\documentclass[twocolumn,twocolappendix]{aastex631}

\usepackage{float}
\usepackage[leftcaption]{sidecap}
\sidecaptionvpos{figure}{c}
\usepackage{placeins}
\usepackage{balance}
\usepackage{amssymb}
\usepackage{pifont}
\usepackage{bm}
\usepackage{soul}
\newcommand{\BalbinotoneNmem}{11}

\newcommand{\BalbinotoneNfeh}{$N_{\rm [Fe/H]}$ = 4}
\newcommand{\Balbinotonemostdistant}{5.3}
\newcommand{\Balbinotoneage}{$\tau = 11.7$ Gyr}
\newcommand{\Balbinotoneisofeh}{$\rm [Fe/H]_{iso} = -2.2$}
\newcommand{\Balbinotonestellarvdisp}{$\sigma_* = 0.27^{+0.10}_{-0.07}$}

\newcommand{\Balbinotonevdisp}{$\sigma_v = 3.7^{+1.5}_{-1.1}~\kms$}
\newcommand{\Balbinotonemlratio}{$M_{1/2}/L_{V,1/2} = 730^{+1140}_{-450}~\MLunit$}
\newcommand{\Balbinotonevdispbayes}{$2\ln\beta = 12.6$}

\newcommand{\Balbinotonefehmaxdiff}{$\Delta \rm [Fe/H]_{max} = 0.76 \pm 0.25$~dex}
\newcommand{\Balbinotonefehmodel}{[Fe/H] =  $-2.59^{+0.21}_{-0.23}$}
\newcommand{\Balbinotonefehdisp}{$\sigma_{\rm [Fe/H]} = 0.41^{+0.27}_{-0.18}$~dex}
\newcommand{\Balbinotonefehdispbayes}{$2\ln\beta = 4.5$}
\newcommand{\BalbinotonevdispbayesJackknife}{$2\ln\beta = 0.7$}
\newcommand{\BalbinotonevdispJackknife}{$\sigma_v = 2.4^{+1.5}_{-1.2}~\kms$}
\newcommand{\BalbinotonemlratioJackknife}{$M_{1/2}/L_{V,1/2} = 290^{+680}_{-220}~\MLunit$}

\newcommand{\BLISSoneNmem}{5}

\newcommand{\BLISSoneage}{$\tau = 7$ Gyr}
\newcommand{\BLISSoneisofeh}{$\rm [Fe/H]_{iso} = -0.9$}

\newcommand{\BLISSonevdisp}{$\sigma_v <\! 5.3~\kms$}

\newcommand{\BLISSonevdispbayes}{$2\ln\beta = -2.8$}

\newcommand{\BLISSonefehmodel}{[Fe/H] =  $-0.83^{+0.14}_{-0.14}$}
\newcommand{\BLISSonefehdisp}{$\sigma_{\rm [Fe/H]} <\! 0.68$~dex}
\newcommand{\BLISSonefehdispbayes}{$2\ln\beta = -3.0$}

\newcommand{\DELVEoneNmem}{10}

\newcommand{\DELVEonemostdistant}{4.8}
\newcommand{\DELVEoneage}{$\tau = 12.5$ Gyr}
\newcommand{\DELVEoneisofeh}{$\rm [Fe/H]_{iso} = -2.2$}

\newcommand{\DELVEonevdisp}{$\sigma_v <\! 3.9~\kms$}

\newcommand{\DELVEonevdispbayes}{$2\ln\beta = -3.2$}
\newcommand{\DELVEonefehbrightest}{[Fe/H]$_{\rm brightest} = -2.68^{+0.14}_{-0.14}$}

\newcommand{\DELVEonefehmaxdiff}{$\Delta \rm [Fe/H]_{max} = 0.32 \pm 0.25$~dex}

\newcommand{\DELVEthreeNmem}{5}

\newcommand{\DELVEthreeage}{$\tau = 13.5$ Gyr}
\newcommand{\DELVEthreeisofeh}{$\rm [Fe/H]_{iso} = -2.2$}

\newcommand{\DELVEthreevdisp}{$\sigma_v <\! 8.7~\kms$}

\newcommand{\DELVEthreevdispbayes}{$2\ln\beta = -0.8$}
\newcommand{\DELVEthreefehbrightest}{[Fe/H]$_{\rm brightest} = -2.98^{+0.29}_{-0.29}$}

\newcommand{\DELVEfourNmem}{8}
\newcommand{\DELVEfourNkin}{7}
\newcommand{\DELVEfourNfeh}{$N_{\rm [Fe/H]}$ = 6}

\newcommand{\DELVEfourage}{$\tau = 13.5$ Gyr}
\newcommand{\DELVEfourisofeh}{$\rm [Fe/H]_{iso} = -1.9$}

\newcommand{\DELVEfourvdisp}{$\sigma_v = 2.5^{+2.0}_{-1.3}~\kms$}

\newcommand{\DELVEfourvdispbayes}{$2\ln\beta = 0.6$}

\newcommand{\DELVEfourfehmaxdiff}{$\Delta \rm [Fe/H]_{max} = 1.94 \pm 0.55$~dex}
\newcommand{\DELVEfourfehmodel}{[Fe/H] =  $-2.17^{+0.27}_{-0.27}$}
\newcommand{\DELVEfourfehdisp}{$\sigma_{\rm [Fe/H]} = 0.62^{+0.22}_{-0.19}$~dex}
\newcommand{\DELVEfourfehdispbayes}{$2\ln\beta = 17.3$}
\newcommand{\DELVEfourvdispbayesJackknife}{$2\ln\beta = -1.7$}
\newcommand{\DELVEfourvdispJackknife}{$\sigma_v <\! 6.6~\kms$}

\newcommand{\DELVEfiveNmem}{3}

\newcommand{\DELVEfiveage}{$\tau = 10$ Gyr}
\newcommand{\DELVEfiveisofeh}{$\rm [Fe/H]_{iso} = -2.2$}

\newcommand{\DELVEsixNmem}{3}

\newcommand{\DELVEsixage}{$\tau = 13.5$ Gyr}
\newcommand{\DELVEsixisofeh}{$\rm [Fe/H]_{iso} = -2.2$}

\newcommand{\DELVEsixfehbrightest}{[Fe/H]$_{\rm brightest} = -1.49^{+0.51}_{-0.51}$}

\newcommand{\DracoIINmem}{28}

\newcommand{\DracoIIage}{$\tau = 13.5$ Gyr}
\newcommand{\DracoIIisofeh}{$\rm [Fe/H]_{iso} = -2.2$}

\newcommand{\DracoIIvdisp}{$\sigma_v <\! 2.6~\kms$}
\newcommand{\DracoIImlratio}{$M_{1/2}/L_{V,1/2} <\! 1040~\MLunit$}

\newcommand{\DracoIIfehmodel}{[Fe/H] =  $-2.94^{+0.32}_{-0.39}$}
\newcommand{\DracoIIfehdisp}{$\sigma_{\rm [Fe/H]} = 0.57^{+0.27}_{-0.27}$~dex}

\newcommand{\EridanusIIINmem}{8}

\newcommand{\EridanusIIIage}{$\tau = 12.5$ Gyr}
\newcommand{\EridanusIIIisofeh}{$\rm [Fe/H]_{iso} = -2.2$}

\newcommand{\EridanusIIIvdisp}{$\sigma_v <\! 8.7~\kms$}

\newcommand{\EridanusIIIvdispbayes}{$2\ln\beta = -0.6$}
\newcommand{\EridanusIIIfehbrightest}{[Fe/H]$_{\rm brightest} = -3.33^{+0.22}_{-0.22}$}

\newcommand{\EridanusIIIfehmaxdiff}{$\Delta \rm [Fe/H]_{max} = 0.11 \pm 0.63$~dex}

\newcommand{\KimoneNmem}{13}

\newcommand{\Kimonemostdistant}{20.5}
\newcommand{\Kimoneage}{$\tau = 13$ Gyr}
\newcommand{\Kimoneisofeh}{$\rm [Fe/H]_{iso} = -2.2$}

\newcommand{\Kimonevdisp}{$\sigma_v = 2.6^{+1.5}_{-1.2}~\kms$}
\newcommand{\Kimonemlratio}{$M_{1/2}/L_{V,1/2} = 1020^{+1870}_{-740}~\MLunit$}
\newcommand{\Kimonevdispbayes}{$2\ln\beta = 0.7$}
\newcommand{\Kimonefehbrightest}{[Fe/H]$_{\rm brightest} = -2.69^{+0.14}_{-0.14}$}
\newcommand{\KimonevdispbayesJackknife}{$2\ln\beta = -3.8$}
\newcommand{\KimonevdispJackknife}{$\sigma_v <\! 2.6~\kms$}

\newcommand{\KimoneJackknifeSourceID}{\textit{Gaia} DR3 2721058972554851968}

\newcommand{\Kimthreemostdistant}{15.7}
\newcommand{\Kimthreeage}{$\tau = 9.5$ Gyr}
\newcommand{\Kimthreeisofeh}{$\rm [Fe/H]_{iso} = -1.6$}

\newcommand{\Kimthreevdisp}{$\sigma_v <\! 2.9~\kms$}
\newcommand{\Kimthreemlratio}{$M_{1/2}/L_{V,1/2} <\! 950~\MLunit$}
\newcommand{\Kimthreevdispbayes}{$2\ln\beta = -3.2$}

\newcommand{\KoposovoneNmem}{7}

\newcommand{\KoposovoneNfeh}{$N_{\rm [Fe/H]}$ = 2}

\newcommand{\Koposovoneage}{$\tau = 7.0$ Gyr}
\newcommand{\Koposovoneisofeh}{$\rm [Fe/H]_{iso} = -0.6$}

\newcommand{\Koposovonevdisp}{$\sigma_v <\! 5.4~\kms$}

\newcommand{\Koposovonevdispbayes}{$2\ln\beta = -2.6$}

\newcommand{\Koposovonefehwavg}{$\rm [Fe/H]_{wavg.} = -0.79 \pm 0.13$}
\newcommand{\Koposovonefehmaxdiff}{$\Delta \rm [Fe/H]_{max} = 0.45 \pm 0.32$~dex}

\newcommand{\KoposovtwoNmem}{18}

\newcommand{\Koposovtwoage}{$\tau = 13.5$ Gyr}
\newcommand{\Koposovtwoisofeh}{$\rm [Fe/H]_{iso} = -2.2$}

\newcommand{\Koposovtwovdisp}{$\sigma_v = 2.7^{+1.7}_{-1.3}~\kms$}
\newcommand{\Koposovtwomlratio}{$M_{1/2}/L_{V,1/2} = 200^{+540}_{-160}~\MLunit$}
\newcommand{\Koposovtwovdispbayes}{$2\ln\beta = 0.9$}
\newcommand{\Koposovtwofehbrightest}{[Fe/H]$_{\rm brightest} = -2.86^{+0.18}_{-0.18}$}
\newcommand{\KoposovtwovdispbayesJackknife}{$2\ln\beta = -1.6$}
\newcommand{\KoposovtwovdispJackknife}{$\sigma_v <\! 5.4~\kms$}

\newcommand{\LaevensthreeNmem}{9}

\newcommand{\LaevensthreeNfeh}{$N_{\rm [Fe/H]}$ = 5}

\newcommand{\Laevensthreeage}{$\tau = 13$ Gyr}
\newcommand{\Laevensthreeisofeh}{$\rm [Fe/H]_{iso} = -2.2$}

\newcommand{\Laevensthreevdisp}{$\sigma_v <\! 4.5~\kms$}
\newcommand{\Laevensthreemlratio}{$M_{1/2}/L_{V,1/2} <\! 350~\MLunit$}
\newcommand{\Laevensthreevdispbayes}{$2\ln\beta = -3.4$}

\newcommand{\Laevensthreefehmodel}{[Fe/H] =  $-1.97^{+0.12}_{-0.12}$}
\newcommand{\Laevensthreefehdisp}{$\sigma_{\rm [Fe/H]} <\! 0.56$~dex}
\newcommand{\Laevensthreefehdispbayes}{$2\ln\beta = -2.5$}

\newcommand{\MunozoneNmem}{6}

\newcommand{\Munozoneage}{$\tau = 13$ Gyr}
\newcommand{\Munozoneisofeh}{$\rm [Fe/H]_{iso} = -1.5$}

\newcommand{\Munozonefehbrightest}{[Fe/H]$_{\rm brightest} = -1.26^{+0.28}_{-0.28}$}

\newcommand{\PSoneNmem}{15}

\newcommand{\PSoneNfeh}{$N_{\rm [Fe/H]}$ = 9}

\newcommand{\PSoneage}{$\tau = 8$ Gyr}
\newcommand{\PSoneisofeh}{$\rm [Fe/H]_{iso} = -1.2$}
\newcommand{\PSonestellarvdisp}{$\sigma_* = 0.48^{+0.15}_{-0.11}$}

\newcommand{\PSonevdisp}{$\sigma_v = 2.1^{+0.8}_{-0.6}~\kms$}
\newcommand{\PSonemlratio}{$M_{1/2}/L_{V,1/2} = 70^{+100}_{-40}~\MLunit$}
\newcommand{\PSonevdispbayes}{$2\ln\beta = 7.0$}

\newcommand{\PSonefehmodel}{[Fe/H] =  $-1.24^{+0.09}_{-0.08}$}
\newcommand{\PSonefehdisp}{$\sigma_{\rm [Fe/H]} <\! 0.35$~dex}
\newcommand{\PSonefehdispbayes}{$2\ln\beta = -3.0$}
\newcommand{\PSonevdispbayesJackknife}{$2\ln\beta = -2.5$}
\newcommand{\PSonevdispJackknife}{$\sigma_v <\! 2.4~\kms$}
\newcommand{\PSonemlratioJackknife}{$M_{1/2}/L_{V,1/2} <\! 120~\MLunit$}
\newcommand{\PSoneJackknifeSourceID}{\textit{Gaia} DR3 6759858716624992768}

\newcommand{\SeguethreeNmem}{35}
\newcommand{\SeguethreeNkin}{19}

\newcommand{\Seguethreemostdistant}{13.3}
\newcommand{\Seguethreeage}{$\tau = 2.6$ Gyr}
\newcommand{\Seguethreeisofeh}{$\rm [Fe/H]_{iso} = -0.55$}

\newcommand{\Seguethreevdisp}{$\sigma_v <\! 2.7~\kms$}

\newcommand{\Seguethreevdispbayes}{$2\ln\beta = -3.6$}
\newcommand{\Seguethreefehbrightest}{[Fe/H]$_{\rm brightest} = -0.88^{+0.21}_{-0.21}$}

\newcommand{\UrsaMajorIIINmem}{16}

\newcommand{\UrsaMajorIIINfeh}{$N_{\rm [Fe/H]}$ = 12}

\newcommand{\UrsaMajorIIIage}{$\tau = 12$ Gyr}
\newcommand{\UrsaMajorIIIisofeh}{$\rm [Fe/H]_{iso} = -2.2$}

\newcommand{\UrsaMajorIIIvdisp}{$\sigma_v <\! 2.5~\kms$}
\newcommand{\UrsaMajorIIImlratio}{$M_{1/2}/L_{V,1/2} <\! 2730~\MLunit$}
\newcommand{\UrsaMajorIIIvdispbayes}{$2\ln\beta = -3.7$}

\newcommand{\UrsaMajorIIIfehmodel}{[Fe/H] =  $-2.65^{+0.11}_{-0.11}$}
\newcommand{\UrsaMajorIIIfehdisp}{$\sigma_{\rm [Fe/H]} <\! 0.35$~dex}
\newcommand{\UrsaMajorIIIfehdispbayes}{$2\ln\beta = -3.0$}

\newcommand{\YMCAoneNmem}{3}

\newcommand{\YMCAoneNfeh}{$N_{\rm [Fe/H]}$ = 3}

\newcommand{\YMCAoneage}{$\tau = 11.7$ Gyr}
\newcommand{\YMCAoneisofeh}{$\rm [Fe/H]_{iso} = -2.2$}

\newcommand{\YMCAonefehmodel}{[Fe/H] =  $-1.93^{+0.19}_{-0.22}$}
\newcommand{\YMCAonefehdisp}{$\sigma_{\rm [Fe/H]} <\! 0.83$~dex}

\newcommand{\UFCSNmemTotal}{212}
\newcommand{\UFCSNfehTotal}{45}

\newcommand{\UFCSNGaiaTotal}{117}
\newcommand{\UFCSNhiresTotal}{20}

\newcommand{\UFCSNdistantfraction}{12.7}
\newcommand{\UFCSNdistant}{27}
\newcommand{\UFCSNsystemsFivePlus}{16}
\newcommand{\UFCSNsystemsTenPlus}{7}
\newcommand{\UFCSNsystemsBeyondFourahalf}{9}

\usepackage{amsmath}
\usepackage{amssymb}
\usepackage{xspace}
\usepackage{xifthen}
\usepackage{eso-pic}

\newcommand{\Gaia}{{\it Gaia}\xspace}

\definecolor{forestgreen}{HTML}{228B22}
\definecolor{urlblue}{HTML}{000000}

\mathchardef\mhyphen="2D

\newlength{\dhatheight}

\newcommand{\unit}[1]{\ensuremath{\mathrm{\,#1}}\xspace}

\newcommand{\degree}{\ensuremath{{}^{\circ}}\xspace}

\newcommand{\e}{\unit{e^{-}}}

\newcommand{\secref}[1]{Section~\ref{sec:#1}}

\newcommand{\tabref}[1]{Table~\ref{tab:#1}}

\newcommand{\figref}[1]{Figure~\ref{fig:#1}}

\newcommand{\bandvar}[2][]{%
  \ifthenelse{\isempty{#1}}{\var{#2}}{\var{#2\_#1}}%
}

\newcommand{\var}[1]{\ensuremath{\texttt{\MakeUppercase{#1}}}\xspace}

\providecommand\physrep{\ref@jnl{Phys.~Rep.}}%
\providecommand\apjs{\ref@jnl{ApJS}}%
\providecommand{\jcap}{\ref@jnl{JCAP}}%
\defcitealias{2025arXiv251002431C}{Cerny, Bissonette, \& Ji et~al.}
\newcommand{\uma}{Ursa~Major~III/UNIONS~1}
\newcommand{\MLunit}{\ensuremath{\mathrm{M_{\odot}}\,\mathrm{L_{\odot}^{-1}}}}
\newcommand{\xmark}{\ding{55}}
\newcommand{\cmark}{\ding{51}}
\newcommand\kms{\mbox{km\,s$^{\rm -1}$}}
\usepackage{float}

\begin{document}

\title{A Chemodynamical Census of the Milky Way's Ultra-Faint Compact Satellites. \\ I. A First Population-Level Look at the Internal Kinematics and Metallicities of \\ 19 Extremely-Low-Mass Halo Stellar Systems}

\correspondingauthor{William Cerny}
\email{william.cerny@yale.edu}

\author[0000-0003-1697-7062]{William~Cerny}
\affiliation{Department of Astronomy, Yale University, New Haven, CT 06520, USA}

\author[0000-0002-9110-6163]{Ting~S.~Li}
\affiliation{David A. Dunlap Department of Astronomy \& Astrophysics, University of Toronto, 50 St. George Street, Toronto, ON, M5S 3H4, Canada}
\affiliation{Dunlap Institute for Astronomy \& Astrophysics, 50 St. George Street, Toronto, ON, M5S 3H4, Canada}

 \author[0000-0002-6021-8760]{Andrew~B.~Pace}
\affiliation{Department of Astronomy, University of Virginia, 530 McCormick Road, Charlottesville, VA 22904, USA}
\affiliation{Galaxy Evolution and Cosmology (GECO) Fellow}

\author[0000-0002-4733-4994]{Joshua~D.~Simon}
\affiliation{Observatories of the Carnegie Institution for Science, 813 Santa Barbara St., Pasadena, CA 91101, USA}

\author[0000-0002-7007-9725]{Marla~Geha}
\affiliation{Department of Astronomy, Yale University, New Haven, CT 06520, USA}

\author[0000-0002-4863-8842]{Alexander~P.~Ji}
\affiliation{Department of Astronomy and Astrophysics, University of Chicago, Chicago, IL 60637, USA}
\affiliation{Kavli Institute for Cosmological Physics, University of Chicago, Chicago, IL 60637, USA}

\author[0000-0001-8251-933X]{Alex~Drlica-Wagner}
\affiliation{Fermi National Accelerator Laboratory, P.O.\ Box 500, Batavia, IL 60510, USA}
\affiliation{Kavli Institute for Cosmological Physics, University of Chicago, Chicago, IL 60637, USA}
\affiliation{Department of Astronomy and Astrophysics, University of Chicago, Chicago, IL 60637, USA}
\affiliation{NSF-Simons AI Institute for the Sky (SkAI),172 E. Chestnut St., Chicago, IL 60611, USA}

\author[0009-0000-9203-1653]{Jordan~Bruce}
\affiliation{ Department of Astronomy, Indiana University, Swain West, 727 E. 3rd Street, Bloomington, IN 47405, USA}

\author[0000-0001-9852-9954]{Oleg~Y.~Gnedin}
\affiliation{Department of Astronomy, University of Michigan, Ann Arbor, MI 48109, USA}

\author[0000-0002-5564-9873]{Eric~F.~Bell}
\affiliation{Department of Astronomy, University of Michigan, Ann Arbor, MI 48109, USA}

\author[0000-0003-3519-4004]{Sidney~Mau}
\affiliation{Department of Physics, Duke University Durham, NC 27708, USA}

\author[0000-0002-9933-9551]{Ivanna Escala}
\affiliation{Space Telescope Science Institute, 3700 San Martin Drive, Baltimore, MD 21218, USA}

\author[0009-0006-5977-618X]{Daisy~Bissonette}
\affiliation{Department of Astronomy and Astrophysics, University of Chicago, Chicago, IL 60637, USA}

\author[0000-0002-1445-4877]{Alessandro Savino}
\affiliation{Department of Astronomy, University of California, Berkeley, CA 94720, USA}

\author[0000-0002-7155-679X]{Anirudh~Chiti} 
\affiliation{Kavli Institute for Particle Astrophysics \& Cosmology, P.O. Box 2450, Stanford University, Stanford, CA 94305, USA}

\author[0000-0001-6196-5162]{Evan~N.~Kirby}
\affiliation{Department of Physics and Astronomy, University of Notre Dame, Notre Dame, IN 46556, USA}

\shortauthors{Cerny et al.}
\shorttitle{A Chemodynamical Census of the Milky Way's Ultra-Faint Compact Satellites. I. }
\begin{abstract}
Deep, wide-area photometric surveys have uncovered a population of compact ($r_{1/2} \approx$ 1--15~pc),  extremely-low-mass ($M_* \approx$ 20--$4000~M_{\odot}$) stellar systems in the Milky Way halo that are smaller in size than known ultra-faint dwarf galaxies (UFDs) and substantially fainter than most classical globular clusters (GCs). Very little is known about the nature and origins of this population of ``Ultra-Faint Compact Satellites'' (UFCSs) owing to a dearth of spectroscopic measurements. Here, we present the first spectroscopic census of these compact systems based on Magellan/IMACS and Keck/DEIMOS observations of 19 individual UFCSs, representing $\sim$2/3 of the known population. We securely measure mean radial velocities for all 19 systems, velocity dispersions for 15 (predominantly upper limits),  metallicities for 17, metallicity dispersions for 8, and \Gaia-based mean proper motions for 18.  This large new spectroscopic sample provides the first insights into population-level trends for these extreme satellites. We demonstrate that: (1) the UFCSs are kinematically colder, on average, than the UFDs, disfavoring very dense dark matter halos in most cases, (2)  the UFCS population is chemically diverse, spanning a factor of $\sim$300 in mean iron abundance ($\rm -3.3 \lesssim [Fe/H] \lesssim -0.8$), with multiple systems falling beneath the ``metallicity floor'' proposed for GCs,  and (3) while some higher-metallicity and/or younger UFCSs are clearly star clusters, the dynamical and/or chemical evidence allows the possibility that up to $\sim$50\% of the UFCSs in our sample (9 of 19) may represent the smallest and least-massive galaxies yet discovered.
\end{abstract}  


\section{Introduction} \label{sec:intro}
Throughout the majority of the 20th century, the known, gravitationally-bound satellite populations of the Milky Way (MW) halo fell relatively neatly into two categories. The numerous, bright globular clusters (GCs), spanning  typical half-mass radii $\sim$1--10~pc, represented archetypal simple stellar populations (SSPs) -- coeval, monometallic systems with stellar masses $\sim$$10^4$--$10^6 \rm \ M_{\odot}$. On the other hand, the eleven dwarf spheroidal galaxies (dSphs) known around the MW were far more spatially extended and displayed more complex stellar populations with age and metallicity spreads suggestive of extended star formation histories \citep[e.g.,][]{1980ApJ...240..804A,1995AJ....109.1628B,1997RvMA...10...29G}.  Radial velocity measurements of stars in the dSphs later revealed their significant mass-to-light ratios indicative of substantial dark matter components \citep{1983ApJ...266L..11A, 1983ApJ...266L..17F,1986AJ.....92..777A,1987AJ.....94..657A,1991AJ....102..914M,1994MNRAS.269..957H,1995AJ....110.2131A}, deepening the empirical divide between these galaxies and the baryon-dominated GCs.  While the discovery of chemical inhomogeneities within GCs -- including light element anticorrelations and small iron abundance spreads -- certainly complicated the picture of GCs as SSPs (\citealt{1978ApJ...223..487C,1979ARA&A..17..309K,1980ApJ...237L..87P,1994PASP..106..553K} and references therein; see \citealt{2012A&ARv..20...50G,2018ARA&A..56...83B,2019A&ARv..27....8G} for recent reviews),  these clusters' compact sizes alone still provided a seemingly infallible means of distinguishing them from the MW dSphs at similar stellar mass \citep{2007ApJ...663..948G}. Indeed, of the 146 GCs cataloged by \citet{1996AJ....112.1487H} near the turn of the century, the largest (Palomar 14, with a half-light radius of $r_{1/2} = 23$~pc) was more than a factor of 10 smaller than the most compact known MW dwarf satellite at the time (the Sextans dSph; now estimated at $r_{1/2} \approx  350$~pc; \citealt{1990MNRAS.244P..16I,2018ApJ...860...66M}).

\par The discovery of the first ultra-faint MW satellites in the late 2000s presented a significant challenge to this simplistic morphological distinction. As vividly exemplified by early debates over the nature of objects such as Willman~1, Segue~1, and Segue~2 -- each discovered in the early years of the Sloan Digital Sky Survey -- the classification of these newfound low-surface-brightness systems as compact dwarf galaxies or extended, faint star clusters was not at all obvious \citep{2005AJ....129.2692W,2007ApJ...654..897B,2009ApJ...692.1464G,2009MNRAS.398.1771N,2009MNRAS.397.1748B,2011AJ....142..128W,2011ApJ...733...46S,2013ApJ...770...16K}. Dedicated spectroscopic measurements of the kinematics and chemistries of stars in these systems served to settle many of these early debates, clearly establishing the existence of the ultra-faint dwarf galaxies (UFDs) as a distinct class of metal-poor, highly dark-matter-dominated galaxies \citep{2005ApJ...630L.141K,2006ApJ...650L..51M,2007MNRAS.380..281M, 2007ApJ...670..313S}. Nonetheless, the classification of compact MW satellites has remained at the forefront of discussion, with many newly-discovered systems remaining at best galaxy \textit{candidates} until their classifications can be established spectroscopically \citep{2019BAAS...51c.409S}. This has particularly been true for systems at $r_{1/2} \approx15$--30~pc where the most extended GCs overlap in size with fainter UFDs \citep[e.g.,][]{2011AJ....142..128W,2014ApJ...786L...3L,2014MNRAS.441.2124B,2021ApJ...910...18C}, leading some authors to refer to this regime of the absolute magnitude vs. half-light radius plane as the ``valley of ambiguity'' \citep{2011AJ....142...88F,2015ApJ...813...44L,2020ApJ...890..136M} or the ``trough of uncertainty'' \citep{2018ApJ...852...68C}.

\par
\begin{figure*}[ht!]
    \centering
    \includegraphics[width = \textwidth]{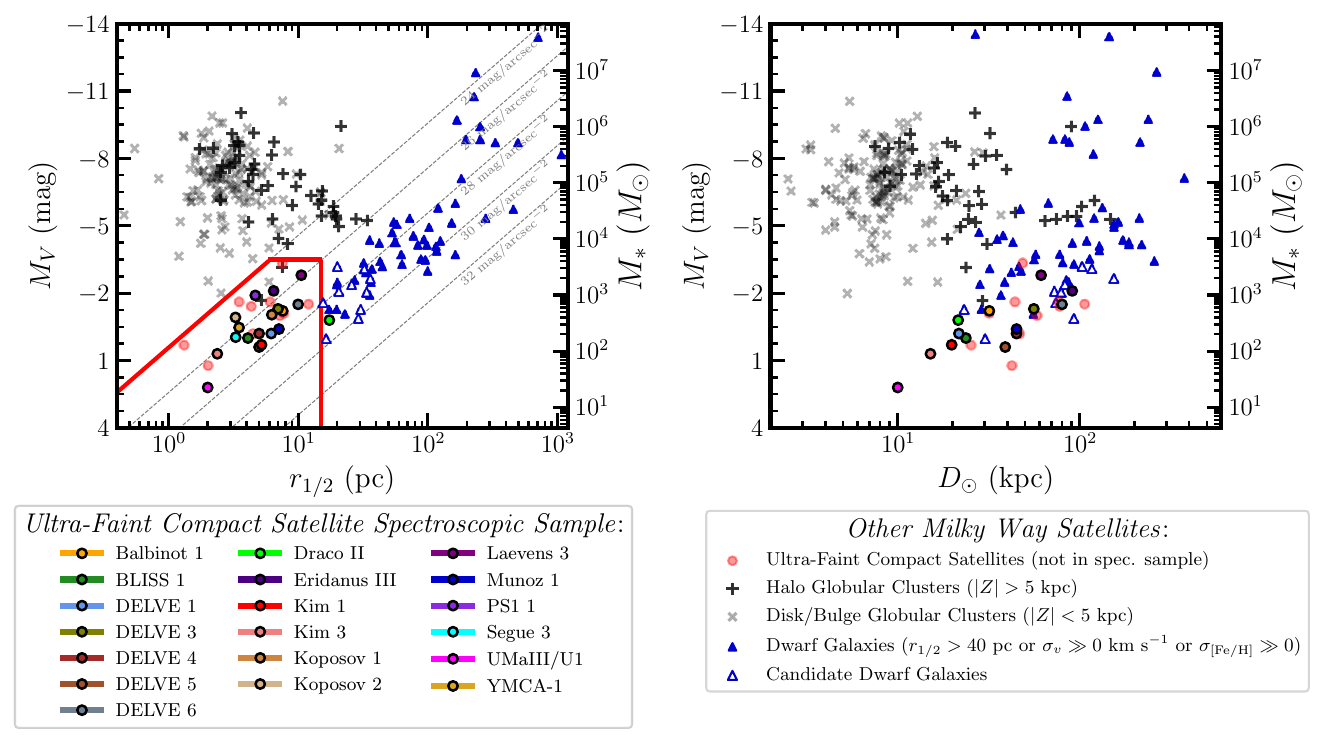}
    \caption{\textbf{A visualization of our UFCS definition and the properties of the systems that comprise our spectroscopic sample.} (Left) Absolute magnitude ($M_{V}$) vs. azimuthally-averaged half-light radius ($r_{1/2}$) for the UFCSs as well as the MW GCs (grey $\times$ or black $+$, depending on whether they inhabit the disk/bulge or halo) and dwarf satellite galaxies (filled blue triangles if confirmed, unfilled if not). The morphological boundaries of our UFCS selection are demarcated by red lines, and the UFCSs in our spectroscopic sample are shown as colored points with black outlines. The remaining UFCSs are shown as filled red circles. Two classical GCs (Palomar 13 and AM~4) fall within our UFCS selection. (Right) Absolute magnitude vs. heliocentric distance ($D_{\odot}$) for the same satellite samples. At any given distance, the UFCSs represent the faintest known halo systems; the lack of UFCSs at larger distances is likely a selection effect. No errorbars are shown in either panel for visual clarity. In both panels, the righthand y-axis displays the equivalent stellar masses ($M_*$) assuming $M_*/L_V = 2$.}
    \label{fig:fig1_population}
\end{figure*}

 Although the UFDs at this $\sim$15--30~pc size scale have received copious attention over the last two decades, the nature of an even more enigmatic population of MW satellites has remained underdiscussed. Recent large-scale photometric surveys have also uncovered a growing population of fainter, more compact ultra-faint systems in the halo ($ -3 \lesssim M_V \lesssim +2 $; $r_{1/2} \approx 1$--15~pc), which have often been regarded as star clusters on the basis of their compact sizes (\figref{fig1_population}). Very little is known about the nature and origins of these compact systems:  even their classifications as star clusters, as opposed to dark-matter-dominated dwarf galaxies, are only assumptions that have rarely been tested. Observationally, many of these systems lie on or near an extrapolation of the sequence delineated by brighter UFDs in the $M_V$--$r_{1/2}$ plane toward lower luminosities (see \figref{fig1_population}). Theoretically, this population appears to match  $\Lambda$CDM predictions for the properties of the smallest galaxies from semi-analytical models tuned to match the properties of brighter satellites \citep{2022MNRAS.516.3944M,2024MNRAS.529.3387A,2025arXiv251115808A,2025arXiv250907066B}. The properties of these faint, compact systems may also be consistent with predictions for the asymptotic remnants of tidally-stripped dwarf galaxies inhabiting cuspy dark matter halos (``microgalaxies''; \citealt{2020MNRAS.491.4591E,2024ApJ...968...89E,2024ApJ...965...20E}), or alternatively the properties of globular-cluster-like dwarf galaxies formed from a single self-quenching star formation event \citep{2025arXiv250909582T,2025arXiv251121824B}. Thus, while the complete body of observational evidence and theoretical predictions remains small, these studies support the possibility that some of these extreme MW satellites may be galaxies at the very threshold of galaxy formation. 
\par Confirmation of some or all of these faint, compact satellites as galaxies would have far-reaching importance for our understanding of the small-scale properties of dark matter and the physics of galaxy formation in low-mass halos. At a population level, the raw number and luminosity function of faint satellite galaxies strongly constrain particle dark matter models that suppress the abundance of low-mass subhalos (see e.g., \citealt{2000PhRvL..85.1158H,2000ApJ...542..622C,2013JCAP...03..014A,2014MNRAS.442.2487K,2018MNRAS.473.2060J,2021PhRvL.126i1101N,2021JCAP...08..062N,2022PhRvD.106l3026D,2024arXiv240816042N,2025PhRvD.111f3079T,2025arXiv251201361L}). These same population-level observables also probe the galaxy occupation fraction in the $\lesssim 10^{7}$--$10^8 \rm \, M_{\odot}$ halo mass regime, which itself is set by the interplay of the gas cooling, star formation, and feedback processes occurring in the least-massive halos to form stars as well as by cosmic reionization \citep[e.g.,][]{2000ApJ...539..517B,2009MNRAS.399L.174O,2010ApJ...710..408B,2010MNRAS.402.1995M,2019MNRAS.488.4585G, 2019ApJ...874...40M,2020MNRAS.498.4887B,2021ApJ...923...35M,2023MNRAS.524.2290N,2025ApJ...983L..23N,2025arXiv250907066B,2025arXiv251115808A}. Lastly, but certainly not exhaustively, the small physical sizes and relative proximity of these compact satellites (often $D_{\odot} < 50~$kpc) position them as among the most promising sites for the indirect detection of dark matter annihilation and/or decay into Standard Model particles if they are indeed dark-matter-dominated (see e.g.,  \citealt{2024PhRvD.109j3007B,2024ChPhC..48k5112Z,2024PhRvD.109h3018C,2024PhRvD.110j3048Z} and especially \citealt{2024arXiv240401181C} for recent analyses including the faintest and most compact satellites). 
\par The key limitation in disentangling the classifications of these ambiguous satellites, and thereby unlocking these broad insights, has been the near-complete lack of spectroscopic measurements of their resolved stars. This dearth partially reflects the greater community interest in targeting more extended satellites that are morphologically favored to be dwarf galaxies (i.e., those with $r_{1/2} \gg$ 15 pc), but the observational challenges associated with spectroscopically characterizing these faint, compact systems have also been a limiting factor. For example, the compactness of these systems (typical angular half-light radii $r_{1/2} \lesssim  1\arcmin$) makes them challenging to observe with fiber-fed multi-object spectrographs due to fiber-separation constraints, and their frequent lack of stars brighter than $g\approx 19$--20 makes them inaccessible to even state-of-the-art spectroscopic surveys (e.g., SDSS-V and DESI, which have nominal magnitude limits of $r \approx 18$ and $r \approx 19$ for stars, respectively; \citealt{2023ApJS..267...44A}, \citealt{2023ApJ...947...37C}). The combination of these factors has resulted in only a small and heterogeneous sample of faint, compact satellites with any published spectroscopic measurements \citep{2011AJ....142...88F,2012ApJ...753L..15M,2018MNRAS.480.2609L,2019MNRAS.490.1498L}.
\par Our goal in this work is to overcome these long-standing difficulties and spectroscopically characterize the properties of these elusive, compact MW satellites \textit{at the population level} for the first time. In so doing, we seek to probe whether any of these systems could represent the smallest and least luminous galaxies yet discovered. These goals are enabled by a large-scale, multi-year effort to collect data on these systems with the Keck II and Magellan-Baade telescopes as well as by a homogeneous reprocessing of a substantial body of archival data from the former facility \citep{geha_paper1}.
\par The remainder of this work is organized as follows. In \secref{sample}, we begin by providing our working definition of the ``Ultra-Faint Compact Satellites'' (UFCSs). In \secref{observations}, we introduce our spectroscopic sample of 19 UFCSs and the new and archival observations used for our analysis. We also describe our data reduction procedure, stellar radial velocity measurements, and Calcium-Triplet-based metallicity measurements. In \secref{membership}, we identify member stars in each system. We subsequently use these member samples to characterize the UFCSs' internal kinematics in \secref{vdispconstraints} and their chemical properties in \secref{metals}.  In \secref{classification}, we leverage these new constraints and a range of other evidence to ascertain the nature of individual UFCSs as dwarf galaxies or star clusters. In \secref{disco}, we discuss the stability of the UFCSs against evaporation, possible evidence for spatially-extended stellar populations in some UFCSs, and the UFCS-GC connection, before providing an outlook for future efforts to definitively classify the most promising galaxy candidates in our sample. We summarize our findings in \secref{summary}. 
\par In a forthcoming second paper in this series (Cerny et al. 2026b, in prep.; hereafter Paper II), we study the orbital histories, accreted origins, and tidal influences of our sample of UFCSs.

\section{Sample Definition, Nomenclature, and Structural Properties}
\label{sec:sample}
\subsection{Defining the UFCSs: A Rough Selection}
One major goal of this work is to ascertain what shared dynamical and chemical properties -- if any -- define the population of objects referred to here as the UFCSs. No comprehensive definitions of the UFCSs exist in the literature at the time of writing, and it remains unclear how these objects are related to the GCs and UFDs. To guide our choice of systems to target spectroscopically and to define the limits of our analysis, we adopted a three-pronged working definition of a UFCS as (1) a gravitationally-bound constituent of the MW stellar halo, as suggested by a significant height above the Galactic plane ($|Z| > 5$~kpc) that distinguishes it from clusters in the disk and bulge, with (2) a low intrinsic luminosity, reflected by a surface brightness within the azimuthally-averaged half-light radius  $\mu_{1/2} > 24~$mag~arcsec$^{-2}$ and an absolute $V$-band magnitude $M_V > -3.5$,  which together separate it from nearly all of the classical GCs\footnote{This corresponds to a stellar mass limit of $M_* < 4300 \rm \ M_\odot$ assuming a stellar mass-to-light ratio of $M_*/L_V = 2$. This cut also separates the UFCSs from the (much brighter) ultra-compact dwarf (UCD) galaxies known around external galaxies.}, and with (3) an azimuthally-averaged half-light radius $r_{1/2} \lesssim 15$~pc -- a threshold that excludes nearly all known UFDs and satellites that have traditionally been regarded as UFD candidates upon discovery \citep{2015ApJ...813..109D,2023ApJ...953....1C}. 
\par The morphological boundaries outlined above are visualized in the left panel of \figref{fig1_population} by the thick red border, where we compare the UFCSs to the known population of MW satellite galaxies and classical MW GCs. In this figure and throughout this work, the reference sample of UFD and GC measurements is taken from the Local Volume Database (\citealt{2025OJAp....8E.142P}, release 1.0.4; see Appendix \secref{comparisonrefs} for a complete reference list). For the GCs in particular, this database's entries are primarily sourced from the \citeauthor{1996AJ....112.1487H} (\citeyear{1996AJ....112.1487H}, 2010 edition)
 catalog of clusters augmented by recent discoveries in the disk/bulge. When available, the $M_V$ values in the database are adopted from the uniform analysis of \citet{2020PASA...37...46B}.

\par We identified an initial sample of 32 MW satellites that satisfy all three criteria of our UFCS definition. Ordered by year of discovery, these are Palomar~13 \citep{1955PASP...67...27W,1955PASP...67..258A}, AM~4 \citep{1982PASP...94...40M},  Koposov~1 and Koposov~2 \citep{2007ApJ...669..337K},  Segue~3 \citep{2010ApJ...712L.103B}, Mu\~{n}oz~1 \citep{2012ApJ...753L..15M}, Balbinot~1 \citep{2013ApJ...767..101B}, Laevens~3 \citep{2015ApJ...813...44L}, Kim~1 \citep{2015ApJ...799...73K}, Eridanus~III \citep{2015ApJ...807...50B,2015ApJ...805..130K}, Kim~2/Indus I \citep{2015ApJ...803...63K, 2015ApJ...807...50B,2015ApJ...805..130K}, Kim~3 \citep{2016ApJ...820..119K}, SMASH~1 \citep{2016ApJ...830L..10M}, DES~1 \citep{2016MNRAS.458..603L}, DES~J0111$-$1341 \citep{2017MNRAS.468...97L},  DES~3 \citep{2018MNRAS.478.2006L}, PS1~1/Prestgard~64\footnote{PS1~1 was first identified by amateur astronomer T. Prestgard prior to its rediscovery in data from \Gaia. Hereafter, we use the published name from \citet{2019MNRAS.484.2181T} for ease of indexing.} and Torrealba~1 \citep{2019MNRAS.484.2181T}, OGLE-CL LMC 863/DES~4, OGLE-CL LMC 874/DES 5, and OGLE-CL LMC 845/\Gaia~3   (\citealt{2016AcA....66..255S,2019MNRAS.484.2181T}; hereafter referred to by their latter names only)\footnote{The systems \Gaia 4--7 from \citet{2019MNRAS.484.2181T} are excluded owing to their proximity to the Galactic plane ($|Z| < 5$~kpc). DES 4, DES 5, and \Gaia 3 are included by our definition, but we note that they may be associated with the LMC disk.
}, BLISS~1 \citep{2019ApJ...875..154M}, DELVE~1 \citep{2020ApJ...890..136M}, HSC~1 \citep{2019PASJ...71...94H}, YMCA-1 \citep{2021RNAAS...5..159G}, DELVE~3, DELVE~4, and DELVE~5 \citep{2023ApJ...953....1C}, DELVE~6 \citep{delve6_discovery}, \uma{} \citep{2024ApJ...961...92S}, and the very recently discovered systems ``Alice'' \citep{2025arXiv250511120P} and DELVE~7 \citep{2025arXiv251011684T}. Among this list, Palomar~13 and AM~4 are best considered classical GCs meeting our UFCS definition, as they have prior dynamical, chemical, and/or stellar population measurements establishing their nature as star clusters (see \secref{connection} for a more detailed discussion of the faintest classical GCs).  DES J0111$-$1341 may be a false-positive detection associated with density fluctuations in the Sagittarius stream (i.e., not a \textit{bona fide} MW satellite; see \citealt{2019ApJ...875...77P,2020ApJ...893...47D}), and thus we exclude this system throughout this work.  Lastly, we appended the satellite Draco~II \citep{2015ApJ...813...44L} to this list. Draco~II falls just outside our size selection boundary ($r_{1/2} = 17^{+4}_{-3} \rm \ pc$) but has a classification that, like the remaining UFCSs, has not yet been clearly established.  After the exclusion of the three aforementioned systems and the inclusion of Draco~II, the total parent sample of UFCSs that we consider here is 30.

\par One important aspect of our UFCS definition is that it is \textit{not} a catch-all for satellites with potentially ambiguous classifications: it is specifically limited to the faintest and most compact halo systems. Accordingly, our selection excludes the MW satellites Laevens~1/Crater \citep{2014ApJ...786L...3L,2014MNRAS.441.2124B} and Sagittarius~II \citep{2015ApJ...813...44L} -- two significantly brighter ($M_V \lesssim -5$) and more extended systems ($r_{1/2} \gtrsim 20$~pc) that were once considered possible dwarf galaxy candidates but are now regarded as likely GCs (\citealt{2016ApJ...822...32W, 2016MNRAS.460.3384V,2021MNRAS.503.2754L,2025arXiv250305927Z}). Moreover, our selection also excludes a small number of ultra-faint MW satellites with somewhat larger sizes of $15  \ \rm pc \lesssim r_{1/2} \lesssim 30$~pc that have been argued to be dwarf galaxies but for which some ambiguity remains. These include Tucana~III \citep{2015ApJ...813..109D,Simon_2017,2017ApJ...838...44H,2018ApJ...866...22L,2019ApJ...882..177M}, DELVE~2 (\citealt{2021ApJ...910...18C}; Cerny et al. 2026c, in prep.), Virgo~II \citep{2023ApJ...953....1C}, Cetus~II \citep{2015ApJ...813..109D,2023ApJ...959..141W}, and Phoenix~III  \citep{2025arXiv251011684T}. These more-extended satellites warrant robust spectroscopic characterization and definitive classification, but that is beyond the scope of this work. 
\par Lastly, we note that our compilation currently excludes the ultra-faint, compact star clusters known to reside in the MW's low-mass satellite galaxies: one each in Eridanus II \citep{2015ApJ...805..130K,2016ApJ...824L..14C} and Ursa~Major~II \citep{Zucker_2006,2022ApJ...926..162E}. The six GCs embedded in the Fornax dSph \citep{1939PASP...51...40B,doi:10.1073/pnas.25.11.565} are all too luminous to be UFCSs according to our definition \citep{10.1046/j.1365-8711.2003.06275.x,2019ApJ...875L..13W}. No UFCSs are presently known around external galaxies, though the faintest GCs in M31 and its satellite galaxies are not far from passing our morphological selections \citep[e.g.,][]{2011MNRAS.414..770H,2014MNRAS.442.2165H,2017PASA...34...39C} and upcoming surveys may yield UFCS discoveries in the M31 environment.

\subsection{A Note on Nomenclature}
We have chosen to introduce a new name for this class of system, namely ``Ultra-Faint Compact Satellites'' (UFCSs), which we believe to be the most concise and descriptive label that avoids assigning a dwarf galaxy or star cluster classification to these systems. In the literature, individual examples of these systems have received a wide range of names broadly split between those that presume a star cluster classification, such as Faint Halo Clusters (e.g., \citealt{2019AJ....157...12B}) or ultra-faint star clusters \citep[e.g.,][]{2010ApJ...712L.103B, 2012ApJ...753L..15M,2020ApJ...890..136M,2023ApJ...953....1C}, and those that instead highlight the ambiguity in the nature of these candidates, such as ultra-faint objects (``UFOs''; \citealt{2017MNRAS.466.1741C}), faint, ambiguous satellites \citep{2024ApJ...961...92S}, and, most similarly to our preferred naming, ultra-faint compact stellar systems \citep{2024arXiv240401181C} and hyper-faint compact stellar systems (HFCSSs; \citealt{2025OJAp....8E.142P,2025arXiv251115808A}). We opt to include ``satellites'' as a specific qualifier, as opposed to merely ``stellar systems,'' to indicate our limited focus on systems inhabiting the MW halo and to emphasize our exclusion of open clusters and GCs in the disk and bulge. Nonetheless, we acknowledge the remarkable progress made in recent years to discover and characterize increasingly faint GCs in highly-reddened regions of the Galaxy \citep[e.g.,][]{2005AJ....129..239K,2008AJ....136.2102S,2011A&A...535A..33M,2018ApJ...860L..27C, 2018PASA...35...25B, 2022A&A...662A..95G,2023MNRAS.526.1075P,2024A&A...687A.201B,2024A&A...689A.115S}, which have contributed significantly to the known faint star cluster population and may in the future yield discoveries with sizes and luminosities in the UFCS regime. 

\par The prevailing convention for naming newly discovered MW satellites has been to refer to dwarf galaxies based on the constellations within which they reside and star clusters based on the survey through which they were discovered or after the investigator who discovered them \citep[e.g.,][]{2019ARA&A..57..375S}; occasionally, dual names have been given to objects to reflect the classification ambiguity at the time of discovery. As our analysis will observationally demonstrate, it is likely that at least a small fraction of the UFCSs -- and potentially as high as $\sim$50\% -- are in fact very-low-mass dwarf galaxies. If dwarf galaxy classifications for these UFCSs are definitively established, it would be appropriate to refer to them as UFDs. Some authors have begun using the term ``hyper-faint'' dwarf galaxies to refer to confirmed dwarf galaxies with $M_V \gtrsim -3; \ L_V \lesssim 1000 \, \rm \, L_{\odot}$ \citep[e.g.,][]{2014ApJ...795L..13H,2015ApJ...808...95S,10.1093/mnras/stx1900,2018MNRAS.479.2853N}, which we believe would also be appropriate (though we note our sample here extends to $M_V = -3.5$). In cases where reclassification as a galaxy is necessary, we advocate for these systems' original published names to be retained to avoid confusion, as has been past practice. 

\begin{deluxetable*}{ccccccccc}
\tablewidth{\textwidth}
\tabletypesize{\small}
\tablecaption{Structural Properties and Distances of the UFCS in our Spectroscopic Sample\label{tab:litproperties}}
\tablehead{
Name & $\alpha$ & $\delta$ & $M_V$ & $M_{*}$ & $r_{1/2}$ & $(m-M)_0$ & $D_{\odot}$ & References \\
 & (deg) & (deg) & (mag) & $(M_{\odot})$ & (pc) &  & (kpc) &  
}
\startdata
Balbinot 1 & $332.680^{+0.001}_{-0.001}$ & $14.940^{+0.001}_{-0.001}$ & $-1.2^{+0.7}_{-0.7}$ & $520^{+420}_{-240}$ & $8^{+2}_{-1}$ & $17.49^{+0.08}_{-0.10}$ & $32^{+1}_{-2}$ & (1) \\
BLISS 1 & $177.510^{+0.003}_{-0.003}$ & $-41.772^{+0.002}_{-0.002}$ & $0.0^{+1.7}_{-0.7}$ & $100^{+150}_{-70}$ & $4^{+1}_{-1}$ & $16.87^{+0.20}_{-0.13}$ & $24^{+2}_{-2}$ & (2) \\
DELVE 1 & $247.725^{+0.002}_{-0.002}$ & $-0.972^{+0.003}_{-0.003}$ & $-0.2^{+0.8}_{-0.6}$ & $180^{+150}_{-90}$ & $6^{+2}_{-1}$ & $16.68^{+0.20}_{-0.20}$ & $22^{+2}_{-2}$ & (3),(4) \\
DELVE 3 & $290.396^{+0.003}_{-0.003}$ & $-60.784^{+0.001}_{-0.001}$ & $-1.3^{+0.4}_{-0.6}$ & $640^{+400}_{-220}$ & $7^{+1}_{-1}$ & $18.73^{+0.09}_{-0.23}$ & $56^{+2}_{-6}$ & (5) \\
DELVE 4 & $230.775^{+0.001}_{-0.001}$ & $27.395^{+0.001}_{-0.001}$ & $-0.2^{+0.5}_{-0.8}$ & $250^{+230}_{-110}$ & $5^{+1}_{-1}$ & $18.28^{+0.19}_{-0.19}$ & $45^{+4}_{-4}$ & (5) \\
DELVE 5 & $222.104^{+0.002}_{-0.002}$ & $17.468^{+0.001}_{-0.001}$ & $0.4^{+0.4}_{-0.9}$ & $160^{+160}_{-70}$ & $5^{+1}_{-1}$ & $17.97^{+0.17}_{-0.17}$ & $39^{+3}_{-3}$ & (5) \\
DELVE 6 & $33.070^{+0.003}_{-0.004}$ & $-66.056^{+0.002}_{-0.002}$ & $-1.5^{+0.4}_{-0.6}$ & $770^{+480}_{-270}$ & $10^{+4}_{-3}$ & $19.51^{+0.11}_{-0.16}$ & $80^{+4}_{-6}$ & (6) \\
Draco II & $238.174^{+0.005}_{-0.005}$ & $64.579^{+0.006}_{-0.006}$ & $-0.8^{+0.4}_{-1.0}$ & $510^{+590}_{-220}$ & $17^{+4}_{-3}$ & $16.67^{+0.05}_{-0.05}$ & $21.5^{+0.4}_{-0.4}$ & (7) \\
Eridanus III & $35.695^{+0.001}_{-0.001}$ & $-52.284^{+0.002}_{-0.002}$ & $-2.1^{+0.5}_{-0.5}$ & $1180^{+680}_{-430}$ & $6.5^{+0.7}_{-0.6}$ & $19.80^{+0.04}_{-0.04}$ & $91^{+4}_{-4}$ & (8) \\
Kim 1 & $332.921^{+0.003}_{-0.003}$ & $7.027^{+0.003}_{-0.003}$ & $0.3^{+0.5}_{-0.5}$ & $130^{+80}_{-50}$ & $5.2^{+0.7}_{-0.6}$ & $16.48^{+0.10}_{-0.10}$ & $20^{+1}_{-1}$ & (9) \\
Kim 3 & $200.688^{+0.001}_{-0.001}$ & $-30.600^{+0.001}_{-0.001}$ & $0.7^{+0.3}_{-0.3}$ & $90^{+30}_{-20}$ & $2.4^{+0.9}_{-0.7}$ & $15.90^{+0.11}_{-0.04}$ & $15.1^{+1.0}_{-0.3}$ & (10) \\
Koposov 1 & $179.825^{+0.001}_{-0.001}$ & $12.261^{+0.003}_{-0.003}$ & $-1.0^{+0.7}_{-0.7}$ & $450^{+390}_{-210}$ & $6^{+2}_{-2}$ & $17.80^{+0.10}_{-0.10}$ & $36^{+2}_{-2}$ & (11), (12) \\
Koposov 2 & $119.572^{+0.001}_{-0.001}$ & $26.257^{+0.002}_{-0.002}$ & $-0.9^{+0.8}_{-0.8}$ & $400^{+440}_{-210}$ & $3.3^{+0.5}_{-0.5}$ & $16.90^{+0.20}_{-0.20}$ & $24^{+2}_{-2}$ & (12); $\S$\ref{sec:membership} \\
Laevens 3 & $316.729^{+0.001}_{-0.001}$ & $14.984^{+0.001}_{-0.001}$ & $-2.8^{+0.2}_{-0.3}$ & $2380^{+680}_{-460}$ & $11^{+1}_{-1}$ & $18.94^{+0.05}_{-0.02}$ & $61^{+1}_{-1}$ & (13) \\
Munoz 1 & $225.450^{+0.004}_{-0.004}$ & $66.969^{+0.002}_{-0.002}$ & $-0.4^{+0.9}_{-0.9}$ & $250^{+320}_{-140}$ & $7^{+2}_{-2}$ & $18.27^{+0.25}_{-0.25}$ & $45^{+5}_{-5}$ & (14) \\
PS1 1 & $289.171^{+0.003}_{-0.003}$ & $-27.827^{+0.003}_{-0.003}$ & $-1.9^{+0.5}_{-0.5}$ & $990^{+580}_{-360}$ & $5^{+1}_{-1}$ & $17.40^{+0.20}_{-0.20}$ & $30^{+3}_{-3}$ & (15) \\
Segue 3 & $320.380^{+0.001}_{-0.001}$ & $19.118^{+0.001}_{-0.001}$ & $-0.0^{+0.8}_{-0.8}$ & $180^{+190}_{-90}$ & $3.3^{+0.6}_{-0.6}$ & $17.35^{+0.08}_{-0.08}$ & $30^{+1}_{-1}$ & (12),(16),(17) \\
UMaIII/U1 & $174.708^{+0.003}_{-0.003}$ & $31.078^{+0.003}_{-0.003}$ & $2.2^{+0.4}_{-0.3}$ & $21^{+7}_{-6}$ & $2.0^{+1.0}_{-0.8}$ & $15.00^{+0.20}_{-0.20}$ & $10^{+1}_{-1}$ & (18) \\
YMCA-1 & $110.838^{+0.003}_{-0.003}$ & $-64.832^{+0.003}_{-0.003}$ & $-0.5^{+0.6}_{-0.6}$ & $260^{+190}_{-110}$ & $3.5^{+0.5}_{-0.5}$ & $18.72^{+0.15}_{-0.17}$ & $55^{+4}_{-4}$ & (19) \\
\enddata
\tablecomments{Targets are ordered alphabetically here and throughout this work. See Appendix \ref{sec:litpropertydetails} for extensive details about assumptions behind the measurements quoted in this table, and see Table \ref{tab:supplementalproperties} for systems not in our spectroscopic sample. References enumerated above are: (1) \citealt{2013ApJ...767..101B}, (2) \citealt{2019ApJ...875..154M},   (3) \citealt{2020ApJ...890..136M}, (4) \citealt{Simon2024}, (5) \citealt{2023ApJ...953....1C}, (6) \citealt{delve6_discovery}, (7) \citealt{2018MNRAS.480.2609L}, (8) \citealt{2018ApJ...852...68C}, (9) \citealt{2015ApJ...799...73K}, (10) \citealt{2016ApJ...820..119K},  (11) \citealt{2014AJ....148...19P}, (12) \citealt{2018ApJ...860...66M},  (13)  \citealt{2019MNRAS.490.1498L},  (14) \citealt{2012ApJ...753L..15M}, (15) \citealt{2019MNRAS.484.2181T}, (16) \citealt{2017AJ....154...57H}, (17) \citealt{2011AJ....142...88F}, (18) \citealt{2024ApJ...961...92S}, and (19) \citealt{2022ApJ...929L..21G}.}\end{deluxetable*}

\subsection{Properties of the Known UFCSs}
For this work, we compiled a new catalog of the properties of the 30 known UFCSs based on literature photometric studies, including their morphological properties (radii, ellipticities, and position angles), absolute magnitudes, and distances. A subset of these measurements is presented in \tabref{litproperties} for systems studied spectroscopically in this work (see following section), and an analogous summary for systems not in our spectroscopic sample is included in Appendix \ref{sec:litpropertydetails}, \tabref{supplementalproperties}.
\par In a substantial number of cases, we found it necessary to recompute derived properties of the UFCSs,  make assumptions about unreported measurements and measurement uncertainties, and/or to sample from asymmetric posterior distributions when computing derived properties. Our procedures for interpreting literature measurements are detailed  in Appendix \ref{sec:litpropertydetails} and frequently result in estimates that diverge at a small level from quoted values in the literature. We particularly emphasize that we use estimates of \textit{azimuthally-averaged} physical half-light radii (which we call $r_{1/2}$) throughout this work. This is most consistent with common dynamical mass estimators that assume spherical symmetry; however, this quantity is often omitted in photometric studies in favor of the elliptical semi-major axis (which we call $a_{1/2}$ and is related to $r_{1/2}$ via $r_{1/2} \equiv a_{1/2}\sqrt{1-\epsilon}$, where $\epsilon$ is the ellipticity). In addition, we highlight that all stellar masses presented here were newly derived from literature $M_V$ estimates assuming a fixed stellar-mass-to-light ratio of $M_*/L_V = 2$ (where $L_V$ is the total $V$-band luminosity of a given system) and a solar absolute magnitude of $M_{V,\odot} = 4.83$ \citep{1976asqu.book.....A}. These estimates explicitly account for the asymmetric uncertainties that arise from the non-linear mapping between $M_V$ and $L_V$, but do not account for potential variations of $M_*/L_V$ with stellar population age or metallicity.

\section{Observations, Data Reduction, and Spectroscopic Measurements}
\label{sec:observations}
We obtained medium-resolution, multi-object spectroscopy of 21 of the 30 known UFCSs through dedicated new Magellan and Keck observations in the 2019A--2024B semesters paired with a re-reduction of existing data available through the Keck Observatory Archive.  Of the 21 systems targeted or re-analyzed using archival data, we were able to successfully recover a reliable velocity detection for 19 systems, which form the basis of our spectroscopic sample in this work. These 19 systems span both celestial hemispheres, cover nearly the full range of distances over which UFCSs are currently known ($\rm 10 \ kpc \lesssim$ $D_{\odot} \rm \lesssim 90 \ kpc$ for our sample), and extend across most of the $M_V$--$r_{1/2}$ parameter space encompassed by our UFCS selection (see \figref{fig1_population}). Thus, we expect our spectroscopic sample to be broadly representative of the known MW UFCS population.
\par In the subsections that follow, we elaborate on the sample of UFCSs targeted with each telescope and detail our spectroscopic observations of these systems. We then describe our procedures for data reduction, our measurements of stellar velocities and equivalent widths, and our derivation of stellar metallicities. These procedures generally differ for each telescope/instrument so as to make use of existing, optimized reduction and measurement pipelines, though we note that all stellar metallicities were derived using the same calibration to ensure homogeneity across our UFCS sample.

\subsection{Magellan/IMACS Sample} 
\label{sec:imacs}
\par We pursued spectroscopy of 12 UFCSs in the southern sky with the Inamori-Magellan Areal Camera and Spectrograph (IMACS) on the 6.5m Magellan-Baade telescope at Las Campanas Observatory in Chile \citep{10.1117/12.670573,Dressler_2011}. Our observations targeted the Balbinot~1, BLISS~1, DELVE~1, DELVE~3, DELVE~6,  Eridanus~III,  Gaia~3, Kim~1, Kim~2, Kim~3, PS1~1, and YMCA-1 systems. \cite{Simon2024} presented a first look at the IMACS observations of DELVE~1 and Eridanus~III; here, we omit their DELVE~1 IMACS data in favor of a fully independent Keck dataset for the system, but we retain the same Eridanus~III IMACS dataset. We failed to identify any clear members in our IMACS data for Balbinot~1 and Kim~1 due to poor observing conditions, Kim~2 due to the lack of suitably bright member stars, and \Gaia~3 due to a combination of the system's compactness paired with significant LMC star contamination. We later identified numerous member stars in Kim~1 and Balbinot~1 with deeper Keck observations (see \secref{deimos}) and therefore based our analysis on those data instead. In summary, our complete sample of UFCSs with useful IMACS data consists of seven systems: BLISS~1, DELVE~3, DELVE~6, Eridanus~III, Kim~3, PS1~1,  and YMCA-1. Basic details of these observations are presented in \tabref{obstable}.

\subsubsection{IMACS Observations and Data Reduction} 
\par Our IMACS observations of the seven UFCSs listed above were carried out with the instrument's $f$/4 camera in one of two configurations, both of which targeted the strong Calcium II near-infrared Triplet (CaT) lines at $8498 {\rm \ \AA},8542 {\rm \ \AA},$ and $8662 {\rm \ \AA}$. The first configuration used the 1200 $\ell$/mm grating blazed at $9000 \rm \ \AA$, providing $R \approx 11000$ across a typical wavelength range of $7500$--$9000 \rm \ \AA$. This configuration has been a staple of past IMACS analyses of UFDs \citep{li17,2018ApJ...857..145L,2020ApJ...892..137S,2022ApJ...939...41C,2023ApJ...950..167B,2024ApJ...961..234H}. The second configuration used the instrument's 600 $\ell$/mm grating blazed at $7500 \rm \ \AA$, yielding a lower spectral resolution of $R \approx 5000$ across a broader wavelength range of $6000$--$9000\rm \ \AA$ that notably covers the $\rm H\alpha$ line at $ 6563\rm \; \AA$. This lower-resolution grating was specifically chosen for observations of our more distant southern-sky UFCS targets (DELVE~3, DELVE~6, and Eridanus~III, each at $D > 50$~kpc), for which we accepted lower velocity precision in order to recover fainter member stars.
\par For six of the seven UFCSs that we targeted successfully with IMACS,  we observed a single multi-object mask composed of $0.7'' \ \rm \times \ 5''$ slits and used just one of the two gratings (see \tabref{obstable}).  For the seventh target, Eridanus~III,  we observed one mask with the 1200 $\ell$/mm grating and a separate mask with the 600 $\ell$/mm grating (as described in more detail in \citealt{Simon2024}). Stellar targets for these masks were generally designed based on isochrone selections applied to photometry from the Dark Energy Camera (DECam; \citealt{2015AJ....150..150F}) paired with selections based on \Gaia DR2 or DR3 photometry and astrometry \citep{2016A&A...595A...1G,2018A&A...616A...1G,2021A&A...649A...2L,2023A&A...674A...1G}.  On a typical observing night, we obtained science exposures through these masks with exposure times of $\sim$15--40 minutes. After every two to three science exposures, we obtained a HeNeArKr arc frame and a quartz flat frame in the same instrument position in order to monitor the time evolution of the wavelength solution.
\par We reduced the IMACS data following the procedure first described in \citet{2017ApJ...838...11S} and used throughout most analyses of UFDs using the same instrument.  Briefly, the 2D spectra were first bias-subtracted. The \texttt{COSMOS} pipeline \citep{2017ascl.soft05001O} was then used to map slits on the 2D spectra using the comparison arc images and slit coordinates and to derive a preliminary wavelength solution. Next, a modified version of the DEEP2 reduction pipeline \citep{2012ascl.soft03003C,2013ApJS..208....5N} was used to carry out flat fielding and sky subtraction. Finally, a precise wavelength solution was derived based on a polynomial fit to the arc lamp calibration frame taken after a given set of science exposures in our observing sequence. Because six of our seven targets were observed for just a single epoch/run, we could safely coadd across all 1D spectra from a given run and then subsequently normalize the coadded spectrum. The sole exception to this coaddition procedure was Eridanus~III, which was observed across three separate runs; for that system, we followed \citet{Simon2024} and kept the distinct epochs separate.

\begin{deluxetable*}{cccccc}
\tablecaption{Summary of the Keck/DEIMOS and Magellan/IMACS Spectroscopic Observations \label{tab:obstable}}
\tablehead{\colhead{UFCS Name} & \colhead{Program P.I.} & \colhead{Instrument (Grating)} & \colhead{UTC Date} & \colhead{MJD} & \colhead{$t_{\rm exp}$ (s)}}
\startdata
BLISS 1 & T.S.~Li & IMACS (1200 $\ell$/mm) & 2019-02-12 & 58526.35 & 6300 \\
Balbinot 1 & W. Cerny & DEIMOS (1200G) & 2022-09-25 & 59847.74 & 5400 \\
DELVE 1  & W. Cerny & DEIMOS  (1200G) & 2024-02-15 & 60355.64 & 5100 \\
DELVE 3 & W. Cerny, J. Simon & IMACS (600 $\ell$/mm) & 2024-07-01,02 & 60492.90 & 9660  \\
DELVE 4 & W. Cerny & DEIMOS  (1200G) & 2023-04-24 & 60058.59 & 3600  \\
& W. Cerny & DEIMOS  (1200G) & 2023-06-12 & 60107.48 & 3900  \\
DELVE 5 & W. Cerny  & DEIMOS (1200G) & 2023-06-12 & 60107.40 & 7200  \\
DELVE 6 & W. Cerny; O. Gnedin & IMACS (600 $\ell$/mm) & 2023-12-14 & 60292.09 & 8400   \\
Draco II & M.~Geha & DEIMOS (1200G) & 2015-07-18 & 57221.38 & 4800 \\
 & M.~Rich & DEIMOS (1200G) & 2016-09-05 & 57636.27 & 3600  \\
& M.~Geha & DEIMOS (1200G) & 2017-02-27 & 57811.60 & 2400  \\
Eridanus III & T.S.~Li, J.~Simon &  IMACS (1200 $\ell$/mm) & 2020-02-02 & 58882.1 & 8400  \\
 & T.S.~Li, J.~Simon &  IMACS (1200 $\ell$/mm) & 2020-12-17 & 59201.2 & 3600  \\
 & T.S.~Li, J.~Simon &  IMACS (600 $\ell$/mm) & 2021-09-12,13 & 59471.0 & 24950 \\
Kim 1 & W.~Cerny &  DEIMOS (1200G) & 2022-09-25 & 59847.51 & 7200  \\
Kim 3 & T.S.~Li &  IMACS (1200 $\ell$/mm) & 2021-06-27 & 59393.05 & 11400  \\
Koposov 1 & M.~Geha & DEIMOS  (1200G) & 2011-05-28 & 55709.66 & 9600  \\
Koposov 2 & E.~Tollerud & DEIMOS  (1200G) & 	2014-10-20 & 56950.57 & 3600  \\
 &  &  & 2015-02-17 & 57070.28 & 3600  \\
 & &  & 2015-02-18 & 57071.27 & 3600  \\
Laevens 3 & D.~Mackey & DEIMOS (1200G) & 2015-09-08 & 57273.32 & 3600  \\
Munoz 1 & R.~Munoz & DEIMOS (1200G)  & 2011-05-28 & 55709.39 & 7800  \\
PS1 1 &  T.S.~Li & IMACS (1200 $\ell$/mm) & 2021-07-11 & 59406.26 & 4800   \\
Segue 3 & M. Geha & DEIMOS (1200G) & 2009-11-16 & 55151.21 & 3600  \\
 &  &  & 2009-11-16 & 55151.27 & 3600  \\
  &  &  & 2010-05-16 & 55332.59 & 3960  \\
UMaIII/U1 & W. Cerny, S. Smith & DEIMOS (1200G) & 2023-04-24 & 60058.29 & 3600 \\
 & W. Cerny, A. Ji  & DEIMOS (1200G) & 2025-04-02 & 60767.32 & 16800 \\
YMCA-1 & W.~Cerny, J.~Simon  & IMACS (1200 $\ell$/mm) & 2024-12-23 & 60667.35 & 2700 \\
\enddata
\tablecomments{Each row corresponds to an individual mask-run of observations; some runs used more than one mask and thus are split here. The MJD and UTC values correspond to the mid-point of observations for a given mask-run. }
\end{deluxetable*}

\subsubsection{Magellan/IMACS Velocity Measurements} 
We measured velocities from each normalized 1D IMACS spectrum following the procedure first introduced in \citet{li17}, which involves shifting a set of empirical spectral templates through a range of velocities and identifying the best-fitting velocity shift through Markov Chain Monte Carlo (MCMC) sampling with \texttt{emcee} \citep{2013PASP..125..306F}.  Each spectrum was fit with several templates spanning different spectral types: for the 1200 $\ell$/mm grating observations, these were template spectra of the very metal-poor Red Giant Branch (RGB) star HD 122563, the metal-poor RGB star HD 26297, the Blue Horizontal Branch (BHB) star HD 161817, the very metal-poor subgiant HD 140283, and the metal-poor Main Sequence (MS) star HD 31128. The latter two templates have not generally been used in prior IMACS analyses but were found to be necessary in our analysis here given the prevalence of subgiant and MS stars in our member samples for certain nearby UFCSs. For the 600 $\ell$/mm grating observations, we used only two templates -- one of HD 122563 and the other of the BHB star HD~86986. This grating was exclusively used  for our more distant UFCS targets where all targetable members were (faint) RGB or BHB stars; thus, these two templates were sufficient for our purposes. 
\par After all template fits were performed for a given star, the template yielding the smallest $\chi^2$ was selected for our primary velocity measurement. The observed velocity ($v_{\rm obs}$) and its uncertainty were derived from the median and 16th/84th percentiles of the MCMC posterior after $5\sigma$ clipping. To correct for the miscentering of stars within their slits, we next derived the radial velocity shift $v_{\rm tel}$ of the observed sky absorption (telluric) lines relative to an empirical template of the B-type star HR 4781 (for observations with the 1200 $\ell$/mm grating) or HR 1244 (for observations with the 600 $\ell$/mm grating). This telluric fit was carried out through a nearly identical MCMC procedure to that used for the stellar velocity but over a distinct wavelength range covering the Fraunhofer
A band. The final velocity for each star (after a heliocentric correction) was then computed according to $v_{\rm hel} = v_{\rm obs} - v_{\rm tel}$ and the associated statistical uncertainty was taken as $\epsilon_{\rm v_{\rm hel}} = \sqrt{\epsilon_{\rm v_{\rm obs}}^2 + \epsilon_{\rm v_{\rm tel}}^2}$. Lastly, we added a 1.0~\kms{} systematic error floor in quadrature to this statistical uncertainty for observations with the 1200 $\ell$/mm grating or a $3.0$~\kms{} floor for observations with the 600 $\ell$/mm grating. The floor for the 1200 $\ell$/mm grating is based on the repeatability of velocity measurements across repeat observations on distinct nights \citep{Simon_2017,li17}, while the floor for the 600 $\ell$/mm grating was derived by \citet{Simon2024} based on the degradation in spectral resolution and rms scatter of the wavelength solution of the 600 $\ell$/mm grating relative to the 1200 $\ell$/mm grating.
\par This velocity-fitting procedure was attempted for all spectra with $S/N > 2$/pixel (for the 1200 $\ell$/mm grating) or  $S/N > 3$/pixel (for the 600 $\ell$/mm grating). Stars above this $S/N$ limit with poor fits (e.g., as reflected in bimodal velocity posteriors) were manually removed. This vetting was particularly necessary for spectra taken with the lower-resolution 600 $\ell$/mm grating in the range $3 \lesssim  S/N \lesssim  5$. The velocity signal in these low-signal-to-noise 600 $\ell$/mm grating spectra was often dominated by the H$\alpha$ line (which falls outside the wavelength range of the 1200 $\ell$/mm grating observations) as opposed to the CaT. Ultimately, our kinematic constraints based on 600 $\ell$/mm data are rather weak, and no qualitative results in this work depend strongly on these marginal spectra. 
\par For the special case of Eridanus~III, for which both 600 $\ell$/mm and 1200 $\ell$/mm observations exist, we directly adopted the velocity measurements from \citet{Simon2024} instead of re-measuring them. The \citet{Simon2024} velocities were derived from a very similar fitting approach and assumed the same systematic uncertainty floors described above.

\subsubsection{Magellan/IMACS Equivalent Width measurements}
We also measured the Equivalent Widths (EWs) of the three CaT lines from our normalized IMACS spectra following the procedure from \citet{2017ApJ...838...11S} and \citet{2018ApJ...857..145L}. Each CaT line was fit with a Gaussian-plus-Lorentzian profile through a non-linear least-squares optimization procedure using \texttt{mpfit} \citep{2009ASPC..411..251M}. The EWs of the three lines were then individually determined by integrating the fitted Gaussian and Lorentzian profiles. Hereafter, we refer to the sum of the three CaT line EWs as $\sum \rm EW_{\rm CaT}$. The statistical uncertainties on the EWs were calculated analytically from the uncertainties on the Gaussian and
Lorentzian integrals derived from the covariance matrix of the fit parameters from \texttt{mpfit}. Based on the repeatability of IMACS EW measurements across nights, we added a 0.2 $\rm \AA$ systematic term in quadrature to the uncertainty on the summed EW (\citealt{li17}).
\par We only report EW measurements derived from IMACS spectra above a hard signal-to-noise cutoff of $S/N > 5$/pixel for both gratings. The fits for all likely UFCS member stars were visually inspected to ensure the reasonableness of the EW measurements and a small fraction were flagged and excluded based on unusual line ratios. For an even smaller number of cases where artifacts associated with poor sky subtraction or chip gaps affected exactly one of the three CaT lines, we used the inter-line relations derived by \citet{2024ApJ...961..234H} to predict the EW of the missing third line and its associated uncertainty from the other two.  For illustration, a subset of the CaT fits yielding useful metallicity measurements for UFCS member stars are shown in Appendix \ref{sec:spectrafits}. As demonstrated there, most EW measurements that yielded metallicities in this work were derived from spectra with significantly higher signal-to-noise than our nominal minimum of $S/N = 5$.

\subsection{Keck II/DEIMOS Sample}
\label{sec:deimos}
In the 2022B--2024A semesters, we pursued spectroscopy of six northern sky UFCS targets --  Balbinot~1, DELVE~1, DELVE~4, DELVE~5, Kim~1, and \uma{} -- with the DEep Imaging Multi-Object Spectrograph (DEIMOS; \citealt{2003SPIE.4841.1657F}) mounted on the 10-m Keck II telescope at the W. M. Keck Observatory on Maunakea, Hawai`i. Complementing these newer observations, we drew archival data for the Draco~II, Laevens~3, Mu\~{n}oz~1, Segue~3, Koposov~1, and Koposov~2 UFCSs from the database of \citet{geha_paper1}, who present a homogeneous reduction of all public DEIMOS spectra available in the Keck Observatory Archive.\footnote{\url{https://koa.ipac.caltech.edu/}}  DEIMOS observations for the former four targets have been published in the literature \citep{2011AJ....142...88F, 2012ApJ...753L..15M, 2016MNRAS.458L..59M, 2018MNRAS.480.2609L, 2019MNRAS.490.1498L}; however, the Koposov~1 and Koposov~2 observations, as well as a single more recent Draco~II mask, have not yet been presented in any dedicated study. Our re-analysis of the four targets with published measurements was warranted by the state-of-the-art DEIMOS reduction utilities developed within recent years and the new astrometric measurements provided by \Gaia Data Release 3 \citep{2023A&A...674A...1G}.  In addition, the newer mask of Draco~II observations significantly expands the sample of known members compared to the two published spectroscopic studies \citep{2016MNRAS.458L..59M,2018MNRAS.480.2609L} and offers an additional epoch to check for binary star radial velocity variations.
\par We successfully recovered a velocity detection for all DEIMOS targets listed above, and thus our spectroscopic sample consists of 12 UFCSs (DEIMOS) plus 7 UFCSs (IMACS) yielding 19 systems in total (see \tabref{obstable}). Our sample encompasses all UFCSs with published spectroscopic studies, which are limited to four systems beyond those targeted as part of our census effort (Draco~II, Laevens~3, Mu\~{n}oz~1, and Segue~3) as well as three systems targeted in association with our census but published first as standalone results: DELVE~1 and Eridanus~III (presented by \citealt{Simon2024} in tandem with high-resolution spectra of their single brightest stars) and \uma{} (presented in tandem with its discovery by \citealt{2024ApJ...961...92S} and recently revisited with additional DEIMOS observations by our team in \citetalias{2025arXiv251002431C}~\citeyear{2025arXiv251002431C}).

\subsubsection{DEIMOS Observations and Data Reduction}
The new and archival DEIMOS observations analyzed here exclusively used the instrument's 1200 line $\rm mm^{-1}$ grating (1200G) and OG550 order-blocking filter, providing a spectral resolution of $R \approx  6000$ across a typical wavelength range 6500--$9000 \rm \ \AA$. For our dedicated new observations of Balbinot~1, DELVE~1, DELVE~4, DELVE~5, Kim~1, and \uma{}, we observed either one or two slitmasks per target. Each mask consisted of $\sim$40--100 slits with widths 0.7$\arcsec$ and minimum lengths of 4.5--5$\arcsec$. Stellar targets for these more recent masks (i.e., those observed since 2022 by our team) were typically selected based on public DECam photometry and the astrometry available from \Gaia DR3. Older masks (i.e., those observed before 2018) were designed more heterogeneously based on a variety of photometric datasets and predate the availability of \Gaia proper motions.  At the beginning of each DEIMOS observing night, we obtained XeNeArKr calibration arc exposures through each mask as well as internal quartz flat frames. Science exposures were typically 20--40 minutes each. The archival observations were predominantly taken under the same configuration albeit with small variations in the chosen central wavelength, slit width, and exposure time.
\par All DEIMOS observations in this work (both new and archival) were reduced using the state-of-the-art utilities provided by the open-source Python package \texttt{PypeIt} \citep{Prochaska2020}, which includes the officially-supported DEIMOS data reduction pipeline. \texttt{PypeIt} reduces the eight-CCD DEIMOS detector  as four mosaic images and automatically carries out flat-fielding, wavelength calibration based on the arc frames, and extraction of the 1D spectra. For our reductions, we disabled \texttt{PypeIt}'s default flexure correction and heliocentric correction and instead computed our own corrections at the velocity measurement stage, as described in the subsection below. Further details about our DEIMOS reductions are provided in \citet{geha_paper1}.

\subsubsection{DEIMOS Velocity Measurements} 
\par We measured stellar velocities from our reduced 1D DEIMOS spectra using the \texttt{dmost} pipeline -- a suite of codes specifically designed to improve and systematize measurements from spectra taken with DEIMOS and its 1200G grating.  All aspects of this pipeline are presented in \citet{geha_paper1}, and we provide only an abbreviated summary here.   
\par \texttt{dmost} measures stellar velocities on an exposure-by-exposure basis by forward-modeling each observed 1D spectrum with both a synthetic stellar model from the PHOENIX library \citep{2013A&A...553A...6H} paired with a telluric model from \texttt{TelFit} \citep{2014AJ....148...53G}. The coadded spectrum of each source across all exposures is used to select the best-fitting stellar template from the coarse PHOENIX grid via $\chi^2$ minimization, while the telluric spectrum is taken to be a constant across all sources on a given mask and determined via a joint fit to the majority of slits on the mask. With templates selected, the \texttt{emcee} sampler is used to derive posterior probability distributions for the stellar velocity and the telluric velocity shift needed to account for slit miscentering. The posteriors from this sampling provide exposure-level velocities, which are then heliocentric-corrected and combined via inverse-variance weighting to yield final mask-level velocity measurements $v_{\rm hel}$ and statistical velocity uncertainties $\epsilon_{\rm v,stat}$. For low $S/N$ cases where reliable velocities could not be derived for a majority of the available exposures, \texttt{dmost} measures the velocity from a coadded 1D spectrum across exposures. 
\par The final step of \texttt{dmost}'s velocity measurement procedure is to compute systematic velocity errors, which are critical to accurate velocity dispersion measurements -- particularly for the least-massive systems. Systematic errors within \texttt{dmost} are split into two components: a signal-to-noise dependent scaling term applied to the statistical velocity errors and a (per-epoch) systematic velocity uncertainty floor. Assuming this scaling-plus-floor functional form, \citet{geha_paper1} derived a prescription to compute the final velocity uncertainty, $\epsilon_{v} = \sqrt{(1.4\epsilon_{\rm v,stat})^2 + 1.1^2}$,  based on the repeatability of velocity measurements across scores of masks in the DEIMOS archive observed with the same 1200G grating. This prescription was validated against larger public radial velocity datasets \citep{2022ApJS..259...35A,2023A&A...674A...5K,2023ApJS..268...19W,2024AJ....168...58D}, confirming the reasonableness of the derived total errors. For the remainder of this work, we only make use of velocity measurements with total uncertainties $\epsilon_v < 15$~\kms{} so as to remain in the regime where this systematic uncertainty prescription applies.
\par The majority of our DEIMOS UFCS targets were observed for only a single epoch, and the procedure above fully describes the production of our final measurements in these cases. However, for three targets (DELVE~4, Draco~II, and Segue~3), two or three well-spaced epochs of velocity measurements were available for a subset of stars (see \tabref{obstable}). For these cases, the velocity estimation procedure described above was independently applied to each mask, and the final velocity measurement that we report is the inverse-variance-weighted average of the mask-level measurements. To probe for velocity variability, \texttt{dmost} additionally performs a $\chi^2$ goodness-of-fit test to estimate the probability that a given star varies with respect to the weighted-mean velocity measured across all masks. We conservatively flagged all stars with logarithmic $p$-values of $\log_{10}(p) < -1$ as suspected binaries and excluded them from our kinematic analyses. For the special case of \uma{} only, we did not average across epochs or apply this binarity test. Instead, we matched the dedicated analysis of \citetalias{2025arXiv251002431C}~\citeyearpar{2025arXiv251002431C} who identified a single confirmed binary based on two epochs of DEIMOS data combined with additional velocity monitoring from Gemini/GMOS, Keck/HIRES, and Magellan/MagE. We did not flag the three additional candidate binaries discussed in that work because their binarity status is not yet well-established.

\subsubsection{DEIMOS Equivalent Width Measurements}
The \texttt{dmost} pipeline also measures CaT EWs through a non-linear least-squares fit to the coadded, normalized 1D spectrum of each star. For stars with high signal-to-noise coadded spectra ($S/N > 15$/pixel), \texttt{dmost} models each CaT line with a Gaussian-plus-Lorentzian profile. For stars with lower signal-to-noise spectra ($S/N < 15$/pixel) or poor Gaussian-plus-Lorentzian fits, \texttt{dmost} instead models the CaT lines with only a Gaussian profile. The EWs of the three lines and their associated statistical uncertainties are computed analytically using the best-fit model parameters and the associated covariance matrices.  Lastly, the pipeline applies a scaling-plus-floor systematic uncertainty derived based on the repeatability of EW measurements, with a different prescription for the Gaussian vs. Gaussian-plus-Lorentzian cases (see \citealt{geha_paper1} for more detail); for reference, the EW uncertainty floor is $0.05 \rm \ \AA$ for Gaussian-plus-Lorentzian fits. If a star was observed on more than one mask, the final value of $\sum \rm EW_{\rm CaT}$ was determined through an inverse-variance-weighted average of the EWs derived from each individual mask.
\par To limit ourselves to the regime where EW measurements are the most reliable, we only present DEIMOS metallicities derived from spectra with $S/N \geq 7$/pixel. This is a slightly higher threshold compared to IMACS owing to DEIMOS' lower spectral resolution. Representative CaT fits to the DEIMOS spectra for UFCS member stars yielding metallicities are shown in Appendix \ref{sec:spectrafits}  (\figref{deimos_ew_fits1}--\ref{fig:deimos_ew_fits2}). As was the case with our IMACS measurements before, most EW measurements from DEIMOS that ultimately yielded metallicities for UFCS members were derived from spectra considerably above this minimum $S/N$ threshold.

\subsection{Stellar Metallicities from the Calcium Triplet}
We derived stellar metallicities ([Fe/H]) for RGB stars from our IMACS and DEIMOS CaT EW measurements using the recent calibration from \citet{2026MNRAS.546ag019N}. This calibration updates the coefficients of the \citet{2013MNRAS.434.1681C} calibration that has regularly been applied in prior analyses of MW UFDs. We specifically used the form of the \citet{2026MNRAS.546ag019N} calibration
that takes as inputs both the summed CaT EW across all three lines ($\sum\rm EW_{\rm CaT}$) and the absolute $V$-band magnitude of each star (hereafter $V_0$); the latter serves as a proxy for surface gravity. The calibration from \citet{2013MNRAS.434.1681C} was originally derived from stars with absolute magnitude $V_0 < 2$; here, we follow the common literature convention and assume that this calibration can reasonably be applied to \textit{all} RGB stars (in practice, $V_0 \lesssim 3$). \citet{geha_paper1} confirm the reasonableness of this extrapolation based on a large body of DEIMOS observations of GCs available in the Keck Observatory Archive; see their Figure 13.
\par Estimates of $V_0$ were derived using public broadband photometry from ground-based surveys and the distance moduli reported in \tabref{litproperties}. For UFCS targets with DESI Legacy Imaging Surveys DR10 (LS DR10; \citealt{2019AJ....157..168D}) or DELVE DR2  \citep{2022ApJS..261...38D} photometry, we first obtained apparent, dereddened $V$-band magnitudes from the extinction-corrected $g,r$-band photometry using the color-dependent piecewise relations derived for DES DR2 and presented in Appendix B of \citet{Abbott_2021}. For our single UFCS target with only Pan-STARRS~1 DR2 \citep{2016arXiv161205560C} photometry, we instead used the quadratic function and coefficients provided in Equation 6 and Table 6 of \citet{2012ApJ...750...99T}. To account for photometric zeropoint uncertainties and scatter in these filter transformations, we added a 0.02 mag error floor in quadrature to the input $g,r$-band magnitude uncertainties in all cases. After the apparent, dereddened $V$ magnitudes were derived, we computed $V_0$ estimates by subtracting the distance modulus of the corresponding UFCS as compiled here in \tabref{litproperties}. We treated the distance modulus error as a component of the total uncertainty budget for $V_0$.
\par To derive full [Fe/H] posteriors for each star, we propagated the posterior samples for each input through the \citet{2026MNRAS.546ag019N} relation.  We assumed Gaussian uncertainties on the coefficients of the \citet{2026MNRAS.546ag019N} relation and on $V_0$. Because our optimization-based CaT fits did not yield posteriors directly, we assumed that $\sum \rm EW_{\rm CaT}$ obeys a truncated normal distribution bounded below at zero.  After clipping nonphysical samples ($\rm [Fe/H] < -6$), our final metallicity estimates and their uncertainties were computed from the median and 16th/84th percentiles of the resultant [Fe/H] posterior. For the DEIMOS metallicities, we followed \citet{geha_paper1} and added a 0.1~dex systematic uncertainty term to account for scatter in the EW--[Fe/H] calibration. We did not apply this additional scatter term to our IMACS-derived metallicity uncertainties following past work with the instrument; however, we note that the systematic uncertainty floor assumed for $\sum \rm EW$ measurements derived from IMACS spectra (0.2$\rm \AA$) is $4\times$ larger than the same floor derived for DEIMOS  (0.05$\rm \ \AA$).  This results in an effective metallicity uncertainty floor of $\sim$0.1 dex for both instruments. 
\par When comparing our final CaT metallicities against those derived from the prior calibration of \citet{2013MNRAS.434.1681C}, we found that the updated \citet{2026MNRAS.546ag019N} calibration yielded metallicities that were typically 0.1--0.25~dex lower, with the largest differences seen at the higher-metallicity end ($\rm [Fe/H] > -1.5$). These differences are consistent with the discussion in the latter work and the analytic expectation from the coefficients of the two calibrations. We made no attempt to bring our (heterogeneously-derived) comparison sample of GCs and UFDs onto this revised metallicity scale, so this small offset relative to prior measurements should be kept in mind.

\subsubsection{Metallicities for \uma{}}
For the special case of \uma{}, which hosts just a single RGB star, we did not make use of CaT-based [Fe/H] measurements from DEIMOS. Instead, we directly adopted the sample of Ca K metallicities for 12 stars presented in \citetalias{2025arXiv251002431C}~\citeyearpar{2025arXiv251002431C} based on low-resolution $R\approx 1000$, blue-optical Keck I/LRIS spectra. These EW-based Ca K measurements were used to derive [Fe/H] measurements via the KP calibration from \citet{1999AJ....117..981B}. As described in the former work, we assumed a substantial $\pm 0.3$~dex global metallicity zeropoint uncertainty to account for possible differences relative to the CaT metallicity scale set by the \citet{2013MNRAS.434.1681C} relation (see also \citealt{2025arXiv250616462L}). Despite the difference in spectral lines and calibrations, we found excellent ($< 1\,\sigma$) consistency between the DEIMOS CaT measurement of \uma{}'s single RGB star (here using the \citealt{2026MNRAS.546ag019N} calibration) and the LRIS Ca~K measurement of the same star. For reference, we include the DEIMOS CaT spectrum of this star in \figref{deimos_ew_fits2}, though we reiterate that it is not used for our analysis.

\section{Stellar Membership Determination}
\label{sec:membership}

To isolate UFCS member stars within our spectroscopic catalogs, we leveraged our newly derived radial velocities, public ground-based photometry, and the proper motion and parallax information available for brighter stars in \Gaia DR3. In \secref{genproc} below, we first outline our general membership selection procedure before providing a significantly more extensive description of our selections for each UFCS in \secref{detailedmembership}.  Readers interested only in an abbreviated summary of our selections are encouraged to skip to \secref{memsummary}.

\subsection{General Procedure}
\label{sec:genproc}
 Our general UFCS membership selection procedure involved making conservative cuts based on the astrometric and photometric information to remove obvious foreground stars, which typically revealed a prominent radial velocity signal associated with the targeted UFCS. From there, a member sample could be identified unambiguously based on a simple velocity selection, or occasionally, both a velocity and proper motion selection. We avoided factoring in spatial information unless absolutely necessary, so as to preserve the possibility of identifying stars at large projected radii that may have been stripped via tides or ejected through two-body interactions. We also avoided incorporating metallicity/CaT EW information so as to avoid biasing ourselves against detecting metallicity spreads that might be expected if any UFCSs are dwarf galaxies. We only invoked spatial or metallicity information to exclude a star if there were several lines of evidence pointing to the same conclusion, for example, a large spatial separation paired with a discrepant proper motion. 
\par In detail, we began by retaining only stars satisfying the astrometric selections 
\[ (\varpi - 3 \sigma_\varpi < 0) \ \wedge\]
\[ (|\mu_{\alpha*}| < 7 \rm \ mas \ yr^{-1}) \wedge (|\mu_{\delta}| < 7 \rm \ mas \ yr^{-1})\]
where $\mu_{\alpha *} \equiv \mu_{\alpha}\cos\delta$ and $\mu_{\delta}$ are the \Gaia DR3 proper motion components in the R.A.\ and Decl.\ directions, respectively, and $\varpi$ is the \Gaia parallax with associated uncertainty $\sigma_\varpi$. These loose selections eliminated nearby foreground stars and stars with tangential velocities such that they would be unbound from the MW at the typical distances of the UFCSs in our sample ($D_\odot > 10$~kpc). We did not consider the \Gaia proper motion uncertainties when making these selections, but the selections are inclusive enough to avoid removing \textit{bona fide} members with large uncertainties.
\par In tandem with these astrometric selections, we applied an isochrone filter in color--magnitude (CMD) space to select UFCS members based on model isochrones from the PAdova and tRieste Stellar Evolutionary Code (PARSEC; \citealt{2012MNRAS.427..127B,2013MNRAS.434..488M,2016ApJ...822...73R}), Version 1.2S, which we accessed through the \texttt{ezpadova} Python package.\footnote{https://github.com/mfouesneau/ezpadova} The isochrone ages and metallicities for the UFCSs were generally adopted from the literature, though in some cases we re-determined them iteratively with our kinematic selections as detailed in the subsection below. We note that the PARSEC V1.2S isochrone grid extends to a minimum metallicity of $Z=0.0001$ ($\rm [Fe/H] \approx -2.19$).  To select members based on these models, we performed 1D filtering assuming a $(g-r)_0$ color tolerance of 0.2 mag about the isochrone RGB/MS and a generous tolerance of 0.4 mag in $g_0$ about the HB. For a few relatively younger UFCSs in our sample ($\tau < 8$~Gyr), we additionally included stars within $0.2$~mag of the isochrone subgiant branch (in $g_0$). 
\par After these initial selections, we identified the approximate mean velocity associated with the UFCS targeted in each spectroscopic field through visual inspection of the heliocentric velocity vs. projected elliptical radius ($r_{\rm ell}$) plane. The velocity signal was found to be unambiguous for all cases in our UFCS sample (though we remind the reader that two targets for which our IMACS observations recovered no clear velocity signal were excluded from our final spectroscopic sample). A loose velocity selection, typically $|\Delta v_{\rm hel}| \lesssim   20$--30~\kms{} about this mean, was then made to encompass all plausible members, including possible binaries. Lastly, in a small number of cases, we applied a more restrictive proper motion selection to hone the resultant pool of radial-velocity-selected member candidates into a final sample of member stars. These selections were determined subjectively, but we found them straightforward to choose for nearly all UFCSs and we took particular care to explore the properties of stars near the selection boundaries.
\par
\begin{figure*}
    \centering
    \includegraphics[width= 0.85\textwidth]{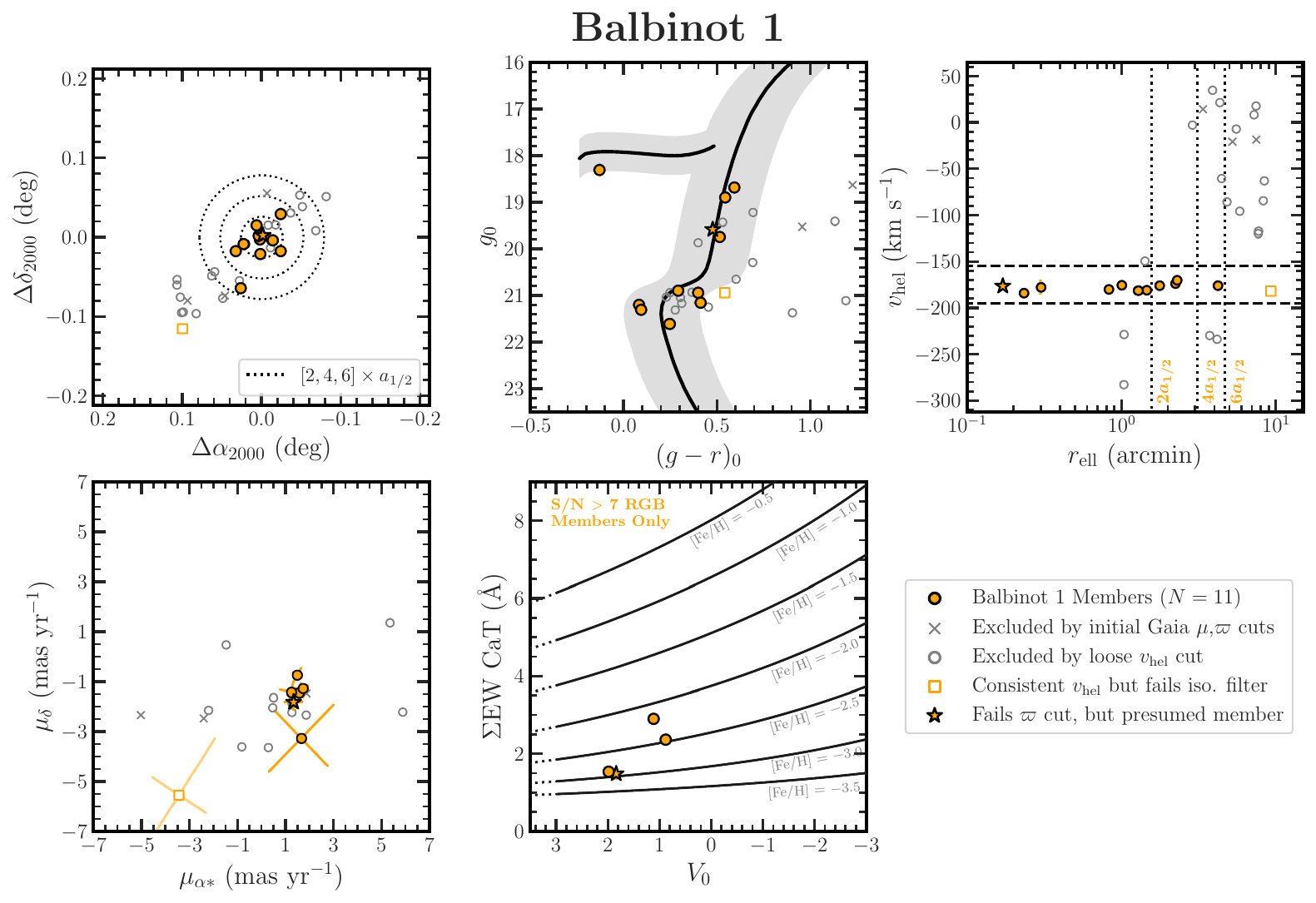}
    \caption{Member sample for Balbinot~1. Members were selected based on a \Balbinotoneage{}, \Balbinotoneisofeh{} isochrone. We identify \BalbinotoneNmem{} total members including four RGB stars, one BHB star, and six MSTO stars. Refer to the main text of \secref{genproc} for a detailed description of each panel. \label{fig:balbinot1}} 
    
    \includegraphics[width= 0.85\textwidth]{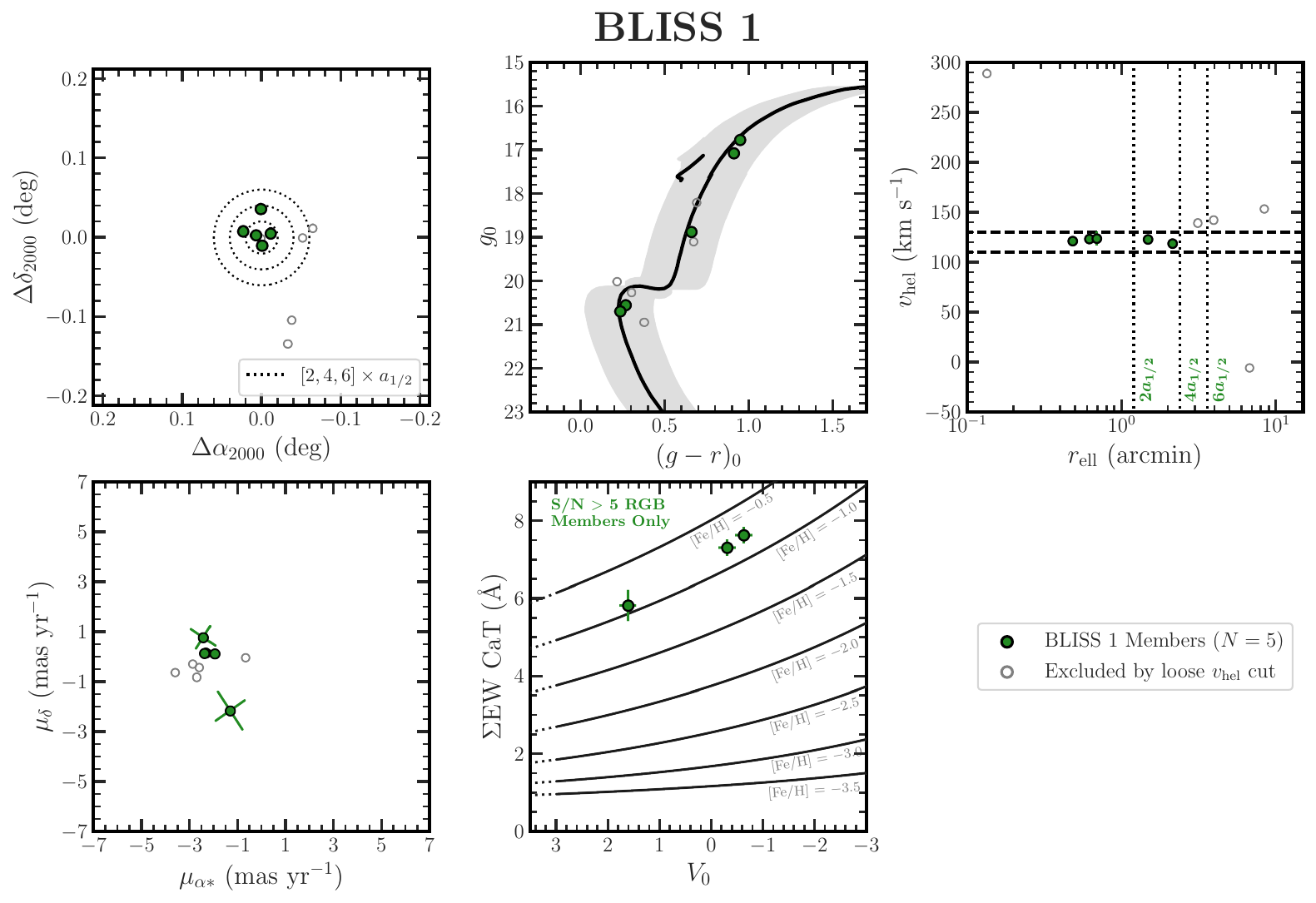}
    \caption{Member sample for BLISS~1. Members were selected based on a \BLISSoneage{}, \BLISSoneisofeh{} isochrone. This is a somewhat younger and more metal-rich model than reported by \citet{2019ApJ...875..154M}, but it better fits the observed slant of the RGB. We identify \BLISSoneNmem{} total members including three bright RGB stars and two stars at the MSTO. \label{fig:bliss1}} 
\end{figure*}

\begin{figure*}
    \centering
    \includegraphics[width= 0.85\textwidth]{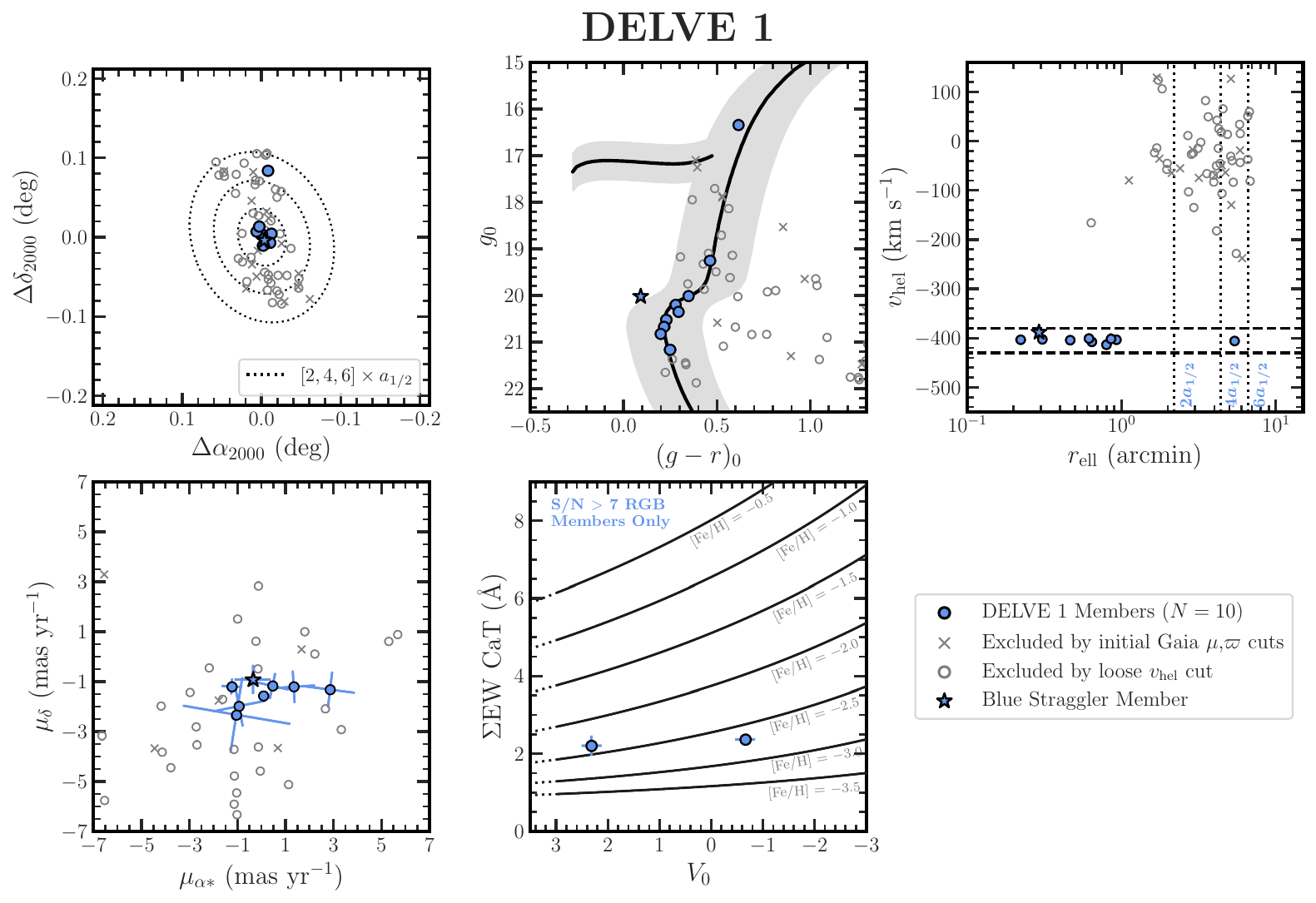}
    \caption{Member sample for DELVE~1. Members were selected based on a \DELVEoneage{}, \DELVEoneisofeh{} isochrone. We identify \DELVEoneNmem{} total members including two RGB stars, seven subgiant/MSTO/MS stars, and one candidate blue straggler.  \label{fig:delve1}} 

   \includegraphics[width= 0.85\textwidth]{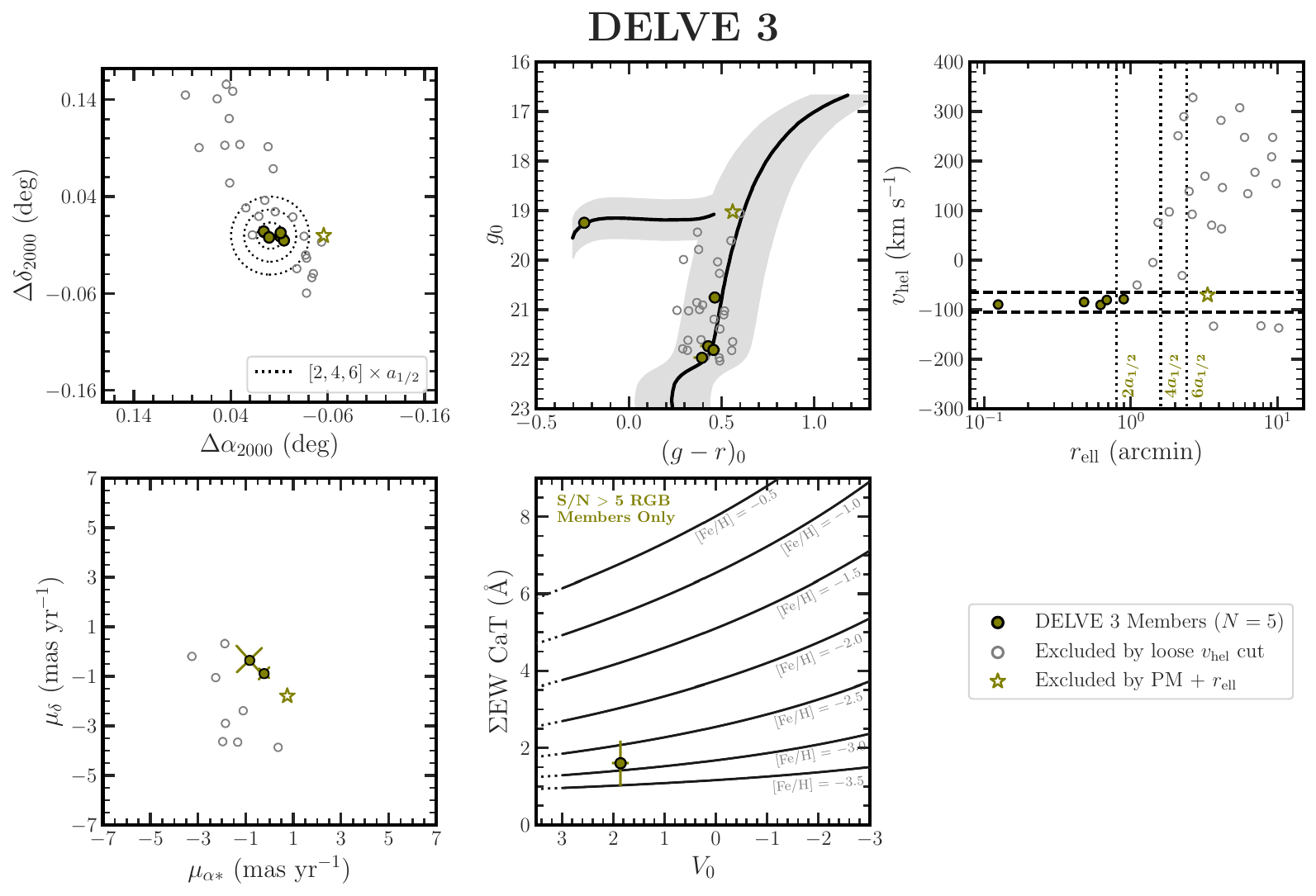}
    \caption{Member sample for DELVE~3. Members were selected based on a \DELVEthreeage{}, \DELVEthreeisofeh{} isochrone. We identify \DELVEthreeNmem{} total members including four RGB stars and one BHB star. \label{fig:delve3} }

\end{figure*}

\begin{figure*}
    \centering
    \includegraphics[width=0.85\textwidth]{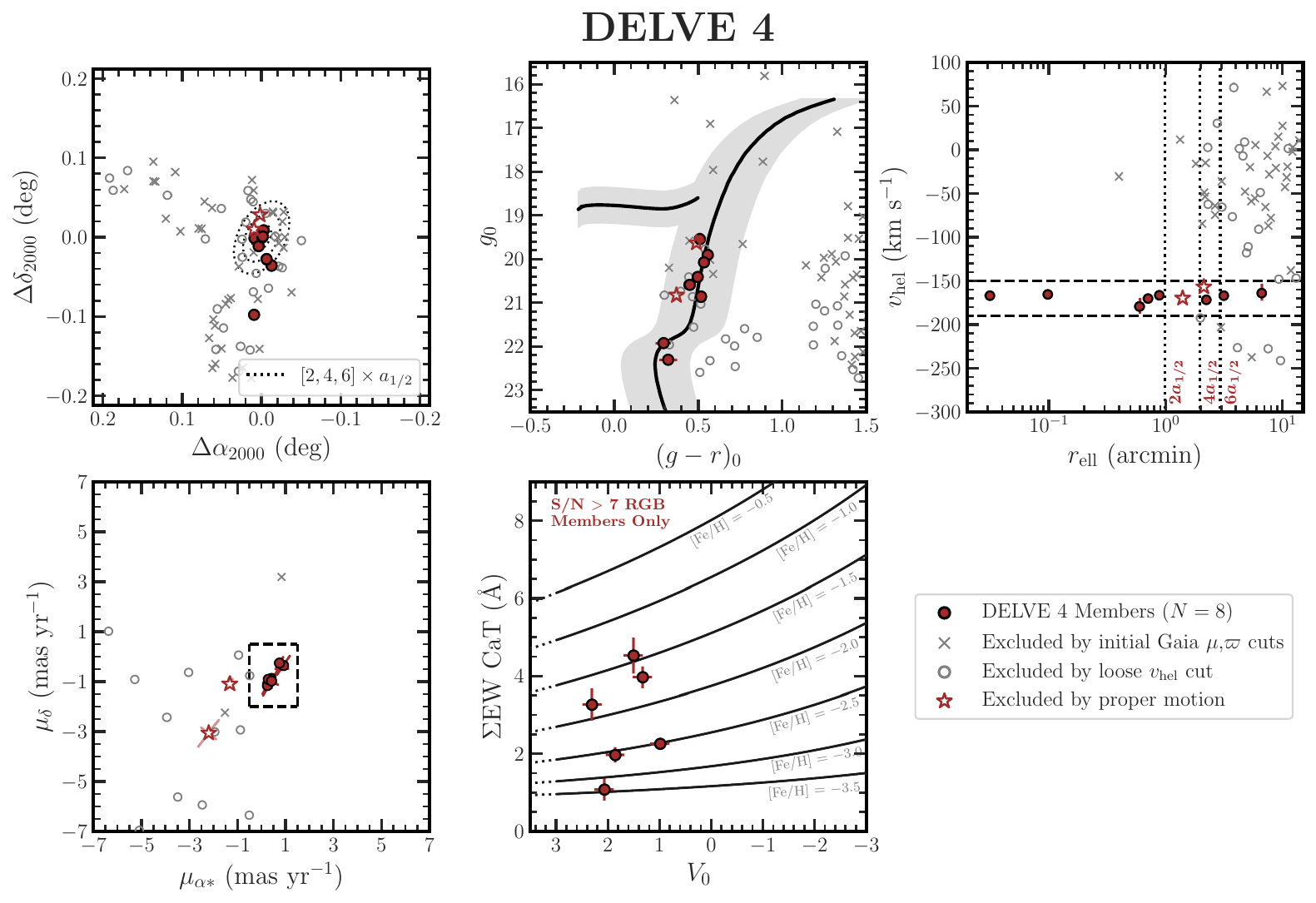}
    \caption{Member sample for DELVE~4. Members were selected based on a \DELVEfourage{}, \DELVEfourisofeh{} isochrone. We identify a total of \DELVEfourNmem{} likely members including six RGB stars and two stars at the MSTO, though see \secref{membership} for caveats.   \label{fig:delve4}}
   \includegraphics[width=0.85\textwidth]{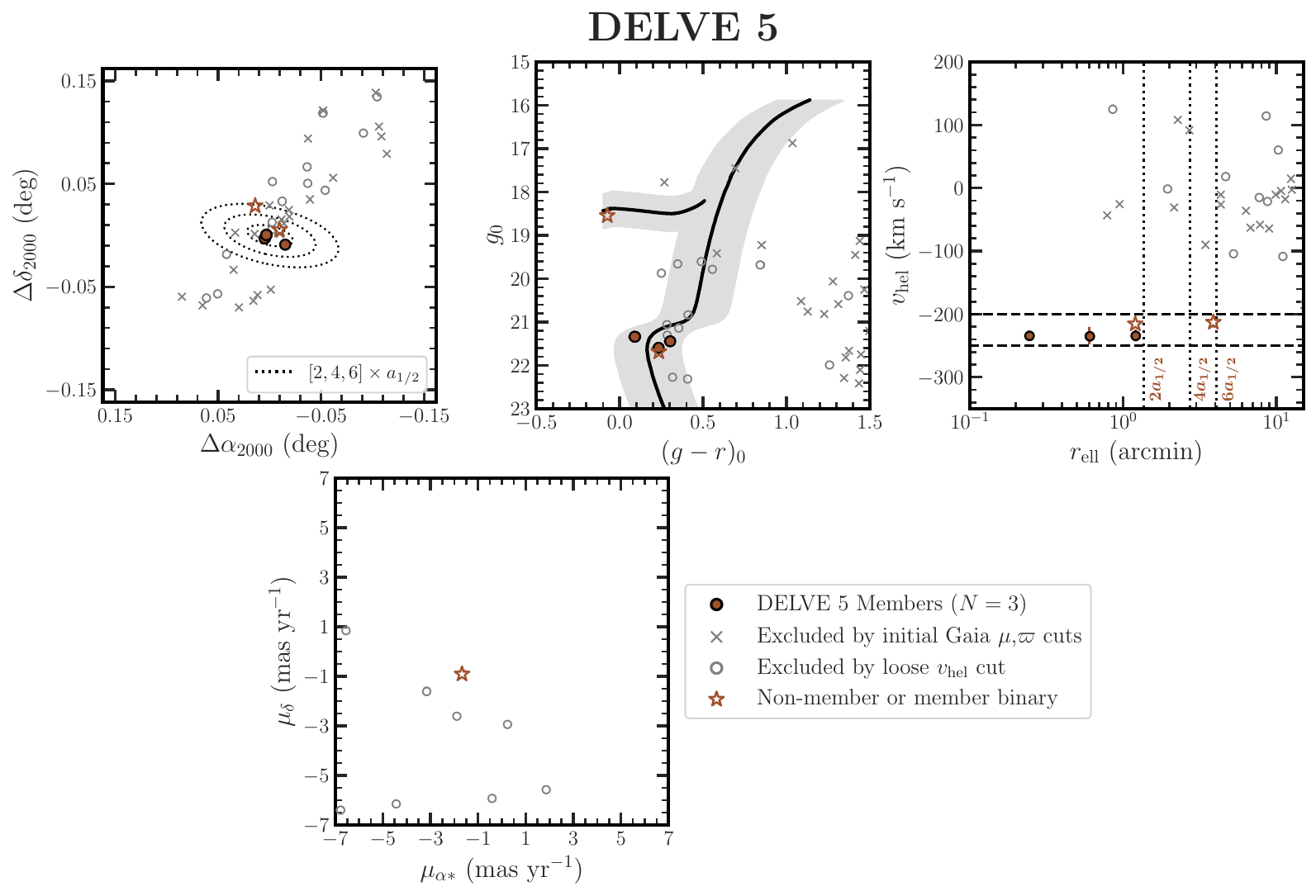}
    \caption{Member sample for DELVE~5. Members were selected based on a \DELVEfiveage{}, \DELVEfiveisofeh{} isochrone. We identify \DELVEfiveNmem{} confident MSTO members each at $v_{\rm hel} \approx -233$~\kms{}. We further highlight two stars at $v_{\rm hel} \approx -214$~\kms{} that are either non-members or member binaries observed far from their center-of-mass velocities. \label{fig:delve5}}
\end{figure*}

\begin{figure*}
    \centering
    \includegraphics[width=0.85\textwidth]{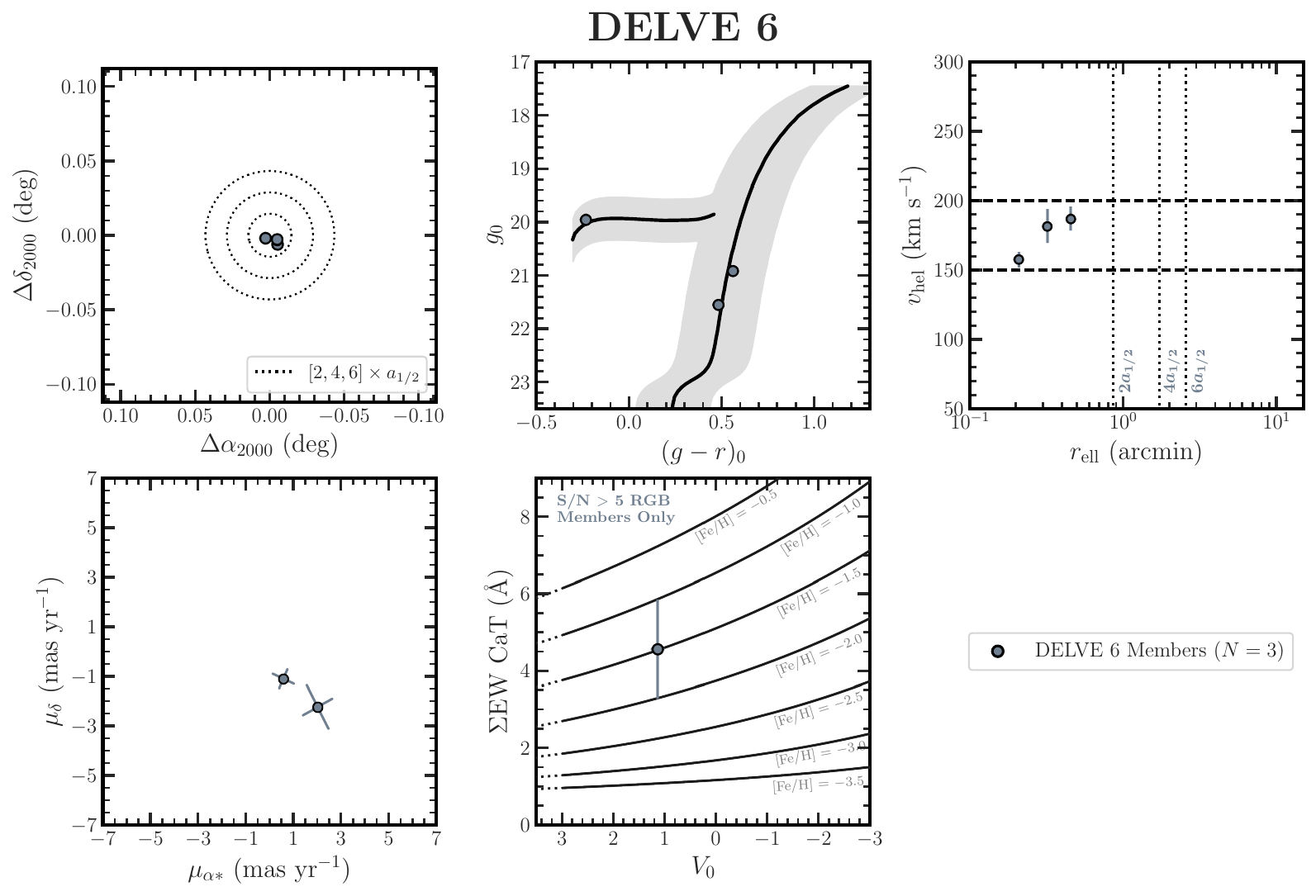}
    \caption{Member sample for DELVE~6. Members were selected based on a \DELVEsixage{}, \DELVEsixisofeh{} isochrone. We identify \DELVEsixNmem{} total members including two RGB stars and one BHB star. No other stars yielded useful velocities.    \label{fig:delve6}} 
     \includegraphics[width=0.85\textwidth]{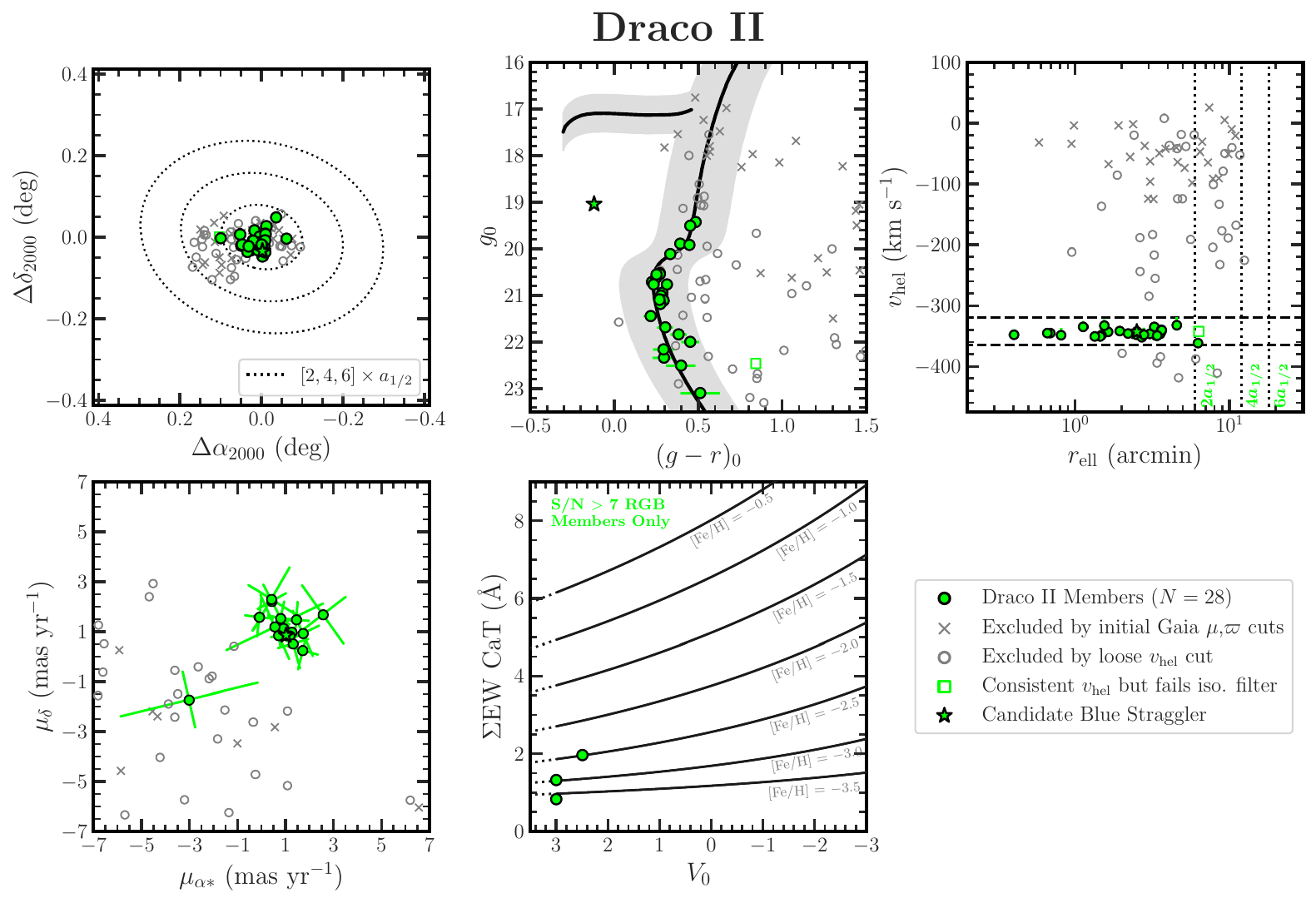}
    \caption{Member sample for Draco~II. Members were selected based on a \DracoIIage{}, \DracoIIisofeh{} isochrone. We identify \DracoIINmem{} members, including four RGB stars,  one subgiant, 22 stars along the MS, and one candidate blue straggler member.}
    \label{fig:dracoII}
    
\end{figure*}

\begin{figure*}
    \centering
    \includegraphics[width=0.85\textwidth]{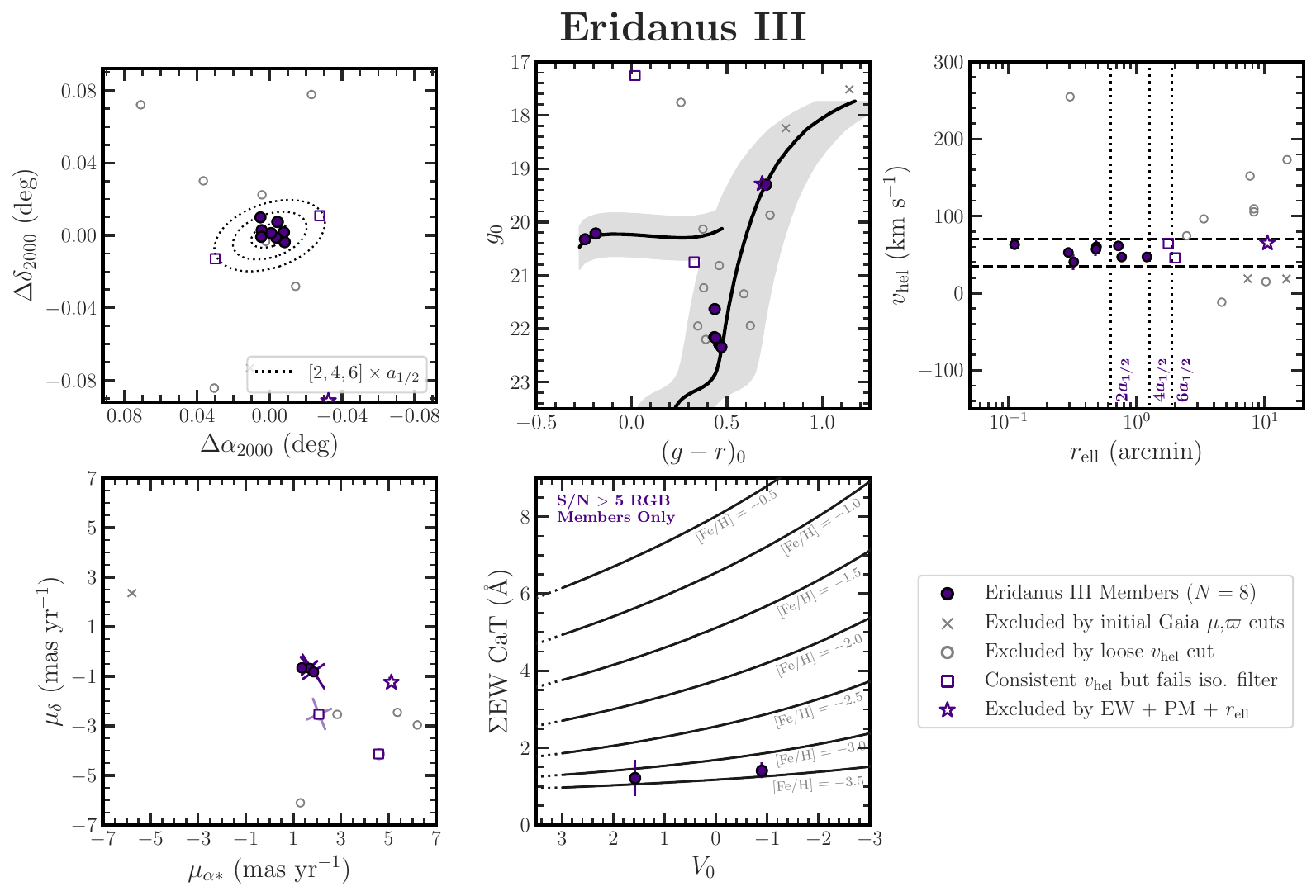}
    \caption{Member sample for Eridanus~III. We adopt the \citet{Simon2024} sample of \EridanusIIINmem{} total members including six RGB and two BHB stars. We overplot a \EridanusIIIage{}, \EridanusIIIisofeh{} isochrone to guide the eye. \label{fig:eridanusIII}}
    \includegraphics[width= 0.85\textwidth]{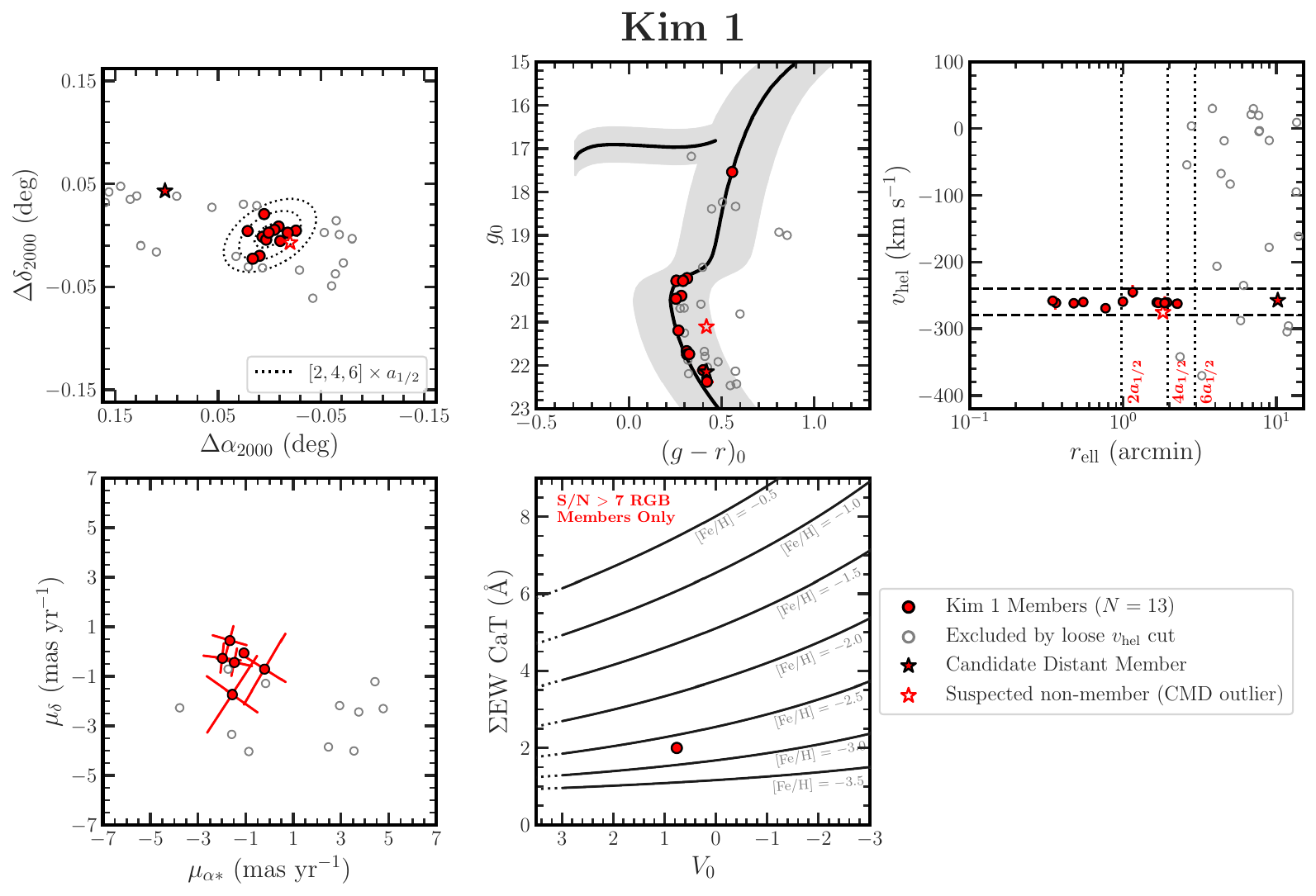}
    \caption{Member sample for Kim~1. Members were selected based on a \Kimoneage{}, \Kimoneisofeh{} isochrone. We identify \KimoneNmem{} confident members including a single RGB star and 12 MSTO or MS stars. This includes one candidate MS member at a large projected radius. \label{fig:kim1}}
\end{figure*}

\begin{figure*}
    \centering
    \includegraphics[width= 0.85\textwidth]{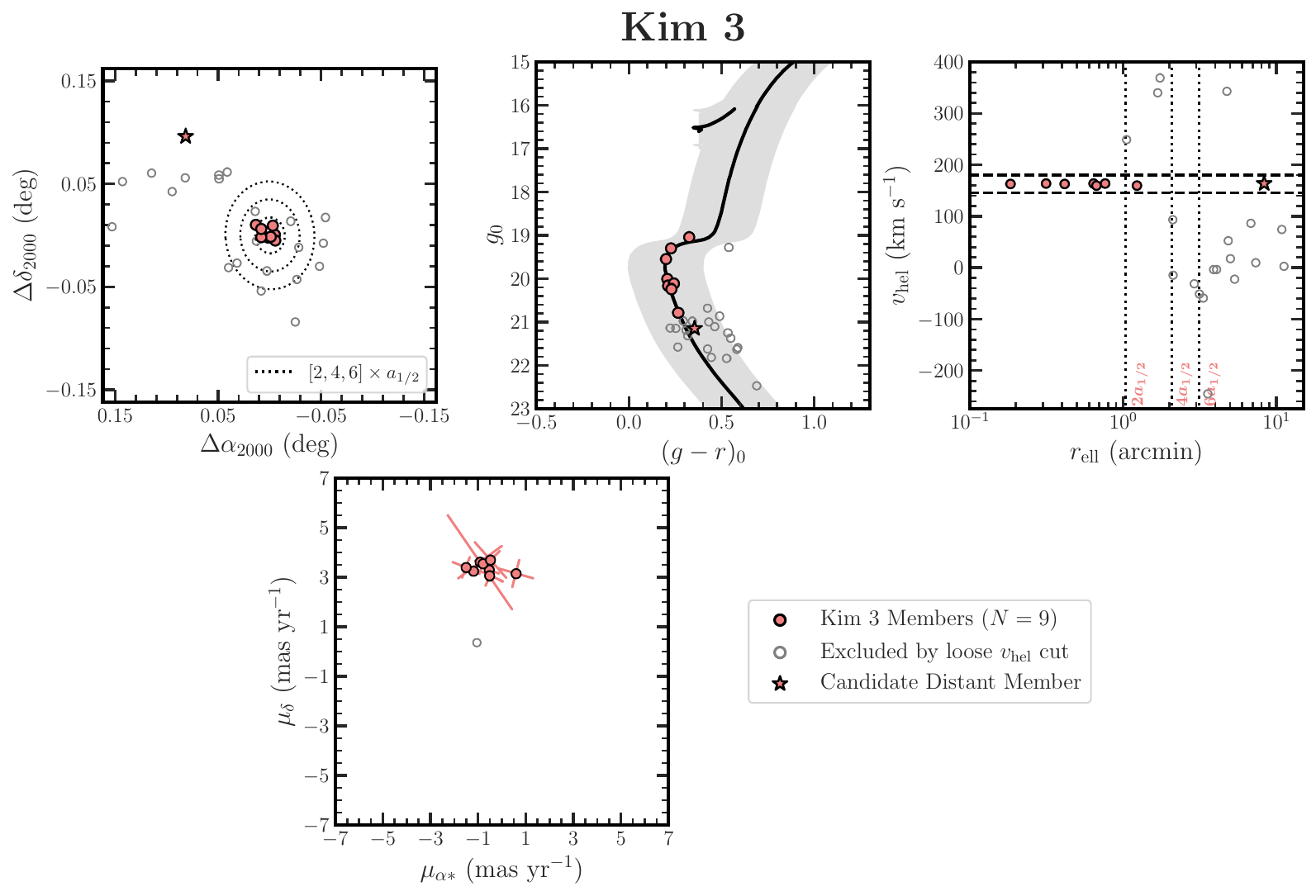}
    \caption{Member sample for Kim~3. Members were selected based on a \Kimthreeage{}, \Kimthreeisofeh{} isochrone. We identify 8 secure members, all at the MSTO or MS, as well as one candidate MS member at a large projected radius.  \label{fig:kim3}}
      \includegraphics[width= 0.85\textwidth]{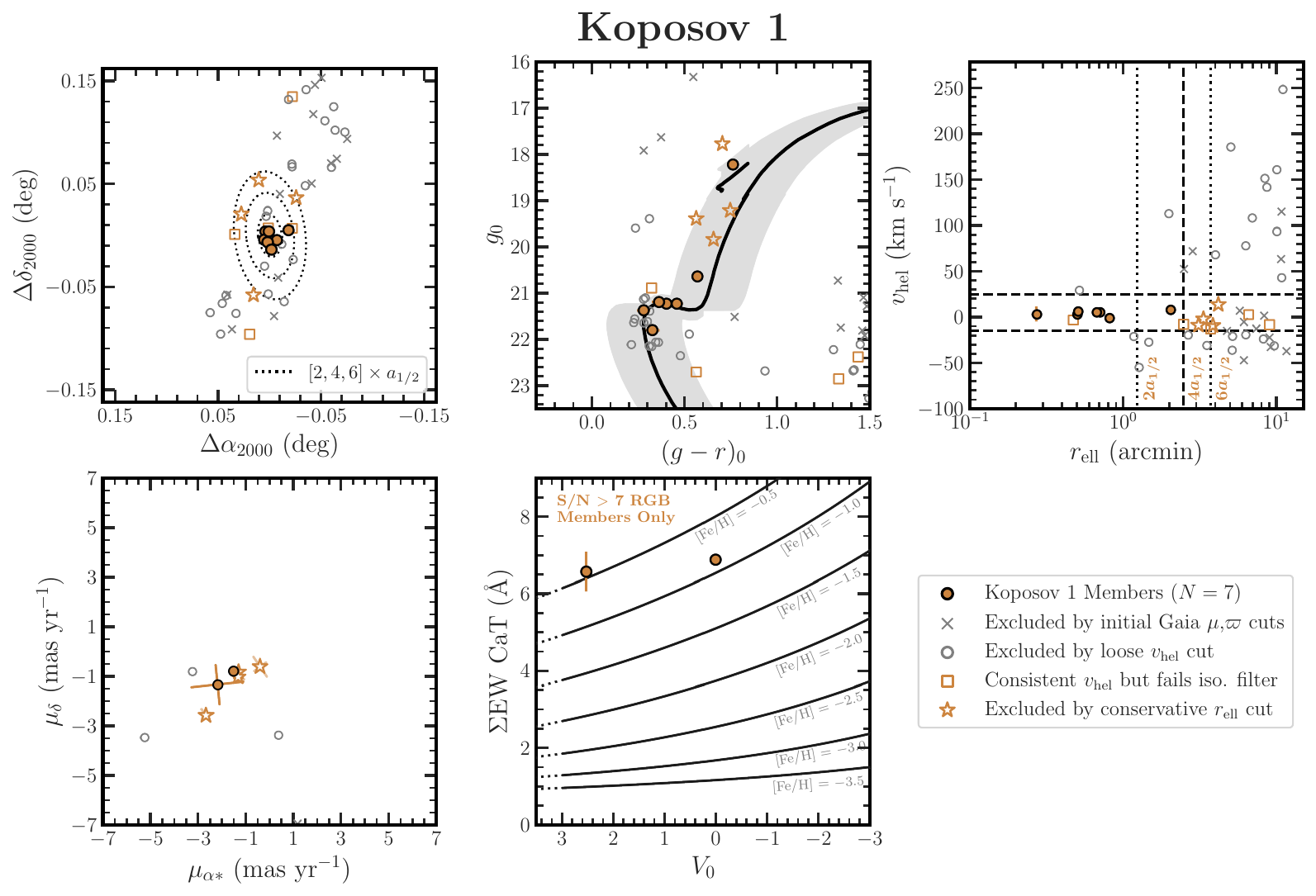}
    \caption{Member sample for Koposov~1. Members were selected based on a \Koposovoneage{}, \Koposovoneisofeh{} isochrone (parameters from \citealt{2014AJ....148...19P}; see \secref{membership}). We identify \KoposovoneNmem{} members within a conservative radial selection of $r_{\rm ell} < 4 \, a_{1/2}$, including two RGB stars and five subgiant or MSTO stars. Beyond this radius, there are several additional velocity-consistent member candidates; we currently exclude them because contamination from the Sagittarius dSph cannot be ruled out. \label{fig:koposov1}}
\end{figure*}

\begin{figure*}
    \centering
    \includegraphics[width= 0.85\textwidth]{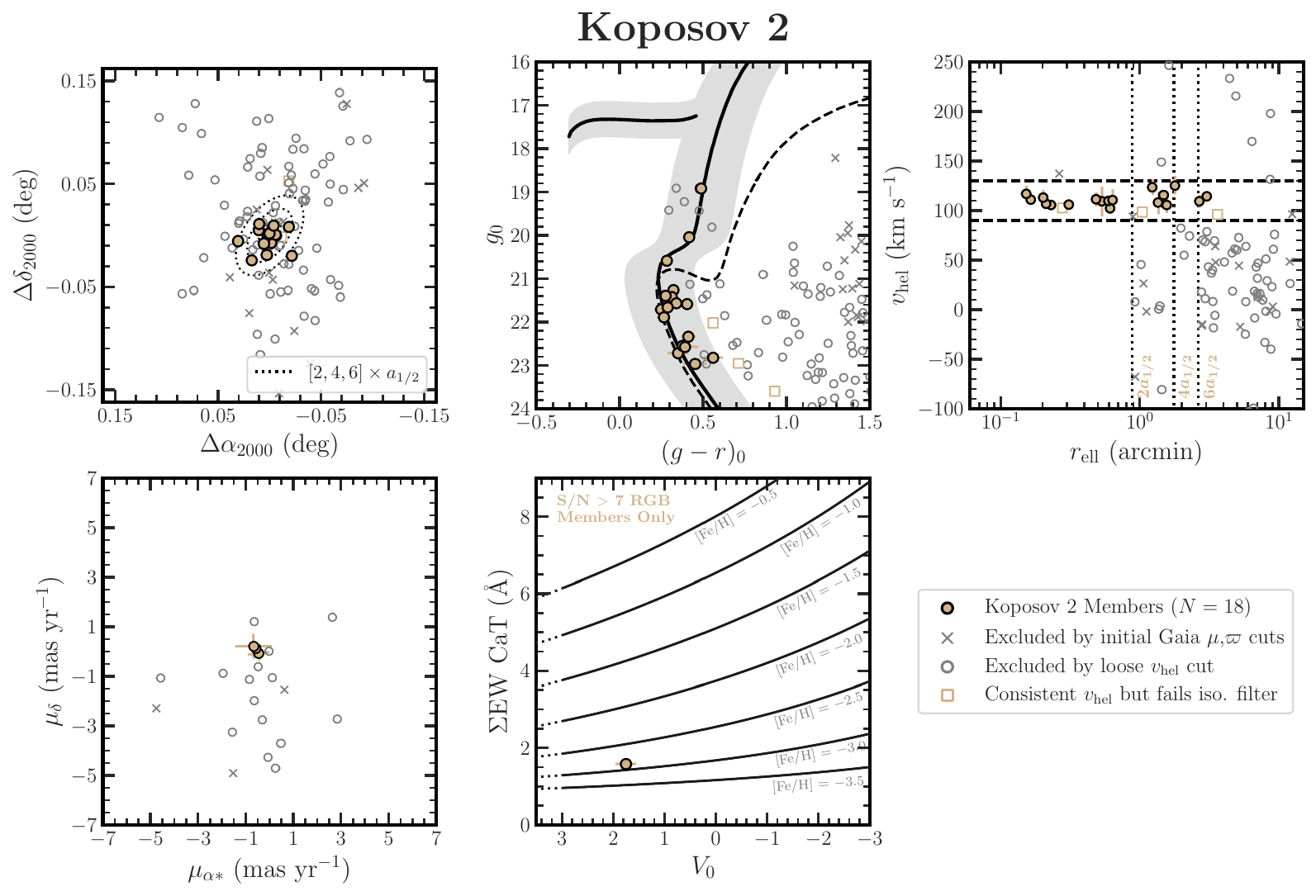}
    \caption{Member sample for Koposov~2. Members were selected based on our proposed isochrone model with \Koposovtwoisofeh{}, \Koposovtwoage{}, $(m-M)_0 = 16.9$. This is a significantly different model than the best-fit isochrone from \citet{2014AJ....148...19P}, shown as a black dashed line in the top-center panel. We identify \KoposovtwoNmem{} members assuming this revised model including two RGB stars, one subgiant, and 15 MS stars. \label{fig:koposov2}}
    
     \includegraphics[width= 0.85\textwidth]{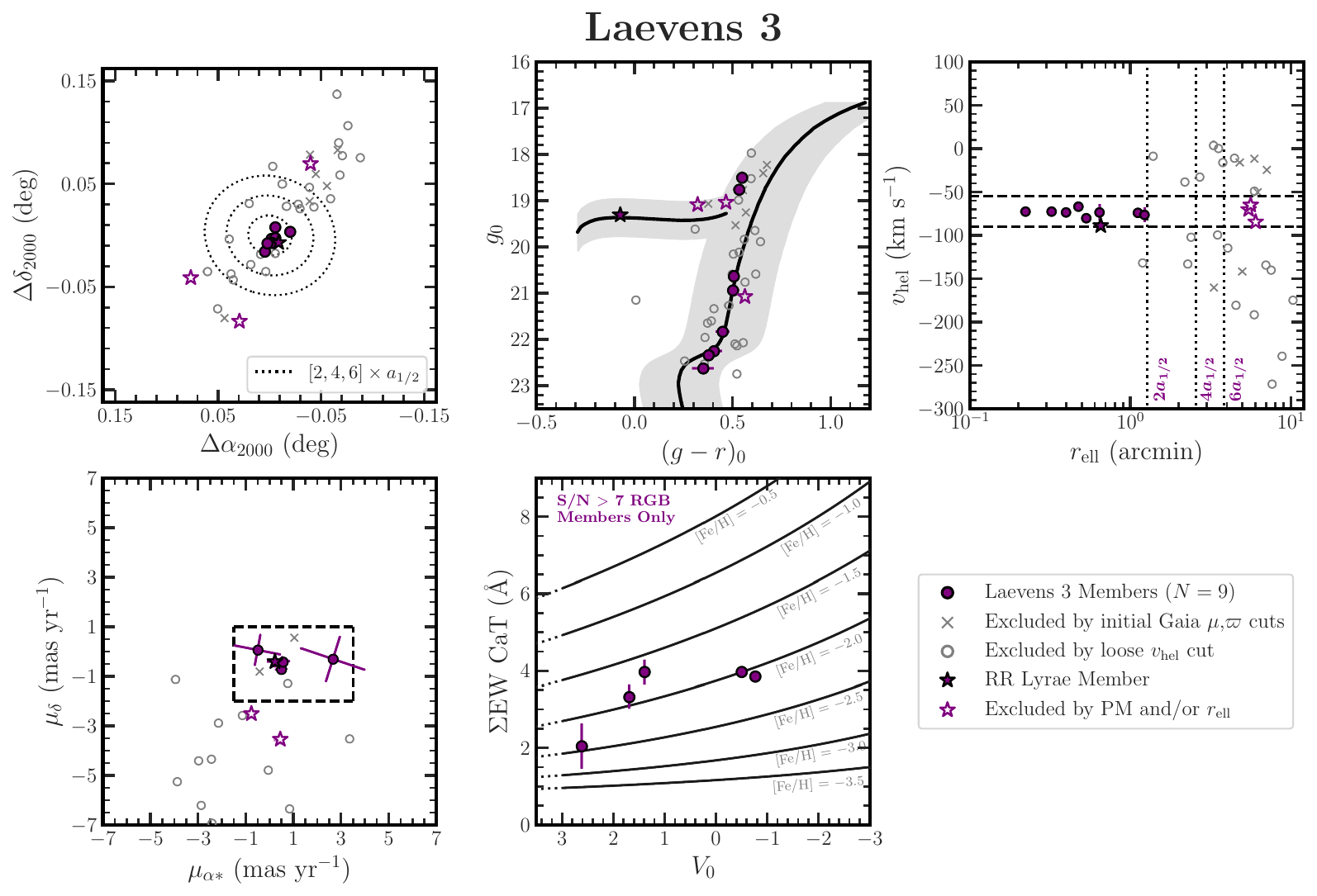}
    \caption{Member sample for Laevens~3. Members were selected based on a \Laevensthreeisofeh{}, \Laevensthreeage{} isochrone. We identify \LaevensthreeNmem{} members in total, including five RGB stars, three subgiants, and one RR Lyrae star.  The two brightest RGB stars appear unusually blue relative to the isochrone, likely due to crowding/blending in the dense central regions of Laevens~3. \label{fig:laevens3}} 
\end{figure*}

\begin{figure*}
    \centering
    \includegraphics[width= 0.85\textwidth]{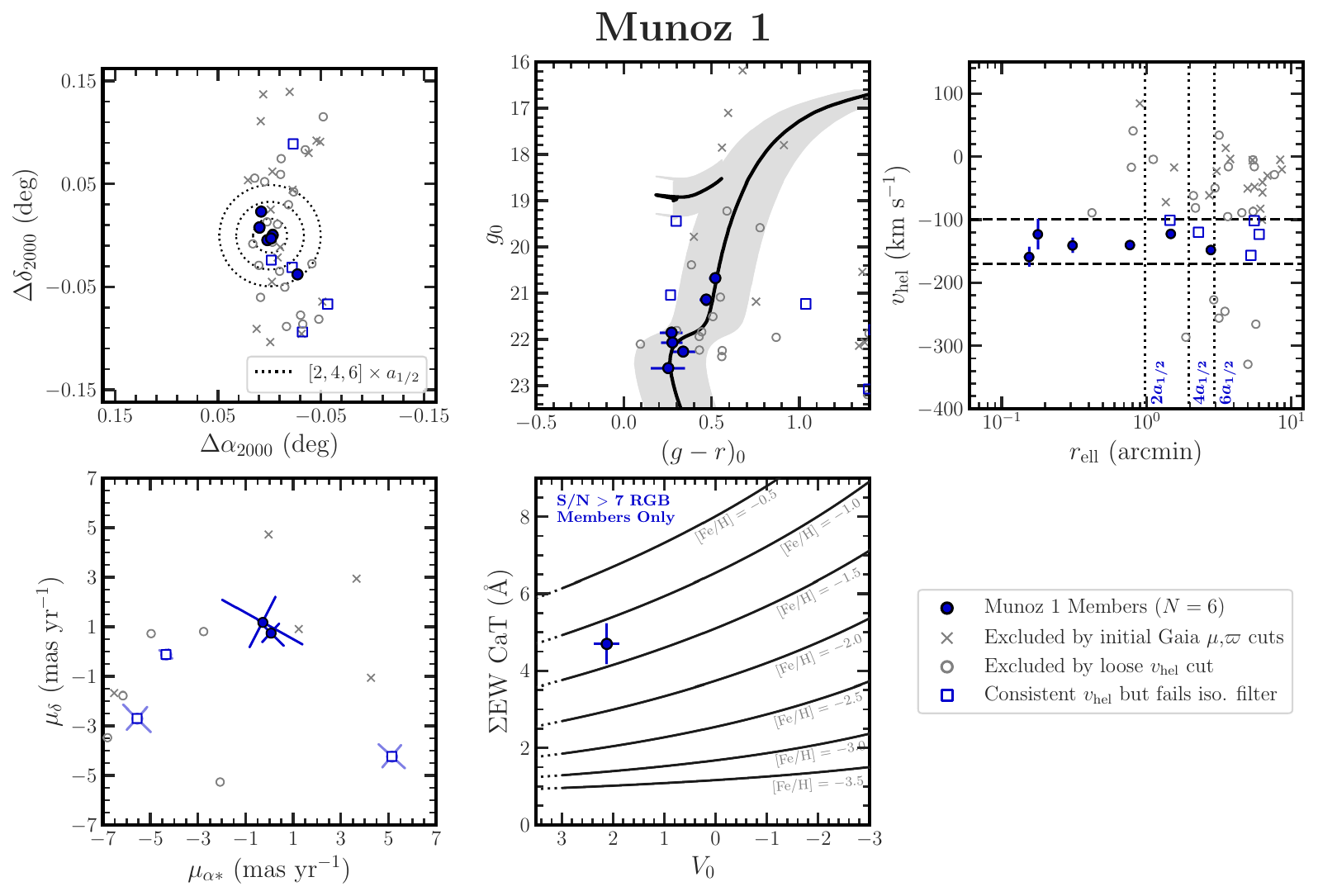}
    \caption{Member sample for Mu\~{n}oz~1. Members were selected based on a \Munozoneisofeh{}, \Munozoneage{} isochrone. We identify \MunozoneNmem{} total members including two RGB stars and four stars near the MSTO. However, due to the large velocity errors for the suspected members and the presence of contaminants at similar velocities, we caution that all membership for the system is tentative. \label{fig:munoz1}}
    
    \includegraphics[width= 0.85\textwidth]{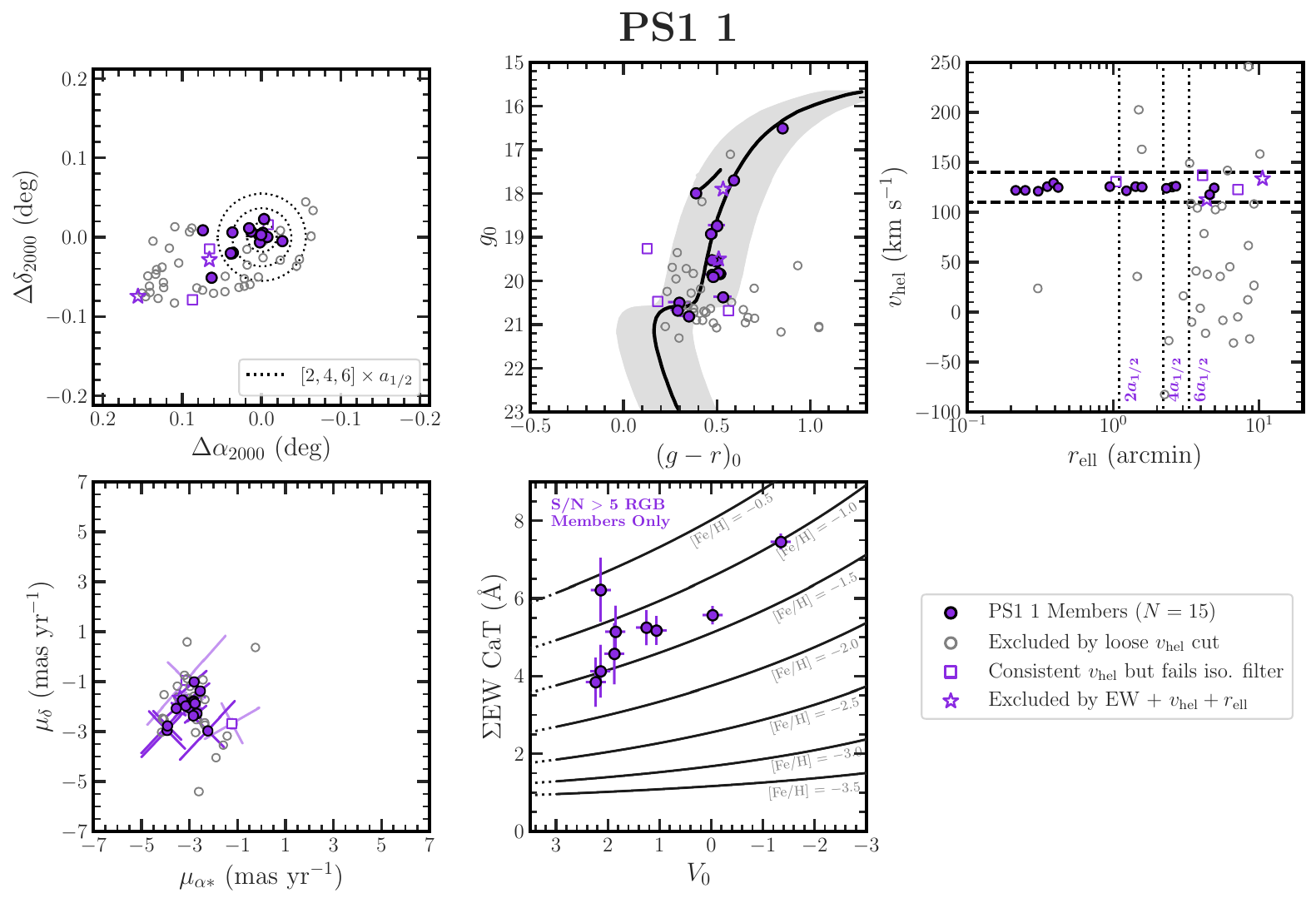}
    \caption{Member sample for PS1~1. Members were selected based on a \PSoneisofeh{}, \PSoneage{} isochrone. We identify \PSoneNmem{} likely members, all but three of which are RGB stars. \label{fig:ps11}}
   
\end{figure*}

\begin{figure*}
    \centering
    \includegraphics[width= 0.85\textwidth]{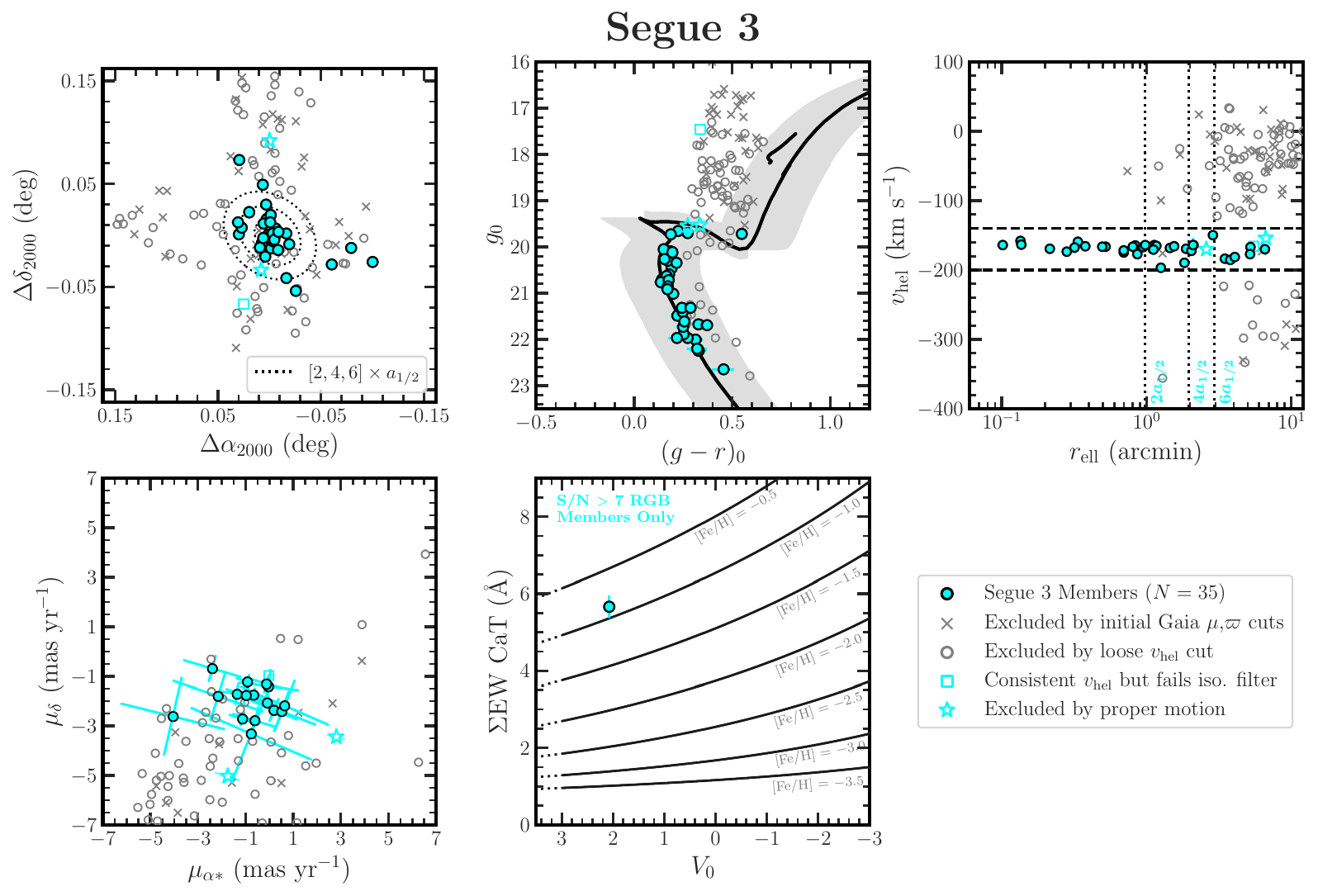}
    \caption{Member sample for Segue~3. Members were selected based on a \Seguethreeage{}, \Seguethreeisofeh{} isochrone. We identify \SeguethreeNmem{} total members, including one RGB star and 34 MS or MSTO stars. \label{fig:segue3}}
    \includegraphics[width= 0.85\textwidth]{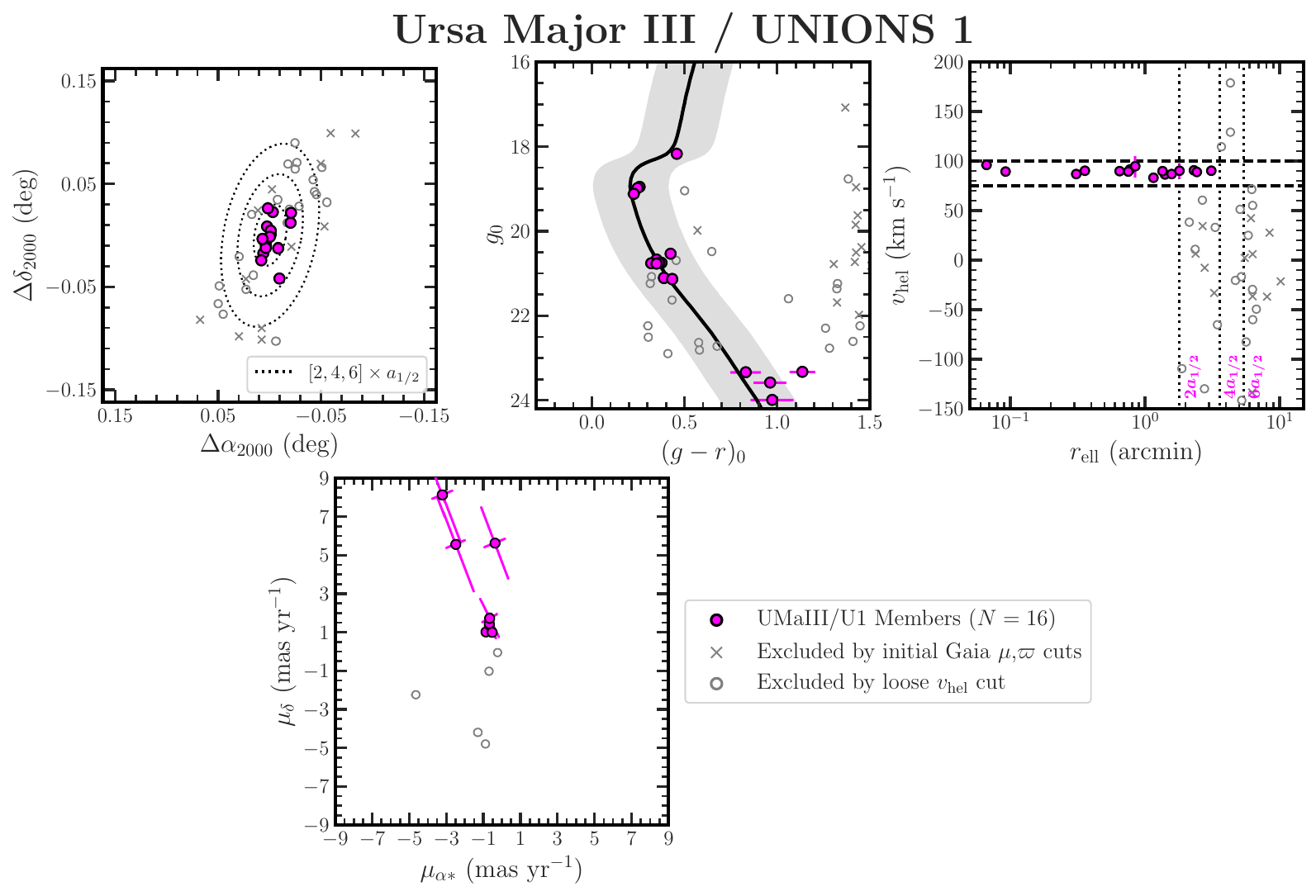}
    \caption{Member sample for \uma{}. We adopt the  \UrsaMajorIIINmem-star member sample from \citetalias{2025arXiv251002431C}~\citeyearpar{2025arXiv251002431C}, which includes 15 MS stars and one RGB star. We display the best-fit isochrone from \citet{2024ApJ...961...92S}  with \UrsaMajorIIIage{}, \UrsaMajorIIIisofeh{}. We note that one faint member falls outside our isochrone window, but we retain it  here following our analysis in \citetalias{2025arXiv251002431C}~\citeyearpar{2025arXiv251002431C}. \label{fig:unions1}}
    
\end{figure*}

\begin{figure*}
    \centering
    \includegraphics[width= 0.85\textwidth]{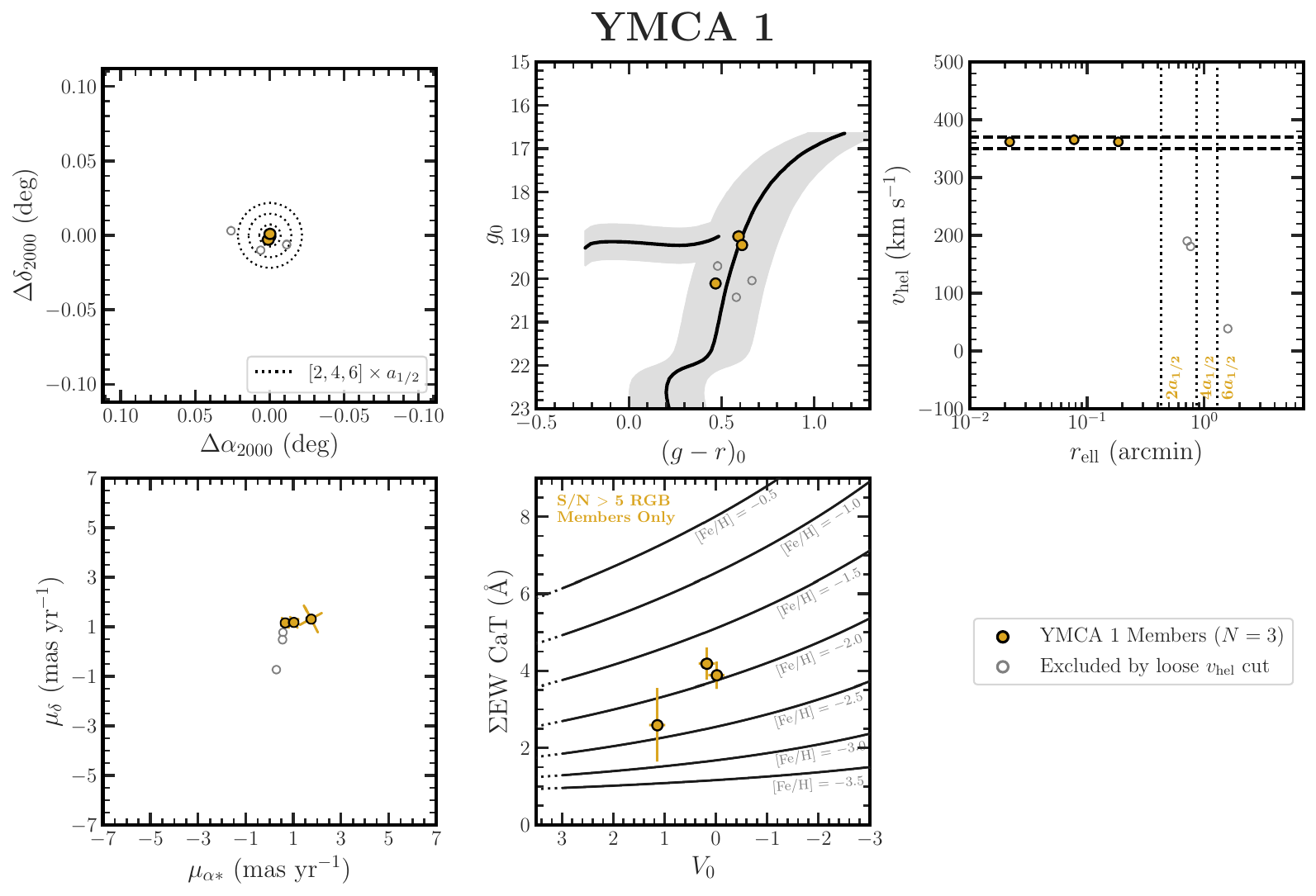}
    \caption{Member sample for YMCA-1. Members were selected based on a \YMCAoneage{}, \YMCAoneisofeh{} isochrone; this uses the age from \citet{2022ApJ...929L..21G} but a lower metallicity to better match the photometric colors and spectroscopic metallicities of the observed RGB members. We identify \YMCAoneNmem{} RGB members in total.  \label{fig:ymca1}}
    
\end{figure*}

  In the following subsection, we describe our membership choices for each UFCS -- including our handling of edge cases -- and summarize the resultant member samples. To accompany these descriptions, we visualize our membership selections in Figures \ref{fig:balbinot1}--\ref{fig:ymca1}. In these figures, we plot a multi-dimensional representation of the spatial, proper motion, velocity, metallicity, and photometric information available for our spectroscopic stellar targets. Specifically, for all UFCSs, these figures show: the projected spatial positions of all spectroscopic targets in a small region around each UFCS, with contours of $[2,4,6] \times a_{1/2}$ overlaid (top left), a CMD of these targets with the PARSEC isochrone used for filtering overplotted in black (top center), the heliocentric velocities $v_{\rm hel}$ for all stars vs. their projected elliptical radii $r_{\rm \rm ell}$ in arcminutes relative to the UFCS centroid (top right), and a \Gaia-based proper-motion vector-point diagram (bottom left). A final panel (bottom right), included only for UFCSs with at least one RGB member star, displays $\sum$EW$_{\rm CaT}$ vs. $V_0$ for confirmed RGB members with spectra passing our metallicity signal-to-noise cut ($S/N > 5$ for IMACS or $S/N > 7$ for DEIMOS). We overplot contours showing lines of constant metallicity based on the \citet{2026MNRAS.546ag019N} calibration. Across these membership panels, we show confirmed members passing all of our selections as filled, colored circles with black outlines. Velocity-consistent non-members are shown as unfilled, colored shapes (squares or stars; see legends for details). Non-members excluded only by their velocities are shown as unfilled grey circles, and non-members excluded by our initial astrometric and/or isochrone selections are shown as grey $\times$ symbols.

%%%%%%%%%% PAGE 1 of 9 %%%%%%%%%% 

%%%%%%%%%% PAGE 2 of 9 %%%%%%%%%% 

%%%%%%%%%% PAGE 3 of 9 %%%%%%%%%% 

%%%%%%%%%% PAGE 4 of 9 %%%%%%%%%% 

%%%%%%%%%% PAGE 5 of 9 %%%%%%%%%% 

%%%%%%%%%% PAGE 6 of 9 %%%%%%%%%% 

%%%%%%%%%% PAGE 7 of 9 %%%%%%%%%% 

%%%%%%%%%% PAGE 8 of 9 %%%%%%%%%% 

%%%%%%%%%% PAGE 9 of 9 %%%%%%%%%% 

\subsection{Description of Membership Selections for Individual Systems}
\label{sec:detailedmembership}
\textbf{Balbinot~1} (\figref{balbinot1}): For Balbinot~1, we adopted the best-fit isochrone age of \Balbinotoneage{} derived by \citet{2013ApJ...767..101B} based on deep CFHT/MegaCam imaging paired with the lowest metallicity in our PARSEC grid, \Balbinotoneisofeh{}. This is a lower metallicity than the $\rm [Fe/H] = -1.57$ model found by \citet{2013ApJ...767..101B} but is more consistent with our eventual spectroscopic mean metallicity for the system (\Balbinotonefehmodel{}; see \secref{metals}). After applying our initial astrometric cuts and a CMD selection based on this isochrone, we identified a clear excess of 10 stars with velocities $v_{\rm hel} \approx -175$~\kms, motivating us to adopt a loose velocity selection of $-195~\kms < v_{\rm hel} < -155$~\kms. To this sample of 10 stars, we added back one additional star (\Gaia DR3 2735227386765138688, denoted by a filled gold star in \figref{balbinot1}) that narrowly missed our initial parallax cut but has a consistent radial velocity and proper motion and lies at the very center of the system. A single star within our loose velocity selection  boundaries (\Gaia DR3 2735213466775943424; the unfilled gold square in \figref{balbinot1}) was excluded because of its discrepant proper motion, large projected radius from Balbinot~1's centroid, and, most importantly, its redder photometric color that placed it outside our isochrone window.
\par In summary, we identified a total of \BalbinotoneNmem{} members including four RGB stars, one BHB star, and six stars at the MSTO. Notably, this member sample extends out to an elliptical radius of $\sim$\Balbinotonemostdistant$\, a_{1/2}$, hinting at a spatially-extended stellar population for Balbinot~1. We discuss this star further in \secref{distantmembers}, where we quantitatively evaluate the expected density of interlopers passing our membership selection criteria. \newline

\textbf{BLISS~1} (\figref{bliss1}): Although our spectroscopic catalog for BLISS~1 is relatively sparse -- containing just 10 stars in total -- we identified a clear velocity signal comprising five stars in the narrow range $118~\kms < v_{\rm hel} < 124~\kms$, all of which fell close to the system centroid ($r_{\rm ell} \lesssim 3.5\, a_{1/2}$). No stars in our catalog fall within $\pm 15$~\kms{} of these five member candidates, making the stellar membership of BLISS~1 straightforward. That being said, we note that one of the five velocity-selected member candidates differs in proper motion with respect to the others at the $\sim$2.5$\sigma$~level. We maintain that this star is most likely a member despite this discrepancy and anticipate that more precise proper motions from \Gaia DR4 will confirm its membership.
\par In summary, we identify five total members of BLISS~1 including three RGB stars and two stars near the MSTO. The three brighter stars are clearly consistent with a redward-sloping RGB sequence characteristic of a more metal-rich and/or intermediate-age population, as first seen by \citet{2020ApJ...890..136M}. While \citet{2020ApJ...890..136M} reported an age $\tau = 9.6^{+1.6}_{-0.8}$~Gyr and metallicity $\rm [Fe/H]_{iso} = -1.4$,  we found an improved fit with a \BLISSoneage{}, \BLISSoneisofeh{} isochrone that is better matched to our mean spectroscopic metallicity for the system (\BLISSonefehmodel{}; see \secref{metals}). Both of these isochrones result in the same member sample when used for our CMD filtering, but we use our revised age estimate throughout this work and display this alternate model in \figref{bliss1}. \newline

\textbf{DELVE~1} (\figref{delve1}): DELVE~1 stands out as a clear phase-space overdensity of 10 stars near $v_{\rm hel} = -400~\kms$ that is separated in velocity by $>150~\kms$ from all other stars in our DEIMOS spectroscopic catalog. The stellar membership for DELVE~1 was therefore found to be unambiguous and can be reproduced with a simple selection of $v_{\rm hel} < -380~\kms$. 9 of the 10 members closely trace our adopted \DELVEoneage{}, \DELVEoneisofeh{} isochrone (assuming the recently-revised distance modulus of $(m-M)_0 = 16.68$ from \citealt{Simon2024}), while the tenth star (\Gaia DR3 4359353150735257984) is photometrically consistent with the expected locus of blue stragglers in our CMD.  This blue straggler candidate has a mildly outlying velocity relative to the remaining candidates that suggests it may be a binary system. This would be consistent with the hypothesis that some or all blue straggler stars in star clusters are the result of mass transfer during binary evolution \citep[e.g.,][]{1964MNRAS.128..147M,2011Natur.478..356G, 2024A&A...685A..33R}. We regard this star as a confident member but later excluded it from our kinematic analyses given its possible binarity.
\par In summary,  our final sample of DELVE~1 members includes 10 stars, including two RGB stars, seven subgiant/MSTO/MS stars, and one candidate blue straggler. Compared to the IMACS sample presented by our team in \citet{Simon2024}, we identified seven overlapping members in addition to three new faint members. The eighth IMACS member was observed with DEIMOS but did not yield a reliable velocity. A joint, multi-epoch analysis of the IMACS and DEIMOS datasets will be presented at a later date, but here we note that there is tentative evidence for radial velocity variation for one star, \Gaia DR3 4359353155032064256, across these two instruments' datasets. Specifically, we observed a velocity difference of $\Delta v_{\rm hel} \approx 13~\kms$ between the DEIMOS and IMACS velocity measurements taken $\sim$325 days apart, which can be compared to the per-epoch uncertainties of $\sim$4~\kms. We flagged this member star as a possible binary and excluded it from our kinematic analyses, as with the blue straggler above. Our kinematic sample for the system therefore includes eight stars.  
\par Lastly, we highlight that one of our 10 members in DELVE~1 is located at a projected radius of $r_{\rm ell} = \DELVEonemostdistant{}\,a_{1/2}$ (as also seen in the \citealt{Simon2024} sample). Given the large separation in velocity between DELVE~1 and the empirical MW foreground star velocity distribution, this star is highly likely to be a member (see \secref{distantmembers} for more discussion). \newline

\textbf{DELVE~3} (\figref{delve3}): We identified five stars within $r_{\rm ell} < 1\arcmin$ of DELVE~3's centroid that are all consistent with an old, metal-poor isochrone (\DELVEthreeage{}, \DELVEthreeisofeh{}; set based on the upper/lower limits from \citealt{2023ApJ...953....1C}) and share velocities $-105~\kms < v_{\rm hel} < -65~\kms$. The brightest two of these stars exhibit self-consistent \Gaia proper motions. We regard these five stars as secure members of DELVE~3. One additional member candidate (\Gaia DR3 6445898050394010240; depicted as an unfilled green star in \figref{delve3}) fell just within our wide velocity selection ($v_{\rm hel} \approx -69~\kms$). While this star is consistent with the RGB of our best-fit isochrone, its large projected radius ($r_{\rm ell} = 8.3\,a_{1/2}$) and discrepant proper motion strongly suggest it is a non-member and thus we excluded it.
\par Our final member sample for DELVE~3 therefore includes \DELVEthreeNmem{} stars, including four RGB stars and one BHB star. \newline

\textbf{DELVE~4} (\figref{delve4}): DELVE~4 presents the most ambiguous membership of any UFCS in our sample. In our DEIMOS spectroscopic catalog, we identified a highly significant clustering of 10 stars at velocities $-180~\kms < v_{\rm hel} < -160~\kms$ that all trace a \DELVEfourage{}, \DELVEfourisofeh{} isochrone (including eight on the RGB).  Interestingly, though, these 10 stars display a large dispersion in CaT EWs and three fall at large radii ($r_{\rm ell} > 4\, a_{1/2}$). In addition, the raw number of RGB stars appears to be in tension with the number expected from the system's stellar mass (as was noticed originally in the course of the photometric analysis presented in \citealt{2023ApJ...953....1C}). The stars bright enough to have \Gaia DR3 proper motion measurements fall within a relatively narrow selection box defined by $-0.5 \ \rm mas \ yr^{-1} < \mu_{\alpha *} < 1.5 \ \rm mas \ yr^{-1}$, $-2 \ \rm mas \ yr^{-1} < \mu_{\delta} < 0.5 \ \rm mas \ yr^{-1}$, with two exceptions that we identified as non-members because of their discrepant $\mu_{\alpha *}$ relative to the remaining set of velocity-filtered candidates. After removing these two stars (both of which fell near the isochrone RGB), the velocities of the eight remaining member candidates were found to be separated by at least $13$~\kms{} from all other stars in our catalog.
\par The key question for the interpretation of DELVE~4's properties is whether all eight remaining member candidates (six RGB and two MSTO stars) are true members, and, by extension, whether the large implied metallicity spread among the RGB stars is a real feature of the system. After extensive exploration of the multi-dimensional data available, we found no clear evidence for internal correlations that would suggest only a subset being members. Specifically, we reiterate that the six RGB stars exhibit self-consistent \Gaia proper motions and their radial velocities span a narrow range of $<10$~\kms{}. To assess whether these RGB candidates could be misidentified foreground MS stars, we used \texttt{dmost} to fit a Gaussian profile to the gravity-sensitive Mg I $8807\rm \; \AA$ line in our DEIMOS spectra. This demonstrated that the EWs for all six of these RGB candidates are below the dividing line between dwarf stars and giants suggested by \citet{2012A&A...539A.123B}, i.e., $\rm EW_{\rm MgI} < 0.3 \rm \AA$; thus, their status as RGB stars appears secure. No metallicity or proper motion information is available for the faintest two stars (at the MSTO), though their velocities alone provide strong support for their membership. We therefore consider all eight of these stars to be likely members of DELVE~4.
\par Despite all of this evidence, we remain cautious about the possibility of membership contamination for two reasons. First, three of the four innermost RGB member candidates ($r_{\rm ell} < 2 \, a_{1/2}$) have relatively higher metallicities ([Fe/H] $\approx -1.5$); this scenario plausibly could arise if DELVE~4 is a monometallic system with contamination at larger radii. Second, the presence of the two proper motion outliers at similar radial velocities implies that contamination by foreground/background halo substructures is a real risk in the DELVE~4 field. In an attempt to identify possible contaminating halo structures that could contribute interloper stars at similar velocities to DELVE~4, we searched the \texttt{galstreams} package \citep{2023MNRAS.520.5225M} for stellar streams that spatially overlapped with DELVE~4's position.  This revealed that DELVE~4 is co-located in projection with the Wukong/LMS-1 dwarf galaxy stellar stream \citep{2020ApJ...898L..37Y,2020ApJ...901...48N} and the tidal tails of the globular cluster M3, the latter of which is the same structure as the independently-reported stellar stream Svol \citep{Yang_2023}. Both of these structures are well in the foreground of DELVE~4 ($D_{\odot} < 20$~kpc vs. $D_{\odot} = 45 \pm 4$ kpc; \citealt{2023ApJ...953....1C}) but nonetheless could not be immediately rejected as contaminating sources because the distances of individual stars are not known.  Considering the most-recently available LMS-1 stream track from \citet{2024ApJ...967...89I} and the M3 track from \citet{Yang_2023}, we found that the former structure is a poor match in radial velocity at DELVE~4's position while the latter is offset by just $\sim$20~\kms{} in radial velocity but is clearly distinct in both proper motion components. These discrepancies suggest that the putative DELVE~4 member stars that we identified are unlikely to originate from these two spatially-superimposed streams. 

\par With no obvious reason to suspect contamination,  we proceeded under the assumption that our DELVE~4 member sample is pure for the remainder of this work.  In summary, then, we identified a total of eight member stars including six RGB members and two MSTO members. One of the more distant RGB members in this sample (\Gaia DR3 1270955387816956928) varied in velocity by $\sim$24~\kms{} between our two epochs of observation, and thus we flagged this star as a binary and excluded it from our kinematic analyses. \newline 

\textbf{DELVE~5} (\figref{delve5}): DELVE~5 is extraordinarily dim and low mass ($M_V = +0.4$; $M_* = 160 \rm \ M_\odot$), somewhat distant ($D_{\odot} = 39$~kpc),  and likely contains no RGB stars, making it a challenging target for multi-object spectroscopy even with Keck. In our DEIMOS catalog, we identified five stars within $r_{\rm ell} < 4\arcmin$ with velocities $-250~\kms < v_{\rm hel} < -200~\kms$. Four of these stars are faint and consistent with being at the main-sequence turnoff (MSTO) of DELVE~5 based on the \DELVEfiveage{}, \DELVEfiveisofeh{}  best-fit isochrone from \citet{2023ApJ...953....1C}. Three of these four MSTO candidates are clearly members based on their statistically-indistinguishable velocities of $v_{\rm hel} \approx -233$~\kms, while the fourth MSTO candidate is more ambiguous because of its mildly discrepant velocity ($v_{\rm hel} = -212.0 \pm 9.2$~\kms). The fifth and final velocity-consistent member candidate (\Gaia DR3 1236270945623558912) lies near the BHB of our adopted isochrone, but its velocity ($v_{\rm hel} = -214.2 \pm 1.4$~\kms) is offset at a highly-significant level from the mean defined by the three confident MSTO members.  The dearth of stars in our spectroscopic catalog with $v_{\rm hel} < -150$~\kms{} suggests that interlopers at DELVE~5's velocity are uncommon, supporting the case that these latter two stars are members. However, explaining their significant velocity offsets would likely necessitate that both are binaries. To be conservative, we treated both of these stars as non-members for the remainder of this work, though this should be revisited once multi-epoch velocity observations become available.
\par Our final sample of DELVE~5 members is therefore limited to three secure members at the MSTO. None of these three stars have \Gaia proper motions or metallicity measurements from our spectroscopy, making DELVE~5 the least well-characterized UFCS in our sample. \newline

\textbf{DELVE~6} (\figref{delve6}): The slitmask for our IMACS observations of DELVE~6 contained just five science targets. Three of these stars were identified as high-probability photometric members based on the joint spatial and CMD fit from \citet{delve6_discovery} and yielded usable radial velocity measurements with IMACS. Our spectra for the remaining two stars did not yield reliable velocities ($S/N < 3$). Of the three photometric member candidates with useful velocities, we found that two (the BHB star and the fainter RGB star in \figref{delve6}) have the same velocity within uncertainties; we therefore consider them secure members despite the relatively poor velocity precision. The third candidate -- the brighter RGB star in \figref{delve6} --  has a radial velocity that is mildly offset from the other two stars (by $\sim$2.9$\sigma$  and $\sim$$1.8\sigma$), though all other properties of this star strongly favor membership: it falls at $<2 \, a_{1/2}$, lies exactly on the predicted isochrone RGB, and has a proper motion consistent with the BHB star at the $<1.5\, \sigma$ level. We consider this star a likely member as well and speculate that its velocity offset may result from binary motion. 
\par Our final member sample for DELVE~6 comprises three stars in total: two RGB stars and one BHB star. Notably, the secure confirmation of the BHB star breaks the age-metallicity degeneracy encountered during the isochrone fitting in \citealt{delve6_discovery} (see their Fig. 3),  favoring an ancient age for the system. Given the particularly low signal-to-noise of our observations of these three stars ($3 < S/N < 6$), though, it would be worthwhile to obtain deeper observations to solidify these membership determinations. \newline

\textbf{Draco~II} (\figref{dracoII}): As was the case with DELVE~1, Draco~II's highly negative radial velocity separates it from the vast majority of MW foreground stars ($v_{\rm hel} \ll -300$~\kms{}).  With a loose selection of $-365~\kms{} < v_{\rm hel} < -320~\kms{}$, we isolated a sample of 29 stars with no additional stars within $\pm 15$~\kms{}. All but two of these stars were consistent with an old, metal-poor isochrone (\DracoIIage{}, \DracoIIisofeh{}). The two exceptions included one clear nonmember falling significantly redder than our isochrone selection as well as one candidate blue straggler star; we excluded the former but retained the latter. Lastly, within the remaining 28 stars, we identified the star \Gaia DR3 1640413452483584640 as an edge case falling $10$~\kms{} below all other members, at the largest projected radius in the velocity-selected sample ($r_{\rm ell} \approx 2\,a_{1/2}$), and with a proper motion that deviated from the remaining candidates at the 2$\sigma$~level. We retained this star as a member given the relative dearth of contaminants at similar velocities and the reasonable proper motion consistency; however, given its mildly outlying velocity, we flagged it as a suspected binary and excluded it from our kinematic analyses.

\par Including the blue straggler and the edge-case member, our final sample for Draco~II comprises \DracoIINmem{} stars in total, nearly all of which lie along the MS. Compared to the prior analysis by \citet{2018MNRAS.480.2609L}, our new sample nearly doubles the sample of spectroscopic members in Draco~II thanks to the addition of new DEIMOS observations with distinct slitmasks. We recover all the previously identified members from that work. \newline

\textbf{Eridanus~III} (\figref{eridanusIII}): For Eridanus~III, we directly adopted the member sample from \citet{Simon2024}, which was determined based on the same set of IMACS velocity measurements presented in this work.  In total, this sample includes eight members: six RGB stars and two BHB stars. Several stars have second-epoch velocity measurements but none display radial velocity variations suggestive of binarity. In \figref{eridanusIII}, we display the \citet{Simon2024} velocity selection of $35~\kms{} < v_{\rm hel} < 70~\kms{}$ and overplot a \EridanusIIIage{}, \EridanusIIIisofeh{} isochrone for visual reference. \newline

\par \textbf{Kim~1} (\figref{kim1}): In our DEIMOS catalog for Kim~1, we identified 14 stars within the wide velocity range $-280~
\kms{} < v_{\rm hel} < -240~\kms{}$ that all pass our CMD selection based on a \Kimoneage{}, \Kimoneisofeh{} isochrone. This is an older and more metal-poor model compared to the isochrone parameters fit in the discovery paper by \citet{2015ApJ...799...73K}; however, it is more consistent with the photometric color and spectroscopic metallicity of the system's brightest RGB star, which we later found to be very metal-poor (\Kimonefehbrightest{}; see \secref{metals}).
\par Of the 14 stars in this velocity window, we found that 12 were clear members based on their velocities, colors, and spatial positions; the 5 of these 12 stars observed by \Gaia further display self-consistent proper motions. The two remaining stars were more ambiguous: the first is significantly redder than the isochrone track and is a mild velocity outlier, while the second is an extreme spatial outlier falling at $r_{\rm ell} \approx \Kimonemostdistant{}\,a_{1/2}$. These stars are shown  in \figref{kim1} as an unfilled red star and a filled red star, respectively. Given the very tight color-magnitude sequence followed by the 12 clear members, we flagged the former star as a suspected non-member even though it formally falls within our isochrone window. On the other hand, we found no reason to reject the more distant star and consider it a tentative member worthy of further follow-up.
\par Our final member sample therefore comprises 13 total stars including a single RGB star and 12 stars at the MSTO or on the MS. While the status of the most distant member remains somewhat uncertain (see \secref{distantmembers}), the inclusion of this star does not significantly impact our ultimate constraints on Kim~1's properties. \newline

\par \textbf{Kim~3}  (\figref{kim3}): Our IMACS catalog for Kim~3 features a velocity peak at $v_{\rm hel} \approx 165$~\kms{} comprised of nine stars, with no other stars within $\pm 30$~\kms{}. All nine of these stars are closely consistent with an old, metal-poor isochrone (\Kimthreeage{}, \Kimthreeisofeh{}). The only ambiguity is whether the faintest star in this sample (\Gaia DR3 6181853910690372096) is a member. This star lies at $r_{\rm ell} \approx \Kimthreemostdistant{}\, a_{1/2}$, while the remaining eight lie at $r_{\rm ell} < 2.5\,a_{1/2}$. This star's velocity is indistinguishable from the remaining eight, but no proper motion or parallax information is available.  Because there was no clear basis to reject this star, we retained it in our sample but consider its membership tentative. A future proper motion measurement would easily clarify this star's membership given the other members'
strong proper motion clustering in \figref{kim3}.
\par Our final member sample for Kim~3 therefore consists of nine total members, including eight secure members and one tentative distant member. The eight secure members include one subgiant and seven MSTO or MS stars; the distant member candidate is a MS star. \newline

\textbf{Koposov~1}  (\figref{koposov1}): Koposov~1 lies in a field that is densely populated with stars from the Sagittarius dSph, making CMD and phase-space contamination a significant issue for both isochrone fitting and identifying spectroscopic members. At the time of discovery,  \citet{2007ApJ...669..337K} found that Koposov~1 is comprised of a somewhat old, metal-poor stellar population ($\tau = 8\rm ~Gyr; \rm [Fe/H]_{\rm iso}  = -2.0$) at $\sim$50~kpc.  However, a more recent analysis by \citealt{2014AJ....148...19P} has instead suggested the system is considerably more metal-rich (\Koposovoneisofeh{}), significantly closer  ($D_{\odot} = 34$~kpc), and of a slightly younger age ($\tau = 7$~Gyr). Through iterative velocity and isochrone selections, we found that the model from \citet{2014AJ....148...19P} is a much better fit to our spectroscopic member candidates, matching the photometry of the numerous stars with velocities $-15~\kms{} < v_{\rm hel} < 25~\kms{}$. 
\par Applying a velocity based on this window and an isochrone selection based on the \citet{2014AJ....148...19P} parameters, we identified a total of 11 plausible members, including six within $r_{\rm ell} < 2a_{1/2}$, one at $\sim 4\,a_{1/2}$, and the remaining four beyond this threshold. The innermost six stars are almost certainly members, and the seventh is a likely member but less clearly so. However, the four remaining stars at $> 4\,a_{1/2}$ exhibit a larger dispersion in velocities and their colors have larger scatter about the best-fit isochrone. These features lead us to believe the four outermost velocity-consistent candidates are non-members, and thus we applied a radial separation criterion of $r_{\rm ell} < 4\,a_{1/2}$ to preserve the purity of our sample. To be maximally conservative, we later performed our kinematics fits over only the six innermost members within $<2\,a_{1/2}$, excluding the seventh, farthest likely member at $\sim$$4\,a_{1/2}$.
\par After these selections, our final sample for Koposov~1 consisted of seven members including two RGB stars, three subgiants, and two stars at the MSTO. We view this sample as both tentative and conservative; re-analyses with improved photometric constraints on the best-fit isochrone age and distance (and/or deeper spectroscopy) may warrant revisions to this member sample. \newline

\textbf{Koposov~2} (\figref{koposov2}): 
Like Koposov~1, Koposov~2's stellar population properties have been debated in the literature. Initially, \citet{2007ApJ...669..337K} found that Koposov~2's CMD was best fit by a $\tau = 8$~Gyr, $\rm [Fe/H]_{\rm iso} = -2.0$ isochrone at a distance $D_{\odot} \approx 40$~kpc. Based on a deeper photometric dataset, \citet{2014AJ....148...19P} revised  these estimates to a younger age $\tau \approx 4$--6~Gyr, higher metallicity $\rm [Fe/H]_{\rm iso} = -0.6$, and closer distance $D_{\odot} = 33.5\, \pm 1.5$~kpc.  This significant disagreement likely reflects the challenge of isochrone-fitting amidst contamination from Sagittarius (though, unlike Koposov~1, we note that Koposov~2 is \textit{not} likely to be physically associated with Sagittarius; see Paper II).
\par While we initially explored using the \citet{2014AJ....148...19P} parameters, we discovered two anomalies that motivated us to investigate alternative models. Assuming an initial velocity selection of $90~\kms <v_{\rm hel} < 130~\kms$ and our astrometric cuts -- but no isochrone selection -- we isolated a sample of 21 stars encompassing all plausible Koposov~2 members in our DEIMOS spectroscopic catalog. While $\sim$18 of these stars would be selected by an isochrone filter based on the \citet{2014AJ....148...19P} parameters, 
the three brightest stars -- which conspicuously follow an RGB-like sequence -- were missed by this filter despite their consistent radial velocities, self-consistent, halo-like proper motions, and close-in positions (two at $r_{\rm ell} < 1\arcmin$). Notably, these three stars were all identified as $P > 0.99$ \Gaia proper-motion members based on the mixture-model analysis of \citet{2021MNRAS.505.5978V}. In addition, we noticed that neither the best-fitting model from \citet{2007ApJ...669..337K}  nor that from \citet{2014AJ....148...19P} was a good match to Koposov~2's MS in color--magnitude space. Using our kinematically-selected sample, we found that both of these anomalies (i.e., the RGB candidates and the MS slope) could be better matched by an ancient, very-metal-poor isochrone model (\Koposovtwoage{}, \Koposovtwoisofeh{}) so long as a $\sim$40\% smaller distance was assumed: $D_{\odot}  \approx 24$~kpc (equivalently, a distance modulus of $(m-M)_0 = 16.90$). The color of the MS track remains an imperfect match under this revised model, and three velocity-consistent candidate MS members that  fall redder than our filter remain excluded. We speculate this color mismatch could reflect either a local color offset in the DECam photometric calibration, the need to explore different levels of $\alpha$-element enhancement compared to our solar-scaled PARSEC isochrones, or even a significant population of equal-mass MS binaries. 
\par A clear prediction of this older, closer, more metal-poor picture of Koposov~2 is that the two brightest stars should have spectra resembling metal-poor giants. This is indeed the case for the brightest star: we determined a metallicity of \Koposovtwofehbrightest{} (see \secref{metals} and the bottom-center panel of \figref{koposov2}). As for its surface gravity, we found that the EW of the Mg I 8807~$\rm \AA$ line for the brightest star was consistent with the locus of giants based on the \citet{2012A&A...539A.123B} criterion (see above description of DELVE~4 for more details). This consistency strengthens the case for Koposov~2 as an old, metal-poor stellar system. 
\par The last important check was to consider whether these stars could be interlopers from the Sagittarius dSph. Comparing to the star particles from the Sagittarius simulations carried out by \citet{2021MNRAS.501.2279V}, we found that the only debris expected within a $\sim 3\deg^2$ of Koposov~2 is at $\gtrsim 80$~kpc. A small ($< 25\%$) fraction of this debris does overlap Koposov~2 in radial velocity; however, we find that the candidate Koposov~2 RGB 
members are also systematically offset in proper motion from this predicted debris. Lastly, we note that stars at [Fe/H] $< -2.5$ in Sagittarius are very rare \citep[see e.g., ][for context]{2017A&A...605A..46M, 2018ApJ...855...83H,2020ApJ...901..164C,2024ApJ...963...95C}, and finding such a metal-poor Sagittarius interloper star within an arcminute of Koposov~2's centroid would be highly unlikely. The combination of all of these factors leads us to believe that these stars are \textit{bona fide} RGB members of Koposov~2,  validating our revised choice of isochrone model. 
\par As with DELVE~4, we chose to carry out the remainder of our analysis under the assumption that our Koposov~2 member sample is accurate and free from contamination, even if this may need to be revisited in the future. Our final sample therefore includes \KoposovtwoNmem{} member stars, all but the brightest two of which fall on the MS. We urge significant caution in the interpretation of Koposov~2's properties until its age and distance can be better established. \newline

\textbf{Laevens~3} (\figref{laevens3}): After our astrometric cuts and an initial velocity selection of $-90~\kms{} < v_{\rm hel} < -55~\kms{}$ (bracketing the mean velocity from \citealt{2019MNRAS.490.1498L}), we obtained a sample of 12 stars including seven with \Gaia proper motions. Two of these seven stars lie at large elliptical radii ($r_{\rm ell} > 6\, a_{1/2}$) and have precise proper motion measurements that significantly diverge from the remaining five. To remove them, we imposed a loose proper motion selection of $-1.5 \rm \ mas \ yr^{-1} < \mu_{\alpha *} < 3.5 \rm \ mas \ yr^{-1}$ and  $-2  \rm \ mas \ yr^{-1}< \mu_{\delta} < 1 \rm \ mas \ yr^{-1}$. We then excluded an additional star, \Gaia DR3 1760242455925564672, because it lies at a very large radius ($r_{\rm ell} \approx 8.7\, a_{1/2}$) and exhibits much larger CaT and Mg I EWs than the other candidate members.
\par These selections resulted in a final sample of \LaevensthreeNmem{} Laevens~3 members, including five RGB stars, three subgiants, and one HB star. The HB star (\Gaia DR3 1760245891900130688) is a known RR Lyrae member of Laevens~3. Its velocity appears as an outlier in our DEIMOS data, likely due to the photospheric velocity shifts associated with its radial pulsations and the lack of dedicated treatment of these stars within our \texttt{dmost} velocity measurement procedure. Consequently, we excluded this variable star for our kinematic analyses.
\par Our sample of nine stars is larger than that of \citet{2019MNRAS.490.1498L} who reported six members and one tentative candidate from their own reduction of the same DEIMOS spectra. Both studies agree that these seven stars are plausible members. The two additional spectroscopic members that we identified include the RR Lyrae variable, which they identify as a photometric member but flag as a spectroscopic non-member in their Table 2. The second -- the lowest signal-to-noise star in our sample ($S/N  = 3.5$/pixel) -- does not appear at all in \citet{2019MNRAS.490.1498L}'s catalog. We speculate that this omission arises from their cut at $S/N = 3$ and the differences in reduction. \newline

\textbf{Mu\~{n}oz~1} (\figref{munoz1}): Our re-reduction of the DEIMOS data obtained by \citet{2012ApJ...753L..15M} 
revealed six isochrone-consistent stars falling in the wide velocity window $-170~\kms < v_{\rm hel} < -100~\kms$, bracketing the published systemic velocity for Mu\~{n}oz~1. Two of these stars are plausible members but fail our nominal DEIMOS velocity error quality cut of $\epsilon_v < 15~\kms{}$ and thus do not factor further into our  analysis. Of the four stars with more trustworthy measurements, two brighter RGB candidates have $S/N > 10$ while the other two have $2 < S/N < 3$. The brighter two stars share similar proper motions and radial velocities; we regard these two stars as confident members. One of the two lower-$S/N$ stars has a similar radial velocity to this pair, while the other is different by $\sim$3--4$\sigma$. We opt to include both these faint stars as tentative members, but the latter star in particular should be re-evaluated when higher-precision observations become available. We note that we do not report a velocity dispersion for Mu\~{n}oz~1 because the sample of available kinematic members is too small (see \secref{vdispconstraints}), nor do we report a metallicity for either faint member, and thus these two low-$S/N$ members have negligible impact on our interpretation of Mu\~{n}oz~1's properties.
\par In short, our final sample for Mu\~{n}oz~1 consists of two RGB stars with good $S/N$ and four faint member candidates at the MSTO with poor $S/N$. We were unable to directly compare this sample with the published sample from \citet{2012ApJ...753L..15M} because they do not individually report members, but we do note that both studies identify a RGB member at $g< 21$ that we strongly suspect is the same star. \newline

\par \textbf{PS1~1} (\figref{ps11}):  In the field of PS1~1, we identified a clear excess of 20 stars in the velocity range $110~\kms{} <  v_{\rm hel} < 140$~\kms{} including 17 that both passed our astrometric selections and were consistent with a \PSoneage{}, \PSoneisofeh{} isochrone.  Of these 17 plausible members, 13 fell within $r_{\rm ell} < 3\, a_{1/2}$ and shared closely consistent proper motions; we view these 13 stars as confident members. The remaining four candidates fell at larger radii ($r_{\rm ell} > 4\, a_{1/2}$) and warranted additional scrutiny. The most distant star among these four (\Gaia DR3 6759859300740397056, at $\sim$$18.9\,a_{1/2}$) exhibited both a larger CaT EW than the inner members as well as a statistically-significant velocity offset ($v_{\rm hel} = 133.8 \pm 1.0$~\kms{} vs. $v_{\rm hel} \approx 124$~\kms{} for the confident members); these properties led us to reject this star as a clear non-member. Among the remaining three, the star \Gaia DR3 6759863462562275328 was again found to be an outlier in both velocity and CaT EW; we excluded this star on the same grounds as the more distant star, albeit more tentatively. Lastly, the final two stars (each at $\sim$$8\,a_{1/2}$)  shared velocities consistent within $2\,\sigma$ of the mean velocity of the innermost members, and we found no reason to doubt their membership. 
\par In summary, we identified a total of \PSoneNmem{} members of PS1~1. 11 of these stars clearly fall along the RGB,  while three fainter members fall on the subgiant branch. The final star lies close to the isochrone model red clump / red horizontal branch. While our sample of spectroscopic members conspicuously extends as far out as $\sim 8 \, a_{1/2}$, we caution that the basic structural properties of PS1~1 are among the least well-characterized across the UFCSs in our sample. This is because the PS1 imaging available at the time of its discovery was relatively shallow.  \newline

\textbf{Segue~3} (\figref{segue3}):  Although the discovery analysis of Segue~3 by \citet{2011AJ....142...88F} favored an old, metal-poor stellar population for the system, more recent isochrone-fitting analyses by \citet{2013MNRAS.433.1966O} and \citet{2017AJ....154...57H} have persuasively established that the system is more likely to be a relatively-younger, metal-rich system in the outer halo. We therefore adopted the isochrone parameters from the latter work, namely an age \Seguethreeage{} and metallicity \Seguethreeisofeh{}. Pairing this isochrone filter with our astrometric cuts and a wide velocity selection of $-200~\kms < v_{\rm hel} < -140~\kms$, we identified an initial sample of 37 candidate members with no other stars falling within $20~\kms{}$ of these boundaries. We subsequently excluded two stars with discrepant proper motions (shown as unfilled stars in \figref{segue3}).  
\par The resultant sample of 35 stars represents a maximally-complete set of all possible Segue~3 members in our spectroscopic catalog -- albeit not necessarily a pure sample. Although there was no direct photometric or astrometric evidence to support excluding any additional stars, the implied velocity dispersion for this complete 35-star sample would be a high $\sigma_v \approx  9~\kms{}$ (after removing sources with velocity variation across epochs) -- well above the expectation for the central velocity dispersions of faint star clusters or UFDs. As seen in \figref{segue3}, this is primarily due to the increased scatter in putative member star velocities at large radii, which plausibly could reflect either contamination or the tidal disruption of the system.  Following a similar approach to the original DEIMOS study of Segue~3 by \citet{2011AJ....142...88F}, we addressed this concern for our kinematic analyses by conservatively placing a radial cut of $r_{\rm ell} < 3\,a_{1/2}$. This yielded a final kinematic member sample of \SeguethreeNkin{} stars.
\par In total, our final sample of secure Segue~3 members includes \SeguethreeNmem{} stars, including a single RGB star and 34 MS stars. The RGB star holds significant weight in our analysis because it enables us to report the first spectroscopic metallicity for Segue~3 (see \secref{metals}). We note that this star (\Gaia DR3 1785664195553222272) was previously excluded from membership by \citet{2011AJ....142...88F} because of their use of an old, metal-poor isochrone model. However, it was later re-identified by \citet{2017AJ....154...57H} as a faint giant/subgiant member (log($g$) $\approx 3.5$) based on spectral-energy-distribution fitting to Washington $CT_1$ and SDSS photometry. The combination of the surface gravity estimate from \citet{2017AJ....154...57H}, our radial velocity measurement, and the Gaia DR3 proper motion of this star provides strong support for its classification as an RGB member of Segue~3. \newline

\textbf{\uma{}} (\figref{unions1}):  For \uma{}, we directly adopted the member sample constructed by \citetalias{2025arXiv251002431C}~\citeyearpar{2025arXiv251002431C} based on two epochs of Keck/DEIMOS spectroscopy. This sample, totaling 16 stars, includes all 11 original members reported in the discovery analysis by \citealt{2024ApJ...961...92S} in addition to 5 (generally fainter) new members. As shown here in \figref{unions1}, these 16 stars all fall in the  velocity range $75~\kms < v_{\rm hel} < 100~\kms$ and trace a \UrsaMajorIIIage{}, \UrsaMajorIIIisofeh{} isochrone. One faint member star (\texttt{S24\_M9}) nominally falls outside our isochrone filter, but we retain it here following \citetalias{2025arXiv251002431C}~\citeyearpar{2025arXiv251002431C}. \newline

\textbf{YMCA-1} (\figref{ymca1}): Similarly to our observations of DELVE~6, we designed a sparse slitmask covering just six primary targets in the field of YMCA-1.  These stellar targets were carefully chosen to be high-probability members based on LS DR10 photometry and \Gaia DR3 proper motions  (where available).  Our IMACS spectra confirm the three most promising candidates from this set of six, which share velocities of $v_{\rm hel} \approx 361~\kms{}$ and are all centrally positioned ($r_{\rm ell} < 1\arcmin$). The other three stars have velocities separated by $\gg$100~\kms{} and thus are clear non-members. 
\par Our final member sample for YMCA-1 therefore consists of three RGB stars. We note that matching the photometry of these stars required a more metal-poor isochrone (\YMCAoneisofeh{}) compared to that quoted by \citet{2022ApJ...929L..21G}, who found $\rm [Fe/H]_{\rm iso} \approx -1.12^{+0.21}_{-0.13}$. While the choice of this alternate isochrone does not affect our spectroscopic member selection due to our wide color tolerance, this  metal-poor model is more consistent with our measured spectroscopic metallicity for the system (\YMCAonefehmodel{}; see \secref{metals}).

\subsection{Summary of Membership Selections}
\label{sec:memsummary}
\par In total, the selections described above yielded \UFCSNmemTotal{} spectroscopic member stars across our sample of 19 UFCSs. The number of member stars per UFCS ranges from just 3 stars (DELVE~5, DELVE~6, and YMCA-1) to \SeguethreeNmem{} stars (Segue~3), with a median of 9 per system. \UFCSNsystemsFivePlus{} (\UFCSNsystemsTenPlus) UFCSs in our sample have at least 5 (10) members. Of these \UFCSNmemTotal{} members, \UFCSNGaiaTotal{} stars have complete five-parameter astrometric solutions from \Gaia (i.e., positions, parallaxes, and proper motions). \UFCSNfehTotal{} of the  \UFCSNmemTotal{} members are RGB stars for which we were able to measure CaT metallicities. We release catalogs summarizing the properties of all  \UFCSNmemTotal{} member stars in a Zenodo repository associated with this work; see \secref{data_avail}.

\par While our membership selections are bespoke and rely on hard cuts, we found that most UFCSs' member samples could be defined relatively unambiguously  -- often thanks to a large velocity separation between the UFCSs and the bulk of the MW foreground star population (see e.g., DELVE~1 and Draco~II, both at $v_{\rm hel} < -300$~\kms{}) and due to careful spectroscopic target selection. The few UFCSs in our sample that posed the most significant challenges were those superimposed in projection over known halo substructures -- for example, Koposov~1 and Koposov~2, which each overlap the Sagittarius dSph and for which even prior isochrone-fitting analyses have struggled to converge to consensus age, metallicity, and distance estimates. Ultimately, as the quality of the photometric, astrometric, and spectroscopic data available for the UFCSs improves, it would be worthwhile to explore more sophisticated probabilistic membership modeling techniques that jointly model the chemokinematics of both the targeted UFCS in each field and the empirical interloper population \citep[see e.g.,][]{2011ApJ...738...55M,2017ApJ...839...20C,2020MNRAS.495.3022P,2026ApJ...998...47S}. These approaches are challenging to apply to sparse datasets like those presented here, but offer the benefit of capturing the impact of membership uncertainty on the downstream analyses that follow. In lieu of applying these techniques, we take care to acknowledge and explore the impact of critical membership edge cases in our kinematic and chemical analyses in the following sections. \pagebreak

\section{Stellar Kinematics of the UFCS\lowercase{s}}
\label{sec:vdispconstraints}
\subsection{Mean Velocity and Velocity Dispersion Fits}
We determined the systemic mean velocity ($v_{\rm sys}$) and radial velocity dispersion ($\sigma_{v}$) of each UFCS in our sample through a two-parameter Bayesian fit. The velocity distribution of each UFCS was modeled as a Gaussian with a spread constituted by a component associated with measurement errors and an intrinsic dispersion component. The corresponding log-likelihood for this model was 
\begin{align}
 \ln \mathcal{L} = -\frac{1}{2} \sum_{i=1}^N \ln \left(2\pi(\epsilon_{v,i}^2+\sigma_{v}^2)\right)-\frac{1}{2} \sum_{i=1}^N \frac{\left(v_{\rm hel, i}- v_{\rm sys}\right)^2}{\left(\epsilon_{v,i}^2+\sigma_v^2\right)} 
\end{align}
\citep[e.g.,][]{2006AJ....131.2114W} where $\epsilon_{v,i}$ refers to the velocity uncertainty of the $i$th star, $v_{\rm hel,i}$. This likelihood neglects any radial dependence and ordered velocity structure such as rotation or tidally-induced gradients; we assume these signatures are undetectable given our small member samples and current velocity precision. By default, these fits were performed over the velocity distribution of all member stars that we selected for each UFCS minus any stars flagged as possible binaries, RR Lyrae, or blue stragglers. The three exceptions to this were Koposov~1 and Segue~3, for which we imposed radial selections of $r_{\rm ell} < 2 \, a_{1/2}$ and $r_{\rm ell} < 3\,a_{1/2}$, respectively (see preceding section), and \uma{}, for which we adopted the 14-star ``Complete Epoch 2'' sample from \citetalias{2025arXiv251002431C}~\citeyearpar{2025arXiv251002431C} directly.

\par To derive posteriors for the two model parameters and simultaneously evaluate the Bayesian evidence (hereafter $\rm \ln\mathcal{Z}_{\sigma}$) of the data given our Gaussian model, we performed dynamic nested sampling \citep{2004AIPC..735..395S,10.1214/06-BA127,2019S&C....29..891H} using the \texttt{dynesty} package \citep{2020MNRAS.493.3132S,sergey_koposov_2024_12537467}. We adopted flat priors of $-500~\kms < v_{\rm sys} < 500~\kms$ and $0~\kms < \sigma_v< 10~\kms$; this dispersion prior was chosen to yield conservative upper limits. We used uniform sampling with 2048 live points and sampling weights set to yield a 65/35 evidence/posterior split, imposing that each sampling run yield a minimum of 30,000 effective posterior samples. This nested sampling was then repeated for an alternative one-parameter model where $\sigma_v$ was fixed to zero, i.e., no intrinsic spread, yielding a separate Bayesian evidence $\ln\mathcal{Z}_{\sigma=0}$.
\par These fits frequently yielded posteriors with modes and/or tails bounded against the lower prior bound of $\sigma_v = 0$~\kms{}. To rigorously decide whether to quote resolved dispersion estimates or upper limits in these cases,  we computed logarithmic Bayes factors from the difference in the \texttt{dynesty} evidence estimates between the free-dispersion and zero-dispersion models, 
\begin{align}
     2\ln\beta \equiv  2(\Delta \ln\mathcal{Z}) = \rm 2(\ln\mathcal{Z}_\sigma - \ln\mathcal{Z}_{\sigma=0}).
\end{align} Under this construction, positive Bayes factors $(2\ln\beta > 0)$ imply a preference for the free-dispersion, two-parameter model (i.e., a resolved velocity dispersion) while negative Bayes factors $(2\ln\beta < 0)$ imply the data are better described by a zero-dispersion model that does not include $\sigma_v$ as a free parameter (i.e., an unresolved velocity dispersion).\footnote{We stress that negative Bayes factors  \textit{do not} demonstrate that the true dispersion is zero; instead, they indicate that the data provide insufficient evidence to favor a model in which the dispersion is included as a free parameter.} Larger values of $|2\ln\beta|$ imply a stronger preference for one model over the other: adapting the qualitative scale from \citet{kassrafferty} to our application here,  Bayes factors $2\ln\beta > 10$ would suggest a ``very strong'' preference for the free-dispersion model, $6 <  2\ln\beta < 10$  would suggest a ``strong'' preference for the free-dispersion model,  $2 < 2\ln\beta < 6$ would suggest a ``positive'' preference for the free-dispersion model,  and finally, $0 < 2\ln\beta  < 2$ would suggest a preference for the free-dispersion model that is ``not worth more than a bare mention''. On the other hand, $2\ln\beta < -10$ would suggest a ``very strong'' preference for the zero-dispersion model, $-10 < 2\ln\beta < -6$ would suggest a ``strong'' preference for the zero-dispersion model, and so on.
\par Based on these Bayesian model comparisons, we adopted a threshold of $2\ln\beta = 0$ below which we quote upper limits at the 95\% credible level (for $2\ln\beta < 0$) and above which we derived resolved credible intervals from the median and 16th/84th percentiles of the posterior  ($2\ln\beta > 0$). We note that this threshold for quoting resolved dispersions is permissive: cases with $0 < 2\ln\beta < 2$ correspond to only weak evidence for a resolved dispersion.  We refer to this subset of cases as ``marginally-resolved'' and interpret them cautiously throughout this work.

\subsection{Results: Velocity Dispersions Constraints for Individual UFCSs}
\begin{figure*}
    \centering
    \vspace{0.4em}
    \includegraphics[width=0.76\textwidth]{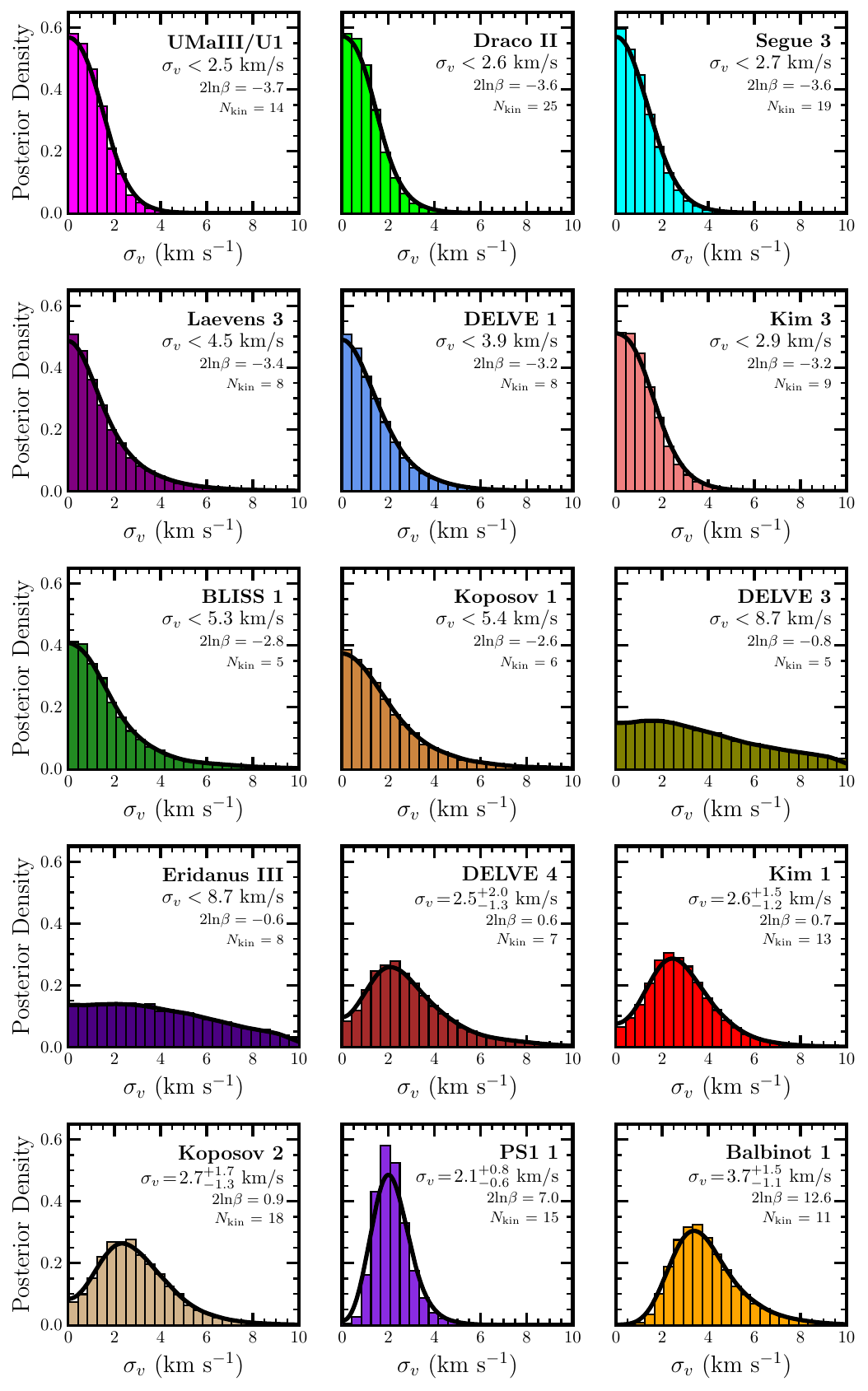}
    \caption{\textbf{Marginalized posterior probability distributions for the UFCS' velocity dispersions}, ordered by increasing Bayes factor from our comparison against zero-spread models (from least resolved to most resolved). We show the posteriors from our \texttt{dynesty} sampling for all UFCSs with $\geq 5$ kinematic members; we do not report dispersions for systems with fewer than five kinematic members. In each panel, we summarize the marginalized posterior with either a 95\% credible upper limit (if $2\ln\beta < 0$) or the median and lower/upper errors derived from the 16th/84th percentile (if $2\ln\beta > 0$). We also overplot a kernel density estimate (Epanechnikov kernel, bandwidth 1~\kms{}, with reflection at zero) as a solid black line. }
    \label{fig:ridgeplot}
\end{figure*}

Our complete set of constraints on the stellar kinematics of the UFCSs is presented in \tabref{dynamicstable}. In total, we were able to meaningfully constrain both the mean velocity and velocity dispersion for the 15 of our 19 UFCSs with $N_{\rm kin} \geq 5$ kinematic members.  The resulting posterior probability distributions for $\sigma_v$ (marginalized over $v_{\rm sys}$) for these 15 cases are shown as 1D histograms in \figref{ridgeplot}. For the four remaining UFCSs with samples of $<$~5 kinematic members, we performed the identical two-parameter Bayesian fits; however, in these cases we treated the dispersion as a nuisance parameter and we report only the derived systemic mean velocities $v_{\rm sys}$. 
\par The outcomes of the 15 well-constrained fits can be split into three broad categories based on Bayes factors: fits where the zero-dispersion model was clearly favored ($2\ln\beta < -2$), i.e., unresolved dispersions, those where the default free-dispersion model was clearly favored ($2\ln\beta > 2$), i.e., resolved dispersions, and lastly, cases where neither model was clearly favored ($-2 \lesssim  2\ln\beta \lesssim 2$). More than half of our sample fell in the first category: we found eight UFCSs (Segue~3, \uma{}, Kim~3, Laevens~3, Draco~II, DELVE~1, BLISS~1, and Koposov~1) with negative Bayes factors ranging from \UrsaMajorIIIvdispbayes{} to \BLISSonevdispbayes{}. As shown in the first three rows of \figref{ridgeplot}, the posteriors for these systems pile up against the lower prior boundary at $\sigma_v =0$~\kms{}, indicating that the data provide no significant evidence for a nonzero dispersion. Using the rough approximation that the frequentist $z-$score obeys $z \approx \sqrt{2\ln\beta}$,  these Bayes factors translate to a $\sim 1.5$--2$\sigma$ preference for the zero-dispersion model over the model with $\sigma_v$ left free. In all eight of these cases, we quote upper limits at the 95\% credible level; these range from a relatively stringent \UrsaMajorIIIvdisp{} and \DracoIIvdisp{} (for \uma{} and Draco~II, respectively) to a comparatively weaker \BLISSonevdisp{}  and \Koposovonevdisp{} (for BLISS~1 and Koposov~1, respectively).  
\par In the second category, we found exactly two UFCSs for which the free-dispersion model was clearly favored: Balbinot~1 (with \Balbinotonevdispbayes{}) and PS1~1 (with \PSonevdispbayes{}). For these two cases, we report resolved dispersions of \Balbinotonevdisp{} and \PSonevdisp{} from the median and 16/84th percentiles of the marginalized posteriors (though we note these uncertainties are significantly non-Gaussian). Both of these estimates considerably exceed these systems' predicted velocity dispersions if they are comprised entirely of stars, \Balbinotonestellarvdisp{} and \PSonestellarvdisp{}; see \secref{masses} for details of these predictions.
\par Lastly, in the third category, we found five UFCSs with Bayes factors in the intermediate range ($-2 \lesssim 2\ln \beta \lesssim  2$) for which there was no clear preference for one velocity dispersion model or the other. The fits for two of these five systems (DELVE~3 and  Eridanus~III) yielded mildly negative Bayes factors (\DELVEthreevdispbayes{} and \EridanusIIIvdispbayes{}) that lead us to report 95\% upper limits of \DELVEthreevdisp{} and \EridanusIIIvdisp{}, respectively. These two limits are weak primarily due to the poor  signal-to-noise and velocity precision of member stars in our IMACS datasets for these UFCSs. On the other hand, our nominal fits to the other three UFCSs in this category (DELVE~4, Kim~1, and Koposov~2) yielded marginally positive Bayes factors of \DELVEfourvdispbayes{}, \Kimonevdispbayes{}, and \Koposovtwovdispbayes{}, respectively,  with corresponding dispersions of \DELVEfourvdisp{}, \Kimonevdisp{} and \Koposovtwovdisp{}. The statistical significances of these three dispersions are sufficiently small that they can be considered inconclusive. Moreover, as we demonstrate in the subsubsection below, these three cases -- as well as our more clearly resolved dispersions for Balbinot~1 and PS1~1 -- should be interpreted cautiously given the strong impact of individual member stars.

\subsubsection{Robustness to Velocity Outliers}
\label{sec:veloutliers}
\par Velocity dispersion estimates derived from small member samples run a significant risk of being biased by outliers in the form of unidentified spectroscopic binaries and/or interloping non-members. In general, binaries and interlopers are expected to inflate velocity dispersion estimates, which would lead us to estimate larger dynamical masses and potentially even erroneously identify the presence of dark matter even in cases when it is not present \citep[e.g.,][]{2010ApJ...722L.209M,2022ApJ...939....3P,2023ApJ...956...91W,2025MNRAS.543.1120G}.  As vividly illustrated by the case of \uma{} (\citetalias{2025arXiv251002431C}~\citeyear{2025arXiv251002431C}), binaries are a particularly acute risk for the UFCSs, where the intrinsic dispersions are expected to be $\mathcal{O}$(0.1--0.5)~\kms{} if they are clusters (see \secref{masses}) and $\mathcal{O}$(1--2)~\kms{} if they are galaxies \citep{2019ARA&A..57..375S,2025arXiv251115808A}. These predicted intrinsic dispersions can be compared with the small number of well-characterized binaries in UFDs,  which have semi-amplitudes as large as $\sim$10--20~\kms{} \citep{2014ApJ...780...91K,2022MNRAS.514.1706B,2024ApJ...968...21H}. Multi-epoch measurements are only available for a handful of our UFCS targets and thus our dispersion measurements presented above are certainly subject to this risk.
\par To evaluate the impact of possible unidentified binaries or interlopers on our dispersion constraints, we carried out a series of jackknife tests. For each UFCS, we iterated through our kinematic member sample, removing a single star at a time and carrying out the identical velocity dispersion fitting procedure described above. This included running the fits with the intrinsic dispersion fixed to zero, enabling us to compute a velocity dispersion Bayes factor associated with each member subsample. Our particular focus was on evaluating whether the two UFCSs with clearly resolved dispersions (Balbinot~1 and PS1~1) as well as the three UFCSs with marginally-resolved dispersions  (DELVE~4, Kim~1, and Koposov~2) would become upper limits ($2\ln\beta < 0$) if a single star was excluded from their kinematic member samples.
\par Importantly, we found that all five of these UFCSs' dispersions were indeed highly sensitive to a single star. Most significantly, we found that PS1~1's clearly-resolved dispersion of \PSonevdisp{} (with \PSonevdispbayes{}) entirely vanished when the star \PSoneJackknifeSourceID{} was removed, yielding a strong upper limit of \PSonevdispJackknife{} (with \PSonevdispbayesJackknife). Similarly, Kim~1's nominal marginally-resolved dispersion of \Kimonevdisp{} (with \Kimonevdispbayes) dropped to become a strong upper limit of \KimonevdispJackknife{} (with \KimonevdispbayesJackknife) when the star \KimoneJackknifeSourceID{} was removed from our kinematic sample. The remaining cases were less dramatic: for Koposov~2, we found that a single star in the wings of the velocity distribution with a small velocity uncertainty drove the resolved dispersion. Removing it shifted the posterior toward zero, yielding a weak upper limit \KoposovtwovdispJackknife{} (with \KoposovtwovdispbayesJackknife{}) that reflects a significant tail to large dispersions. Likewise, removing one star for DELVE~4 shifted our nominal resolved estimate of \DELVEfourvdisp{} to an upper limit of \DELVEfourvdispJackknife{}  (with \DELVEfourvdispbayesJackknife), though in this case the very small sample size  ($N =\DELVEfourNkin{}$ kinematic members) makes these jackknife tests harder to interpret. For the fifth and final UFCS, Balbinot~1,  the removal of its most outlying star reduced our nominal clearly-resolved dispersion of \Balbinotonevdisp{} (with \Balbinotonevdispbayes) to a smaller and only marginally significant dispersion of \BalbinotonevdispJackknife{} (with \BalbinotonevdispbayesJackknife). 
We present diagnostic figures associated with these jackknife tests in Appendix \ref{sec:jackknifediagnostics}, \figref{jackknifeposteriors}.

\par Collectively, these jackknife tests emphasize that the five resolved and marginally-resolved dispersions are tentative at best.  Without additional data -- particularly multi-epoch velocity monitoring -- it is not possible to assess whether the dispersions derived from these jackknife tests are more accurate constraints than our nominal results reported in \tabref{dynamicstable}. In the absence of a concrete reason to remove any of the stars isolated above, we continue to report the original results derived from our complete kinematic member samples. Nonetheless, given the  general trend that the initially-reported velocity dispersions of faint MW satellites are commonly revised \textit{downward} when revisited with expanded member samples and/or multi-epoch observations \citep{2011ApJ...743..167B,2013ApJ...770...16K,2017ApJ...838...83K,2021MNRAS.503.2754L,2022MNRAS.514.1706B,2023ApJ...950..167B,geha_paper1}, we view it as reasonably likely that the jackknife estimates will prove to be closer approximations to the true velocity dispersions of these UFCSs.

\subsubsection{Dynamical Masses and Mass-to-Light Ratios}
\label{sec:masses}
Leveraging our set of 15 new velocity dispersion estimates, we computed dynamical masses within the half-light radius ($M_{1/2}$) for the UFCSs in our sample 
with the mass estimator from \citet{2010MNRAS.406.1220W}, 
\begin{align}
 M_{1/2} = 930 \ M_{\odot} \left(\frac{\sigma_v}{\kms}\right)^2 \left(\frac{r_{1/2}}{\rm pc}\right). 
 \label{eq:wolfmass}
\end{align}
The posteriors on $\sigma_v$ and $r_{1/2}$ were directly propagated into the equation above to yield full posterior distribution on $M_{1/2}$; we then computed either upper limits at the 95\% credible level (for UFCSs with unresolved velocity dispersions) or the median and 16th/84th percentiles (for those with resolved dispersions) from this $M_{1/2}$ posterior.  Mass-to-light ratios were subsequently derived by dividing our $M_{1/2}$ posterior by our posterior for $L_{V,1/2} \equiv L_{V}/2$. Lastly, to contextualize our $\sigma_v$ estimates, we computed the expected  velocity dispersions of the UFCSs if they are comprised only of stars, $\sigma_*$, by inverting Equation \ref{eq:wolfmass}, i.e., $\sigma_* \equiv \sqrt{M_*/(930\,r_{1/2})}$.
\par The resulting estimates of $M_{1/2}$, $M_{1/2}/L_{V,1/2}$, and $\sigma_*$ are presented in \tabref{dynamicstable}. We emphasize that these estimates each assume that the UFCSs are in dynamical equilibrium -- a requirement for the applicability of the \citet{2010MNRAS.406.1220W} estimator. The degree to which this assumption applies will depend on the UFCSs' densities, and, relatedly, whether they inhabit dark matter halos. This remains an open question, but we remark that there is already tentative evidence for disequilibrium in a subset of UFCSs in our sample in the form of spatially-extended stellar populations (see \secref{distantmembers}).

\begin{figure*}
    \centering
    \includegraphics[width=\textwidth]{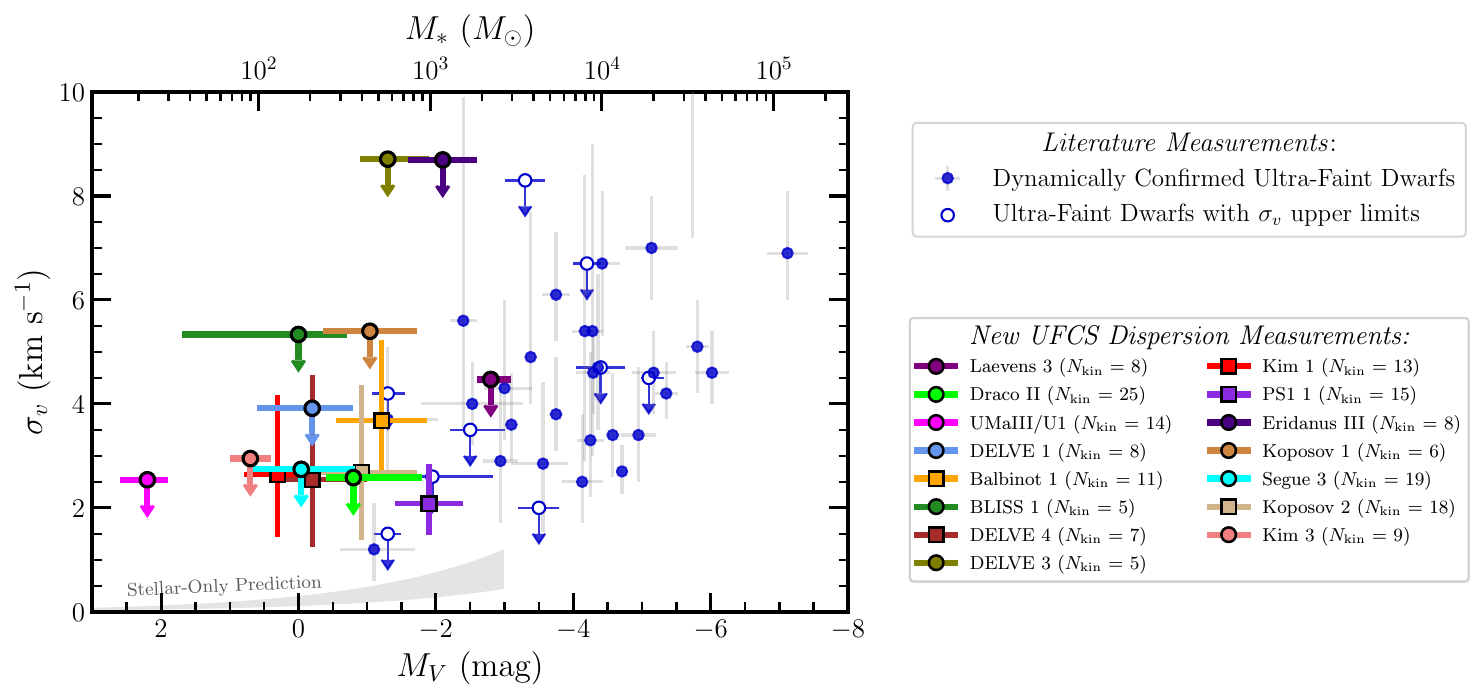}
    \caption{\textbf{Comparing our new UFCS velocity dispersion measurements to the population of ultra-faint MW satellite galaxies.} Our sample of 15 new velocity dispersion constraints are shown as colored markers  to the left of the plot; this includes 10 upper limits (circles) and 5 resolved/marginally-resolved dispersions (squares; see \secref{veloutliers} for caveats). The background UFD sample is split into dynamically-confirmed galaxies (filled blue circles) and presumed galaxies with upper limits on their velocity dispersions (unfilled blue circles). The high frequency of upper limits for UFCSs in our sample suggests that they are kinematically colder than the MW UFDs. For illustration, we highlight the predicted range of  $\sigma_v$ for stellar-only systems in grey (assuming $M_*/L_V = 2$~\MLunit{} and radii $r_{1/2} = 2$--15~pc; see \secref{masses} for details). Note that the radius dependence of these predicted dispersions prohibits extrapolation beyond the UFCS regime.  \label{fig:vdispcomparison} }
    
    \includegraphics[width=\textwidth]{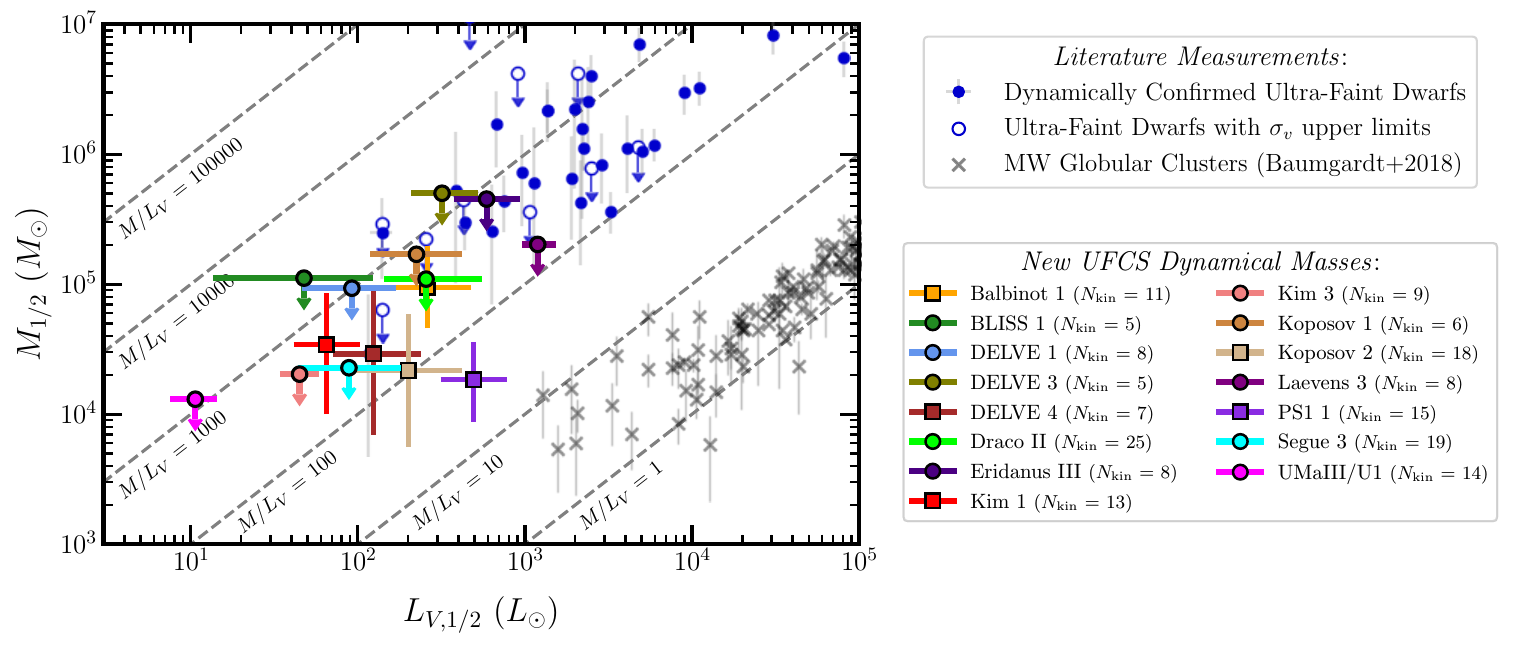}
    \caption{\textbf{Constraints on the dynamical masses of the UFCSs within their half-light radii ($M_{1/2}$) vs. their $V$-band luminosities within their half-light radii ($L_{V,1/2}$)}. As with \figref{vdispcomparison}, we include only UFCSs with  $\geq$~5 kinematic members; UFCSs with resolved (unresolved) dispersions are shown as colored squares (circles). We compare our new measurements to similar mass estimates for the MW UFDs (filled and unfilled blue circles) computed in the Local Volume Database based on the same \citet{2010MNRAS.406.1220W}, as well as to the sample of GCs with half-mass measurements from \citet{2018MNRAS.478.1520B}. No luminosity uncertainties are shown for these comparison samples.  In grey, we overplot lines of constant mass-to-light ratio (in units of $M_{\odot}/L_{\odot}$). While insufficient to rule out dark matter for individual objects, our measurements and upper limits suggest that the UFCSs have mass-to-light ratios that cannot be significantly larger than the UFD population average. \label{fig:mlratios}}
\end{figure*}

\subsection{Results: A Kinematically Cold, Low-Mass Population of UFCSs}
We present the first population-level view of the stellar kinematics of the UFCSs in \figref{vdispcomparison}, where we plot the velocity dispersions of all 15 UFCSs with velocity dispersion constraints (including upper limits) as a function of their absolute magnitudes / stellar masses. In grey, we shade the predicted velocity dispersions for stellar-only systems in the UFCS regime ($M_V > -3.5$; $2 \rm \ pc < r_{1/2} < 15 \ pc$). As a point of comparison, we also plot the observed velocity dispersions of the UFD satellites of the MW in blue (filled points if the velocity dispersion has been resolved by prior spectroscopic measurements, or unfilled points for upper limits). 
\par This population-level view makes clear that the UFCSs are generally kinematically cold: our data yielded upper limits on the velocity dispersions for 2/3 of our sample (10 of 15 systems). While the strength and robustness of these limits vary considerably from target to target, several of our upper limits are as low as $\sigma_v \lesssim 2.5$--3~\kms{} -- sufficiently strong to rule out the dispersions seen in the majority of UFDs (albeit still well above these UFCSs' predicted dispersions of $\sim$0.1--1~\kms in the stellar-only scenario). Even when considering the five systems for which we do report resolved or marginally-resolved velocity dispersions, the largest is a mere \Balbinotonevdisp{} and the remainder are $< 3$~\kms{}  (noting the caveats from \secref{veloutliers}). This preponderance of upper limits and low dispersions was by no means a foregone conclusion: the hypothesis that the UFCSs are ubiquitously kinematically hot -- comparable to the well-known case of the extreme dwarf galaxy Segue 1 ($M_V = -1.5$; $\sigma_v = 3.7$~\kms{}) -- had not been previously tested based on a statistical sample of systems.  Our results now rule out this scenario for nearly half of the sample and disfavor it for the remainder. This result has potentially significant implications for dark matter models that predict the substantial dynamical heating of the faintest galaxies (e.g., self-interacting dark matter after gravothermal collapse or fuzzy dark matter; see e.g., \citealt{2019PhRvD.100f3007Z,2022PhRvD.106f3517D,2023ApJ...949...67Y,2023ApJ...949...68D,2025PhRvD.112f3008Z}). 

\begin{deluxetable*}{lccccccc|ccc}
\tablewidth{\textwidth}
\tabletypesize{\small}
\tablecaption{Dynamical Properties of the UFCSs derived in this work\label{tab:dynamicstable}}
\tablehead{Name & $N_{\rm mem}$ & $N_{\rm kin}$ & $v_{\rm sys}$ & $\sigma_v$ & $\sigma_{\ast,\rm pred.}$ & $M_{1/2}$ & $M_{1/2}/L_{V,1/2}$ & $N_{\rm 6D}$ & $\bar{\mu}_{\alpha\ast}$ & $\bar{\mu}_{\delta}$ \\
 &  &  & $\rm (km\ s^{-1})$ & $\rm (km\ s^{-1})$ & $\rm (km\ s^{-1})$ & $(10^5\ M_{\odot})$ & $\rm M_{\odot}L_{\odot}^{-1}$ &  & $\rm (mas\ yr^{-1})$ & $\rm (mas\ yr^{-1})$ 
}
\startdata
Balbinot 1* & 11 & 11 &  $-175.9^{+1.3}_{-1.5}$ & $3.7^{+1.5}_{-1.1}$ & $0.27^{+0.10}_{-0.07}$ & $0.9^{+1.0}_{-0.5}$ & $730^{+1140}_{-450}$ & 6 & $1.48^{+0.12}_{-0.12}$ & $-1.29^{+0.13}_{-0.13}$ \\
BLISS 1 & 5 & 5 &  $122.6^{+1.0}_{-1.1}$ & $< 5.3$ & $0.16^{+0.10}_{-0.08}$ & $< 1.1$ & $< 9490$ & 5 & $-2.34^{+0.02}_{-0.02}$ & $0.13^{+0.02}_{-0.02}$ \\
DELVE 1 & 10 & 8 &  $-402.8^{+1.1}_{-1.1}$ & $< 3.9$ & $0.18^{+0.07}_{-0.05}$ & $< 0.9$ & $< 2650$ & 8 & $0.02^{+0.07}_{-0.07}$ & $-1.54^{+0.05}_{-0.05}$ \\
DELVE 3 & 5 & 5 &  $-84.8^{+3.0}_{-2.8}$ & $< 8.7$ & $0.32^{+0.09}_{-0.06}$ & $< 5.0$ & $< 3510$ & 2 & $-0.31^{+0.27}_{-0.27}$ & $-0.82^{+0.28}_{-0.28}$ \\
DELVE 4* & 8 & 7 &  $-167.7^{+1.4}_{-1.4}$ & $2.5^{+2.0}_{-1.3}$ & $0.23^{+0.10}_{-0.06}$ & $0.3^{+0.6}_{-0.2}$ & $440^{+1220}_{-350}$ & 6 & $0.42^{+0.08}_{-0.08}$ & $-0.75^{+0.11}_{-0.11}$ \\
DELVE 5 & 3 & 3 &  $-233.5^{+6.3}_{-6.2}$ & \ldots & $0.19^{+0.08}_{-0.05}$ & \ldots & \ldots & 0 & \ldots & \ldots \\
DELVE 6 & 3 & 3 &  $171.4^{+6.6}_{-6.2}$ & \ldots & $0.28^{+0.10}_{-0.07}$ & \ldots & \ldots & 2 & $0.92^{+0.39}_{-0.39}$ & $-1.28^{+0.38}_{-0.38}$ \\
Draco II & 28 & 25 &  $-342.4^{+0.7}_{-0.7}$ & $< 2.6$ & $0.18^{+0.09}_{-0.05}$ & $< 1.1$ & $< 1040$ & 17 & $1.01^{+0.09}_{-0.09}$ & $0.96^{+0.10}_{-0.10}$ \\
Eridanus III & 8 & 8 &  $54.0^{+2.5}_{-2.1}$ & $< 8.7$ & $0.44^{+0.12}_{-0.09}$ & $< 4.5$ & $< 1820$ & 3 & $1.36^{+0.08}_{-0.07}$ & $-0.64^{+0.08}_{-0.08}$ \\
Kim 1* & 13 & 13 &  $-260.8^{+1.2}_{-1.1}$ & $2.6^{+1.5}_{-1.2}$ & $0.16^{+0.04}_{-0.03}$ & $0.3^{+0.5}_{-0.2}$ & $1020^{+1870}_{-740}$ & 6 & $-1.12^{+0.10}_{-0.10}$ & $-0.09^{+0.09}_{-0.09}$ \\
Kim 3 & 9 & 9 &  $163.2^{+0.7}_{-0.8}$ & $< 2.9$ & $0.20^{+0.05}_{-0.04}$ & $< 0.2$ & $< 950$ & 8 & $-0.86^{+0.16}_{-0.16}$ & $3.34^{+0.12}_{-0.12}$ \\
Koposov 1 & 7 & 6 &  $4.7^{+1.3}_{-1.4}$ & $< 5.4$ & $0.28^{+0.13}_{-0.09}$ & $< 1.7$ & $< 1860$ & 2 & $-1.54^{+0.12}_{-0.11}$ & $-0.80^{+0.09}_{-0.09}$ \\
Koposov 2* & 18 & 18 &  $109.8^{+1.5}_{-1.3}$ & $2.7^{+1.7}_{-1.3}$ & $0.37^{+0.16}_{-0.12}$ & $0.2^{+0.4}_{-0.2}$ & $200^{+540}_{-160}$ & 3 & $-0.58^{+0.16}_{-0.16}$ & $0.10^{+0.10}_{-0.10}$ \\
Laevens 3 & 9 & 8 &  $-72.8^{+1.0}_{-1.0}$ & $< 4.5$ & $0.49^{+0.07}_{-0.05}$ & $< 2.0$ & $< 350$ & 5 & $0.44^{+0.15}_{-0.14}$ & $-0.53^{+0.11}_{-0.11}$ \\
Munoz 1 & 6 & 4 &  $-138.7^{+4.5}_{-4.0}$ & \ldots & $0.20^{+0.11}_{-0.07}$ & \ldots & \ldots & 2 & $-0.03^{+0.60}_{-0.61}$ & $0.83^{+0.56}_{-0.56}$ \\
PS1 1* & 15 & 15 &  $124.5^{+0.7}_{-0.7}$ & $2.1^{+0.8}_{-0.6}$ & $0.48^{+0.15}_{-0.11}$ & $0.2^{+0.2}_{-0.1}$ & $70^{+100}_{-40}$ & 15 & $-2.91^{+0.03}_{-0.03}$ & $-1.85^{+0.03}_{-0.03}$ \\
Segue 3 & 35 & 19 &  $-165.5^{+0.7}_{-0.8}$ & $< 2.7$ & $0.24^{+0.11}_{-0.07}$ & $< 0.2$ & $< 710$ & 16 & $-0.72^{+0.17}_{-0.17}$ & $-1.87^{+0.11}_{-0.11}$ \\
UMaIII/U1 & 16 & 14 &  $89.0^{+0.6}_{-0.6}$ & $< 2.5$ & $0.11^{+0.04}_{-0.03}$ & $< 0.1$ & $< 2730$ & 8 & $-0.76^{+0.09}_{-0.08}$ & $1.16^{+0.11}_{-0.12}$ \\
YMCA-1 & 3 & 3 &  $361.8^{+1.7}_{-1.6}$ & \ldots & $0.29^{+0.09}_{-0.07}$ & \ldots & \ldots & 3 & $0.88^{+0.15}_{-0.15}$ & $1.19^{+0.16}_{-0.16}$ \\
\enddata
\tablecomments{Upper limits are quoted at the 95\% credible level. Mass-to-light ratios are rounded to the nearest ten. Proper motion estimates were derived exclusively from spectroscopically-confirmed members. UFCSs marked with a * were found to have velocity dispersions strongly affected by a single star; see \secref{veloutliers} for details. Reported $M_{1/2}$ estimates were derived from the \citet{2010MNRAS.406.1220W} estimator under the assumption of dynamical equilibrium, which may not apply for disrupting systems.}\end{deluxetable*}

\par In \figref{mlratios}, we extend these comparisons to the UFCSs' dynamical masses. Our spectroscopic velocity dispersion measurements rule out dynamical masses within the half-light radius of $M_{1/2} > 2\times \rm 10^5 \ M_{\odot}$ for all but the two least-well-constrained UFCSs in our sample (DELVE~3 and Eridanus~III). More than 80\% of the MW UFDs with published velocity dispersion estimates have dynamical masses above this threshold (including \textit{all} UFDs with resolved dispersions); thus, we conclude that the UFCSs are a lower-mass population than the UFDs. While this is somewhat of a trivial conclusion given the lower observed stellar masses of the UFCSs, we emphasize again that the scenario in which the majority of the UFCSs had Segue-1-like dynamical masses ($M_{1/2} \approx 6 \times 10^5 \rm \ M_{\odot}$) had not been ruled out prior to this work. In addition to being lower than the UFDs, our dynamical mass upper limits also begin to suggest that the UFCSs' total dynamical masses within $r_{1/2}$ are smaller than those of the MW's classical GCs (black crosses in \figref{mlratios}). This would imply that the UFCSs, regardless of their (unknown) dark matter contents, have lower mean densities within $r_{1/2}$ than GCs given the similar physical sizes. The low densities of the UFCSs have major implications for their long-term survival against two-body relaxation and tidal disruption (see \secref{stellaronly}).

\par Finally, posed in terms of their dark matter contents,  we find that the majority of the UFCSs in our sample have mass-to-light ratio upper limits or resolved estimates in the range $100$~\MLunit{} $\lesssim M_{1/2}/L_{V,1/2} \lesssim 1000$~\MLunit{}. This range is similar to the range spanned by the MW UFDs and rules out the possibility that the UFCSs are even more highly dark-matter-dominated than these galaxies within their half-light radii. There is no unambiguous evidence from our sample that the population of UFCSs is \textit{less} dark-matter-dominated on average, though we do observe that the two individual UFCSs in our sample with the largest stellar masses (Laevens~3 and PS1~1) have lower mass-to-light ratios that begin to diverge from the UFD population average (\Laevensthreemlratio{} and \PSonemlratio{}, respectively). This divergence is even clearer if the velocity outlier we identified in PS1~1 is excluded, in which case the system's mass-to-light ratio becomes \PSonemlratioJackknife{} -- an upper limit stronger than the measured value for any known UFD. Neglecting these two cases, we expect that establishing a systematic difference in mass-to-light ratios between the UFCS population and the MW UFDs (if one exists) would require typical velocity dispersion upper limits roughly $\sim$2$\times$ smaller than our current measurements, reducing the UFCSs' estimated masses and mass-to-light ratios by a factor of $\sim$4. Achieving these stronger limits should be observationally feasible with current instruments, even if robustly resolving the UFCSs' dispersions at high statistical significance may not be (see \secref{outlook} for a more extended outlook). 
\par This all being said, we note that these comparisons of enclosed masses and mass-to-light ratios between the UFCS and UFD populations are complicated by the poorly-understood halo masses and scale radii of the UFCSs (if they are galaxies). If the UFCSs inhabit dark matter halos identical to those of the UFDs, then the dynamical masses of the UFCSs within $r_{1/2}$ are computed from a significantly smaller volume than the mass estimates for UFDs; this would lead us to infer smaller estimates of $\sigma_v$ within $r_{1/2}$,  $M_{1/2}$, and $M_{1/2}/L_{V,1/2}$.  Conversely, if the UFCSs inhabit \textit{lower-mass} halos than the UFDs -- as might be expected cosmologically from the stellar-halo mass relation -- then one would expect the mass density within $r_{1/2}$ to be larger for these very-low-mass systems assuming the ratio of $r_{1/2}$/$r_s$ does not vary substantially with mass (see discussion in \citealt{2023MNRAS.525..325K}). As larger and more precise spectroscopic datasets become available and begin to test for systematic differences between these populations, it will become increasingly important to carry out principled comparisons that take into account this difference in stellar component sizes and to explore the possible radial variation of $\sigma_v$.

\subsection{Mean Proper Motions}
\label{sec:propermotion}
Nearly all UFCSs in our sample have two or more stars bright enough to have 5D astrometric solutions (i.e., position, parallax, and proper motion measurements) reported in \Gaia DR3. To enable studies of the UFCSs' orbital histories, we derived new systemic proper motion estimates for each UFCS through simple Bayesian fits constraining their systemic mean proper motions in the right ascension and declination directions ($\bar{\mu}_{\alpha*}$ and $\bar{\mu}_{\delta}$). Following \citet{2019ApJ...875...77P}, the observed proper motions of member stars in each UFCS were modeled as draws from a bivariate Gaussian centered on the systemic mean. The corresponding log likelihood was
\begin{align}
 \ln \mathcal L = -\frac{1}{2} (\boldsymbol{\mu} - \bar{\boldsymbol{\mu}})^{T} \mathcal{C}^{-1}(\boldsymbol{\mu} - \bar{\boldsymbol{\mu}}) - \frac{1}{2} \ln \left( (2\pi)^2 \det \mathcal{C} \right)
 \end{align}
where $\boldsymbol{\mu} \equiv \left\{\mu_{\alpha*}, \mu_{\delta}\right\}$ is the data vector and $\bar{\boldsymbol{\mu}} \equiv \left\{\bar{\mu}_{\alpha*}, \bar{\mu}_{\delta}\right\}$ is the systemic mean proper motion vector of the UFCS that we sought to constrain. The covariance matrix $\mathcal{C}$ was defined as 
\begin{align}
 \mathcal{C} = \begin{pmatrix}
\sigma_{\mu_{\alpha*}}^2 &  \rho\sigma_{\mu_{\alpha*}} \sigma_{\mu_{\delta}}\\
\rho \sigma_{\mu_{\alpha*}} \sigma_{\mu_{\delta}} & \sigma_{\mu_{\delta}}^2 
\end{pmatrix} 
\end{align}
where $\rho$ is the correlation coefficient between the two proper motion components (recorded in the \Gaia DR3 source catalogs as \texttt{pmra\_pmdec\_corr}). 
\par To derive posterior probability distributions for $\mu_{\alpha*}$ and $\mu_{\delta}$, we performed MCMC sampling with \texttt{emcee} assuming non-informative priors. Prioritizing membership purity over completeness, we limited these fits to include only spectroscopically-confirmed member stars. We also excluded stars with poor-quality \Gaia astrometric fits indicated by \texttt{astrometric\_excess\_noise\_sig} $> 3$ or \texttt{ruwe} $> 1.4$.  At the conclusion of the MCMC sampling, we derived estimates of the two proper motion components and their uncertainties for each UFCS from the median and 16th/84th percentile of the resultant marginalized posterior probability distributions.

\par Our final proper motion estimates are reported in the final columns of \tabref{dynamicstable}, where we also report the number of confirmed member stars with proper motions as $N_{\rm 6D}$. In total, we were able to derive the mean proper motion for 18 of the 19 UFCSs in our sample based on a median of $N_{\rm 6D} = 6$ member stars per system.  This includes, to the best of our knowledge, the first proper motion measurements for two UFCSs: Kim~1 and PS1~1. For the remaining 16 UFCSs, we generally found good agreement between our estimates and those available in the literature \citep[e.g., those from][for a non-exhaustive list of studies covering overlapping targets]{2018MNRAS.480.2609L,2021MNRAS.505.5978V,2022A&A...657A..54B,2022MNRAS.515.4005P,2022ApJ...940..136P,2023ApJ...953....1C,2025arXiv251024849W}.
\par The sole UFCS for which we were unable to measure a proper motion was DELVE~5. Previously, \citet{2023ApJ...953....1C} reported a proper motion for DELVE~5 based on four possible member candidates including one star with a claimed membership probability of $p =1.0$. Our radial velocity for this star (\Gaia DR3 1236270945623558912; the BHB candidate shown in \figref{delve5}) implies it is more likely a non-member (see \secref{membership}). While it could also be a member binary observed far from its center-of-mass velocity, we currently advise that the \citet{2023ApJ...953....1C} measurement should not be used until this BHB star's membership can be more securely confirmed or disputed. Spectroscopic confirmation of the other three (low-probability) \Gaia member candidates in the field of DELVE~5 could also yield an independent proper motion measurement for the system that is not subject to the uncertainty in the BHB star's membership.
\par For the remainder of this work, we do not make use of our new mean proper motion measurements. However, they form an essential component of our orbit analyses in Paper II.

\section{Mean Metallicities and Metallicity Spreads of the UFCS\lowercase{s}}
\label{sec:metals}
\subsection{Metallicity and Metallicity Dispersion Estimates}
\par Although the number of RGB stars with measurable CaT metallicities is very limited in most UFCSs, even a single stellar metallicity measurement offers valuable information about the classification of a given system. We therefore attempted to derive a metallicity for all UFCSs in our sample with at least one RGB star, adopting a three-tiered approach with increasing complexity as a function of the number of available metallicity measurements (hereafter $N_{\rm [Fe/H]}$).
\par As our simplest and most frequently-available measure, we report the metallicity of the brightest RGB star in each UFCS ($\rm [Fe/H]_{\rm brightest}$). Measurements based on one star run the risk of significantly misestimating the true mean metallicity of a system with an intrinsic dispersion; nonetheless, this was the only available option for six UFCSs in our sample.  Next, for all UFCSs with $\geq 2$ metallicity measurements, we also report the inverse-variance-weighted average metallicity of all available stars ($\rm [Fe/H]_{\rm wavg.}$) as well as the maximum metallicity difference between any pair of stars ($\Delta \rm [Fe/H]_{\rm max}$, with an uncertainty set by the quadrature sum of the individual metallicity errors of the two stars). Lastly, for UFCSs with at least three RGB stars with spectroscopic metallicity measurements, we attempted to fit for the mean metallicity and metallicity dispersion through a Bayesian procedure analogous to that used for our velocity dispersion measurements. Specifically, we fit the observed metallicity distribution with a Gaussian model with two free parameters: the systemic mean metallicity of the UFCS ([Fe/H]) and the intrinsic metallicity dispersion, $\sigma_{\rm [Fe/H]}$. We assumed a uniform prior of $-4 < \rm [Fe/H] < 0$ 
 for the mean metallicity and $0 \rm \ dex < \sigma_{\rm [Fe/H]} < 2\rm \ dex$ for the metallicity dispersion; the latter was chosen to yield conservative limits.  We sampled the posterior probability distribution of each parameter with \texttt{dynesty} using the same configuration as for our velocity dispersion fits.  This sampling was then repeated for a model with $\sigma_{\rm [Fe/H], model} = 0$~dex to enable us to compute metallicity dispersion Bayes factors and rigorously decide whether to quote resolved estimates ($2\ln\beta > 0$) from the median and 16th/84th percentiles of the posterior or upper limits at the 95\% credible level ($2\ln\beta < 0$).

\begin{deluxetable*}{lccccccc|cc}
\tablewidth{\textwidth}
\tabletypesize{\small}
\tablecaption{Metallicity Properties of the UFCS derived in this Work\label{tab:metallicitytable}}
\tablehead{
Name & $N_{\rm [Fe/H]}$ & $\rm [Fe/H]_{\rm model}$ & $\sigma_{\rm [Fe/H]}$ & $\rm [Fe/H]_{\rm w.avg.}$ & $\rm \Delta [Fe/H]_{max}$ & $\rm [Fe/H]_{\rm brightest}$ \\
 & (stars) &  & (dex) &  & (dex) &  
}
\startdata
Balbinot 1 & 4 & $-2.59^{+0.21}_{-0.23}$ &  $0.41^{+0.27}_{-0.18}$ & $-2.51^{+0.08}_{-0.08}$ & $0.76^{+0.25}_{-0.25}$ & $-2.47^{+0.14}_{-0.14}$ \\
BLISS 1 & 3 & $-0.83^{+0.14}_{-0.14}$ &  $< 0.68$ & $-0.82^{+0.08}_{-0.08}$ & $0.13^{+0.24}_{-0.24}$ & $-0.79^{+0.13}_{-0.13}$ \\
DELVE 1 & 2 & \ldots &  \ldots & $-2.58^{+0.12}_{-0.12}$ & $0.32^{+0.25}_{-0.25}$ & $-2.68^{+0.14}_{-0.14}$ \\
DELVE 3 & 1 & \ldots &  \ldots & \ldots & \ldots & $-2.98^{+0.29}_{-0.29}$ \\
DELVE 4 & 6 & $-2.17^{+0.27}_{-0.27}$ &  $0.62^{+0.22}_{-0.19}$ & $-2.16^{+0.08}_{-0.08}$ & $1.94^{+0.55}_{-0.55}$ & $-2.51^{+0.15}_{-0.15}$ \\
DELVE 5 & 0 & \ldots &  \ldots & \ldots & \ldots & \ldots \\
DELVE 6 & 1 & \ldots &  \ldots & \ldots & \ldots & $-1.49^{+0.51}_{-0.51}$ \\
Draco II & 3 & $-2.94^{+0.32}_{-0.39}$ &  $0.57^{+0.27}_{-0.27}$ & $-2.74^{+0.12}_{-0.12}$ & $1.33^{+0.48}_{-0.48}$ & $-2.49^{+0.16}_{-0.16}$ \\
Eridanus III & 2 & \ldots &  \ldots & $-3.32^{+0.21}_{-0.21}$ & $0.11^{+0.63}_{-0.63}$ & $-3.33^{+0.22}_{-0.22}$ \\
Kim 1 & 1 & \ldots &  \ldots & \ldots & \ldots & $-2.69^{+0.14}_{-0.14}$ \\
Kim 3 & 0 & \ldots &  \ldots & \ldots & \ldots & \ldots \\
Koposov 1 & 2 & \ldots &  \ldots & $-0.79^{+0.13}_{-0.13}$ & $0.45^{+0.32}_{-0.32}$ & $-0.88^{+0.14}_{-0.14}$ \\
Koposov 2 & 1 & \ldots &  \ldots & \ldots & \ldots & $-2.86^{+0.18}_{-0.18}$ \\
Laevens 3 & 5 & $-1.97^{+0.12}_{-0.12}$ &  $< 0.56$ & $-1.97^{+0.08}_{-0.08}$ & $0.74^{+0.44}_{-0.44}$ & $-2.09^{+0.13}_{-0.13}$ \\
Munoz 1 & 1 & \ldots &  \ldots & \ldots & \ldots & $-1.26^{+0.28}_{-0.28}$ \\
PS1 1 & 9 & $-1.24^{+0.09}_{-0.08}$ &  $< 0.35$ & $-1.24^{+0.07}_{-0.07}$ & $0.93^{+0.46}_{-0.46}$ & $-1.01^{+0.16}_{-0.16}$ \\
Segue 3 & 1 & \ldots &  \ldots & \ldots & \ldots & $-0.88^{+0.21}_{-0.21}$ \\
UMaIII/U1 & 12 & $-2.65^{+0.11}_{-0.11}$ &  $< 0.35$ & $-2.65^{+0.10}_{-0.10}$ & $0.72^{+0.48}_{-0.48}$ & $-2.47^{+0.33}_{-0.33}$ \\
YMCA-1 & 3 & $-1.93^{+0.19}_{-0.22}$ &  $< 0.83$ & $-1.91^{+0.11}_{-0.11}$ & $0.51^{+0.51}_{-0.51}$ & $-1.95^{+0.15}_{-0.15}$ \\
\enddata
\tablecomments{All estimates refer to spectroscopic metallicities. $\rm [Fe/H]_{\rm model}$ and its uncertainties refer to the mean metallicity derived from our MCMC fits for UFCSs with $>3$ stars. $\rm [Fe/H]_{\rm w.avg.}$ is an inverse-variance weighted average of all stellar metallicities for a given system, and $\rm [Fe/H]_{\rm brightest}$ is the metallicity of the brightest star. We recommend using the leftmost available [Fe/H] estimate per row.}\end{deluxetable*}

\subsection{Results: A Chemically Diverse Population of Satellites Extending Below $\rm[Fe/H] = -2.5$}
We summarize our tiered metallicity estimates in \tabref{metallicitytable}. In total, we were able to measure a metallicity for at least one star in 17 of our 19 UFCS targets. The two UFCSs for which we report no measurements are DELVE~5 and Kim~3, both of which intrinsically have no RGB members for which the CaT EW--[Fe/H] relationship is calibrated. Prior to the initiation of our census, the only UFCSs with published spectroscopic metallicity measurements were Laevens~3,  Mu\~{n}oz~1, and Draco~II; our new measurements more than triple the sample with spectroscopic metallicities available.

\begin{figure*}[htp]
    \centering    \includegraphics[width=\textwidth]{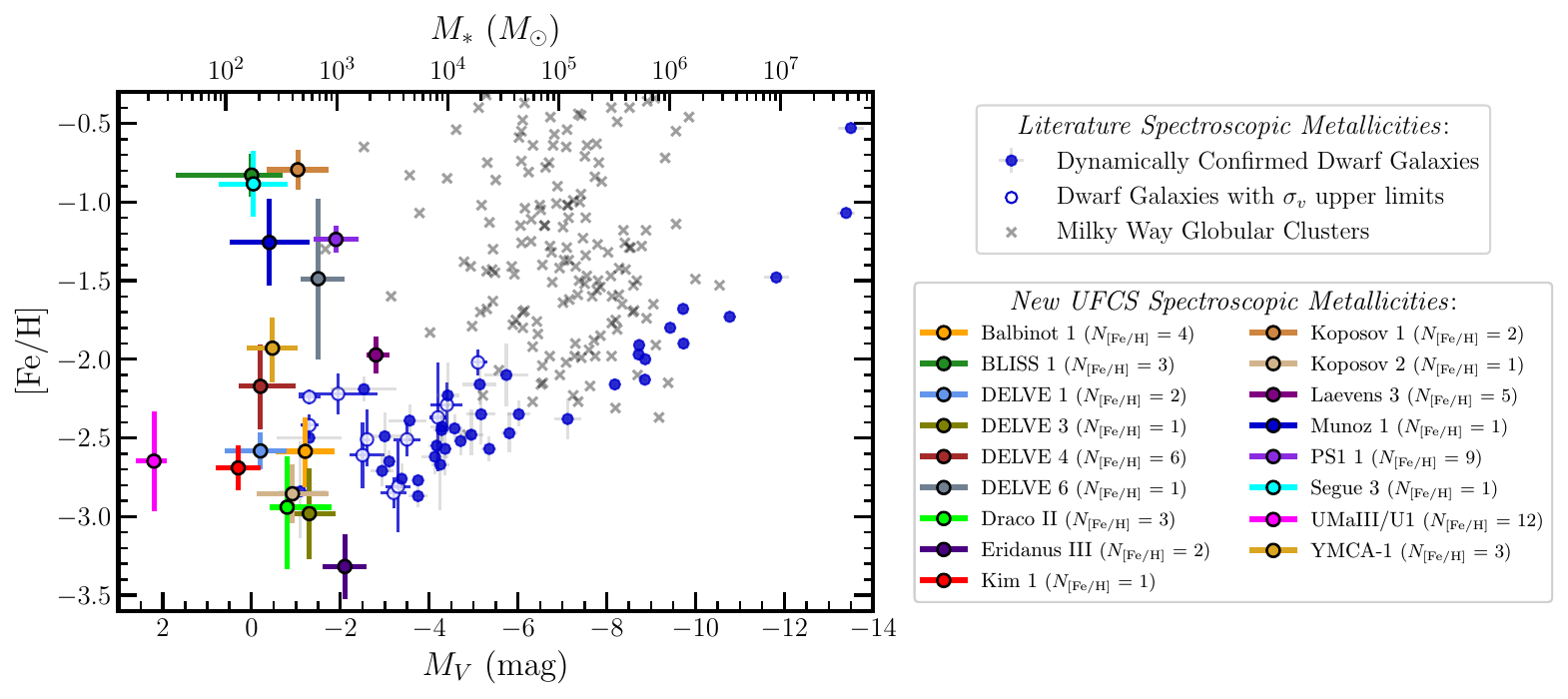}
    \caption{\textbf{The mass-metallicity plane for all MW satellites with spectroscopic measurements.} Our new UFCS metallicities are shown here as colored points; the errorbars are the true observational errors (\textit{not} the metallicity dispersions). Comparison dwarf galaxy and GC spectroscopic metallicities drawn from the Local Volume Database \citep{2025OJAp....8E.142P}. The legend denotes the number of metallicity measurements per system ($N_{\rm [Fe/H]}$). Notably, we identify eight UFCSs that fall at metallicities below the lowest-metallicity known MW GCs ([Fe/H] $\lesssim -2.5$): Balbinot~1, DELVE~1, DELVE~3, Draco~II, Eridanus~III,  Kim~1, Koposov~2, and \uma{}. The remaining UFCSs exhibit diverse metallicities up to a maximum of [Fe/H] $\approx -0.8$.  Note that if any of these objects are indeed dwarf galaxies, their mean metallicities may differ somewhat from the estimates here owing to our small-$N$ sampling of their underlying metallicity distributions. }
    \label{fig:lzr}
\end{figure*}

\par In \figref{lzr}, we visualize the mean metallicities of the UFCSs in the [Fe/H]--$M_V$ plane and compare them to those of the MW GC and UFD populations. We draw three main conclusions from this figure. First, it is immediately clear that the UFCSs are a \textit{chemically diverse} class of systems:  our observations suggest that the UFCSs span as large as a factor of $\sim$300 in mean iron abundance ($-3.3 \lesssim \rm [Fe/H] \lesssim -0.8$) even at roughly fixed stellar mass. Second, we report the discovery of a \textit{population} of compact ($r_{1/2} \lesssim 15$~pc) MW satellites with metallicities [Fe/H] $\lesssim -2.5$. Third, we observe an apparent scarcity of UFCSs at $\rm [Fe/H] < -3$, with only one UFCS in our sample (Eridanus~III) clearly below this threshold and two just above (DELVE~3 and Draco~II).
\par The interpretation of these three features is closely linked to the UFCSs' classifications. Notably, the metallicity range spanned by our spectroscopic sample of UFCSs is larger than the range spanned by the entire population of MW satellite galaxies (excluding the LMC), strongly suggesting that at least some of the UFCSs are not likely to be dwarf galaxies. More specifically, the highest-metallicity UFCSs clearly deviate from the empirical dwarf galaxy mass-metallicity/luminosity-metallicity sequence seen in \figref{lzr} and are more likely star clusters. On the other hand, the lowest-metallicity UFCSs resemble both the faintest GCs and the MW UFDs and are more ambiguous. We extend these arguments to classify individual UFCSs in \secref{classification}. 
\par If the UFCSs at $\rm [Fe/H] < -2.5$ are in fact star clusters, they would significantly expand the population of MW GCs in this metallicity regime, with broader implications for the physics of GC formation and destruction. It has frequently been suggested that GCs exhibit a metallicity floor of $\rm [Fe/H] \approx -2.5$, potentially arising from a combination of a host galaxy mass-metallicity relation and the lower specific frequency of GCs in low-mass, low-metallicity hosts \citep[e.g.,][]{2019MNRAS.487.1986B}. Recently, however, the confirmation of a single Local Group GC unambiguously known below this proposed floor -- namely, the cluster EXT8 in M31, at [Fe/H] = $-2.9$ \citep{2020Sci...370..970L} -- has challenged this picture. Moreover, the discovery of stellar streams with metallicities $\rm [Fe/H] < -2.5$ that likely originated from GC progenitors (e.g., Sylgr, Phoenix, and C-19) has pointed to the existence of a population of ancient MW GCs that formed below the purported metallicity floor but have since been destroyed \citep{2019ApJ...883...84R,2020Natur.583..768W,2022Natur.601...45M,2022ApJ...928...30L,2024ApJ...967...89I}. Confirming UFCSs in this metallicity range as star clusters would provide further evidence against a strict metallicity floor and may support the idea that extremely metal-poor GCs are rare today because they are more efficiently destroyed than their higher-metallicity counterparts -- potentially due to their lower formation masses (e.g., \citealt{2026enap....4..500K}) or more eccentric orbits.
\par The existence of GCs at these metallicities would also be consistent with theoretical and empirical expectations. Based on the observed metallicity distribution function of MW field stars, \citet{2020MNRAS.492.4986Y} predicted the existence of $\sim$10 MW GCs at $\rm [Fe/H] < -2.5$. In addition, the GC formation and evolution model of \citet{2024MNRAS.527.3692C} predicts that the MW should host at least four surviving GCs with $\rm [Fe/H] < -2.5$ and masses $>10^3\,\rm M_\odot$. If some of the most metal-poor UFCSs are confirmed to be star clusters, they could help reconcile the observed MW GC population with these predictions -- albeit only if the UFCSs formed at significantly higher masses than presently observed.

\par Alternatively, if the UFCSs at $\rm [Fe/H] \lesssim -2.5$ are  galaxies, their metallicities would provide strong constraints on the behavior of the galaxy mass/luminosity-metallicity relation at the faint extreme. The metallicity distribution of the UFCSs seen in \figref{lzr} -- with its notable scarcity of systems at [Fe/H] $<-3$ -- is consistent with prior observations of a metallicity plateau in the MW UFDs \citep[e.g.,][]{2019ARA&A..57..375S,2023ApJ...958..167F,geha_paper2}. Provided that this feature of the UFCS/UFD population is established based on a clearer understanding of the UFCSs' classifications and metallicity distributions, this plateau may point to a floor in the metallicity of the intergalactic medium, enrichment by Population III stars, and/or strong outflows in the faintest dwarf galaxies \citep[e.g.,][]{2022MNRAS.516.3944M,2024MNRAS.529.3387A,2025OJAp....8E.151W,2025arXiv250907066B,2025arXiv251005232R}.

\subsection{Results: Constraints on Metallicity Spreads}
\par Across our sample of 19 UFCSs, we were able to perform full Bayesian metallicity dispersion fits for just eight systems (BLISS~1, Balbinot~1, DELVE~4, Draco~II, Laevens~3, PS1~1, \uma{}, and YMCA-1) after imposing a relaxed minimum sample size of $\geq3$ stars with measured metallicities. For an additional three systems (DELVE~1, Koposov~1, and Eridanus~III) with exactly two metallicity measurements, we computed only the difference statistic $\Delta \rm [Fe/H]_{max}$. Finally, for the remaining eight UFCSs in our sample, we were unable to place any constraints on  internal metallicity spreads.
\par The UFCSs in our sample with the best-sampled metallicity distributions are PS1~1 (with \PSoneNfeh{} CaT metallicities) and \uma{} (with \UrsaMajorIIINfeh{} Ca K metallicities from \citetalias{2025arXiv251002431C}~\citeyear{2025arXiv251002431C}). For these two cases, our fits yielded negative Bayes factors of \PSonefehdispbayes{} and \UrsaMajorIIIfehdispbayes{}, respectively, implying a preference for models with $\sigma_{\rm [Fe/H]} = 0$~dex. Accordingly, we report 95\% credible upper limits of \PSonefehdisp{} and \UrsaMajorIIIfehdisp{} (respectively) for these two systems. These Bayes factors and upper limits suggest the absence of large internal metallicity variation but are not yet strong enough to rule out the possibility of the smallest spreads seen in UFDs (see, e.g., \citealt{2015ApJ...811...62K} and \citealt{2024ApJ...961..234H}). 
\par After PS1~1 and \uma{}, the next best-sampled metallicity distributions for UFCSs in our sample are those of DELVE~4 (\DELVEfourNfeh), Laevens~3 (\LaevensthreeNfeh{}), and Balbinot~1 (\BalbinotoneNfeh{}). In the case of DELVE~4, we found a highly significant resolved metallicity dispersion of \DELVEfourfehdisp{} (with \DELVEfourfehdispbayes{}) that is robust to the removal of any single star. Notably, the metallicity distribution of DELVE~4 also spans nearly two orders of magnitude (\DELVEfourfehmaxdiff{}), though see \secref{membership} for membership caveats.  For Balbinot~1, we found a smaller dispersion than DELVE~4 (\Balbinotonefehdisp{}) that is moderately significant (\Balbinotonefehdispbayes{}). This latter dispersion arises from a single star in our four-star metallicity sample  (\Gaia DR3 2735226970153175808) that falls at a higher metallicity and thereby drives a statistically-significant difference relative to the most metal-poor RGB member star (\Balbinotonefehmaxdiff). Finally, for Laevens~3, we placed a weak upper limit of \Laevensthreefehdisp{} (with \Laevensthreefehdispbayes{}). This upper limit is consistent with the results of \citet{2019MNRAS.490.1498L}, who reported $\sigma_{\rm [Fe/H]} < 0.5$~dex at the 95\% credible level based on the same DEIMOS spectra.

\par The remaining UFCSs in our sample have too few stars to offer meaningful constraints. Nonetheless, for the three systems with exactly $N_{\rm [Fe/H]} = 3$, we found a resolved dispersion for Draco~II (\DracoIIfehdisp{}) and weak upper limits for BLISS~1 and YMCA-1 (\BLISSonefehdisp{} and \YMCAonefehdisp{}, respectively).  For an additional three UFCSs (DELVE~1, Eridanus~III, and Koposov~1) with exactly two RGB stars with measured metallicities, we quantified only the statistical difference between each star's measurement, $\Delta \rm [Fe/H]_{\rm max}$. In these cases, we found no clear evidence for internal metallicity variation: \DELVEonefehmaxdiff{}, \EridanusIIIfehmaxdiff{}, and \Koposovonefehmaxdiff{}, respectively. Observations of additional member stars in these six UFCSs will be required before any definitive conclusions regarding their metallicity distributions can be drawn.

\section{Dwarf Galaxies or Star Clusters? \newline The Nature of the Ultra-Faint Compact Satellites}
\label{sec:classification}
With our large sample of new chemodynamical measurements, augmented by a wealth of constraints from prior photometric studies, we are now positioned to make inferences about the nature of the UFCSs both individually and as a population. In \secref{diagnostics} below, we begin by reviewing observational diagnostics suitable for separating dwarf galaxies and star clusters. Then, in the subsequent sections (\secref{starclusters} to \secref{unconstrained}), we apply these diagnostics to classify the UFCSs in our sample. We summarize our population-level findings in \secref{landscape} and \tabref{qualitative}.

\subsection{A Review of Observational Diagnostics}
\label{sec:diagnostics}
To classify the 19 UFCSs in our sample, we considered six main observational diagnostics:  \newline
\par \textit{(1) Mass-to-Light Ratios}: Dwarf galaxies and star clusters are definitionally distinguished by the presence or absence of a dark matter halo. In practice, the dark matter content of faint stellar systems is best observationally diagnosed by their mass-to-light ratios. The MW UFDs with well-constrained velocity dispersions have typical mass-to-light ratios within their half-light radius of $\mathcal{O}(10^2 - 10^{3.5})$~\MLunit{} \citep[e.g.,][]{2019ARA&A..57..375S,2022NatAs...6..659B}, whereas GCs exhibit mass-to-light ratios of just $\sim$2~\MLunit{} on average \citep[e.g.,][and see \figref{mlratios}]{2017MNRAS.464.2174B}. Unfortunately, because upper limits on velocity dispersions are typically found for systems in the $M_V \gtrsim -3$ regime, classifying ultra-faint systems based on this gold standard metric has become uncommon.  \newline

\par \textit{(2) Stellar population ages}: Without exception, the UFDs known around the MW and other hosts are dominated by ancient stellar populations ($\tau > 10$~Gyr). The lack of young stars in these ``fossil'' systems reflects their rapid star formation in the early universe followed by quenching during the epoch of reionization \citep{2005ApJ...629..259R,2009ApJ...693.1859B,2012ApJ...744...96O,2014ApJ...796...91B,2021ApJ...920L..19S,2025arXiv250518252D}. On the other hand, the MW halo is known to host GCs spanning a wide range in age, including a significant population of ``young halo'' clusters that likely accreted within more massive dwarf galaxies with extended star formation histories such as the LMC, SMC, or Sagittarius \citep[e.g.,][]{1993ASPC...48...38Z,2000ApJ...530..777V,2005ApJ...621L..61C,2008AJ....136.1407D,2009AJ....137.3809C,2016ApJ...822...32W,2016MNRAS.460.3384V}. Therefore, ultra-faint systems with significantly younger or intermediate ages (as determined through isochrone fitting) are empirically much less likely to be dwarf galaxies and instead are more likely to be halo star clusters.  Beyond mean ages, the age \textit{distributions} within satellites may also inform their classifications, as dwarf galaxies can exhibit longer star formation durations compared to their GC counterparts. However, such measurements are (usually) not feasible for the least-massive systems \citep{2025arXiv250518252D}.  \newline

\par \textit{(3) Mean Metallicities}: Spectroscopic studies of the MW and M31 UFDs have demonstrated that these systems are universally very metal-poor, again reflecting their early and inefficient star formation \citep[e.g.,][]{2008ApJ...685L..43K,2009MNRAS.395L...6S,2013ApJ...779..102K,2015ApJ...807..154B}. The metallicities of these faint dwarf galaxies are closely correlated with their luminosities through the luminosity-metallicity relation (LZR), which is commonly parameterized by a power law \citep{2013ApJ...779..102K} but may plateau near $\rm [Fe/H] \approx -2.6$ at the faint end \citep[e.g.,][] {2019ARA&A..57..375S,2023ApJ...958..167F,geha_paper2}. 
In contrast, star clusters form with the metallicity of their birth galaxies, and there is little correlation between mass and metallicity.
If the UFCSs ($L_V < 10^3 \rm \ L_{\odot}$) are dwarf galaxies, the \citet{2013ApJ...779..102K} LZR predicts that these systems should exhibit typical mean metallicities of [Fe/H] $< -2.4$ -- comparable to, or lower than, the lowest-metallicity MW GCs known.  UFCSs with significantly higher metallicities that place them off the LZR (e.g., $\rm [Fe/H] >  -2.0$) are unlikely to be galaxies, even when considering the possible plateau. Thus, metal-rich UFCSs are very likely to be star clusters, while metal-poor UFCSs are more ambiguous. \newline

\par \textit{(4) Metallicity Dispersions}: In addition to their low mean metallicities, the UFDs exhibit significant star-to-star iron abundance variations with typical dispersions of $\sigma_{\rm [Fe/H]} \approx 0.3$--0.7~dex \citep{2007ApJ...670..313S,2019ARA&A..57..375S,2023ApJ...958..167F}. These dispersions signify the presence of potential wells deep enough to allow the UFDs to retain supernova ejecta and thereby self-enrich, indirectly suggesting the presence of dark matter in these systems \citep{2012AJ....144...76W}. Only a single UFD candidate has been observed to have an iron abundance spread upper limit stringent enough to suggest it could be monometallic: Tucana III, with $\sigma_{\rm [Fe/H]} < 0.2$~dex \citep{Simon_2017,2018ApJ...866...22L}. Given that major doubts persist as to whether Tucana III is truly a galaxy (see \citealt{2025arXiv250305927Z} and Appendix C of \citetalias{2025arXiv251002431C}~\citeyearpar{2025arXiv251002431C} for an extended discussion), the empirical evidence implies that metallicity spreads are a universal feature of the UFDs (though see \citealt{2025arXiv250909582T} for a possible counterpoint from simulations).   By contrast, nearly all known MW GCs display negligible iron abundance spreads $\sigma_{\rm [Fe/H]} \lesssim 0.05$~dex \citep{2012AJ....144..183L,2019ApJS..245....5B}. Even the two clearest exceptions among the GC population, M54 and $\omega$~Centauri, are believed to be the stripped nuclear remnants of dwarf galaxies \citep[e.g.,][]{1999Natur.402...55L,2003MNRAS.346L..11B,2008AJ....136.1147B}; these cases thus support the use of iron abundance dispersions as a powerful diagnostic for classifying the morphologically-overlapping populations of GCs and dwarfs.\footnote{The massive Galactic bulge clusters Liller 1 and Terzan 5 also exhibit complex stellar populations with iron spreads, but their mean metallicities are high ($\rm [Fe/H] > -1$) and their orbital kinematics and elemental abundance patterns favor an in-situ formation channel \citep{2009Natur.462..483F,2014ApJ...795...22M,2023ApJ...951...17C,2024A&A...690A.139F,2025A&A...697A..19O}. Similar systems are unlikely to be mistaken for UFDs.}  \newline

\par \textit{(5) Stellar mass segregation:} Mass segregation refers to the dynamical process in which stars of different masses exhibit different radial distributions within a collisional stellar population: due to equipartition of energy, higher-mass stars should sink to smaller radii, while lower-mass stars should migrate to the population's outskirts \citep[e.g.,][]{1969ApJ...158L.139S,1997A&ARv...8....1M,2008gady.book.....B}. If the UFCSs are comprised solely of stars, their implied relaxation timescales (typically $\lesssim$1--2~Gyr) are significantly shorter than their observed ages and thus they are effectively certain to display mass segregation (see \secref{stellaronly}). On the other hand, the presence of significant amounts of dark matter in the UFCSs would increase the relaxation timescale of these systems by orders of magnitude, and thus mass segregation will not occur in the UFCSs if they are highly-dark-matter-dominated dwarf galaxies (though see \citealt{2025ApJ...993..160E} for the case of compact systems with \textit{low} dark matter densities). This approach has been tested as an empirical classifier in several individual UFCS-focused studies \citep{2016ApJ...820..119K,2018ApJ...852...68C,2018MNRAS.480.2609L,2019MNRAS.490.1498L} and was more recently systematized and carried out across a large sample of targets by \citet{Baumgardt2022}.  \newline

\par \textit{(6) Chemical Abundance Patterns:} High-resolution spectroscopic study of stars in the MW UFDs has suggested that they display distinct chemical abundance patterns from most GCs and field halo stars. Specifically, UFD stars display systematically lower abundances of the neutron-capture elements strontium (Sr) and barium (Ba) compared to stars in their GC counterparts (e.g., $\rm [Ba/Fe] < -1$ and $\rm [Sr/Fe] < -1$; \citealt{2010ApJ...716..446S,2010ApJ...708..560F,2013A&A...554A...5K,2015ARA&A..53..631F,2019ApJ...870...83J,2023ApJ...959..141W}). This deficiency is related to the rarity of early neutron-capture element sources \citep{2019ApJ...870...83J}. Additionally, the most metal-poor stars in UFDs display a high frequency of carbon enhancement \citep{2014ApJ...786...74F,Ji2020,2024A&A...686A.266L} that is not seen in the most metal-poor GCs \citep{2022Natur.601...45M,Casey2021}. Such carbon-enhanced metal poor (CEMP) stars with no neutron-capture element enhancement ($\rm [Ba/Fe] < 0$) are thought to be direct descendants of Population~III stars \citep{2015ARA&A..53..631F,Klessen2023}, and have never been observed in a GC \citep{Simon2024}; thus, the presence of these so-called ``CEMP-no'' stars is a  strong classifier for ultra-faint systems. 
\par On the other hand, a significant fraction of stars in nearly all known GCs display light-element anticorrelations indicative of multiple stellar populations -- specifically, enhancements in nitrogen, sodium, and aluminum paired with depletions in carbon, oxygen, and magnesium \citep[see e.g.,][for reviews]{2004ARA&A..42..385G,2018ARA&A..56...83B} -- that are associated with high stellar densities and thus not seen in the UFDs. The detection of these anti-correlations would provide compelling positive evidence for a GC classification, though non-detections would not be surprising in a low-mass cluster as the fraction of enriched stars (i.e., those with these second-population signatures) is correlated with the  mass of GCs \citep[e.g.,][]{2010A&A...516A..55C,2017MNRAS.464.3636M}.\footnote{The faintest GC currently known to display these features is the tidally-disrupted GC ESO280-SC06 ($M_V \approx -4.3$; see \citealt{2025arXiv250615664U}). While this cluster is nearly a magnitude brighter than the boundary of our UFCS definition, its existence suggests that detecting multiple-population signatures in the UFCSs may be plausible if they are heavily-stripped systems.} High-resolution spectroscopy is available for just two UFCSs in our sample (DELVE~1 and Eridanus~III), and thus the UFCSs' detailed abundance patterns have limited bearing on our classification analysis here. \newline

\par We emphasize that these six diagnostics are not an exhaustive list, and they may not all be strict signatures of one classification or another. Moreover, some degree of difference between the UFCSs and the UFDs/GCs would be reasonable to expect, particularly if the UFCSs are a population of satellites that have experienced substantial mass loss. For example, if the UFCSs are dwarf galaxies, it is conceivable that stripping could  result in the UFCSs exhibiting both smaller mass-to-light ratios than the known UFDs as well as mean metallicities that place them off the dwarf galaxy luminosity-metallicity relation. If they are star clusters, it is possible that the UFCSs formed in very-low-mass dark matter that have since been stripped \citep[e.g.,][]{1984ApJ...277..470P,2005ApJ...619..258M,2009ApJ...706L.192B,2022A&A...667A.112V,2024ApJ...971..103G}, potentially imprinting detectable kinematic signatures. Given these complexities, we anticipate that a comprehensive picture of the stellar kinematics,  chemistries, and orbits of the most ambiguous systems will be necessary to achieve high-confidence classifications. 

\subsection{Metal-Rich and Intermediate-Metallicity UFCSs: \newline Likely Star Clusters}
\label{sec:starclusters}
\par Based on the diagnostics outlined above, we identified eight UFCSs in our sample that are more likely to be star clusters: BLISS~1, Kim~3, Koposov~1, Laevens~3,  Mu\~{n}oz~1, PS1~1, Segue~3, and YMCA-1. We detail each case individually below and summarize the available evidence in \tabref{qualitative}. The most recurring line of evidence throughout our discussion is that these eight systems share metallicities and/or ages that are inconsistent with the known MW UFD population.
\subsubsection{BLISS~1}
\par The most straightforward case on this list is BLISS~1, for which we measured a remarkably high mean metallicity of \BLISSonefehmodel{} from $N_{\rm [Fe/H]} = 3$ RGB stars. This metallicity ($\sim$$16\%$ solar) is clearly inconsistent with the known population of MW UFDs, which all exhibit mean metallicities $\rm [Fe/H] < -2$. As expected from a star cluster, we found no evidence for metallicity differences between the three RGB stars (with a weak formal upper limit of \BLISSonefehdisp{} owing to the small $N$) nor any evidence for a significant velocity dispersion in BLISS~1 (\BLISSonevdisp{}, with a Bayes factor of \BLISSonefehdispbayes{} favoring the zero-dispersion model). Although these dispersion limits are quite weak, we conclude that BLISS~1 is overwhelmingly more likely to be a star cluster on the basis of its high spectroscopic metallicity alone.
\par We highlight the similarity between BLISS~1 (\BLISSoneage) and the young GC Palomar 1 ($\tau \approx 6$--8~Gyr; \citealt{1998AJ....115..648R}),  which shares a similar metallicity ($\rm [Fe/H] = -0.6$; \citealt{1996AJ....112.1487H}, 2010 edition) and also falls near the faint end of the MW GC luminosity distribution. Palomar 1 narrowly misses our UFCS selection due to its low height above the plane ($|Z| < 5$~kpc), whereas BLISS~1 falls just above ($|Z|  \approx 8$~kpc; see Paper II). Palomar~1 has long been enigmatic on the basis of its age, metallicity, low alpha abundance, mass function, and orbital kinematics (see \citealt{2020ApJ...901...48N} and references therein), and we expect that further comparing Palomar~1 and BLISS~1 may elucidate their origins. 

\subsubsection{Segue~3}
\par The next most straightforward case is that of Segue~3. As discussed in \secref{membership}, the original interpretation of Segue~3 as an old, metal-poor system \citep{2010ApJ...712L.103B, 2011AJ....142...88F} has been overturned in favor of a picture in which the system is among the youngest GCs in the Galactic halo (\Seguethreeage{}; see \citealt{2013MNRAS.433.1966O,2017AJ....154...57H}). Here, we add further support to this paradigm with our spectroscopic confirmation of a new RGB member star with a metallicity \Seguethreefehbrightest{}. This star was missed in the original spectroscopic analysis by \citet{2011AJ....142...88F} but later re-identified photometrically by \citet{2017AJ....154...57H}; see \secref{membership}. As we illustrate in Paper II, this revised metallicity brings Segue~3 closer toward the empirical age-metallicity trend exhibited by the MW's in-situ star cluster population. Given its young age, high metallicity, and our spectroscopic non-detection of a velocity dispersion in its inner regions (\Seguethreevdisp{} within $r_{\rm ell} < 3\,a_{1/2}$, with \Seguethreevdispbayes{}), we conclude that Segue~3 is a star cluster. This conclusion is independently supported by the detection of stellar mass segregation in the system reported by both \citet{2017AJ....154...57H} and \citet{Baumgardt2022}.
\subsubsection{Koposov~1}
\par Next under consideration is Koposov~1, for which we measured a weighted-average metallicity of \Koposovonefehwavg{} from \KoposovoneNfeh{} likely RGB members. The membership status of each of these two RGB stars is individually somewhat tentative due to the presence of interlopers from Sagittarius in the Koposov~1 field. However, we note that the template fitting carried out at the velocity-measurement stage also supports a higher metallicity for the more secure MSTO member stars as well ($ -1 \lesssim \rm [Fe/H] \lesssim -0.5$).  This high mean metallicity, paired with Koposov~1's intermediate age (\Koposovoneage{}), again strongly disfavors a galaxy classification. Further informed by
our non-detection of a velocity dispersion in the system (\Koposovonevdispbayes{}, with \Koposovonevdisp{}) and by the weakly-significant detection of stellar mass segregation in the system ($P_{\rm seg.} = 74.4\%$) from \citet{Baumgardt2022},  we conclude that Koposov~1 is very likely a star cluster. Given its close proximity to the Sagittarius stream, it is likely that Koposov~1 is an intermediate-age GC that formed within Sagittarius and subsequently was accreted onto the MW (see Paper II for a deeper exploration of this connection based on its orbit).
\subsubsection{PS1~1}
\par For PS1~1, our relatively large sample of \PSoneNmem{} members (of which \PSoneNfeh{} have metallicities) suggested an intermediate metallicity of \PSonefehmodel{} with little to no internal variation (\PSonefehdisp, with \PSonefehdispbayes{}). 
This metallicity distribution alone strongly suggests that PS1~1 is a star cluster. As with Koposov~1, PS1~1's spatial position immediately suggests an association with the Sagittarius dSph: it falls within $\sim$$3\,a_{1/2}$ of the Sagittarius core at a stream latitude of just $B_{\rm Sgr} = -1.8\degree$ (in the coordinate system of \citealt{2010ApJ...714..229L}). On the basis of these properties, we conclude that PS1~1 is likely another intermediate-age (\PSoneage{}) GC that accreted with the Sagittarius dSph. Given this probable classification, the small but statistically significant velocity dispersion that we measured for the system (\PSonevdisp{}, with \PSonevdispbayes{}) likely arises from undetected stellar binaries or membership contamination. If that is indeed the case, it would be reasonable to infer that our alternative jackknife velocity dispersion limit of \PSonevdispJackknife{} and corresponding mass-to-light ratio limit of \PSonemlratioJackknife{} are more appropriate descriptions of the system's stellar kinematics.
\subsubsection{Mu\~{n}oz~1}
\par Resembling PS1~1 above, we found that  Mu\~{n}oz~1 is moderately metal-poor (\Munozonefehbrightest{}), in agreement with the prior measurement of [Fe/H]= $-1.46 \pm 0.32$ by \citet{2012ApJ...753L..15M} based on the same DEIMOS data. Stars of this metallicity are exceptionally rare in UFDs, though not unprecedented (see e.g., the star SDSS J100714+160154 in Segue 1; \citealt{2014ApJ...786...74F}). The high metallicity of this star therefore favors a star cluster classification for  Mu\~{n}oz~1, though the overall evidence is significantly less strong relative to the cases above because we were unable to measure  Mu\~{n}oz~1's velocity dispersion or metallicity dispersion. Moreover, unlike Koposov~1 or PS1~1, the system's age from isochrone fitting (\Munozoneage, from \citealt{2012ApJ...753L..15M}) is fully consistent with the typical ages of both UFDs and ancient MW GCs. Therefore, while we argue a star cluster classification is more likely based on the brightest star's metallicity alone, an expanded spectroscopic member sample will be necessary to definitively test this hypothesis.
\subsubsection{Laevens~3}
\par Laevens~3 is significantly more metal-poor than each of the UFCSs above, with a mean metallicity from our DEIMOS observations approaching 1/100th solar (\Laevensthreefehmodel{}, from \LaevensthreeNfeh{} members with metallicities). This metallicity would be consistent with known GCs but unusually high for a dwarf galaxy of Laevens~3's luminosity. We found no evidence for a velocity dispersion (\Laevensthreevdisp; \Laevensthreevdispbayes{}) nor any evidence for a metallicity dispersion (\Laevensthreefehdisp; \Laevensthreefehdispbayes{}) in the system. While tighter limits will be required to completely rule out the dwarf galaxy scenario, we concur with the prior analysis by \citet{2019MNRAS.490.1498L} that Laevens~3 is most likely a globular cluster. This is consistent with the $\sim$2$\sigma$ detection of mass segregation in the system, $P_{\rm seg.} = 96.6\%$, from \citet{Baumgardt2022}.
\subsubsection{YMCA~1}
\par As with Laevens~3 above, we found that YMCA-1 is a metal-poor system, with a mean metallicity \YMCAonefehmodel{} based on \YMCAoneNfeh{} bright RGB members. From these three stars, we also determined a weak metallicity dispersion upper limit of \YMCAonefehdisp{} (limited by the small sample size) for YMCA-1 but were unable to measure a velocity dispersion. This metallicity is significantly higher than most UFDs at YMCA~1's absolute magnitude, suggesting that YMCA-1 is more likely a star cluster. However, we again cannot yet rule out a dwarf galaxy classification -- particularly given that YMCA~1's ancient age (\YMCAoneage{}, from \citealt{2022ApJ...929L..21G}) resembles both the MW UFDs and ancient GCs. Obtaining spectra for fainter RGB stars in the system would be highly beneficial for confirming (or refuting) the star cluster scenario.
\par Regardless of its classification, we note that YMCA-1's 3D position and kinematics strongly imply a connection to the LMC/SMC system. \citet{2022MNRAS.515.4005P} predicted that YMCA-1 would be bound to the LMC for radial velocities $v_{\rm hel} > 300~\kms{}$; here, we measure $v_{\rm hel} \approx 362~\kms{}$ and thereby confirm this association (see also Paper II). If it is later confirmed that YMCA~1 is a GC, our mean metallicity measurement would place the system as one of the oldest and most metal-poor GCs in the LMC-SMC system (see e.g., the recent homogeneous compilation from \citealt{2024MNRAS.529.3998S}, who report a range of $\rm [Fe/H] = -2.17 \ to \ [Fe/H] = -1.46$ for old LMC GCs). This would make YMCA-1 a particularly valuable tracer of the Clouds' early assembly histories.

\subsubsection{Kim~3}
\par Lastly, and most uncertain among the pool of UFCSs that we identified as star cluster candidates, is Kim~3. Our spectroscopic constraints yielded  a relatively strong upper limit on Kim~3's velocity dispersion (\Kimthreevdisp{}, with \Kimthreevdispbayes{}) with a relatively weaker limit on its mass-to-light ratio of \Kimthreemlratio{} (owing in part to Kim~3's extraordinarily low luminosity). Further suggesting a limited or absent dark matter content, \citet{2016ApJ...820..119K} detected the presence of stellar mass segregation in Kim~3  ($P_{\rm seg.} = 87\%$). No metallicity information is available for the system due to the lack of RGB member stars suitable for CaT-based measurements; however, we do observe that the system is relatively younger than the known MW UFDs. Specifically, \citet{2016ApJ...820..119K} determined an intermediate age of $\tau = 9.5^{+3.0}_{-1.7}$~Gyr for Kim~3 from isochrone-fitting. While the formal uncertainties on this age are large, our spectroscopic members closely trace the \Kimthreeage{} isochrone at the MSTO, with little room for a significantly older model matching a UFD-like stellar population. Therefore, while the lack of metallicity information available for Kim~3 prohibits us from making any confident judgments about its nature, the combination of features described above tilts the balance toward a GC classification. Obtaining metallicity measurements for stars in Kim~3 is a critical next step for testing this hypothesis.

\subsection{The Very-Metal-Poor UFCSs: The Smallest, Least-Massive Galaxies?}
\label{sec:galaxycandidates}
The UFCSs Balbinot~1, DELVE~1, Draco~II, DELVE~3, Eridanus~III, and Kim~1 all feature ancient ($\tau > 10$~Gyr) stellar populations with one or more stars at a metallicity $\rm [Fe/H] \lesssim -2.5$. If the metallicities of these stars are representative of their host's mean metallicities, then these six UFCSs rank among the most metal-poor stellar populations discovered to date in the local universe.  Given the absence of MW GCs in this metallicity regime, these measurements suggest the tantalizing possibility that these six UFCSs -- representing roughly a third of our total sample -- could be the first examples of MW satellite galaxies with sizes $r_{1/2} \lesssim 15$~pc. We consider each case below, loosely ordered by increasing likelihood of a galaxy classification.  Throughout, we permissively refer to these systems as ``galaxy candidates'' on the basis of their low metallicities, but we stress that our discussion below stops well short of confirming any of these objects as galaxies at this time. We return to the question of what will be necessary to make definitive classifications for these candidates in \secref{outlook}.
\subsubsection{DELVE~3}
\par  The most tenuous candidate on this list is DELVE~3, for which the only available evidence supporting a dwarf galaxy classification is the metallicity of the system's brightest star (\DELVEthreefehbrightest{}). For this star, noise and sky-subtraction residuals around the 8662 $\rm \AA$ CaT line in our IMACS spectrum complicated our $\sum \rm EW_{\rm CaT}$ measurement (see \figref{imacs_ew_fits}), leading us to use the relations from \citet{2024ApJ...961..234H} to predict the EW of this line from the other two CaT lines' EWs. With this caveat in mind, DELVE~3's metallicity is well below the mean metallicity of any known MW GC, positioning the system as a high-priority target for more detailed follow-up. As for the system's stellar kinematics, our velocity dispersion limit for DELVE~3 is too weak to be discriminating (\DELVEthreevdisp{}), and no mass segregation constraints are available.
\par The clearest next step for testing DELVE~3's classification would be to obtain CaT metallicity measurements of additional members from deeper observations. Our current IMACS data confirm four RGB member stars in DELVE~3 (from just 2.7 hours of exposure time), but only the brightest had sufficient signal-to-noise for a reliable metallicity measurement. Obtaining metallicities for the three remaining faint RGB stars would reduce the sizable uncertainty on the system's mean metallicity and would enable straightforward tests of whether DELVE~3 exhibits an internal metallicity spread consistent with a dwarf galaxy classification.  
\subsubsection{Kim~1}
\par Kim~1 is similar to DELVE~3 in the sense that the primary evidence in support of a possible dwarf galaxy classification is the metallicity of its single RGB star (\Kimonefehbrightest{}). For Kim~1,  though, the evidence is significantly more robust due to the exquisite signal-to-noise of our DEIMOS spectrum of this star ($S/N > 100$; see \figref{deimos_ew_fits2}).\footnote{The very low metallicity of Kim~1's brightest star is further supported by a high-resolution Keck/HIRES spectrum obtained by our team, to be presented in future work.} This metallicity suggests that Kim~1 deviates from nearly all known GCs: Kim~1 must either be the lowest metallicity MW GC known by a small margin (below the current record-holder, ESO280-SC06, at $\rm [Fe/H] = -2.54\pm0.06$;  \citealt{2018MNRAS.477.4565S,usman2025}), or instead, the most compact known MW satellite galaxy. 
\par While \citet{2015ApJ...799...73K} originally suggested that Kim~1 is likely to be a disrupting star cluster based on its low central concentration and significant ellipticity, our new spectroscopic measurements also hint at the UFD scenario kinematically. We found a marginally-resolved velocity dispersion of \Kimonevdisp{} for the system, implying a mass-to-light ratio of \Kimonemlratio{}. However, this dispersion comes with two significant caveats. First, the associated Bayes factor of \Kimonevdispbayes{} is too small to provide meaningful evidence in favor of the free-dispersion model over a zero-dispersion model. Second, our jackknife tests suggested that removing a single star from Kim~1's kinematic member sample entirely removed the evidence for a non-zero velocity dispersion (yielding instead a tight upper limit of \KimonevdispJackknife{} with a Bayes factor \KimonevdispbayesJackknife{}). We therefore cannot draw firm conclusions about Kim~1's dark matter content until multi-epoch spectroscopic observations and/or mass segregation measurements become available.  Based exclusively on its brightest star's metallicity, then, we conclude that Kim~1 is a galaxy candidate worthy of further observations, though it remains very possible that the system is instead an unusually low-metallicity star cluster.

\subsubsection{Balbinot~1}
\par With a statistically-significant resolved velocity dispersion (\Balbinotonevdisp, with \Balbinotonevdispbayes) and an appreciable metallicity dispersion (\Balbinotonefehdisp), Balbinot~1 is one of just two systems in our UFCS sample satisfying both classical diagnostics of a dwarf galaxy classification laid out by \citet{2012AJ....144...76W}. Balbinot~1's low metallicity (\Balbinotonefehmodel) is also consistent with the population of MW UFDs at similar luminosity and lower than the metallicity of any known MW GC. These features, if confirmed, would imply that Balbinot~1 is a highly dark-matter-dominated dwarf galaxy (\Balbinotonemlratio{}). 
\par Challenging this conclusion, \citet{Baumgardt2022} detected stellar mass segregation in the system at high confidence ($P_{\rm seg.} = 99.8\%$). This result suggests that Balbinot~1's dynamical timescale is short enough to have allowed for relaxation, disfavoring the possibility that the system inhabits a dark matter halo. The tension between these mass segregation and velocity dispersion results is partially alleviated if the most significant velocity outlier was removed from our kinematic sample (see \secref{veloutliers}). In that case, only a marginally-resolved dispersion persisted (\BalbinotonevdispJackknife{}, with \BalbinotonevdispbayesJackknife{}), and the inferred mass-to-light ratio dropped to \BalbinotonemlratioJackknife{}. This test illustrates that our resolved dispersion for Balbinot~1 is more tentative than our nominal Bayes factor might suggest. By extension, we argue that the star cluster scenario for Balbinot~1 remains viable if our velocity dispersion estimates are currently biased by \textit{two or more} undetected stellar binaries. Given that this is not an especially unlikely scenario, we reserve further interpretation of Balbinot~1's stellar kinematics until multi-epoch velocity monitoring becomes available.
\par Independently, our measurement of a non-zero metallicity dispersion for Balbinot~1  supports the possibility that the system's potential well was deep enough to retain supernova ejecta and allow for self-enrichment, favoring a dwarf galaxy classification. However, we again consider this result tentative given the small sample of stars with measured metallicities (\BalbinotoneNfeh), one of which has a $\sim$3$\sigma$ \Gaia DR3 parallax detection that makes its membership somewhat uncertain (see \secref{membership}). In summary, while several lines of spectroscopic evidence clearly position Balbinot~1 as a viable dwarf galaxy \textit{candidate}, it remains possible that the system is a very-metal-poor star cluster. 
\subsubsection{DELVE~1 and Eridanus~III}
\par  DELVE~1 and Eridanus~III are the only two UFCSs with both medium-resolution and high-resolution spectroscopy available. Each system features a single bright RGB star with a very low metallicity: \DELVEonefehbrightest{} for DELVE~1 and \EridanusIIIfehbrightest{} for Eridanus~III, as inferred here from the CaT. High-resolution Magellan/MIKE spectra of these same two members, presented by our team in \citet{Simon2024}, independently confirm these metallicity estimates and further reveal that both stars display the CEMP-no abundance pattern. The low metallicities of these stars and their CEMP-no abundance patterns would be unprecedented among all known MW GC stars but would be relatively typical for UFD members. These chemical properties thus make DELVE~1 and Eridanus~III highly promising galaxy candidates. 
\par As for metallicity spreads, we found no statistically significant difference between the spectroscopic metallicities of the brightest two stars in each system (\DELVEonefehmaxdiff{} and \EridanusIIIfehmaxdiff{}, respectively). However, an independent \textit{HST} photometric metallicity analysis by \citet{2023ApJ...958..167F} found evidence for a $\sigma_{\rm [Fe/H]_{\rm phot}} = 0.49^{+0.22}_{-0.18}$~dex spread among 13 RGB members of Eridanus~III. At face value, this large photometric dispersion would be sufficient to declare Eridanus~III as a galaxy. However, we note an apparent discrepancy between the metallicity distributions derived photometrically and spectroscopically: \citet{2023ApJ...958..167F}  report no stars below $\rm [Fe/H] < -3.0$ in Eridanus~III and a much higher metallicity ($\rm [Fe/H] = -2.0$), while here (matching \citealt{Simon2024}) we report two members below  $\rm [Fe/H] < -3.0$ that set our mean metallicity for the system. These results are not formally in contradiction because the stellar samples used in both studies are small and cover entirely non-overlapping members, but they suggest that further observations are needed before a metallicity-based classification for Eridanus~III can be established with certainty. 
\par While the collection of chemical features described above is highly suggestive that both objects are galaxies, the dark matter contents of DELVE~1 and Eridanus~III remain ambiguous.  We failed to resolve a velocity dispersion in either system, with a modest limit of \DELVEonevdisp{} (with \DELVEonevdispbayes{}) for DELVE~1 and a non-constraining limit of \EridanusIIIvdisp{} (with \EridanusIIIvdispbayes{}) for Eridanus~III.  \citet{Baumgardt2022} reported tentative evidence for stellar mass segregation in Eridanus~III ($P_{\rm Seg.} = 91.7\%$), which would disfavor the presence of substantial amounts of dark matter. No such mass segregation measurements exist for DELVE~1. Given the strong chemical evidence suggesting that these systems are galaxies, it would be worthwhile to pursue significantly-higher-precision radial velocity observations of both systems.

\begin{deluxetable*}{cccccccc}
\tablewidth{\textwidth}
\tabletypesize{\small}
\tablecaption{Summary of the Evidence For/Against Galaxy or Star Cluster Classifications for the UFCSs \label{tab:qualitative}}
\tablehead{
Name & $\tau > 10\rm \ Gyr$ & $\rm [Fe/H] \lesssim -2.5$ & BF($\sigma_v > 0$) & BF($\sigma_{\rm [Fe/H]} > 0$) & Mass Segregated?\tablenotemark{a} & CEMP-no stars?\tablenotemark{b} & Classification \\
 &  &  & $2\ln\beta$ = & $2\ln\beta$ = & $P_{\rm Seg.}$ = &  & 
}
\startdata
BLISS 1 & \xmark & \xmark &  $-2.8$ & $-3.0$ & \ldots & \xmark & \footnotesize{Definite Star Cluster} \\
Segue 3 & \xmark & \xmark &  $-3.6$ & \ldots & \textsuperscript{\textdagger}75.9\% & \xmark & \footnotesize{Definite Star Cluster} \\
PS1 1 & \xmark & \xmark &  $7.0$ & $-3.0$ & \ldots & \xmark & \footnotesize{Very Likely Star Cluster} \\
Laevens 3 & \cmark & \xmark &  $-3.4$ & $-2.5$ & 96.6\% & \ldots & \footnotesize{Likely Star Cluster} \\
Koposov 1 & \xmark & \xmark &  $-2.6$ & \ldots & \textsuperscript{\textdagger}74.4\% & \xmark & \footnotesize{Very Likely Star Cluster} \\
YMCA-1 & \cmark & \xmark &  \ldots & $-1.8$ & \ldots & \ldots & \footnotesize{Leans Star Cluster} \\
Kim 3 & \xmark & \ldots &  $-3.2$ & \ldots & \textsuperscript{\textdagger}87\% & \ldots & \footnotesize{Leans Star Cluster} \\
Munoz 1 & \cmark & \xmark &  \ldots & \ldots & \ldots & \ldots & \footnotesize{Leans Star Cluster} \\
\hline
Draco II & \cmark & \cmark &  $-3.6$ & $3.2$ & 88.4\% & \ldots & \footnotesize{Likely Galaxy} \\
DELVE 1 & \cmark & \cmark &  $-3.2$ & \ldots & \ldots & \cmark & \footnotesize{Galaxy Candidate} \\
Eridanus III & \cmark & \cmark &  $-0.6$ & \ldots & 91.7\% & \cmark & \footnotesize{Galaxy Candidate} \\
Kim 1 & \cmark & \cmark &  $0.7$ & \ldots & \ldots & \ldots & \footnotesize{Galaxy Candidate} \\
DELVE 3 & \cmark & \cmark &  $-0.8$ & \ldots & \ldots & \ldots & \footnotesize{Galaxy Candidate} \\
Balbinot 1 & \cmark & \cmark &  $12.6$ & $4.5$ & \textsuperscript{\textdagger}99.8\% & \ldots & \footnotesize{Galaxy Candidate} \\
\hline
UMaIII/U1 & \cmark & \cmark &  $-3.7$ & $-3.0$ & \ldots & \ldots & \footnotesize{Ambiguous / Leans Cluster} \\
DELVE 4 & \cmark & \xmark &  $0.6$ & $17.3$ & \ldots & \ldots & \footnotesize{Ambiguous / Leans Galaxy} \\
Koposov 2 & \cmark? & \cmark? &  $0.9$ & \ldots & \textsuperscript{\textdagger}89.5\% & \ldots & \footnotesize{Ambiguous / Gal. Cand.?} \\
\hline
DELVE 5 & \xmark? & \ldots &  \ldots & \ldots & \ldots & \ldots & \footnotesize{Unconstrained} \\
DELVE 6 & \cmark & \xmark &  \ldots & \ldots & \ldots & \ldots & \footnotesize{Unconstrained} \\
\hline
\enddata
\tablecomments{Here, we divide the UFCSs by suspected classification: first likely star clusters, next dwarf galaxy candidates, then systems with status that is particularly ambiguous despite high-quality observations, and finally systems for which the available data does not constrain a classification. }
\tablenotetext{a}{Probabilities in the mass segregation column refer to the results of \cite{Baumgardt2022}; the sole exception is Kim 3 for which we adopt the result from \cite{2016ApJ...820..119K}. We flag constraints derived from samples of $<150$ stars with a \textdagger; these may be less robust.}
\tablenotetext{b}{Only DELVE~1 and Eridanus~III have carbon and neutron-capture measurements; however, we presmuptively exclude UFCSs at [Fe/H] $>-1$ from CEMP-no classification because of their high metallicities.}
\end{deluxetable*}

\subsubsection{Draco~II}
\par Lastly, and perhaps most likely to be a dwarf galaxy, is Draco~II. Draco~II is the best-studied object in our sample, with two prior spectroscopic studies, two independent analyses leveraging photometric metallicities, and two independent mass-segregation analyses. Our measurements confirm the low metallicity of Draco~II (\DracoIIfehmodel{}) and tentatively suggest a metallicity spread of \DracoIIfehdisp{} from $N_{\rm [Fe/H]} = 3$~RGB stars. This sample is too limited to discern Draco~II's true nature; however, more insightfully,  \citet{2023ApJ...958..167F}  presented the system's metallicity distribution based on deep \textit{HST} narrow-band photometry. From a sample of 38 (primarily MS) stars, many of which overlap with our velocity-confirmed member sample, they reported a mean metallicity of $\rm [Fe/H] = -2.72^{+0.10}_{-0.11}$ with a statistically-significant metallicity dispersion of $\sigma_{\rm [Fe/H]} = 0.40 \pm 0.12$~dex. This metallicity spread alone strongly favors a dwarf galaxy classification for the system. As with Eridanus~III above, spectroscopic measurements of the system's MDF based on a significantly larger member sample would be beneficial as a check on the photometric results. This is particularly true since the preceding photometric metallicity analysis of Draco~II based on shallower, ground-based CaHK imaging from the \textit{Pristine} survey yielded a contrasting result -- namely, an upper limit of $\sigma_{\rm [Fe/H]} < 0.24$~dex  \citep{2018MNRAS.480.2609L}.

\par Taking the resolved photometric metallicity dispersion from \textit{HST} at face value, the primary remaining source of classification uncertainty lies in Draco~II's stellar kinematics. Here, we report a strong velocity dispersion limit of \DracoIIvdisp{} at the 95\% credible level based on a large sample of \DracoIINmem{} stars. This limit is nearly twice as strong as the limit previously reported by \citet{2018MNRAS.480.2609L}, and translates to a mass-to-light ratio limit of \DracoIImlratio{}. While still insufficient to rule out the presence of dark matter, our updated mass-to-light ratio now clearly establishes that Draco~II is significantly less dark-matter-dominated than the most extreme dwarfs currently known (e.g., Segue~1). Interestingly, \citet{Baumgardt2022} reported the presence of stellar mass segregation in Draco~II, adding independent evidence for a low or absent dark matter content. These constraints together could support a scenario in which the system is a dwarf galaxy with a low dark matter density. This could arise if Draco~II has lost the majority of its mass through tidal stripping -- consistent with the tentative evidence for spatially-extended stellar features reported by \citet{2018MNRAS.480.2609L} and \citet{2024MNRAS.527.4209J}. 
\par In summary, given its low metallicity, the evidence for a metallicity spread from both spectroscopy and photometry,  and the plausible dynamical picture of Draco~II as a tidally-stripped satellite, we conclude that Draco~II is most likely a UFD. \newline

\subsection{UFCSs with the Most Ambiguous Classifications}
In addition to the six most promising galaxy candidates above, we identified three additional UFCSs that have features that plausibly resemble galaxies but for which the star cluster scenario is equally or perhaps even more likely: \uma{}, DELVE~4, and  Koposov~2. 
\subsubsection{\uma{}}
\par \uma{} stands out among our UFCS sample -- and the entire MW satellite population -- because of its extremely low stellar mass ($M_* = 21^{+8}_{-6} \; M_{\odot}$ assuming $M_*/L_V =2$). Initially, \citet{2024ApJ...961...92S} reported a large velocity dispersion of $\sigma_v = 3.7^{+1.4}_{-1.0}~\kms$ for the system from 11 stars, implying a high dark matter content of $M_{1/2}/L_{V,1/2} = 6500^{+9100}_{-4300}$~\MLunit{}. Contemporaneous dynamical modeling by \citet{2024ApJ...965...20E} further suggested that the system is not expected to survive in the MW tidal field for much longer than a single orbital period ($\sim 0.4$~Gyr) if it is comprised solely of stars, suggesting the need for a dark matter halo to stabilize the system. These lines of evidence positioned \uma{} as a promising galaxy candidate. However, \citet{2024ApJ...961...92S} also cautioned that the resolved velocity dispersion that they measured was sensitive to the inclusion of two velocity outliers, which, when removed, caused the dispersion to drop to an upper limit.
\par New kinematic and chemical data from DEIMOS and LRIS, presented in \citetalias{2025arXiv251002431C}~\citeyearpar{2025arXiv251002431C} and reproduced here, have significantly challenged the dwarf galaxy scenario for \uma{}. While we refer the reader to that work for a significantly more extended discussion, the latest Bayesian constraints on the system's velocity dispersion yielded a strong upper limit of \UrsaMajorIIIvdisp{} (with \UrsaMajorIIIvdispbayes{}) under conservative assumptions and priors. This lower dispersion result arises, in part, due to the confirmation and subsequent removal of the suspected spectroscopic binary identified by \citet{2024ApJ...961...92S}; the newer kinematic data also expanded the total member sample and substantially improved the per-star velocity precision. On the metallicity front, the sample of 12 stars with Ca II K metallicities from Keck/LRIS yielded an upper limit on the system's metallicity dispersion of \UrsaMajorIIIfehdisp{}.
\par Based on the diagnostics of \citet{2012AJ....144...76W}, these results suggest that there are no observed features of \uma{} that currently provide positive evidence for the presence of a dark matter halo. This tilts the balance toward a star cluster classification for the system. In further support of this possibility, both \citet{2025MNRAS.tmp..546D} and \citet{2025ApJ...989L..14R} presented dynamical models of \uma{} that could explain the long-term survival of the system in the star cluster scenario through the presence of a substantial population of stellar remnants arising through normal stellar evolution. 
\par This all being said, we stress that the dwarf galaxy scenario is not ruled out for \uma{}.  In fact, the system's low metallicity from the recent Keck/LRIS observations, \UrsaMajorIIIfehmodel{}(stat.) $\pm 0.3$ (zeropoint),  supports its viability as a galaxy \textit{candidate} along the same lines as the other UFCSs discussed above, though this line of evidence would benefit from a more secure determination of the Ca II K metallicity zeropoint. Kinematically, mass-to-light ratios as large as \UrsaMajorIIImlratio{} remain permitted, and several predictions of the dynamical modeling in the star cluster scenario (e.g., the presence of tidal tails) remain untested. Significant further investigation of the system's classification is still warranted.

\subsubsection{DELVE~4}
Again meriting a category of its own, DELVE~4 is a system for which we have relatively high-quality spectroscopic observations yet no clear classification picture due to an underlying uncertainty in the stellar membership. Should all seven suspected members we observe in the system at $v_{\rm hel} \approx -170$~\kms{} prove to be \textit{bona fide} RGB stars bound to the system, then the statistically-significant metallicity dispersion we measured (\DELVEfourfehdisp, with \DELVEfourfehdispbayes{} and a \textit{range} of \DELVEfourfehmaxdiff{}) would strongly favor a dwarf galaxy classification for the system.  We also found a marginally-resolved velocity dispersion for the system (\DELVEfourvdisp{}) that would support this conclusion, though this dispersion was much less significant (\DELVEfourvdispbayes) and vanished upon removal of a single star.
\par If it is the case that DELVE~4 is a dwarf galaxy, it is notable that DELVE~4's mean metallicity, \DELVEfourfehmodel{}, is somewhat higher than the galaxy candidates identified above ([Fe/H] $< -2.5$).  We speculate that this higher metallicity could imply DELVE~4 is the stripped remnant of a more massive dwarf; this hypothesis could plausibly also explain the presence of members at large radii and  the apparent excess of RGB stars in the system compared to expectation for its observed stellar mass.  In the alternative scenario in which only the three innermost stars at  $\rm [Fe/H] \approx -1.5$ are members, then DELVE~4 might represent yet another intermediate-metallicity accreted star cluster albeit one with no obvious intact progenitor galaxy. We maintain that this latter (star cluster) scenario seems less likely, as we found no evidence for interloping halo structures that could explain the velocities and proper motions of the putative member stars in DELVE~4 (\secref{membership}). 
\par Evaluating these two scenarios for DELVE~4 would strongly benefit from improved data on all fronts: deeper photometry to better establish the system's total stellar mass and mass function, more precise astrometry from \Gaia DR4 that will test whether the observed member proper motions are truly self-consistent, and an expanded set of spectroscopic observations that better sample the system's observed metallicity distribution. Spectroscopy of stars in the dense innermost region of DELVE~4 (within $1\arcmin$) with an integral-field unit would be particularly valuable for evaluating DELVE~4's metallicity distribution, as stars positioned near the system's centroid are presumably much less likely to be contaminants than those at larger radii.

\subsubsection{Koposov~2}
\label{sec:ko2}
Despite the availability of high-quality \textit{spectroscopic} data covering a sizable member sample ($N = \KoposovtwoNmem{}$), Koposov~2's classification remains unclear primarily due to uncertainty in its \textit{photometrically-derived} isochrone properties. Specifically, as discussed at length in \secref{membership}, our analysis suggested that estimates of the  age, metallicity, and distance of the system from prior isochrone fitting analyses yielded a poor fit to our velocity-selected member sample and notably missed several candidate RGB members. We subsequently determined that an older and more metal-poor stellar population isochrone (\Koposovtwoage{}, \Koposovtwoisofeh{}) better matches the observed color--magnitude sequence of these candidate velocity-selected members. Under this revised model, we identified two candidate RGB member stars, the brighter of which yielded a metallicity of \Koposovtwofehbrightest{} (assuming it is a member RGB star). The second star did not pass our EW-fit quality cuts despite its reasonable $S/N \approx 15$ spectrum, though we remark that the CaT lines are weak in this star's spectrum as well. 
\par If the membership and this low metallicity for its putative brightest RGB star were confirmed, then Koposov~2 would be a viable galaxy candidate along the same lines as our other targets with a single very-metal-poor RGB star (Kim~1 and DELVE~3). A galaxy classification for the system is potentially further supported by the tentative velocity dispersion in the system that we measured (\Koposovtwovdisp{}, with \Koposovtwovdispbayes{}), which implied a mass-to-light ratio of \Koposovtwomlratio{}. However, like Kim~1 before, this dispersion was found to drop to an upper limit of \KoposovtwovdispJackknife{} (with \KoposovtwovdispbayesJackknife{}) if a single member star was removed. \citet{Baumgardt2022} further reported the detection of stellar mass segregation in the system ($P_{\rm Seg.} = 89.5\%$).  Given the uncertainty in the underlying membership due to the poorly-constrained isochrone models, the possibility that any given star -- whether it be the single star driving the velocity dispersion or the putative brightest member star contributing the low metallicity -- is a non-member remains relatively high. We therefore conclude that neither the kinematic nor the chemical evidence for a galaxy classification is secure, and by extension, there remains a significant likelihood that the system is a star cluster. Rather than pursuing additional spectroscopy, we advocate that the photometric properties of Koposov~2 be re-investigated in significantly more detail, making use of the sample of stellar velocities presented here to mitigate the impact of contamination.

\subsection{UFCSs with Unconstrained Classifications: DELVE~5 and DELVE~6} 
\label{sec:unconstrained}
For two UFCSs -- DELVE~5 and DELVE~6 -- we were unable to draw any meaningful inferences about the relative likelihood of the dwarf galaxy vs. star cluster scenarios. In both cases, we are limited primarily by the small sample size and data quality (in contrast to the three UFCSs described in the preceding subsection, which are data-rich). For DELVE~6, we confirmed just three total members and measured a highly uncertain metallicity from its brightest star (\DELVEsixfehbrightest{}). This metallicity measurement has too large an uncertainty to permit meaningful conclusions and is consistent with both the most metal-rich UFD stars and many MW GCs. The system's ancient age from isochrone fitting, \DELVEsixage, is also consistent with both populations. For DELVE~5, even less information is available: the only stars that we identified as members were three faint MSTO stars from which we could not measure a metallicity nor a velocity dispersion; instead, we report only the system's mean radial velocity. Moreover, unlike for DELVE~6, the age of DELVE~5 is not well constrained from photometry, again preventing us from drawing conclusions about the system's classification from that angle. In each of these cases, deeper observations with the same instruments would go a long way toward resolving the existing ambiguity in their properties.

\subsection{Summary of the UFCS Classification Landscape}
\label{sec:landscape}
In \tabref{qualitative}, we synthesize the individual descriptions above by presenting age, metallicity, velocity dispersion, metallicity dispersion, mass segregation, and chemical abundance information in one place. 
\par The first eight UFCSs in \tabref{qualitative} are more likely to be halo star clusters than ultra-faint dwarf galaxies: BLISS~1, Segue~3, Koposov~1, PS1~1,  Mu\~{n}oz~1, Laevens~3, YMCA-1, and Kim~3.  Our classifications for these UFCSs are primarily based on their higher mean metallicities and/or relatively younger ages ($\rm -0.8 \lesssim [Fe/H] \lesssim -2$ and $3 \rm \ Gyr \lesssim \tau \lesssim 10$~Gyr; first two columns of \tabref{qualitative}). This is because our upper limits on the velocity and metallicity dispersions (when available) are insufficient to definitively classify these systems. These likely halo star clusters represent $\gtrsim 40\%$ (8/19) of our sample.  If we further include the classical GCs Palomar 13 and AM 4, both of which meet our UFCS definition, then we find that at least half (10/21) of the known MW UFCSs with spectroscopic data available are likely to be star clusters.

\par The next six UFCSs in \tabref{qualitative} are those that we identified as the most promising galaxy candidates: Balbinot~1, DELVE~1, DELVE~3, Draco~II, Eridanus~III, and Kim~1. All of these systems (${\sim}1/3$ of our sample) share low metallicities ($\rm [Fe/H] \lesssim -2.5$) that place them on or near the dwarf galaxy LZR but substantially below nearly all GCs. Their ancient stellar population ages from isochrone fitting ($\tau > 10$~Gyr) match the known MW UFDs. Three of these systems (Balbinot~1, Draco~II, and Eridanus~III) exhibit metallicity spreads, either from our spectroscopy or from the photometric metallicity analysis by \citet{2023ApJ...958..167F}. However, only Balbinot~1 has a resolved velocity dispersion; the other five galaxy candidates have only upper limits on their velocity dispersions, dynamical masses, and mass-to-light ratios. This emphasizes the difficulty of using stellar kinematics to classify the UFCSs with current facilities.

\par The last five UFCSs in \tabref{qualitative} remain particularly ambiguous. The first three, \uma{}, DELVE~4, and Koposov~2, are viable galaxy candidates based on either their very low mean metallicities (\uma{}, and, tentatively, Koposov~2) or metallicity spreads (DELVE~4). Including these three UFCSs with the six galaxy candidates above would raise the fraction of UFCSs in our sample that are plausibly galaxies to $\sim$50\% (9 of 19).  However, these systems exhibit other lines of evidence or alternative stellar membership interpretations that permit -- or even favor -- the star cluster scenario. Finally, the last two UFCSs in \tabref{qualitative} are those for which the current spectroscopic and photometric evidence is insufficient to constrain a classification: DELVE~5 and DELVE~6. Our spectroscopic datasets for these systems are particularly poor, with just $\sim$3 members each. We were unable to measure even a single stellar metallicity at useful precision in either case, and our spectroscopic member samples were too small to enable measurements of these systems' velocity or metallicity dispersions. 

\par In summary, we find that $\gtrsim 40$--50\% of the UFCSs are likely low-mass star clusters, $\sim$30--50\% are viable UFD candidates, and the remainder are ambiguous. Of the six classification criteria outlined in \secref{diagnostics}, our census results indicate that the age and mean (spectroscopic) metallicity of a given UFCS are currently the most discriminating. While resolved velocity and metallicity dispersions would normally serve as the gold standard for classification, such measurements remain challenging for UFCSs due to their low masses. Similarly, the mass segregation detections summarized in \tabref{qualitative} do not clearly correlate with the UFCSs we identified as most likely to be galaxies, suggesting that this dynamical diagnostic may be less reliable (see \secref{outlook} for a more extended discussion). Viewed more optimistically, the ability to identify likely star clusters based on their age and metallicity alone implies that even shallow reconnaissance observations can effectively exclude targets when searching for the least-massive galaxies among UFCSs discovered in ongoing and upcoming photometric surveys.

\par Finally, we note that there are $\sim$11 additional UFCSs in the literature with no spectroscopy whatsoever (see \tabref{supplementalproperties}). While these systems have similar absolute magnitudes and distances as our spectroscopic sample (see \figref{fig1_population}), we find no clear correlations between the heliocentric distances, sizes, or luminosities of the UFCSs and our assigned classifications. We thus emphasize the critical need for spectroscopy to classify these remaining $\sim$11 systems and all future UFCS discoveries.

\section{Discussion}
\label{sec:disco}

\subsection{The Dynamical Stability of the UFCSs in the Stellar-Only Scenario}
\label{sec:stellaronly}
The survival of self-gravitating systems evolving in the MW potential is governed both by internal two-body interactions and by the Galactic tidal field \citep[e.g.,][]{1942psd..book.....C,1957ApJ...125..451V,1997ApJ...474..223G,2000MNRAS.318..753F,2003MNRAS.340..227B,2011MNRAS.413.2509G}. Given these processes, the long-term survival of the UFCSs may be surprising if they truly lack dark matter: low-mass, compact star clusters have shallow potential wells and short crossing times and are therefore expected to be particularly vulnerable to evaporation and tidal destruction. On the other hand, if the UFCSs are instead dwarf galaxies, then their compact sizes and higher densities could promote their survivability compared to more spatially-extended UFDs.  

\par While exploring the tidal influences on the UFCSs requires their full orbits and is reserved for Paper II, the stability of the UFCSs against relaxation and evaporation from two-body interactions can be estimated at the order-of-magnitude level from their stellar masses and radii alone. Following a similar approach to past works studying UFCSs \citep[e.g.,][]{2007ApJ...669..337K,2024ApJ...965...20E}, we computed the timescale over which the UFCSs are expected to dynamically relax ($t_{\rm relax}$) if they are truly collisional, self-gravitating systems comprised solely of stars; these calculations assumed that the UFCSs are in isolation and thus are independent of the MW potential and neglect tidal effects.  We first computed the crossing time of each UFCS, 
\begin{align}
    t_{\rm cross} \approx (G\rho_{1/2,*})^{-1/2},
\end{align}
where $\rho_{1/2,*} \equiv (M_*/2)/(4/3 \pi R_{\rm 3D}^3)$ under the assumption that each system is spherical, uniform density, and characterized by its deprojected 3D half-light radius $R_{\rm 3D} \equiv  1.305 r_{1/2}$ (for a Plummer profile; see \citealt{2010MNRAS.406.1220W}). The relaxation time was then computed from $t_{\rm cross}$ according to the relation 
\begin{align}
t_{\rm relax} \approx \left(\frac{N_*}{8 \ln N_*}\right) t_{\rm cross}
\end{align}
where $N_*$ is the total number of stars in each UFCS \citep[e.g.,][]{2008gady.book.....B}. We assumed a fixed ratio $N_* = M_*/(0.3 M_{\odot})$, consistent with simulated SSPs generated with the \texttt{artpop} \citep{2022ApJ...941...26G} package assuming a Kroupa initial mass function \citep{2001MNRAS.322..231K}; the age dependence of this ratio was found to be minimal for the range covered by our sample.
\par
\begin{figure*}
    \centering
    \includegraphics[width=0.9\textwidth]{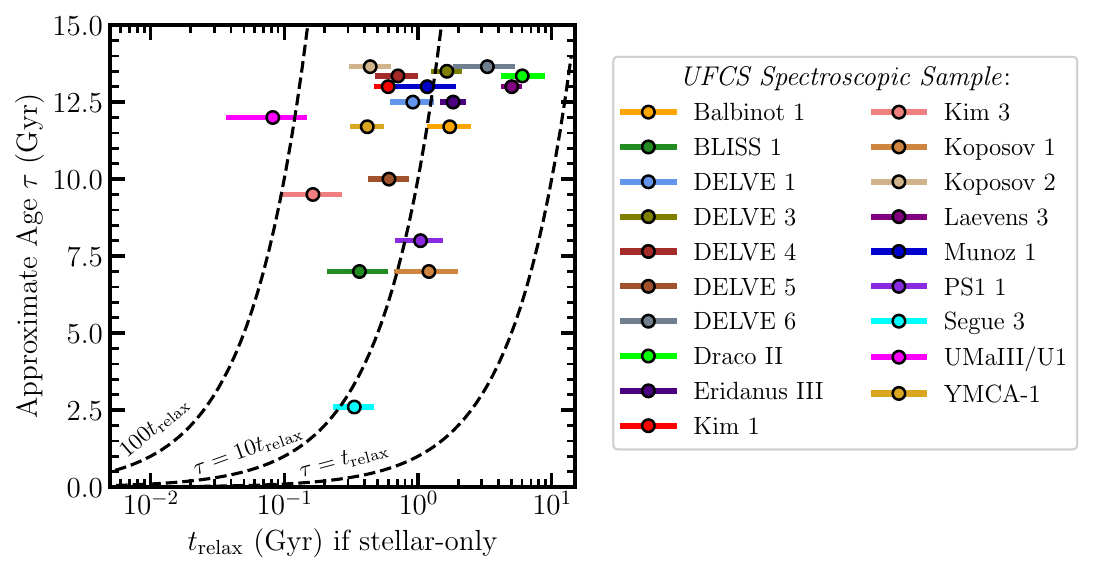}
    \caption{\textbf{Comparing the stellar population ages of the UFCSs to their expected relaxation timescales if they are comprised solely of stars}. Age estimates are taken from literature isochrone fitting analyses and generally lack reliable uncertainties; we omit age errorbars here. For visualization purposes only, systems with reported isochronal ages of exactly 13.5 Gyr have been jittered $\pm 0.15$ Gyr in age to avoid spatial overlap. In black, we overplot contours of $\tau = [1,10,100] \times t_{\rm relax}$. The UFCSs' ages span $4\,t_{\rm relax} \lesssim \tau \lesssim 200 \,t_{\rm relax}$ in this stellar-only scenario, implying that they are dynamically evolved and predicting that they should ubiquitously exhibit stellar mass segregation if they are not dark-matter-dominated dwarf galaxies. Moreover, because the majority of UFCS ages are $\tau \approx 8$--$30\,t_{\rm relax}$, it is likely that most UFCSs are actively evaporating if they lack dark matter halos (see discussion in \secref{stellaronly} for details of these calculations).}
    \label{fig:evaporation}
\end{figure*}

 In \figref{evaporation}, we summarize these calculations by comparing the relaxation timescales of the UFCSs to their ages derived from isochrone fits in the literature.  In black, we overplot contours of  $\tau = [1, 10, 100] \times t_{\rm relax} $ to guide the eye. It is immediately clear from this comparison that all UFCSs should be dynamically relaxed, as they cluster around $\sim$8--$30\,t_{\rm relax}$ and extend to $\sim$$100\,t_{\rm relax}$ in two cases (\uma{} and Kim~3). 
The median $t_{\rm relax}$ for our sample of UFCSs is just 0.9 Gyr, which can be compared with the age of the youngest UFCS in our sample, \Seguethreeage{} (for Segue~3). This large difference emphasizes that age measurement uncertainties -- expected to be of order $\sim$2~Gyr for most UFCSs in our sample -- cannot explain this pattern. In fact, our calculations suggest that even the UFCS in our sample with the longest relaxation time, Draco~II, has $t_{\rm relax} \approx 5$~Gyr -- less than half its reported age of \DracoIIisofeh{}.
\par The most immediate consequence of this dynamical relaxation is mass segregation: all the UFCSs should exhibit clear differences in the radial distribution of lower and higher-mass stars if they lack dark matter. A secondary consequence of this relaxation is expansion, which, in the presence of an external tidal field, leads to gradual mass loss and eventual evaporation. A very rough heuristic for the timescale over which self-gravitating stellar systems lose stars due to two-body interactions is the evaporation timescale, $t_{\rm evap} \approx 140 \, t_{\rm relax}$ \citep{2008gady.book.....B}. Because $t_{\rm relax}$ decreases over time as stars are re-distributed outwards and lost to tides (recall that $t_{\rm relax} \propto \frac{N_*}{\ln N_*}$), applying this heuristic to the calculated $t_{\rm relax}$ from above effectively sets an upper limit on the remaining lifetimes of the UFCSs. Consequently, the UFCSs in our sample -- which commonly exhibit $10 \, t_{\rm relax} < \tau < 100 \, t_{\rm relax}$ -- are likely already losing mass through these processes and in some cases (\uma{} and Kim~3) may be close to complete evaporation if they are purely stellar. The possibility of core collapse and the retention of stellar-mass black holes, neither of which we considered in the calculations above, could further accelerate relaxation and evaporation.  \par While we expect that a more accurate picture of the UFCSs' dynamical evolution in the purely-stellar scenario will require direct $N$-body modeling, we explore whether observational signatures of mass loss are present in our data in the following subsection.

\subsection{Distant Member Candidates: \\ Evidence for Spatially-Extended Stellar Populations, \\ or Foreground Contamination?}
\label{sec:distantmembers}
Stars in the outskirts of the MW's star clusters and satellite galaxies contain rich information about their host populations' dynamical evolution and assembly histories. For the MW GCs in particular, a combination of matched-filter searches, \Gaia astrometry, metallicity-sensitive photometry, multi-object spectroscopy, and RR Lyrae variable stars has led to the discovery of tidal tails, streams, and ``runaway'' stars associated with numerous globular clusters \citep[e.g.,][]{1995AJ....109.2553G, 2000A&A...359..907L, 2001ApJ...548L.165O, 2002AJ....124..349R,2010A&A...522A..71J,2018ApJ...862..114S,2019NatAs...3..667I,2019MNRAS.488.1535P,2024ApJ...967...89I, 2025A&A...697A...8M,2025arXiv251014924C, 2025AJ....170..294C}. For the MW's dwarf satellite galaxies, a similar range of techniques has been used to detect spatially-extended stellar populations suggestive of tidal influences as well as extended dark matter halos \citep[e.g.,][]{2006ApJ...649..201M,2010AJ....140..138M, 2012ApJ...756...79S,2020ApJ...902..106M,2021NatAs...5..392C,2021ApJ...923..218F,2023MNRAS.525.3086L, 2024ApJ...966...33O, 2024MNRAS.527.4209J,2023MNRAS.525.2875S,2024AJ....167...57T}.

\par Detecting these features around the UFCS is especially challenging owing to these systems' extraordinarily sparse stellar populations, which are already overwhelmed by the MW foreground in most current survey datasets. Illustratively, \citet{2023RNAAS...7..127S} ran particle-spray models of Segue~3 in a realistic orbit and found that the implied surface density of member stars expected in the magnitude range $21 < G < 23$ if the system exhibits tidal tails is $\sim$20~$\rm deg^{-2}$ -- to be compared to a density of contaminating field stars of $\sim$20,000~$\rm deg^{-2}$. In qualitative agreement, \citet{2015ApJ...799...73K} performed simplistic $N$-body simulations of Kim~1 in the MW potential (prior to the availability of proper motions). These simulations suggested that the system should exhibit sparse tidal tails that span hundreds of parsecs ($\mathcal{O}(1^{\circ})$  on-sky) and are completely drowned out by the higher-surface-density foreground population.
\par Despite these challenges, there are promising hints of spatially-extended stellar populations in our spectroscopic datasets -- both statistically and for individual UFCSs. Across our sample of 19 UFCSs, \UFCSNsystemsBeyondFourahalf{} systems host at least one likely or confirmed member star at $r_{\rm ell} \geq 4a_{1/2}$, with a collective \UFCSNdistant{} member stars falling beyond this threshold out of our complete sample of \UFCSNmemTotal{} member stars (\UFCSNdistantfraction{}\%). This is roughly double the number expected if one assumes each UFCS obeys a standard \citet{1911MNRAS..71..460P} radial profile and stacked the radial distribution of members across our 19-UFCS sample ($\sim$$6\%$). While our spectroscopy generally oversamples the UFCSs' outer regions due to slit collision constraints near their centers, and while UFCS size measurements carry appreciable uncertainties, we view this high frequency of distant members as an important clue pointing toward larger spatial extents. 
\par Close scrutiny of individual UFCSs' distant members further supports the possibility of spatially-extended stellar populations. The most conspicuous example in our sample is Segue~3, which features a diffuse cloud of stars with consistent velocities that extends as far as $r_{\rm ell} = \Seguethreemostdistant{} \, a_{1/2}$ (with 10 stars at $>4 \, a_{1/2}$). This cloud is sufficiently well-populated and separated from the bulk of the foreground velocity distribution that it likely cannot be solely attributed to foreground contamination. Interestingly, this extended cloud also displays a larger velocity dispersion than the inner population within $3\,a_{1/2}$; this is potentially consistent with expectations for a disrupting cluster.  We refer the reader to \citet{2011AJ....142...88F} for a significantly more detailed discussion of this feature of Segue~3. Another prominent case is  that of DELVE~4, for which we identified a trail of three velocity-consistent stars at $>4\,a_{1/2}$ extending to the south of the system.  This feature resembles a possible tidal tail; however, as we discussed at length in \secref{membership}, the membership of these three stars is somewhat tentative. Finally, we highlight the case of DELVE~1, which hosts a \textit{single} distant member at $\sim$$\DELVEonemostdistant{}\, a_{1/2}$ (\figref{delve1}). While the membership of such a star might normally be ambiguous, the large velocity separation of all DELVE~1 member stars from the empirical MW foreground population sampled by our spectroscopy ($>\!150$~\kms{}) largely rules out the possibility that this star is an interloper.

\par The interpretation of the remaining six UFCSs with candidate member stars at $\geq \! 4a_{1/2}$ (Balbinot~1, Kim~1, Kim~3, Koposov~2, Mu\~{n}oz~1, and PS1~1) was less clear-cut.  Focusing on a subset of three of these cases for which the membership in the central regions was relatively unambiguous, namely Kim~1, Kim~3, and Balbinot~1, we found candidate distant members at $\sim$$\Kimonemostdistant{}\, a_{1/2}$, $\sim$$\Kimthreemostdistant{}\, a_{1/2}$, and $\sim$$\Balbinotonemostdistant{}\, a_{1/2}$ (see \secref{membership}). Unlike the star in DELVE~1, these individual distant member candidates are less clearly separated from the empirical velocity distribution of MW foreground/background stars in each case.  Therefore, to assess the significance of these features, we queried the web interface to the Besancon Galaxy Model version m1612 \citep{2003A&A...409..523R} to sample the expected interloper population within a square region of total solid angle $5$~deg$^2$ centered on each system. We then replicated our isochrone selections using the SDSS photometry provided by the model assuming magnitude limits approximately matched to our spectroscopic datasets. Next, we applied our loose velocity selections as described in \secref{membership}; no parallax or proper motion selections were used because many of our member stars fall below the \Gaia magnitude limit.
After applying these filters, we estimated the number of interlopers expected within an area equivalent to the observed DEIMOS or IMACS fields-of-view ($\sim$64$ \rm \ arcminutes^2$ or $\sim$100$ \rm \ arcminutes^2$, respectively) and limited to the apparent magnitude range spanned by the spectroscopic member samples for each target.
\par For Kim~1, the model predicts $\sim$1.1 MW field stars meeting our selection function. Thus, while stars with properties similar to our distant member candidate are clearly rare,  we cannot yet rule out the possibility that our member candidate at \Kimonemostdistant{}$\, a_{1/2}$ is an interloper contaminant. For Balbinot~1,  the predicted number of stars in our DEIMOS field is $\sim$2.3, indicating an appreciable risk that we could mistake a foreground contaminant as a member; speculatively, this implies one or more interlopers could be driving our observed velocity and metallicity dispersions. Lastly, the expected number within the IMACS FOV for Kim~3 is $\sim2.0$, implying the same conclusion as for Balbinot~1. If we instead narrowed the velocity range used to downselect stars from the Besancon model to $\pm 10$~\kms{} about our measured mean velocity for each system, the number of expected MW field stars was found to be $[0.5,1.1,1.2]$ for the three systems. Under this stronger assumption, the feature in Kim~1 becomes more significant at a level worthy of attention while for the other targets the  qualitative conclusions outlined above remain largely the same. 
\par Clearly, this exercise demonstrates that we cannot yet link these distant member candidates to their putative UFCS hosts with high confidence.  We nonetheless stress that this does not mean that these candidate members are false positive detections of spatially-extended stellar populations; instead, our analysis here suggests that these stars must be further scrutinized before they can be interpreted confidently as signatures of disequilibrium. Metallicity measurements for each of the stars would be a useful tool for confirming or disputing their membership status, as would a larger sample of velocity measurements in each field from which we could evaluate the level of contamination empirically (as we could with DELVE~1). Complementarily, proper motions from wide-field, space-based observatories (\textit{Roman} or \textit{Euclid}) or from deep, multi-epoch ground-based surveys such as the Rubin Observatory LSST would also quickly clarify whether our distant candidates have proper motions consistent with membership. Finally, a more extensive exploration of different radial profile models for the UFCSs (beyond the Plummer models considered here) with deeper photometry would be highly beneficial for understanding the behavior of their stellar populations at large radii. 
\par The uncertainty in some of these features notwithstanding, we argue that both the elevated occurrence rate of distant member candidates across all 19 UFCS member datasets, as well as the specific cases discussed here more extensively, lend new credence to the existing body of evidence for evaporation and/or disruption of the UFCSs reported throughout the literature \citep[e.g.,][]{,2011AJ....142...88F,2015ApJ...803...63K,2015ApJ...799...73K,2024MNRAS.527.4209J}.  By extension, these observations support a scenario in which the UFCSs were born with higher masses than are presently observed. We explore the possible connection between these features and the orbital properties of the UFCSs in Paper II. 

\subsection{Connecting the UFCSs to the Population of  Classical Milky Way GCs}
\label{sec:connection}
Thus far, we have primarily centered our analysis around the possibility that a subset  of the UFCSs could represent dwarf galaxies approaching the low-mass threshold of galaxy formation. However, our results have also suggested that approximately half or more of the UFCSs in our sample are more likely star clusters (\secref{landscape}). Provided this is the case, the origins of these very-low-mass star clusters remain poorly understood. In this subsection, we briefly contextualize these UFCSs within our current picture of the MW GC Luminosity Function / GC Mass Function (GCLF/GCMF) and explore what it implies for their formation and dynamical evolution. 
\par The present-day GCLF of the MW and other massive hosts is observationally characterized by a peaked logarithmic distribution with a turnover at an absolute magnitude of $M_V \approx -7.5$ (mass $M_* \approx 2 \times  10^5 \ M_{\odot}$). This form of the GCLF/GCMF is believed to be the product of cluster formation obeying a Schechter-like or power-law cluster initial mass function modified by cluster dissolution through two-body interactions, tidal stripping and shocking, dynamical friction, and stellar evolution \citep[e.g.,][]{2001ApJ...561..751F,2008ApJ...689..919P,2009MNRAS.394.2113G,2023MNRAS.522.5340G}. The direct expectation from this picture is that the number of clusters at the faint, low-mass end should continuously decrease below the turnover, and that the (surviving) systems found in this regime should commonly display signatures of mass loss.
\par
\begin{figure}
    \centering
    \includegraphics[width=0.45\textwidth]{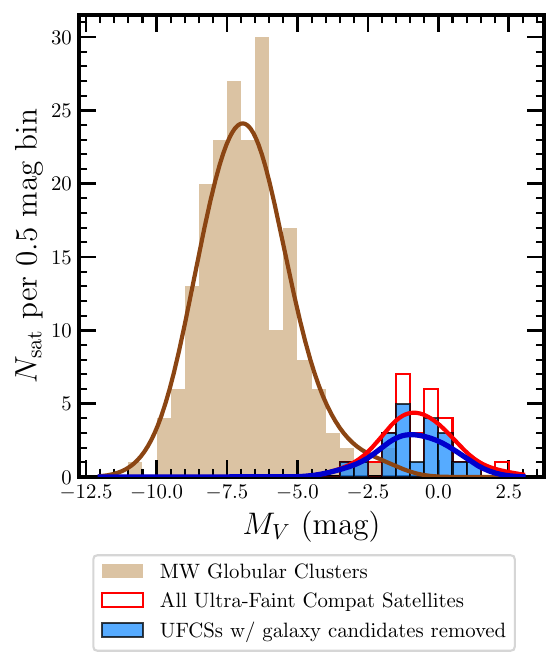}
    \caption{\textbf{A UFCS-driven excess of low-mass clusters in the MW GCLF.} In tan, we plot a histogram of the absolute magnitudes of MW GCs from the Local Volume Database (in 0.5 mag bins). In red outline, we display a similar histogram for the 30 known UFCSs compiled in \secref{sample}. Finally, in blue, we show the same UFCS histogram again but with the permissive set of nine systems that we labeled as galaxy candidates excluded (including the three most ambiguous cases; see \tabref{qualitative}). A Gaussian kernel density estimate (bandwidth 0.75 mag) is overlaid on each histogram. The excess of faint satellites persists even after removing the plausible galaxies, indicating that this is likely a real feature of the MW star cluster population. }
    \label{fig:gclf}
\end{figure}

 The predicted continuously-decreasing luminosity function is in tension with observations of the UFCSs. As we highlight in \figref{gclf}, the presence of more than two dozen systems clustered around an absolute-magnitude (mass) of  $M_V \approx -1$ ($M_* \approx  400 \rm \ M_\odot$) is clearly inconsistent with the tail of a log-normal distribution. This clear excess hints at the possibility that the MW GCLF is bimodal or has a plateau at the faint end, though we do not attempt to parametrize it here because the underlying distribution is likely multi-dimensional.  While this feature has arguably been visible for years in the $M_V$--$r_{1/2}$ plane (see e.g., \figref{fig1_population}), our new classifications uniquely allow us to demonstrate that this excess persists even if we generously assume that all nine of the UFCSs that we flagged as plausible galaxy candidates (six more reliable candidates plus the three ambiguous cases; see \tabref{qualitative}) are indeed galaxies and excise them. In other words, the observed deviation from a lognormal GCLF cannot be explained by simply assuming some fraction of the UFCSs are galaxies. Selection effects, which are expected to primarily depend on $M_V$ for the faintest satellites with a relatively weak size dependence in this regime \citep[e.g.,][]{2025OJAp....8E..89T}, cannot explain the ``gap'' at intermediate luminosities and would point to an even larger population of faint clusters. We therefore conclude that this excess appears to be a real feature of the MW star cluster population.

\par The origin of this excess of low-mass clusters is unclear and points to a gap in our understanding of GC formation and/or destruction.  Undoubtedly, some of the objects at the faint, compact end of the MW satellite population represent the dissolving/disrupting GCs expected from the canonical picture of the GCLF. For one clear example, the metal-poor GC Palomar 13 ($M_V = -3.3$; \citealt{2020PASA...37...46B}) -- which lies within our UFCS selection -- has been demonstrated to have chemical abundances consistent with halo GCs \citep{2019A&A...632A..55K,2021ApJ...908..220T} and displays numerous lines of evidence pointing to substantial stellar mass loss including unambiguous tidal tails \citep{2001AJ....121..935S,2011ApJ...743..167B,2013AJ....146..116H,2020AJ....160..244S}. Another clear example is the very low mass, compact halo cluster ESO37-1/E3 (\citealt{1976A&A....52..309L}; $M_V = -3.6$, \citealt{2020PASA...37...46B}) that is well established to be a GC on the basis of its high (and homogeneous) iron abundance and again displays multiple signatures of disruption including tidal tails \citep{1980ApJ...239..112V,McClure_1985,2015A&A...581A..13D,2020MNRAS.499.2157C}. Palomar~1 ($M_V = -1.7$) and Whiting~1 ($M_V = -4.2$) may be similar examples, though the evidence for disruption in these cases is perhaps less clear \citep{2007A&A...466..181C,2010MNRAS.408L..66N,2022MNRAS.513.3136Z,2025arXiv251014924C}. Broadly speaking, these systems clearly demonstrate a mass-loss-driven evolutionary pathway for GCs that yields UFCS-like systems at present day. The short relaxation times of the UFCSs and the possible presence of spatially-extended stellar populations (see preceding subsections) further support the notion that mass loss processes are indeed operating for (some of) the UFCSs in our sample. 
\par Though this destruction-based explanation can explain some of the star-cluster UFCSs, the deviation from the expected lognormal GCLF tail suggests that the mass-loss-driven channel cannot be a complete explanation for the UFCSs' origins. Instead, we speculate that this GCLF excess could arise if the UFCS region of the size-luminosity plane includes a significant number of star clusters formed with low initial masses within massive dwarf galaxies ($M_* \approx 10^6$--$10^{10} \rm \ M_{\odot}$) that have recently accreted onto the MW. Our results here and in Paper II suggest that $\gtrsim 20\%$ of the UFCSs in our sample plausibly originated with the LMC/SMC (YMCA~1 and DELVE~6) or the Sagittarius dSph (Koposov~1 and PS1~1), with several more accreted with now-disrupted dwarf galaxies.\footnote{These connections align with the substantial body of evidence suggesting that $\gtrsim$40\% of the MW GCs were formed ex-situ in low-mass galaxies that were subsequently accreted onto the Galactic halo \citep[e.g.,][]{2010MNRAS.404.1203F,2019A&A...630L...4M,2024MNRAS.528.3198B,2024OJAp....7E..23C}.}  These three massive dwarf galaxies experienced extended star formation histories, plausibly yielding a population of clusters that form later and are presently observed to be younger. Crucially, by virtue of their ex-situ formation and their recent accretion, the UFCSs born in these galaxies should generally have experienced a gentler tidal field that could have enabled their survival against dissolution and disruption. By contrast, comparably-low-mass clusters that formed in-situ or accreted with earlier mergers would have preferentially been destroyed (see e.g., Section 3.3 of \citealt{2019MNRAS.482.5138B}).  Full orbital histories and more precise ages for the UFCSs in our sample,  as well as the discovery of more UFCSs, will enable more detailed tests of this proposed scenario. It would also be interesting to search for analogs around nearby, isolated, low-mass hosts directly, as has recently become possible with surveys such as \textit{Euclid} \citep[see e.g.,][for state-of-the-art efforts]{2025arXiv250910440H,2025A&A...703A.113L}.
\par In any case,  the presence of this apparent excess of halo clusters at the faint end of the MW GCLF strongly motivates joint modeling of the masses, distances, orbits, and chemical properties of the low-mass cluster population in the MW halo. Clusters in this mass regime have often been omitted from simulation-based work due to resolution and memory limits and/or uncertainties in the physics of low-mass GC evolution \citep[e.g.,][]{10.1093/mnras/stx3124,2024MNRAS.527.3692C}, and few predictions exist for the total number of MW GCs down to the faint extreme (see \citealt{2017MNRAS.466.1741C} and \citealt{2021MNRAS.502.4547W} for two of the only such predictions).  Finally, as this excess relates to efforts to complete the census of MW satellite galaxies and probe the threshold of galaxy formation, we emphasize that the clear confirmation of a sizable population of low-mass clusters removes any hope of morphologically distinguishing dwarf galaxies from clusters in the UFCS regime at the time of discovery. Spectroscopic follow-up will continue to be essential for the classification of the faintest systems, as we elaborate on in the subsection below.

\subsection{Outlook for Classifying the Metal-Poor UFCSs}
\label{sec:outlook}
Although our spectroscopic metallicity measurements provide strong support for the possibility that a subset of the UFCSs are plausibly galaxies, the available kinematic measurements are far from directly discriminating these UFCSs' classifications through their dark matter contents. Indeed, for UFCSs with upper limits on their velocity dispersions, even our strongest constraints permit mass-to-light ratios as high as $M_{1/2}/L_{V,1/2} \approx$ 350--1000$ \; M_{\odot}/L_{\odot}$ at the 95\% credible level. Even the small number of UFCSs with resolved or inconclusive/marginally-resolved dispersions in our data cannot (yet) be unambiguously interpreted as evidence for the presence of dark matter given the strong impact of individual member stars that could be binaries or interlopers (\secref{veloutliers}). Given these limitations, a natural question is: \textit{what will it take to definitively classify the metal-poor UFCSs?} In the subsections below, we describe the prospects for both dynamical and chemical inferences with current and future instrumentation and ultimately provide recommendations for follow-up efforts to classify the UFCSs.

\subsubsection{Possibilities and Pitfalls for Dynamical Constraints}
\par Further radial velocity observations with higher precision are a natural first avenue to consider, as dynamical mass measurements are the gold standard for classifying ultra-faint satellites. At present, we have far from saturated the capabilities of DEIMOS and IMACS owing to our relatively short integration times for many targets. This reflects our prioritization of obtaining a first mean velocity and metallicity measurement for as many systems as possible as opposed to focusing on any specific target.  Additional, deeper observations with IMACS and DEIMOS would certainly help strengthen the current velocity dispersion constraints by pushing toward these instruments' systematic floors of $\sim$1~\kms{} for a large fraction of members. These data would also allow the identification of binaries, potentially offering an immediate test of whether the resolved/marginally-resolved dispersions that we identified are truly robust.
\par Observations with fiber-fed multi-object spectrographs on 8m-class telescopes,  which do not suffer from slit-miscentering effects, would provide for a more substantial improvement. For example, analyses with VLT/FLAMES+GIRAFFE \citep{2000SPIE.4008..129P,2002Msngr.110....1P},  MMT/Hectochelle \citep{2011PASP..123.1188S}, and Magellan/M2FS \citep{2012SPIE.8446E..4YM} have demonstrated the possibility of measuring velocities down to a floor of $\sim$600$ \rm \ m \ s^{-1}$ at high $S/N$ \citep[e.g.,][]{2011ApJ...736..146K,2018MNRAS.481..645Z,2021ApJ...920...92J,2023ApJS..268...19W,2023A&A...675A..49T} and the next-generation multi-object spectrograph ViaSpec\footnote{\url{https://via-project.org/}} (expected first light 2027) is anticipated to achieve precisions of $\sim \!100 \rm \ m \ s^{-1}$. One cautionary note is that these high precision measurements have historically been achieved only for RGB stars, whereas the accessible populations of UFCS members are dominated by MSTO and MS stars. The weak lines in these stars -- particularly for those at low-metallicity and those with warmer effective temperatures -- may make achieving these precisions challenging in practice. Fiber density constraints may also make achieving the requisite precision across large member samples particularly costly given the small angular sizes of the UFCSs.
\par Outside of pursuing higher-precision stellar velocities, an expanded, deeper census of stellar mass segregation with \textit{HST} and/or \textit{JWST} could provide meaningful constraints on the dark matter contents of the UFCSs. Space-based mass segregation constraints already exist for two of our metal-poor galaxy candidates -- Draco~II and Eridanus~III -- and favor mass segregation, though the measurement for Draco~II is based on a sample of $150$ stars \citep{Baumgardt2022} -- insufficient to yield definitive conclusions. Similarly, the existing mass segregation constraint for Balbinot~1 (from the same work) was based on ground-based CFHT/MegaCam photometry of just $141$ stars. Deeper observations of these targets would not only increase the member sample size but also would expand the range of stellar masses probed, yielding more robust constraints on mass segregation.  The other three UFCSs that we identified as most likely to be low-mass galaxies (DELVE~1, Kim~1, and DELVE~3) have no existing space-based imaging, nor deep enough ground-based imaging to achieve the requisite number of stars; however, they are within reach of space-based facilities and could be prioritized in the  future.\footnote{Observations of DELVE~1 are planned for February--March 2026 as part of  \textit{HST} Program \#17787 \citep{2024hst..prop17787C} and several of the UFCSs that we identify as likely star clusters have recently been imaged by \textit{HST} program \#17435 \citep{2023hst..prop17435M,2025arXiv250201741M}. } 
\par As a byproduct of mass segregation, stellar systems experiencing mass loss should also exhibit a shallower/flatter stellar luminosity/mass function owing to the preferential loss of low-mass stars from their outskirts.   This flattening is prominent in many classical MW GCs' mass functions \citep[e.g.,][]{1991AJ....102.1026P,1999A&A...343L...9D,2001AJ....122.3231G,2017MNRAS.472..744B,2024A&A...690A.371L} but should be less significant (or entirely absent) in highly dark-matter-dominated dwarf galaxies, which are less vulnerable to mass segregation and tides. The low-mass stellar luminosity/mass function of individual UFCSs can  therefore potentially serve as a classification diagnostic that again relies on photometric data alone. As quantified by \citet{2025MNRAS.tmp..546D} based on simulated observations, a key advantage of this approach over mass segregation alone is that significantly fewer (photometric) member stars are required to measure a mass function slope. Thus, while this classification technique has yet to be put into practice, \textit{HST} and \textit{JWST} should be able to carry out this measurement for every UFCS currently known.

\par While all three of the above signatures (velocity dispersions, mass segregation, and mass function slopes) provide critical context and should be pursued observationally, we caution that they each become significantly more challenging to interpret for faint, compact systems undergoing mass loss. For example, tidal evolution models applied to the known MW satellite galaxy population predict a continuous decrease in central velocity dispersion and size as these galaxies disrupt \citep{2008ApJ...673..226P,2022MNRAS.511.6001E}. In the asymptotic limit, this stripping process can yield galaxies with mass-to-light ratios that approach those expected from systems comprised only of stars \citep{2024ApJ...965...20E}. Observationally, this prediction finds support from the growing body of measurements suggesting that the MW satellite galaxies experiencing tidal disruption have systematically \textit{lower} central velocity dispersions than their non-disrupting counterparts \citep{Kirby_2013,Simon_2017,2017ApJ...839...20C,2018MNRAS.479.5343K,2021ApJ...921...32J,2024ApJ...966...33O,2025arXiv251202177L}. This implies that even the \textit{absence} of a significant dispersion cannot unambiguously signify a star cluster classification. As first argued in \secref{galaxycandidates}, tidal stripping could be one explanation for our non-detection of a velocity dispersion in Draco~II -- a system for which the likely presence of a stellar metallicity spread favors a dwarf galaxy classification -- though insufficient velocity precision undoubtedly remains a contributing factor.
\par Similar complications potentially apply to mass segregation and mass function classifications. Recent dynamical modeling has shown that mass segregation can potentially arise in the UFCSs even if they do have small dark matter contents ($M/L \lesssim 10$--100) because of their short dynamical times \citep{2025arXiv250522717E}. Mass-to-light ratios in this regime would not be unprecedented for UFDs (see for example the case of Hydrus I; \citealt{2018MNRAS.479.5343K}), and such low mass-to-light ratios are a defining characteristic of the proposed population of globular-cluster-like dwarfs found in the EDGE simulations by \citet{2025arXiv250909582T}. Low mass-to-light ratios could provide one explanation for the detection of mass segregation in Balbinot~1, Draco~II, and Eridanus~III -- all of which we have identified as plausible galaxy candidates (despite their low velocity dispersions / velocity dispersion upper limits) based primarily on their stellar metallicities. 
\par Finally, with regard to mass function slope tests, we warn of the strong dependence on the orbital history of a given object which may complicate efforts to classify UFCSs. As also noted by \citet{2025MNRAS.tmp..546D}, UFCSs that have experienced gentler tidal fields (e.g., recently-accreted systems and/or those in the outer halo) will display less significant depletion of their stellar mass functions; in the limiting case of a first-infall UFCS, it is likely the case that the absence of this depletion cannot be used to rule out a star cluster classification. Indeed, the current (limited) body of observational mass function constraints available for the faintest GCs/UFCSs suggests that the population does not universally exhibit depleted stellar-mass functions -- even among clear star clusters. Specifically, for the UFCS-like GCs Palomar 13 and AM~4, the former shows the expected depletion of low-mass stars while the latter's mass function is nearly consistent with an unevolved, steeper \citet{1955ApJ...121..161S} IMF \citep{2013AJ....146..116H}. Koposov~1 and Koposov~2, while not yet definitively established as star clusters, likewise were found to have Salpeter-like mass functions from ground-based photometry  that would imply minimal mass loss \citep{2014AJ....148...19P}. Palomar 13 is the closest to the Galactic center of all four of these systems, and, unlike the other three systems, is known to be heavily disrupted \citep{2001AJ....121..935S,2020AJ....160..244S}. This ensemble of results suggests that the signature of stellar mass function depletion may only be clearly detected in systems with the most severe mass loss.

\subsubsection{Insights from Chemical Abundances}
\par As an alternative avenue independent of stellar dynamics, chemical abundance measurements from high-resolution spectroscopy  ($\mathcal{R} \gtrsim  20000$) for  UFCS member stars could also provide rich information about these systems' classifications and origins. For example, this type of data would yield higher-precision metallicities and probe whether the UFCSs show the CEMP-no abundance pattern seen in many UFD stars, or, alternatively, the light element correlations seen in many GC stars (see \secref{diagnostics} for a more extended discussion). This approach is primarily limited by the scarcity of bright UFCS members: only $\sim$\UFCSNhiresTotal{} of our  \UFCSNmemTotal{} UFCS members are bright enough for efficient high-resolution spectroscopy with current instruments (adopting a magnitude limit of $g_0 = 19.0$ as a threshold, which corresponds to the rough limit achievable in a single night of 8-m class telescope time).  These \UFCSNhiresTotal{} stars are not distributed evenly across our sample of UFCSs, and a sizable subset of our targeted UFCSs -- as well as the majority of the UFCSs \textit{not} in our sample here -- have no targets amenable for high-resolution spectroscopy before the era of the extremely large telescopes. 
\par Pursuing the brightest individual members, where possible, would nonetheless be valuable for downselecting the most promising galaxy candidates from among our sample of metal-poor UFCSs.  One word of caution for this chemical approach is that targeting only the single brightest star in each UFCS cannot necessarily distinguish between a scenario in which a UFCS is a low-mass galaxy or whether it simply formed in one; this is because clusters are generally expected to track their host galaxies' chemical enrichment patterns. Thus, observations of \textit{multiple} stars are required to test for chemical homogeneity. This underpins our hesitation in definitively classifying DELVE~1 and Eridanus~III as galaxies: while their brightest stars clearly exhibit the low neutron-capture abundances that are the hallmark of UFDs \citep{Simon2024}, it remains possible that both systems are chemically homogeneous star clusters comprised exclusively of stars with this pattern. 

\subsubsection{Searching for Metallicity Variations with Low-Resolution Spectroscopy}
Given the limitations of the approaches above, we advocate for a third, distinct approach to definitively classifying the metal-poor UFCSs in our sample: searching for metallicity and/or chemical abundance variations between stars from \textit{low-resolution, blue-optical} spectroscopy.  The detection of a statistically-significant metallicity difference between even just two UFCS member stars would be sufficient to confirm a galaxy classification so long as the stellar membership is unambiguous. By contrast, a much larger sample of stars with high-precision velocity measurements -- across multiple epochs -- would be required to measure a robust, statistically-significant velocity dispersion that can be  confidently interpreted as the signature of an unseen dark matter halo.
\par While we favored medium-resolution observations covering the red-optical CaT region to maximize velocity precision with DEIMOS and IMACS, this spectral region is information-poor for MSTO and MS stars. By moving to the blue and covering the strong, broad Calcium H\&K lines at $\sim$3930$ \rm \AA $, it is possible to recover iron abundances for faint member stars in most UFCSs at a useful precision ($\sim$$0.3$~dex) with equivalent-width-based calibrations \citep[e.g.,][]{1985AJ.....90.2089B,1999AJ....117..981B,2008A&A...484..721C}. Importantly, this is feasible at \textit{low} spectral resolution ($R\approx 1000$), alleviating the major challenge of achieving the requisite $S/N$ (see e.g., \citealt{2018ApJ...856..142C}; \citealt{2025arXiv250616462L}; Bissonette et al., in prep for recent applications in Local Group dwarf galaxies). Low-resolution, blue-optical spectra would also enable carbon abundance measurements from the CH $G$-band at $\sim$$4300\rm \AA$; this may prove to be a useful secondary approach if more UFCSs are found to host CEMP stars (see discussion in \citealt{Simon2024}). The main foreseeable challenge for this technique is the relatively weak lines expected from metal-poor stars near the MSTO and the corresponding need for high signal-to-noise spectra ($S/N > 25$) to ensure reliable abundances. Nonetheless, the efficiency gained from moving to low resolution puts this within grasp, and we have demonstrated the feasibility of this approach for the case of \uma{} (\citetalias{2025arXiv251002431C} \citeyear{2025arXiv251002431C}). We have begun pursuing this Ca K-based approach for an expanded range of UFCSs and expect it to become an increasingly relevant technique for classifying UFCSs (and other MW satellites) discovered with the Vera C. Rubin Observatory's Legacy Survey of Space and Time, the \textit{Nancy Grace Roman Space Telescope}, and the \textit{Euclid} Wide Survey. 

\section{Summary and Conclusions}
\label{sec:summary}
We have presented the first dedicated spectroscopic census of the Milky Way's Ultra-Faint Compact Satellites (UFCSs) -- an enigmatic population of extremely low mass, small stellar systems which we defined as halo satellites ($|Z| > 5$~kpc) with $r_{1/2} < 15$~pc, $M_V > -3.5$, and $\mu > 24~$mag~$\rm arcsec^{-2}$ (as illustrated in \figref{fig1_population}). Using both new and archival medium-resolution Keck/DEIMOS and Magellan/IMACS spectroscopy, we identified \UFCSNmemTotal{} member stars across 19 UFCSs (\figref{balbinot1}--\figref{ymca1}). Fifteen of these UFCSs lacked any published spectroscopic velocity and metallicity measurements prior to the initiation of our census, and our sample encompasses (and in most cases, improves upon) all existing medium-resolution spectroscopy of the UFCSs. Using this substantial new observational dataset, we characterized the stellar kinematics and chemistries of the 19 systems.  Our primary conclusions are as follows: \newline

\textbf{(1)} We successfully measure the mean radial velocity of all 19 UFCSs in our sample and derive new \Gaia-based mean proper motions for 18 of these systems. Collectively, these measurements significantly expand the number of ultra-faint Milky Way satellites with complete 6D phase-space information, enabling study of their orbital histories in Paper II and beyond. \\

\textbf{(2)} We place constraints on the velocity dispersions of 15 UFCSs in our sample with $\geq 5$ kinematic member stars. For 10 of these 15 satellites, we find unresolved velocity dispersions yielding upper limits of $\sigma_v < 2.5$--$8.7$~\kms{} (at the 95\% credible level). These upper limits demonstrate that the UFCSs are kinematically colder, on average, than most ultra-faint dwarf galaxies. However, even mass-to-light ratios of $1000$~\MLunit{} are not yet excluded for most of our sample, and higher-precision velocity measurements will be necessary before the stellar kinematics of the UFCSs can be used to constrain their dark matter contents and classifications. \\

\textbf{(3)} We find resolved velocity dispersions at the $\sim$2--4~\kms{} level for just two UFCSs in our sample (Balbinot~1 and PS1~1) and marginally-resolved dispersions for three additional UFCSs (DELVE~4, Kim~1, and Koposov~2).  While these resolved dispersions could potentially signify the presence of dark matter, jackknife tests suggest that these five cases are significantly affected by individual member stars. These resolved dispersions may therefore be biased by undetected stellar binaries or membership contamination. Multi-epoch monitoring and expanded spectroscopic member samples will be critical before these kinematic measurements can be reliably interpreted as evidence for the presence of dark matter. \\

\textbf{(4)} We derive mean iron abundances for 18 UFCSs, finding a remarkably wide range of $\rm -3.3 \lesssim  [Fe/H] \lesssim -0.8$.  This significant chemical diversity provides strong evidence that the UFCS population is comprised of more than one class of objects.  We identify six UFCSs that could represent unprecedentedly-low-mass galaxies based primarily on very low metallicities that place them at or below the metallicity floor for GCs ($\rm [Fe/H] \lesssim -2.5$): Balbinot~1, DELVE~1, DELVE~3, Draco~II, Eridanus~III, and Kim~1. We also find that \uma{} and Koposov~2 are similarly metal-poor, though the former is more likely a star cluster and our metallicity measurement for the latter is tentative due to membership uncertainty. \\

\textbf{(5)} We detect internal spectroscopic metallicity variations in just three UFCSs in our sample -- Balbinot~1, DELVE~4, and Draco~II -- with particularly strong evidence in DELVE~4 (subject to caveats in its stellar membership). These dispersions identify the three systems as particularly promising galaxy candidates, though efforts to classify these systems would strongly benefit from additional spectroscopic metallicity measurements.  Our tentative detection of a metallicity dispersion in Draco~II is consistent with an independent photometric estimate from  \citet{2023ApJ...958..167F}. \citet{2023ApJ...958..167F} also reported a photometric metallicity dispersion for Eridanus~III, but we were unable to test this spectroscopically due to our small metallicity sample for the system. \\

\textbf{(6)} We argue that eight UFCSs in our sample are more likely star clusters than galaxies based primarily on their relatively higher metallicities ($\rm -2 \lesssim [Fe/H] \lesssim -0.8$) and/or younger stellar population ages ($3~\rm Gyr \lesssim \tau \lesssim 10~Gyr$): BLISS~1, Kim~3, Koposov~1, Laevens~3, Mu\~{n}oz~1, PS1~1, Segue~3, and YMCA-1. We present preliminary evidence linking several of these systems to the LMC/SMC and the Sagittarius dSph; we explore the accretion histories and origins of these satellites further with their full orbits in Paper II.  \\

\textbf{(7)} Through both an analysis of our new spectroscopic member samples and a review of prior photometric studies, we present tentative evidence that $>1/3$ of the UFCSs in our sample show spatially-extended stellar populations. These features may have arisen from evaporation driven by two-body interactions or tidal disruption and could indicate that the UFCSs formed at higher initial masses than observed today. This conclusion holds regardless of the UFCSs' classifications as dwarf galaxies or star clusters. \\

\textbf{(8)} We draw attention to an excess of halo star clusters at the faint end of the Milky Way GC luminosity function, centered near $M_V \approx -1$, which emerges when the UFCSs are interpreted as star clusters. This excess is inconsistent with the Milky Way GC luminosity function having a lognormal tail. Although this feature has been noted before,  we demonstrate here for the first time that it cannot be explained solely by assuming that some fraction of the faintest compact satellites are galaxies. The origin of this feature is presently unknown; however, we speculate that it may indicate a class of recently-accreted low-mass star clusters that have evolved in relatively gentle tidal fields throughout their lifetimes.  \\

\textbf{(9)} We advocate for \textit{low-resolution, blue-optical
multi-object spectroscopy} focused on resolving iron abundance dispersions as the most promising approach for classifying the UFCSs. The existence of significant metallicity spreads -- even across small samples of secure member stars -- is an unambiguous signifier of a galaxy classification. By contrast, dynamical classifications based on velocity dispersions, stellar mass segregation, or mass function slopes are challenging for the faintest and most compact systems. \\

\par In broad terms, our census strongly suggests that the UFCSs are a complex superposition of both UFDs and faint star clusters in the $M_V$--$r_{1/2}$ plane. Disentangling these two populations based on morphology alone is clearly infeasible, but our results illustrate that significant progress toward understanding the natures of the faintest Milky Way satellites can be made through dedicated follow-up spectroscopy complemented by proper motions and deep ground-based photometry. While it remains the case that there are no \textit{confirmed} Milky Way satellite galaxies smaller than a half-light radius of 15 pc at the conclusion of our analysis, there is a clear path forward toward resolving the status of many of these systems, and thus we optimistically foresee the reclassification of at least some UFCSs as galaxies in the near future.

\section*{Data Availability}
\label{sec:data_avail}
This work is accompanied by a Zenodo repository through which data and posterior chains will be released on a staggered basis: \href{https://doi.org/10.5281/zenodo.18612485}{doi:10.5281/zenodo.18612485}.  The initial public version of this repository accompanying the release of our \texttt{arxiv} preprint (v1.1) includes our complete catalogs of spectroscopic members in each UFCS along with the associated documentation. The next repository update, anticipated at the time of journal publication, will include non-member catalogs as well as posterior chains for our key analyses. Interested users of these products are encouraged to contact the corresponding author. 

\section*{Acknowledgments}
The authors thank the staffs of the W.M. Keck Observatory and the Las Campanas Observatory, without whose expertise and support this work would not have been possible. We also thank the Principal Investigators of DEIMOS observing programs which contributed data that we have re-reduced here (\tabref{obstable}) as well as members of the \texttt{PypeIt} team for their enabling contributions to the reduction of our DEIMOS data. We also thank Ian Roederer and Alice Luna for their contributions to Magellan data collection associated with this long-term project. W.C. thanks Maya Waarts for her patience and support throughout the course of this long-term project.
\par W.C. gratefully acknowledges support from a Gruber Science Fellowship at Yale University. This material is based upon work supported by the National Science Foundation Graduate
Research Fellowship Program under Grant No. DGE2139841. Any opinions, findings, and conclusions or
recommendations expressed in this material are those
of the author(s) and do not necessarily reflect the views
of the National Science Foundation. T.S.L. and J. B. acknowledge financial support from Natural Sciences and Engineering Research Council of Canada (NSERC) through grant RGPIN-2022-04794. OYG was supported in part by National Aeronautics and Space Administration through contract NAS5-26555 for Space Telescope Science Institute program JWST-GO-03433. IE acknowledges financial support from programs HST GO-15891, GO-16235, and GO-16786, provided by NASA through a grant from the Space Telescope Science Institute, which is operated by the Association of Universities for Research in Astronomy, Inc., under NASA contract NAS 5-26555.

\par Some of the data presented herein were obtained at Keck Observatory, which is a private 501(c)3 non-profit organization operated as a scientific partnership among the California Institute of Technology, the University of California, and the National Aeronautics and Space Administration. The Observatory was made possible by the generous financial support of the W. M. Keck Foundation. The authors wish to recognize and acknowledge the very significant cultural role and reverence that the summit of Maunakea has always had within the Native Hawaiian community. We are most fortunate to have the opportunity to conduct observations from this mountain. 
\par This research has also made use of the Keck Observatory Archive (KOA), which is operated by the W. M. Keck Observatory and the NASA Exoplanet Science Institute (NExScI), under contract with the National Aeronautics and Space Administration. This paper also includes data gathered with the 6.5 meter Magellan Telescopes located at Las Campanas Observatory, Chile. 
\par This project used data obtained with the Dark Energy Camera, which was constructed by the Dark Energy Survey (DES) collaboration. Funding for the DES Projects has been provided by the DOE and NSF (USA),   MISE (Spain),   STFC (UK), HEFCE (UK), NCSA (UIUC), KICP (U. Chicago), CCAPP (Ohio State), MIFPA (Texas A\&M University),  CNPQ, FAPERJ, FINEP (Brazil), MINECO (Spain), DFG (Germany), and the collaborating institutions in the Dark Energy Survey, which are Argonne Lab, UC Santa Cruz, University of Cambridge, CIEMAT-Madrid, University of Chicago, University College London, DES-Brazil Consortium, University of Edinburgh, ETH Z{\"u}rich, Fermilab, University of Illinois, ICE (IEEC-CSIC), IFAE Barcelona, Lawrence Berkeley Lab, 
LMU M{\"u}nchen, and the associated Excellence Cluster Universe, University of Michigan, 
NSF's National Optical-Infrared Astronomy Research Laboratory, University of Nottingham, 
Ohio State University, OzDES Membership Consortium
University of Pennsylvania, University of Portsmouth, 
SLAC National Lab, Stanford University, 
University of Sussex, and Texas A\&M University.
 
\par The Legacy Surveys consist of three individual and complementary projects: the Dark Energy Camera Legacy Survey (DECaLS; Proposal ID \#2014B-0404; PIs: David Schlegel and Arjun Dey), the Beijing-Arizona Sky Survey (BASS; NOAO Prop. ID \#2015A-0801; PIs: Zhou Xu and Xiaohui Fan), and the Mayall z-band Legacy Survey (MzLS; Prop. ID \#2016A-0453; PI: Arjun Dey). DECaLS, BASS and MzLS together include data obtained, respectively, at the Blanco telescope, Cerro Tololo Inter-American Observatory, NSF’s NOIRLab; the Bok telescope, Steward Observatory, University of Arizona; and the Mayall telescope, Kitt Peak National Observatory, NOIRLab. Pipeline processing and analyses of the data were supported by NOIRLab and the Lawrence Berkeley National Laboratory (LBNL). 
The Legacy Surveys project is honored to be permitted to conduct astronomical research on Iolkam Du’ag (Kitt Peak), a mountain with particular significance to the Tohono O’odham Nation.

\par NOIRLab is operated by the Association of Universities for Research in Astronomy (AURA) under a cooperative agreement with the National Science Foundation. LBNL is managed by the Regents of the University of California under contract to the U.S. Department of Energy.

BASS is a key project of the Telescope Access Program (TAP), which has been funded by the National Astronomical Observatories of China, the Chinese Academy of Sciences (the Strategic Priority Research Program “The Emergence of Cosmological Structures” Grant \# XDB09000000), and the Special Fund for Astronomy from the Ministry of Finance. The BASS is also supported by the External Cooperation Program of Chinese Academy of Sciences (Grant \# 114A11KYSB20160057), and Chinese National Natural Science Foundation (Grant \# 12120101003, \# 11433005).

The Legacy Survey team makes use of data products from the Near-Earth Object Wide-field Infrared Survey Explorer (NEOWISE), which is a project of the Jet Propulsion Laboratory/California Institute of Technology. NEOWISE is funded by the National Aeronautics and Space Administration.

\par The Legacy Surveys imaging of the DESI footprint is supported by the Director, Office of Science, Office of High Energy Physics of the U.S. Department of Energy under Contract No. DE-AC02-05CH1123, by the National Energy Research Scientific Computing Center, a DOE Office of Science User Facility under the same contract; and by the U.S. National Science Foundation, Division of Astronomical Sciences under Contract No. AST-0950945 to NOAO.

\par The Pan-STARRS1 Surveys (PS1) and the PS1 public science archive have been made possible through contributions by the Institute for Astronomy, the University of Hawaii, the Pan-STARRS Project Office, the Max-Planck Society and its participating institutes, the Max Planck Institute for Astronomy, Heidelberg and the Max Planck Institute for Extraterrestrial Physics, Garching, The Johns Hopkins University, Durham University, the University of Edinburgh, the Queen's University Belfast, the Harvard-Smithsonian Center for Astrophysics, the Las Cumbres Observatory Global Telescope Network Incorporated, the National Central University of Taiwan, the Space Telescope Science Institute, the National Aeronautics and Space Administration under Grant No. NNX08AR22G issued through the Planetary Science Division of the NASA Science Mission Directorate, the National Science Foundation Grant No. AST–1238877, the University of Maryland, Eotvos Lorand University (ELTE), the Los Alamos National Laboratory, and the Gordon and Betty Moore Foundation.

\par This work has made use of data from the European Space Agency (ESA) mission
{\it Gaia} (\url{https://www.cosmos.esa.int/gaia}), processed by the {\it Gaia}
Data Processing and Analysis Consortium (DPAC,
\url{https://www.cosmos.esa.int/web/gaia/dpac/consortium}). Funding for the DPAC
has been provided by national institutions, in particular the institutions
participating in the {\it Gaia} Multilateral Agreement.

\par This research uses services or data provided by the Astro Data Lab \citep{2014SPIE.9149E..1TF,2020A&C....3300411N}, which is part of the Community Science and Data Center (CSDC) Program of NSF NOIRLab. NOIRLab is operated by the Association of Universities for Research in Astronomy (AURA), Inc. under a cooperative agreement with the U.S. National Science Foundation.

\par This work has made use of the Local Volume Database\footnote{\url{https://github.com/apace7/local_volume_database}} \citep{2025OJAp....8E.142P}.

% \newpage
\facilities{Keck: II (DEIMOS), Magellan:Baade (IMACS), \Gaia, Blanco (DECam)}
\software{\texttt{astropy} \citep{2013A&A...558A..33A,2018AJ....156..123A}, \texttt{numpy} \citep{2011CSE....13b..22V,2020Natur.585..357H}, \texttt{scipy} \citep{2020NatMe..17..261V}, \texttt{ugali} \citep{2015ApJ...807...50B}, \texttt{ezpadova}, \texttt{matplotlib} \citep{2007CSE.....9...90H}, \texttt{pandas} \citep{2022zndo...3509134T}, \texttt{dynesty} \citep{2020MNRAS.493.3132S,sergey_koposov_2024_12537467}, \texttt{emcee} \citep{2013PASP..125..306F}}

\bibliography{main}{}
\bibliographystyle{aasjournal}

\appendix 

\onecolumngrid
\section{Compilation of Literature Properties}
\label{sec:litpropertydetails}
\subsection{Procedures for Computing Posteriors and Summary Statistics}
As introduced in \secref{sample}, we compiled a new catalog of the properties of the UFCSs that we make use of throughout our figures, tables, and analyses in this work (and in Paper II). One major goal in creating this compilation, rather than simply using that presented in the Local Volume Database \citep{2025OJAp....8E.142P}, is to provide a means of approximating to the \textit{full posterior distributions} for certain parameters with asymmetric uncertainties. These posteriors could then be propagated into our downstream dynamical and chemical measurements. In addition, our analyses often required making assumptions about missing measurements and/or measurement uncertainties and thus we explicitly state these assumptions below. \newline

\par \textbf{Centroid RA/DEC Positions}:  UFCS centroid positions (i.e., right ascensions $\alpha_{2000}$ and declinations $\delta_{2000}$) are relevant to the determination of stellar membership (e.g., for the calculation of $r_{\rm ell}$) and for our orbit analyses in Paper II. Uncertainties on these position components are commonly omitted in the literature. Therefore, if no uncertainties were provided, we assumed a $\pm 0.003\degree$ error in each component. For all other cases, we imposed a minimum uncertainty of $\pm 0.001\degree$, though quoted uncertainties are often larger than this. For the special case of Balbinot~1, the reported centroid from \citet{2013ApJ...767..101B} is known to be offset from the system's true center, and we therefore adopted the revised value from the Local Volume Database  with our default uncertainty of $0.003\degree$. \newline

\par \textbf{Half-Light Radii and Ellipticities}: Two different radius definitions are frequently used throughout the literature: the elliptical half-light radius / semi-major axis length, which we call $a_{1/2}$, and the azimuthally-averaged/circularized half-light radius, which we call $r_{1/2}$. In the frequently-used \citet{2008ApJ...684.1075M} formalism, the former ($a_{1/2}$) is often referred to as the half-light radius and $r_{1/2}$ (in our nomenclature) is not provided. In these cases -- representing the majority of literature studies -- we recomputed the azimuthally-averaged $r_{1/2}$ as $r_{1/2} \equiv a_{1/2}\sqrt{1-\epsilon}$.  These two definitions become equivalent for $\epsilon = 0$. Many UFCSs were found to have no reported ellipticities in the literature (e.g., because only circular profile models were fit) or only upper limits; in these cases we assumed a fixed $\epsilon = 0$ with no uncertainty. For BLISS~1, where no uncertainty was quoted \textit{and} the ellipticity was very small ($\epsilon = 0.06$), we rounded down to $\epsilon = 0$ to be consistent with our treatment of upper limits. If no uncertainty on $a_{1/2}$ was provided, as was the case for the UFCSs from \citet{2019MNRAS.484.2181T}, we assumed a fixed 25\% fractional uncertainty. \newline

\par \textbf{Distances}:  Many photometric studies of the UFCSs derive distance moduli through isochrone fitting and then transform these distance moduli to heliocentric distances, reporting both. This is a non-linear transformation, and thus the posterior distributions on distances are expected to be non-Gaussian. By default, we adopted the reported distance moduli and carried out the transformation to distances ourselves. If only a distance was quoted, we derived the distance modulus and its uncertainty ourselves for the sake of membership selection and used the resultant posterior based on our general sampling procedure described above. If no distance modulus uncertainties nor distance uncertainties were provided (as was the case for the satellites reported by \citealt{2019MNRAS.484.2181T}), we assumed a $\pm 0.2$ mag Gaussian distance modulus uncertainty. In \tabref{litproperties} and \tabref{supplementalproperties} and throughout the main text, we report distances \textit{exactly as reported in the literature}; it is only at the posterior level that our estimates vary.  \newline

\par \textbf{$V$-band Magnitudes, Luminosities, and Stellar Masses}: Although estimates of absolute $V$-band magnitudes for the UFCSs are inhomogeneously derived throughout the literature, we took literature absolute magnitudes $M_V$ and uncertainties at face value. If no uncertainty was provided, we assumed a  $\pm 0.5$ mag Gaussian uncertainty. In all cases, we used the $M_V$ posteriors to homogeneously recompute integrated $V$-band luminosities from reported absolute magnitudes according to 
\[
L_V = 10^{0.4(4.83-M_V)}
\] (even if $L_V$ estimates were directly provided). Stellar masses were then derived from these $L_V$ estimates assuming a stellar mass-to-light ratio of $M_*/L_V = 2$, appropriate for an old stellar population with a Chabrier or Kroupa IMF \citep[e.g.,][]{2017MNRAS.464.2174B}. We caution that each of these quantities is subject to significant systematics that may not be fully captured by the quoted uncertainties.  \newline

\par \textbf{Ages from Isochrone-Fitting}: 
We make use of isochrone-derived ages for determining stellar membership, informing UFCS classifications, and contextualizing the UFCSs' relaxation times.  In a number of cases (e.g., Koposov~1, Koposov~2, and Segue~3), we found it necessary to critically evaluate constraints from literature studies with discrepant results; see \secref{membership} for discussion of individual objects.  For other cases, if ages were quoted as lower limits (typically due to isochrone grid boundaries), we assumed a fiducial 13.5 Gyr.  Our analysis did not make use of age uncertainties; however, we estimate that $\sim$1--2~Gyr precision is realistic across our sample.  \newline

\par \textbf{General Treatment of Asymmetric Uncertainties:}
Rather than assuming Gaussian distributions when sampling based on measurements with asymmetric uncertainties, we instead sampled from a split-normal distribution using the Python package \texttt{twopiece}\footnote{https://github.com/quantgirluk/twopiece}. Split-normal distributions arise from joining two normal distributions with the same mode but different variances, with a probability density function defined as 
\[
p(x; \mu, \sigma_1, \sigma_2) =
\begin{cases} 
\frac{\sqrt{2}}{\sqrt{\pi} (\sigma_1 + \sigma_2)} \exp\left(-\frac{(x - \mu)^2}{2\sigma_1^2}\right), & x < \mu, \\
\frac{\sqrt{2}}{\sqrt{\pi} (\sigma_1 + \sigma_2)} \exp\left(-\frac{(x - \mu)^2}{2\sigma_2^2}\right), & x \geq \mu.
\end{cases}
\]
For quantities that are strictly positive and for which there were asymmetric uncertainties, we discarded negative samples and re-drew rather than combine split-normal and truncated normal distributions. \\

\subsection{Compiled properties of UFCSs Not in Our Spectroscopic Sample}
In \tabref{supplementalproperties} below, we report the properties of the 13 UFCSs not included in our spectroscopic sample. Below the divider, we include the two classical GCs meeting our UFCS definition (AM 4 and Palomar 13) for completeness.

\begin{deluxetable}{ccccccccc}[!htpb]
\tablewidth{\textwidth}
\tabletypesize{\small}
\tablecaption{Structural Properties and Distances of the UFCS in our Spectroscopic Sample\label{tab:supplementalproperties}}
\tablehead{
Name & $\alpha$ & $\delta$ & $M_V$ & $M_{*}$ & $r_{1/2}$ & $(m-M)_0$ & $D_{\odot}$ & References \\
 & (deg) & (deg) & (mag) & $(M_{\odot})$ & (pc) &  & (kpc) &  
}
\startdata
Alice & $220.774^{+0.003}_{-0.003}$ & $-28.028^{+0.003}_{-0.003}$ &  &  & $10^{+2}_{-2}$ & --- & $74^{+10}_{-10}$ & (1) \\
DELVE 7 & $304.107^{+0.004}_{-0.005}$ & $-50.330^{+0.002}_{-0.002}$ & $1.2^{+0.7}_{-2.1}$ & $130^{+510}_{-90}$ & $2^{+4}_{-1}$ & $18.10^{+0.30}_{-0.30}$ & $42^{+6}_{-5}$ & (2) \\
DES 1 & $8.499^{+0.002}_{-0.002}$ & $-49.039^{+0.001}_{-0.001}$ & $-1.4^{+0.5}_{-0.5}$ & $630^{+360}_{-240}$ & $4.3^{+0.6}_{-0.6}$ & $19.40^{+0.12}_{-0.12}$ & $76^{+4}_{-4}$ & (3) \\
DES 3 & $325.055^{+0.001}_{-0.001}$ & $-52.542^{+0.001}_{-0.001}$ & $-1.6^{+0.5}_{-0.3}$ & $660^{+270}_{-220}$ & $6.0^{+1.1}_{-1.0}$ & $19.41^{+0.08}_{-0.11}$ & $76^{+3}_{-3}$ & (4) \\
DES 4 & $82.095^{+0.003}_{-0.003}$ & $-61.724^{+0.003}_{-0.003}$ & $-1.1^{+0.5}_{-0.5}$ & $470^{+270}_{-170}$ & $8^{+2}_{-2}$ & $17.50^{+0.20}_{-0.20}$ & $32^{+3}_{-3}$ & (5) \\
DES 5 & $77.504^{+0.003}_{-0.003}$ & $-62.580^{+0.003}_{-0.003}$ & $0.3^{+0.5}_{-0.5}$ & $130^{+80}_{-50}$ & $1.3^{+0.3}_{-0.3}$ & $17.00^{+0.20}_{-0.20}$ & $25^{+2}_{-2}$ & (5) \\
Gaia 3 & $95.059^{+0.003}_{-0.003}$ & $-73.414^{+0.003}_{-0.003}$ & $-3.3^{+0.5}_{-0.5}$ & $3580^{+2100}_{-1310}$ & $7^{+2}_{-2}$ & $18.40^{+0.20}_{-0.20}$ & $48^{+5}_{-4}$ & (5) \\
HSC 1 & $334.309^{+0.003}_{-0.003}$ & $3.480^{+0.003}_{-0.003}$ & $-0.2^{+0.6}_{-0.8}$ & $240^{+230}_{-110}$ & $4^{+1}_{-1}$ & $18.30^{+0.20}_{-0.20}$ & $46^{+4}_{-4}$ & (6) \\
Kim 2 & $317.208^{+0.003}_{-0.003}$ & $-51.163^{+0.003}_{-0.003}$ & $-1.5^{+0.5}_{-0.5}$ & $680^{+390}_{-250}$ & $11.9^{+0.8}_{-0.8}$ & $20.10^{+0.10}_{-0.10}$ & $105^{+4}_{-4}$ & (7) \\
SMASH 1 & $95.250^{+0.003}_{-0.003}$ & $-80.396^{+0.003}_{-0.003}$ & $-1.0^{+0.9}_{-0.9}$ & $430^{+540}_{-240}$ & $7^{+4}_{-2}$ & $18.80^{+0.20}_{-0.20}$ & $57^{+5}_{-5}$ & (8) \\
Torrealba 1 & $56.083^{+0.003}_{-0.003}$ & $-69.423^{+0.003}_{-0.003}$ & $-1.6^{+0.5}_{-0.5}$ & $740^{+430}_{-270}$ & $3.4^{+0.8}_{-0.9}$ & $18.20^{+0.20}_{-0.20}$ & $44^{+4}_{-4}$ & (5) \\\hline
AM 4 & $209.088^{+0.001}_{-0.001}$ & $-27.163^{+0.001}_{-0.001}$ & $-1.8^{+0.5}_{-0.5}$ & $910^{+540}_{-330}$ & $6^{+1}_{-1}$ & $18.30^{+0.20}_{-0.20}$ & $46^{+4}_{-4}$ & (9), (10), (11) \\
Palomar 13 & $346.686^{+0.001}_{-0.001}$ & $12.771^{+0.001}_{-0.001}$ & $-3.3^{+0.5}_{-0.5}$ & $3490^{+1970}_{-1300}$ & $8^{+1}_{-1}$ & $16.86^{+0.04}_{-0.04}$ & $23.5^{+0.4}_{-0.4}$ & (9), (10), (12) \\
\enddata
\tablecomments{Targets are ordered alphabetically. References enumerated above are: (1) \citealt{2025arXiv250511120P}, (2) \citealt{2025arXiv251011684T}, (3) \citealt{2018ApJ...852...68C}, (4) \citealt{2018MNRAS.478.2006L}, (5) \citealt{2019MNRAS.484.2181T}, (6) \citealt{2019PASJ...71...94H}, (7) \citealt{2015ApJ...803...63K}, (8) \citealt{2016ApJ...830L..10M},  (9) \citealt{2018ApJ...860...66M}, (10) \citealt{2021MNRAS.505.5957B}, (11) \citealt{2013AJ....146..116H}, and (12) \citealt{2020PASA...37...46B}}\end{deluxetable}

\section{References for Literature Dwarf Galaxy and Globular Cluster Measurements}
\label{sec:comparisonrefs}
\par For the comparison sample of MW satellite galaxies and GCs, we relied on the Local Volume Database \citep{2025OJAp....8E.142P}. The specific references for the dwarf galaxy sample within the LVDB (version 1.0.4) include, in approximate chronological order:  \citet{2007ApJ...670..313S,2009MNRAS.397L..26C,2009ApJ...702L...9C,2011ApJ...736..146K,2011AJ....142..128W,2012AJ....144....4M,2013ApJ...770...16K,2015ApJ...813..109D,2015ApJ...805..130K,2015ApJ...811...62K,2015ApJ...810...56K,2015ApJ...808...95S,2015ApJ...808..108W,2016ApJ...824L..14C,2016ApJ...833...16K,2016ApJ...833L...5D,2016MNRAS.459.2370T,2016MNRAS.463..712T,2017ApJ...838....8L,2017ApJ...838...83K,2018MNRAS.475.5085T,2018MNRAS.479.5343K,2018ApJ...857..145L,2018ApJ...860...66M,2018ApJ...863...25M,2019A&A...623A.129F,2019ARA&A..57..375S,2020ApJ...892..137S,2021ApJ...910...18C,2021ApJ...916...81C,2021ApJ...920...92J,2021ApJ...921...32J,2021NatAs...5..392C,2022ApJ...939...41C,2023ApJ...942..111C,2023ApJ...950..167B,2023ApJ...953....1C,2023AJ....165...55C,2023AJ....166...76S,2024ApJ...961..234H,2024PASJ...76..733H,2024ApJ...967...72R,2025ApJ...979..164C} and \citet{2025ApJ...979..176T}. 
\par For the comparison sample of MW GCs, the measurements within the LVDB are predominantly sourced from three key references: \citet{2018MNRAS.478.1520B} for structural properties (sizes) and kinematic measurements,  \citet{2020PASA...37...46B} for absolute magnitudes, and \citet{2021MNRAS.505.5957B} for distances. Additional measurements were taken from \citet{1996AJ....112.1487H}, \citet{1998AJ....115..648R}, \citet{2005AJ....129..239K}, \citet{2007MNRAS.381L..45B}, \citet{2008AJ....136.2102S}, \citet{2008A&A...489..583K}, \citet{2011A&A...535A..33M}, \citet{2011A&A...527A..81M}, \citet{2011MNRAS.416..465L}, \citet{2013AJ....146..116H}, \citet{2016ApJ...822...32W}, \citet{2017MNRAS.470.2702K}, \citet{2017RNAAS...1...16M}, \citet{2018ApJ...860...66M}, \citet{2018MNRAS.477.4565S}, \citet{2018ApJ...863L..38R}, \citet{2018ApJ...860L..27C}, \citet{2018ApJ...866...12M}, \citet{2019ApJ...875..120J}, \citet{2019ApJ...870L..24B}, \citet{2019MNRAS.484L..90C}, \citet{2020A&A...642L..19G}, \citet{2021NatAs...5..957G}, \citet{2021A&A...649A..86G}, \citet{2021A&A...654A..39O}, \citet{2021A&A...652A.129M}, \citet{2021A&A...650L..11M}, \citet{2022A&A...662A..95G}, \citet{2022MNRAS.509.4962G}, \citet{2022A&A...657A..67D}, \citet{2022MNRAS.513.3993O}, \citet{2023ApJ...942..104D}, \citet{2023ApJ...950..138P}, \citet{2023MNRAS.526.1075P}, \citet{2024ApJ...967...72R}, \citet{2024A&A...688A.133L}, \citet{2024A&A...689A.115S}, \citet{2024ApJ...963L..33B}, \citet{2025A&A...695A.156L}, \citet{2025A&A...695A..47A}.

\par References for the UFCSs were independently compiled and were not drawn from the Local Volume Database. Our compilation of literature measurements for these systems is presented in \tabref{litproperties} and \tabref{supplementalproperties}; in both these tables, the original references are provided in the final column. 

\FloatBarrier
\clearpage
\pagebreak

\section{Calcium Triplet Fits for Bright RGB Members in Each UFCS}
\label{sec:spectrafits}
In \figref{imacs_ew_fits}, \ref{fig:deimos_ew_fits1}, and \ref{fig:deimos_ew_fits2} below,  we display representative fits to the CaT lines in our IMACS and DEIMOS (rest-frame) spectra. For each UFCS, we specifically show the brightest 1–3 RGB member stars.
\FloatBarrier
\begin{figure*}
    \centering
    \includegraphics[width=0.73\textwidth]{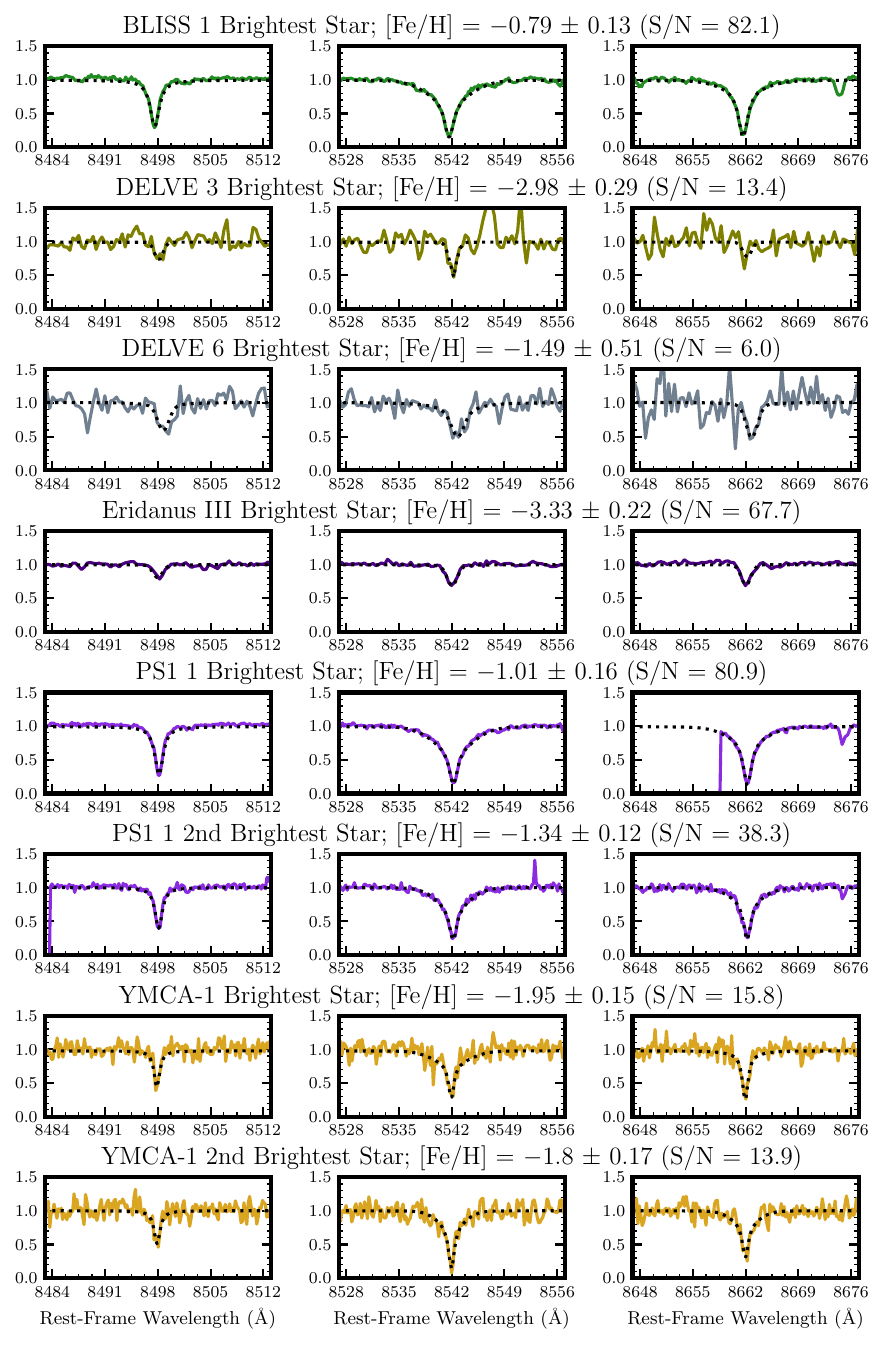}
    \caption{\textbf{CaT fits for a subset of our IMACS targets for which we derived metallicities}. Our normalized, 1D IMACS spectra are shown in color, while the model fits are shown as black dotted lines.}
    \label{fig:imacs_ew_fits}
\end{figure*}
\vspace{-4em}

\begin{figure*}
    \centering
    \includegraphics[width=0.8\textwidth]{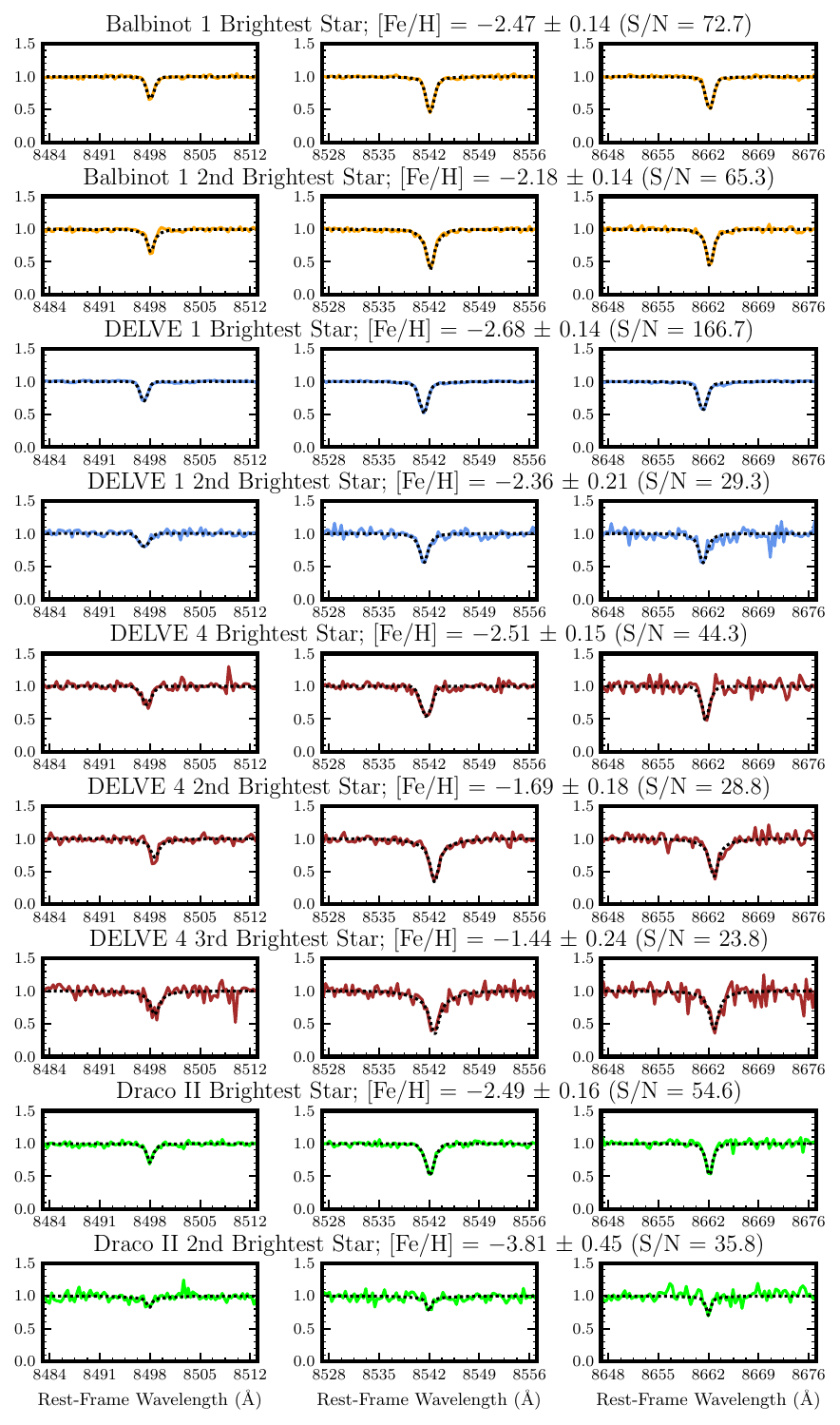}
    \caption{\textbf{CaT fits for stars in our DEIMOS UFCS targets}, analogous to the IMACS fits shown in \figref{imacs_ew_fits}.}
    \label{fig:deimos_ew_fits1}
\end{figure*}

\begin{figure*}
    \centering
    \includegraphics[width=0.8\textwidth]{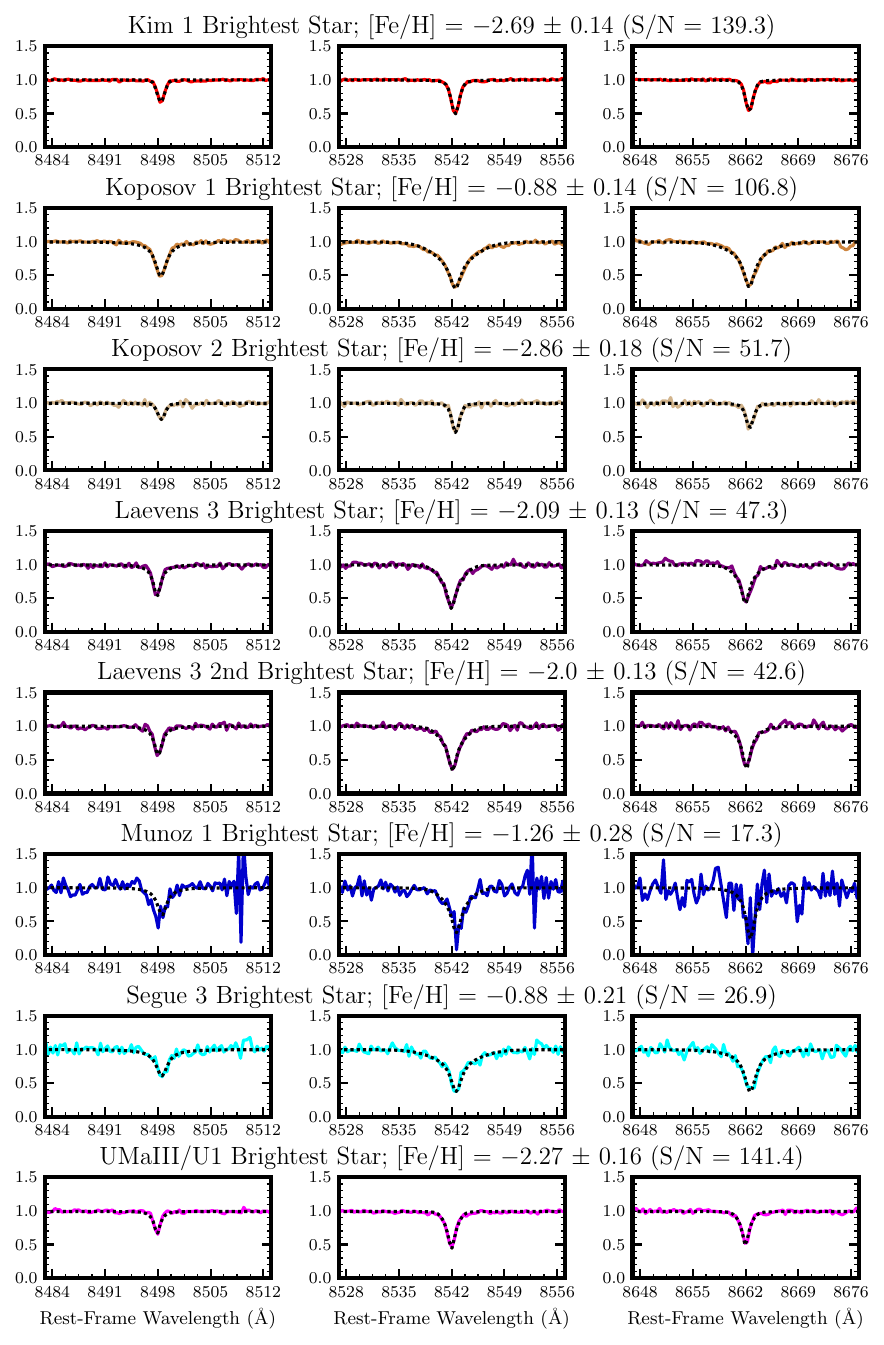}
    \caption{\textbf{A continuation of \figref{deimos_ew_fits1}}, showing DEIMOS spectra of bright UFCS members. Note that our metallicity measurement for the brightest star in Koposov~2 is tentative given significant uncertainty in its membership (and therefore its distance and $V_0$ magnitude). In addition, the \uma{} metallicities presented in this work are based on Ca II K spectra from \citetalias{2025arXiv251002431C}~(\citeyear{2025arXiv251002431C})---not the CaT spectrum shown here.}
    \label{fig:deimos_ew_fits2}
\end{figure*}

\FloatBarrier

\clearpage

\section{Jackknife Diagnostics for UFCSs with Resolved Dispersions}
\label{sec:jackknifediagnostics}
In \secref{veloutliers}, we determined that the five resolved or marginally-resolved velocity dispersions that we report in this work are sensitive to individual member stars. In \figref{jackknifeposteriors} below, we present a diagnostic figure to accompany those determinations. We note that not all of these cases involve outliers in the traditional sense: even stars toward the center of the velocity distribution can have a sufficient impact to alter the interpretation of low-significance dispersions so long as their velocity errors are small and/or the total number of kinematic members is small.

\FloatBarrier

\begin{figure*}
    \centering
    \includegraphics[width=0.74\textwidth]{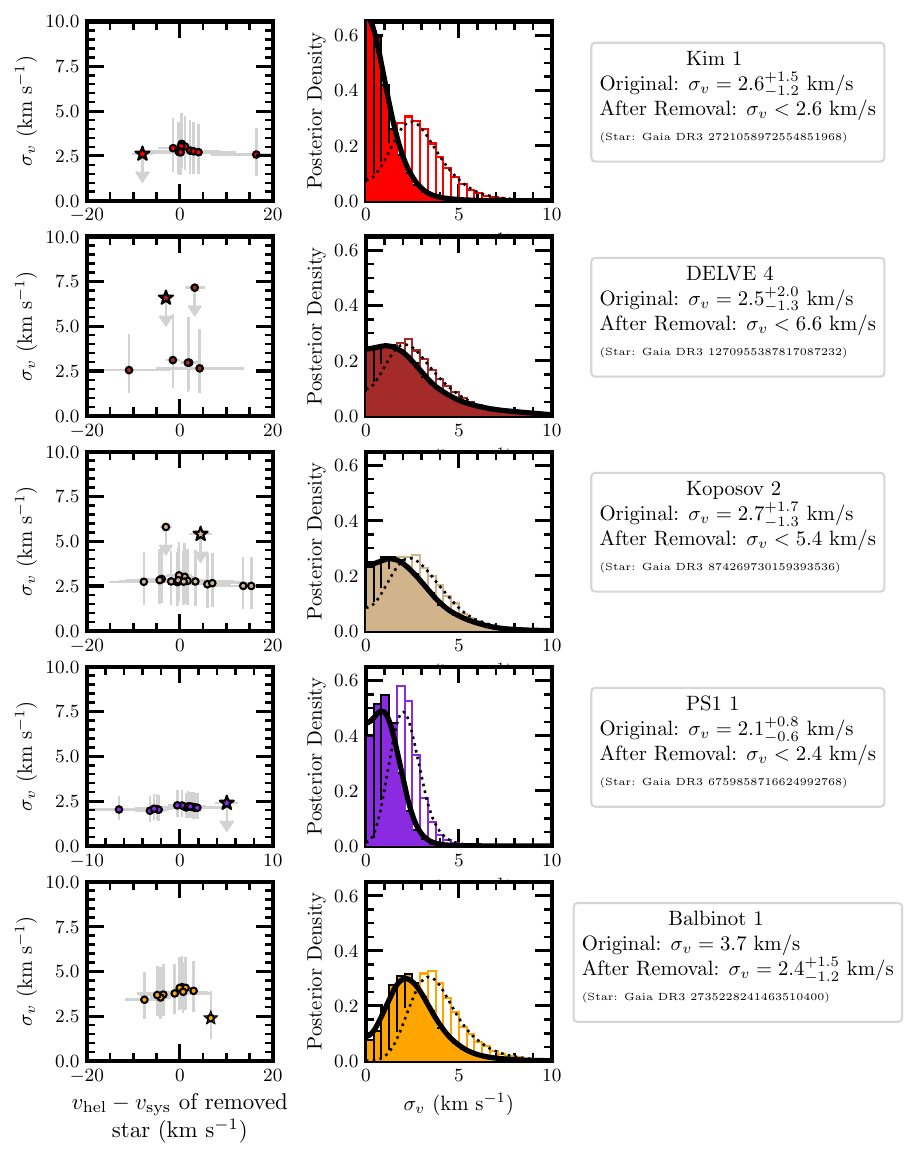}
    \vspace{-2em}
    \caption{\textbf{Diagnosing the impact of individual member stars in the five UFCSs with resolved or marginally-resolved velocity dispersions.} In the left-hand panels, we show the velocity dispersion we inferred after removing each star from our kinematic member sample one by one. The most impactful individual member is shown with a star symbol. In the right-hand panels, we show the posteriors derived from our complete kinematic samples (unfilled histogram with dotted KDE) and the posterior derived from the subsample with the most impactful member star removed (filled histogram with solid black KDE); the former posterior and KDE match those shown in \figref{ridgeplot}.}
    \label{fig:jackknifeposteriors}
\end{figure*}

\end{document}